\documentclass[pdftex,twocolumn,epjc3]{svjour3}
\usepackage{amsmath, amssymb, graphics}
\usepackage{pdflscape}
\usepackage{textcomp}
\usepackage{eurosym}
\usepackage{widetext}
\usepackage{amsfonts}
\usepackage{array}
\usepackage{bm}
\usepackage{helvet}
\usepackage{graphicx}
\usepackage{subfigure}
\usepackage{bm, array}
\usepackage[T1]{fontenc}
\usepackage{color}
\usepackage{enumerate}
\usepackage{soul}
\usepackage[normalem]{ulem}
\usepackage{color}
\usepackage{color}
\usepackage{mathtools, cuted}

\RequirePackage[T1]{fontenc}
\RequirePackage{graphicx}
\RequirePackage{mathptmx}      
\RequirePackage[numbers,sort&compress]{natbib}
\RequirePackage[colorlinks,citecolor=blue,urlcolor=blue,linkcolor=blue]{hyperref}
\journalname{Eur. Phys. J. C}
\newcommand{\mathsym}[1]{{}}
\newcommand{\unicode}[1]{{}}

\def\be{\begin{equation}}
\def\ee{\end{equation}}
\def\beq{\begin{eqnarray}}
\def\eeq{\end{eqnarray}}

\newcommand{\ben}{\begin{eqnarray}}
\newcommand{\een}{\end{eqnarray}}

\def\beq{\begin{equation}}
\def\eeq{\end{equation}}
\def\bea{\begin{eqnarray}}
\def\eea{\end{eqnarray}}
\def\bean{\begin{eqnarray*}}
\def\eean{\end{eqnarray*}}

\newcommand{\bear}{\begin{eqnarray}}
\newcommand{\eear}{\end{eqnarray}}

\newbox\pippobox

\def\be{\begin{equation}}
\def\ee{\end{equation}}
\def\ba{\begin{eqnarray}}
\def\ea{\end{eqnarray}}

\newtheorem{thm}{Theorem}
\newtheorem{prop}{Proposition}
\newtheorem{defn}{Definition}

\begin{document}

\title{Extended phase-space analysis of the Ho\v{r}ava-Lifshitz cosmology}

\author{Genly Leon\thanksref{e1,addr1}
        \and
        Andronikos Paliathanasis\thanksref{e2,addr2,addr3}.
}
\thankstext{e1}{e-mail: genly.leon@ucn.cl}
\thankstext{e2}{anpaliat@phys.uoa.gr}
\institute{Departamento  de  Matem\'aticas,  Universidad  Cat\'olica  del  Norte, Avda. Angamos  0610,  Casilla  1280  Antofagasta,  Chile\label{addr1}
          \and
          Instituto de Ciencias F\'{\i}sicas y Matem\'aticas, Universidad Austral de Chile, Valdivia, Chile\label{addr2}
          \and
          Institute of Systems Science, Durban University of Technology, PO Box 1334, Durban 4000, South
Africa\label{addr3}				
}

\date{Received: date / Accepted: date}

\maketitle

\begin{abstract}
We examine the phase space of Ho\v{r}ava-Lifshitz cosmology for a wide range
of self-interacting potentials for the scalar field under the
detailed-balance condition and without imposing it, by means of the powerful
method of $f$-devisers. A compactification approach is performed for the
exponential potential and for potentials beyond the exponential one,
extending the previous findings in the literature. By using this approach it
is possible to describe the finite region of the phase space and the region
where the phase-space variables becomes infinity. Furthermore, we present
several results concerning the stability of the \emph{de Sitter} solution in
Ho\v{r}ava-Lifshitz cosmology using Center Manifold theory. The advantages
of this procedure are unveiled immediately when it is compared with the
Normal Forms Calculations presented before in the literature.
\end{abstract}




\section{Introduction}

Ho\v{r}ava-Lifshitz (HL) gravity is a power-counting renormalizable theory
with consistent ultra-violet (UV) behavior exhibiting an anisotropic,
Lifshitz scaling between time and space at the UV limit and contains General
Relativity (GR) as an infrared (IR) fixed point \cite{hor3}. HL theory is
closely related to the Einstein-\ae ther theory \cite%
{Jacobson:2000xp,Eling:2004dk,DJ,kann,
Zlosnik:2006zu,CarrJ,Jacobson,Carroll:2004ai,Garfinkle:2011iw}, which is an
effective field theory, preserving locality and covariance, consisting of GR
coupled at second derivative order to a dynamical time-like unit vector
field, the \ae ther. In the IR limit of extended Ho\v{r}ava gravity \cite%
{hor3,Jacobson:2000xp}, the aether vector is assumed to be
hypersurface-orthogonal. That implies that every hypersurface-orthogonal
Einstein-\ae ther solution is a Ho\v{r}-ava solution \cite{TJab13}. However
while the latter equivalence is true in terms of exact solution it is not a
general true for other results which follow from the direct form of the
field equations, like the PPN constraints and other \cite{sot01}. 
In the recent years there has been a lot of interest in some aspects of the HL theory, and
several analyzes have been carried on, e.g., in \cite%
{Cai:2009ar,Charmousis:2009tc,Bogdanos:2009uj,Bellorin:2018blt,Bellorin:2017gzj, Bellorin:2017gab,Bellorin:2016hcu,Bellorin:2016nvh,Bellorin:2016wsl,Bellorin:2015oja, Bellorin:2014qca,Restuccia:2014zda,Bellorin:2013zbp,Bellorin:2012di,Bellorin:2011ff, Bellorin:2010je,dan1,dan2}%
. To reduce the number of terms in the Lagrangian formulation one may impose
the detailed-balance condition \cite{hor3}. Under detailed -balance, the $%
(D+1)$-dimensional theory acquires the renormalization properties of the $D$%
-dimensional one, that is, the theory has a quantum inheritance principle.
However, there is a discussion whether the detailed
-balance condition leads to reliable results or if it is able to reveal the
full information of Ho\v{r}ava-Lifshitz gravity \cite{Kiritsis:2009sh}.
Therefore, one needs to investigate also the phenomenology when the
detailed-balance condition is abandoned. One important application of HL
gravity is that of the HL cosmology \cite%
{Kiritsis:2009sh,Lu:2009em,Leon:2009rc,Mukohyama:2009zs,Mukohyama:2009gg,Brandenberger:2009yt,Danielsson:2009gi,Saridakis:2009bv,Blas:2009qj,Dutta:2009jn,Kim:2009dq,Cai:2009qs,Christodoulakis:2011np,dan111}%
. In \cite{Bramberger:2017tid} it was proposed a possible solution to the
flatness problem, in which it is assumed that the initial condition of the
Universe is set by a small instanton respecting the same scaling. In \cite%
{Bellorin:2018wst} it is analyzed the electromagnetic-gravity interaction in
a pure HL framework formulated in 4+1 dimensions and it is performed a
Kaluza-Klein reduction to 3+1 dimensions. In \cite{Abreu:2018wjg} it was
studied a noncommutative version of the Friedmann-Robertson-Walker (FRW)
cosmological models within the gravitational HL theory. The matter content
of the models is described by a perfect fluid and the constant curvature of
the spatial sections may be positive, negative or zero. In \cite%
{Maier:2017dtb} it is examined the Hamiltonian dynamics of bouncing Bianchi
IX cosmologies with three scale factors in HL gravity. It is considered a
positive cosmological constant plus non-interacting dust and radiation as
the matter content of the models and the dynamics is presented in a
six-dimensional phase space.

In the review \cite{Wang:2017brl} are discussed some recent developments of
the HL theory: (i) universal horizons, black holes and their thermodynamics;
(ii) non-relativistic gauge/gravity duality; and (iii) quantization of the
theory. In the reference \cite{Nilsson:2018knn} was tested HL cosmological
scenarios against the new observational constraints: an updated cosmological
data set from Cosmic Microwave Background (Planck CMB), expansion rates of
elliptical and lenticular galaxies, JLA compilation (Joint Light-Curve
Analysis) data for Type Ia supernovae (SneIa), Baryon Acoustic Oscillations
(BAO) and priors on the Hubble parameter with an alternative parametrization
of the equations. The authors considered the curvature parameter as a free
parameter in the analysis, and they found that the detailed-balance scenario
exhibits positive spatial curvature to more than $3\sigma$, whereas for
further theory generalizations it was found evidence for positive spatial
curvature at $1\sigma$. Moreover, HL gravity is consistent with
the recent observation of the gravitational-wave event GW170817 \cite{gw01}.

The thermodynamics of cosmological models in the HL theory of gravity for a 
perfect fluid was studied in detail in \cite{Wang:2009rw}. The cosmological 
evolution was studied, including the periods of
deceleration and/or acceleration, and the existence of big bang, big crunch,
and big rip singularities, and bouncing universe \cite{Wang:2009rw}. 

Though HL gravity received a lot of attention a decade ago, on its
first stages of the theory, the difficulty to overpass some theoretical
problems have lead the scientists to study other Lorentz violated theories
such as the Einstein- \ae ther theory, but still there are various groups
which try to solve some of the problems of HL theory. For instance the
problem to restore the Lorentz invariance on the low-energy limit has been
the subject of various studies. The most simple scenarios are that
Lorentz invariance is recovered because the renormalization group flow of
the system leads to emergent infrared Lorentz invariance, that supersymmetry
protects infrared Lorentz invariance or a classically Lorentz invariant
matter sector with controlled quantum corrections may simply coexist with HL
gravity under certain conditions \cite{hl001,hl002,hl003,hl004}. 

While there has been various attempts in the study of the
renormalizability HL theory, its complete renormalizability has not yet been
proven \cite{rg01,rg04,rg02,rg03}. However, only recently, in \cite{rg02} the
renormalizability of the projectable HL gravity was proven. Projectable HL
gravity does not provide the limit of GR, however it has many common
physical properties with GR and that makes it an interesting case of study. Other
open subjects of HL gravity are the renormalization group flow \cite{rg05},
the coupling with matter and the effects of Lorentz violation \cite%
{ef01,ef02}. 

Even though the theory has some drawbacks enumerated in Section \ref{cosmequations}, still
there are found several attempts to cure some of these problems. In \cite%
{dan1} the authors reconsidered the \textquotedblleft detailed
balance\textquotedblright\ as a principle that can be used to restrict the
proliferation of couplings in HL gravity, and for the usual shortcomings
-that have been usually associated with it in the literature- were found
easy resolutions for all of them within the framework of detailed balance, 
but the most persistent is related to the projectability. The fact some of
these issues are very difficult to be answered, however, does not spoil the
interest in the theory. Recently, it was increased the interest in finding
solutions in HL gravity and in Einstein-\ae ther theory. Recently in 
\cite{Vernieri:2017dvi} it was presented a procedure to
construct anisotropic interior solutions in Ho\v{r}ava gravity and Einstein-%
\ae ther theory and in \cite{Vernieri:2019vlh} it was investigated interior 
solutions with anisotropic fluids in the context HL gravity and Einstein-
\ae ther theory with non-static \ae ther field, with the field equations 
becoming solvable. In \cite{Paliathanasis:2019qch} we have found exactly 
solvable models for a prototype of HL cosmology imposing the so called 
projectability condition, and considering the \textquotedblleft detailed 
balance\textquotedblright \, condition \cite{Leon:2009rc}. In particular, 
we performed a detailed study of the integrability of the HL scalar field 
cosmology in a FLRW background spacetime. More specifically, we have tested 
if the gravitational field equations possesses the Painlev\'{e} property. 
For the exponential potential of the scalar field we are able to perform 
an analytic explicit integration of the field equations and write the solution 
in terms of a Right Painlev\'{e} series. Relativistic polytropic equations 
of state in the context of HL gravity and Einstein-\ae ther theory were 
examined in \cite{Vernieri:2018sxd}%
. 

In this paper we will be focused more about the general evolution of the solution space for the model. By this it is understood: i) to obtain a global visual picture 
of the solutions space, by characterizing, locally and globally, the so-called attractor solution; ii) to
prove issues concerning asymptotic past and future behavior, iii) generalizing and
simplifying earlier proofs in the literature, and iv) using the Center Manifold Theory to 
rigorously analyze the stability of several cosmological
solutions that are of interest for inflation in HL gravity.

In order to investigate the generic features of a general cosmological
scenario one can apply the powerful method of dynamical system analysis,
which allows to bypass the complexity of the involved equations and extract
information about the global features and behavior of the cosmological
scenario \cite{WE,LeBlanc:1994qm,Heinzle:2009zb,Coley:1999uh,Coley:2003mj}.
In the case of HL cosmology an initial investigation in this direction was
performed in \cite{Leon:2009rc} for the case of exponential potential, in
which the authors have found stable dark-energy-dominated solutions, as well
as bouncing solutions. In \cite{Carloni:2009jc} it was investigated the
properties of cosmological models based on the HL gravity by using the
dynamical system approach but using different normalization procedures in comparison with 
\cite{Leon:2009rc}. Lyapunov method has been used in some gravitational
scenarios, say \cite%
{Setare:2010zd,Cardoso:2008bp,Lavkin:1990gu,Charters:2001hi,Aref'eva:2009xr}%
. In the present work we provide a complete dynamical system analysis of HL
cosmology keeping the potentials arbitrary, which is a major improvement
since it allows for the extraction of information that is related to the
foundations of the cosmological model and not to the specific potential
form. In particular, we apply the method of $f$-devisers \cite%
{Escobar:2013js,Fadragas:2013ina,an1,an2,an3}, in which one first performs
the analysis without the need of an a priori specification of the potential,
and in the end one just substitutes the specific potential form in the
results, instead of having to repeat the whole dynamical elaboration from
the start. As we will see, the results will be richer and more general,
revealing the full capabilities of HL cosmology.

The plan of the work is as follows. In Section \ref{cosmequations} we
briefly review the gravitational field equations in HL cosmology where for
the underlying geometry we consider that of a isotropic and homogeneous
universe. In section \ref{Case1} it is studied the spatially flat FRW\
universe ($k=0$) without a cosmological constant term, i.e. $\Lambda =0,$
under the detailed-balance condition. In section \ref{Case1A} the phase
space analysis for arbitrary potentials for the latter cosmological model is
presented. Moreover, the case of the Exponential potential $V(\phi
)=V_{0}e^{-s\phi }$, with $s$ constant is detailed studied in section \ref%
{Case1B}. In section \ref{Case2C} it is studied a Powerlaw potential $V(\phi
)=\frac{1}{2n}(\mu \phi )^{2n},\mu >0,n=1,2,\ldots $ discussed previously in 
\cite{Alho:2015cza} for a conventional scalar field while in section \ref%
{Case1D} it is studied the E-model with potential $V(\phi )=V_{0}\left(
1-e^{-\sqrt{\frac{2}{3\alpha }}\phi }\right) ^{2n}$ discussed in \cite%
{Alho:2017opd} for a conventional scalar field cosmology. The methodology we
use is different from that employed in \cite{Alho:2015cza,Alho:2017opd}, and
we apply the potentials in a different context. However, we complimentary
use analogous variables as in \cite{Alho:2015cza,Alho:2017opd} to make
comparisons with the relativistic case. The analogous analysis is done in
Section \ref{Case2} for the universe nonzero spatial curvature and without
cosmological constant term under the detailed-balance condition.
Furthermore, in Sections \ref{Case3} and \ref{Case4} we consider that $%
\Lambda \neq 0$ and we perform similar analysis for $k=0$ and $k\neq 0$
respectively. In section \ref{Case5} is is studied the HL model beyond the
detailed-balance condition. With present several results concerning the
stability of the de Sitter solution in Ho\v{r}ava-Lifshitz cosmology using
the Center Manifold theory, summarized in \ref{sectionCM}. The main results 
of this research are proved in sections \ref{PROPOSITION1}, \ref{PROPOSITION2}, \ref{PROPOSITION3}%
, \ref{PROPOSITION4}, \ref{PROPOSITION5}, \ref{PROPOSITION6}, \ref%
{PROPOSITION7}, \ref{PROPOSITION8}, \ref{PROPOSITION9}, \ref{PROPOSITION10},
and in Section \ref{stabilityP26}. We find that in all the cases, with the
exception of the model studied in Section \ref{stabilityP26}, the de Sitter
solution is unstable: saddle or center-saddle. In Section \ref{stabilityP26}
we prove that the de Sitter solution is locally asymptotically stable. The
advantages of this procedure are unveiled immediately when it is compared
with the Normal Forms Calculations presented before in the literature.
Finally, Section \ref{HLconclusions} summarizes our results where also we
draw our conclusions.

\section{The cosmological equations}

\label{cosmequations}

In the simple version of HL gravity \cite{Kiritsis:2009sh} the line-element
can be written as: 
\begin{eqnarray}  \label{metriciit}
ds^2 = - N^2 dt^2 + g_{ij} (dx^i + N^i dt ) ( dx^j + N^j dt ) ,
\end{eqnarray}
where the lapse and shift functions are respectively $N$ and $N_i$. The
spatial metric is given by $g_{ij}$, and roman letters indicate spatial
indices. The scaling transformation is $ t \rightarrow l^3 t~~~\mathrm{and}\
\ x^i \rightarrow l x^i$.

Under the detailed-balance condition  the full action of Ho\v{r}ava-Lifshitz
gravity is given by 
\begin{eqnarray}
S_g &=& \int dt d^3x \sqrt{g} N \left\{ \frac{2}{\kappa^2} (K_{ij}K^{ij} -
\lambda K^2) \right.  \notag \\
&+& \left. \frac{\kappa^2}{2 w^4} C_{ij}C^{ij} -\frac{\kappa^2 \mu}{2 w^2} 
\frac{\epsilon^{ijk}}{\sqrt{g}} R_{il} \nabla_j R^l_k + \frac{\kappa^2 \mu^2%
}{8} R_{ij} R^{ij} \right.  \notag \\
&-&\left. \frac{\kappa^2 \mu^2}{8( 3 \lambda-1)} \left[ \frac{1 - 4 \lambda}{%
4} R^2 + \Lambda R - 3 \Lambda ^2 \right] \right\},  \label{acct}
\end{eqnarray}
where $K_{ij} = \left( {\dot{g_{ij}}} - \nabla_i N_j - \nabla_j N_i
\right)/2N $  is the extrinsic curvature and $C^{ij} = \epsilon^{ijk}
\nabla_k \bigl( R^j_i - R \delta^j_i/4 \bigr)/\sqrt{g} $ the Cotton tensor,
and the covariant derivatives are defined with respect to the spatial metric 
$g_{ij}$. $\epsilon^{ijk}$ is the totally antisymmetric unit tensor, $\lambda
$ is a dimensionless constant and the quantities $\kappa$, $w$ and $\mu$ are
constants. Furthermore, in order to avoid phantom-like solutions we assume $%
\gamma>\frac{1}{3}$, while for simplicity we select to work with units where 
$\kappa^2=8\pi G=1$.

The matter source we consider the following action term \cite%
{Kiritsis:2009sh} 
\begin{eqnarray}
S =\int dt d^3x \sqrt{g} N\left[\frac{3\lambda -1}{4}\frac{\dot\phi^2}{N^2}%
-V(\phi)\right],
\end{eqnarray}
where by simplicity we have not considered the higher derivative terms $%
\propto \phi \nabla^2 \phi, $ $\propto \phi \nabla^4 \phi, $ $\propto \phi
\nabla^6 \phi, \ldots$.

Cosmological Principle, of the homogeneous and isotropic universe, leads to
the consideration of the FRW metric 
\begin{eqnarray}
g_{ij}=a^2(t)\gamma_{ij},~~N^i=0, ~~\gamma_{ij}dx^idx^j=\frac{dr^2}{1- k r^2}%
+r^2d\Omega_2^2,
\end{eqnarray}
where $k=-1,0,+1$ corresponding to open, flat, and closed universe
respectively. Without loss of generality in the following we select $N=1$.

Hence, for that specific line element, the equations are reduced to:
\begin{align} 
& H^2= \frac{1}{6(3\lambda-1)}\left[\frac{3 \lambda -1}{4}\dot\phi^2+V(\phi)%
\right]  \notag \\
& +\frac{1}{16(3\lambda-1)^2}\left[-\frac{\mu^2 k^2}{a^4} -{\mu^2\Lambda ^2}+%
\frac{2\mu^2\Lambda k}{a^2} \right],  \label{Fr2fluid}\\
&\dot{H}+\frac{3}{2}H^2= -\frac{1}{4(3\lambda-1)}\left[\frac{3 \lambda -1}{4}%
\dot\phi^2-V(\phi)\right]  \notag \\
& -\frac{1}{32(3\lambda-1)^2}\left[-\frac{\mu^2 k^2}{a^4} +{3\mu^2\Lambda ^2}%
-\frac{2\mu^2\Lambda k}{a^2} \right], \label{Fr1fluid}\\
& \ddot\phi +3 H \dot \phi+\frac{2 V^{\prime }(\phi)}{3 \lambda -1}=0. \label{Kleina}
\end{align}

If the detailed -balance condition is removed, the gravitational action can
be schematically written as \newline
$S_g = \int dt d^3x \sqrt{g} N (\mathcal{L}_K+\mathcal{L}_V)$, where the
gravitational action have been expanded in its kinetic and potential part.
By varying $N$ and $g_{ij}$, we extract the Friedmann equations  \cite%
{Charmousis:2009tc,Leon:2009rc}: 
\begin{align}  
& H^2 =\frac{2}{(3\lambda-1)}\left[\sigma_0 \left(\frac{3 \lambda -1}{4}%
\dot\phi^2+V(\phi)\right) \right.  \notag \\
& + \frac{\sigma_1}{6} + \left. \frac{\sigma_2 k}{6 a^2}+\frac{\sigma_3 k^2}{%
6a^4} + \frac{\sigma_4 k}{6a^6} \right],\label{Fr2Yc} \\
&\dot{H}+\frac{3}{2}H^2 = -\frac{3}{(3\lambda-1)}\left[\sigma_0 \left(\frac{%
3 \lambda -1}{4}\dot\phi^2-V(\phi)\right) \right.  \notag \\
& \left. -\frac{\sigma_1}{6} -\frac{\sigma_2 k}{6 a^2} +\frac{\sigma_3 k^2}{%
18a^4} +\frac{\sigma_4 k}{6a^6} \right], \label{Fr1Yc}\\
& \ddot\phi +3 H \dot \phi+\frac{2 V^{\prime }(\phi)}{3 \lambda -1}=0. \label{Kleinb}
\end{align}
The constants $\sigma_i$ satisfy $\sigma_0\equiv \kappa^2/12$, $\sigma_2<0$, 
$\sigma_4>0$.

To study the system \eqref{Fr2fluid}, \eqref{Fr1fluid}, \eqref{Kleina}, corresponding to HL cosmology under the
detailed-balance condition, it is more convenient to introduce the auxiliary
variables \cite{Leon:2009rc}: 
\begin{align}
&x=\frac{\dot \phi }{2 \sqrt{6} H}, \quad y=\frac{\sqrt{V(\phi)}}{\sqrt{6}H 
\sqrt{3 \lambda -1}},  \notag \\
& z=\frac{\mu }{4 (3 \lambda -1) a^2 H}, \quad u=\frac{\Lambda \mu }{4 (3
\lambda -1) H},  \notag \\
& s=-\frac{V^{\prime }(\phi)}{V(\phi)}, \quad f\equiv\frac{V^{\prime \prime
}(\phi)}{V(\phi)}-\frac{{V^{\prime}(\phi)}^2}{V(\phi)^2},  \label{HLauxilliaryu}
\end{align}
together with $N=\ln a$ and assuming that $f$ can be written as an explicit
function of $s$ we obtain a closed dynamical system. For the exponential
potential $s$ is constant and $f\equiv 0$. Thus, the system's dimensionality
is reduced in one dimension. Therefore, we extent the phase-space analysis
of \cite{Leon:2009rc} by considering potentials beyond the exponential with
non-trivial $f(s)$. Given $f(s)$, the stability of the hyperbolic
equilibrium points is given by analyzing the signs of the reals parts of the
eigenvalues of the matrix of linear perturbations $\mathbf{Q}$ evaluated at
each equilibrium point. For studying the more general system \eqref{Fr2Yc}, \eqref {Fr1Yc}, \eqref{Kleinb},
we can introduce the auxiliary variables \cite{Leon:2009rc}: 
\begin{align}
& x=\frac{\dot \phi }{2 \sqrt{6} H}, \quad y=\frac{\sqrt{V(\phi)}}{\sqrt{6}H 
\sqrt{3 \lambda -1}},  \notag \\
& x_1=\frac{\sigma_1}{3 (3 \lambda -1) H^2}, \quad x_2=\frac{k \sigma_2}{3
(3 \lambda -1) a^2 H^2},  \notag \\
&x_3=\frac{\sigma_3}{3 (3 \lambda -1) a^4 H^2}, \quad x_4=\frac{2 k \sigma_4%
}{(3 \lambda -1) a^6 H^2}.  \label{HLauxilliaryc}
\end{align}

For the common scalar field potentials the function $f(s)$ is found as
follows. The monomial potential $V(\phi)=\frac{1}{2n}(\mu
\phi)^{2n}$, $\mu >0, n=1,2,\ldots$ \cite{Alho:2015cza} has $f(s)=-\frac{s^2}{%
2 n}$. The so-called E-model studied from the dynamical systems point of
view in \cite{Alho:2017opd} has potential $V(\phi)=V_0\left(1-e^{-\sqrt{%
\frac{2}{3 \alpha}} \phi}\right)^{2 n}$. The corresponding $f$-deviser is $%
f(s)=-\frac{s \left(s-\sqrt{6} \mu \right)}{2 n}$, where $\mu=\frac{2 n}{3 
\sqrt{\alpha}}$. The exponential potential plus a cosmological constant $V(\phi)=V_{0}e^{-l\phi}+V_1$ \cite%
{Yearsley:1996yg,Pavluchenko:2003ge,Cardenas2003} has $f(s)=-s(s-l)$. The hyperbolic potentials:   $V(\phi)=V_{0}\left(\cosh\left( \xi \phi \right)-1\right)$ \cite%
{Ratra:1987rm,Wetterich:1987fm,Matos:2009hf,Copeland:2009be,Leyva:2009zz,
Pavluchenko:2003ge,
delCampo:2013vka,Sahni:1999qe,Sahni:1999gb,Lidsey:2001nj,Matos:2000ng} with $%
f(s)=-\frac{1}{2}(s^2-\xi^2)$ and $V(\phi)=V_{0}\sinh^{-\alpha}(\beta\phi)$ 
\cite%
{Ratra:1987rm,Wetterich:1987fm,Copeland:2009be,Leyva:2009zz,Pavluchenko:2003ge, Sahni:1999gb, UrenaLopez:2000aj}
with   $f(s)=\frac{s^2}{\alpha}-\alpha\beta^2$. The double exponential potential $%
V(\varphi)=V_{0}\left(e^{\alpha\phi}+e^{\beta\phi}\right)$ \cite%
{Barreiro:1999zs,Gonzalez:2007hw,Gonzalez:2006cj} has $f(s)=-(s+\alpha)(s+%
\beta)$. 

The above basic models of HL cosmology proves to have very
interesting cosmological behavior \cite%
{Kiritsis:2009sh,Lu:2009em,Leon:2009rc,Mukohyama:2009zs,
Mukohyama:2009gg,Brandenberger:2009yt,
Danielsson:2009gi,Saridakis:2009bv,Dutta:2009jn,Kim:2009dq, Cai:2009qs}.
However, the gravitational sector itself proves to have instabilities that
cannot be cured by simple tricks such as analytic continuation \cite{Charmousis:2009tc,Bogdanos:2009uj}. 

There are several physical drawbacks in the theory, say, the
projectable HL does not provide the limit of GR, and when the projectability condition {it is applied}-where the lapse
function is a global quantity with no variation over a constant time
hypersurface- the theory has not a local Hamiltonian constraint \cite%
{Mukohyama:2009mz}. Additionally, in the projectable theory, there are
instability problems and strong coupling for cosmological solutions \cite%
{Charmousis:2009tc,Li:2009bg,Blas:2009yd}. On the other hand, the theory
with detailed balance in 3+1 dimensions also it suffers from some problems
(see \cite{dan1}, and references therein): there is a parity violating term
which is of fifth order in derivatives \cite{Sotiriou:2009gy,Sotiriou:2009bx}%
; the squared Cotton tensor, which appears in the theory, is traceless and
vanishes for conformally flat three-dimensional spaces. Therefore, it does not
contribute to the propagator of the scalar graviton. Hence, the scalar mode
does not satisfy the usual dispersion relation and is not power-counting
renormalizable, unlike the spin-2 mode. This spoils the overall UV
properties of the theory \cite{hor3}. The infrared behavior of the scalar
mode is plagued by instabilities and strong coupling at unacceptably low
energies \cite{Charmousis:2009tc,Blas:2009yd}. The (bare) cosmological
constant has the opposite sign from the observed value \cite%
{Sotiriou:2009gy,Sotiriou:2009bx} and has to be large, much larger than the
observed value \cite{Appignani:2009dy}.

Therefore, it is necessary to try to construct suitable extensions
that are free of such problems. A quite general power-counting
renormalizable action is \cite{Kiritsis:2009vz}:  
\begin{equation}
S=S_{kin}+S_{1}+S_{2}+S_{new} ,  \label{q10}
\end{equation}
with  
\begin{align}
& S_{kin}=\alpha \int dtd^3x\sqrt{g}N\!\!\left[(K_{ij}K^{ij}\!-\!\lambda K^2)%
\right],  \label{q4} \\
& S_{1}=\int dtd^3x\sqrt{g}N\left[\gamma_0 {\frac{\epsilon^{ijk}}{\sqrt{g}}}%
R_{il}\nabla_j{R^l}_{k} \!+\!\zeta R_{ij}R^{ij} \right.  \notag \\
& \left. \!+\!\eta R^2\!+\!\xi
R\!+\!\sigma\!\right],  \label{q5} 
\\
&S_2=\int dtd^3x\sqrt{g}N\left[ \beta_0 C_{ij}C^{ij}+\beta_1 R\square
R+\beta_2R^3\right.  \notag \\
& \left.+\beta_3RR_{ij}R^{ij}+\beta_4 R_{ij}R^{ik}{R^{j}}_k\right],
\label{q6} \end{align}
and
\begin{align}
& S_{new}=\int dtd^3x\sqrt{g}N\left[a_1 (a_ia^i)+a_2
(a_ia^i)^2+a_3R^{ij}a_ia_j \right.  \notag \\
& \left.+a_4R\nabla_i a^i+a_5\nabla_ia_j\nabla^ia^j+ a_6\nabla^i a_i
(a_ja^j)+\cdots \right].  \label{q7}
\end{align}

Thus, apart from the known kinetic, detailed-balance and
beyond-detailed-balance combinations that constitute the HL gravitational
action, in (\ref{q7}) it is added a new combination, based on the term 
 $a_{i}\equiv {\frac{\partial _{i}N}{N}}$, which breaks the
projectability condition, and the ellipsis in \eqref{q7} refers to dimension
six terms involving $a_{i}$ as well as curvatures \cite%
{Blas:2009qj}. Such a new combination of terms seems to alleviate the problems of
HL gravity, although there could still be some ambiguities. Therefore, one
should repeat all the relevant investigations of the literature for this
extended version of the theory.

In this paper, however, we will be focused in
the general evolution of the solution space for the model \eqref{Fr2fluid}, \eqref{Fr1fluid}, \eqref{Kleina} (under the detailed-balance condition) for the situations: i): Flat universe
with $\Lambda=0$. ii) Non-flat universe
with $\Lambda=0$. iii) Case 3: Flat
universe with $\Lambda \neq 0$. For
completeness, we perform the dynamical systems analysis of the model: $k\neq
0, \Lambda \neq 0$ under the detailed-balance condition for arbitrary
potentials at the finite region of the phase space. Furthermore, we study the solution space at the finite region of the phase space for the system \eqref{Fr2Yc}, \eqref {Fr1Yc}, \eqref{Kleinb}
without detailed-balance.

\section{Case 1: Flat universe with $\Lambda=0$ under the detailed-balance
condition}

\label{Case1}

\begin{table*}[t]
\caption{Case 1: Equilibrium points at the finite region of the system 
\eqref{GenHLeqxcase1}, \eqref{GenHLeqzcase1}, \eqref{GenHLeqscase1}.}
\label{HLcrit}\centering
\begin{tabular*}{\textwidth}{@{\extracolsep{\fill}}lrrrl}
\hline
Equil. & \multicolumn{1}{c}{$(x,z,s)$} & \multicolumn{1}{c}{Existence} & 
\multicolumn{1}{c}{Eigenvalues} & \multicolumn{1}{c}{Stability} \\ 
Points &  &  &  &  \\ \hline
$P_1(\hat{s})$ & $(1, 0,\hat{s})$ & $f(\hat{s})=0$ & $6-2 \sqrt{6} \hat{s},
1, -2 \sqrt{ 6} f^{\prime }\left(\hat{s}\right)$ & nonhyperbolic for $%
f^{\prime }\left(\hat{s}\right)=0$, \\ 
&  &  &  & or $\hat{s}=\sqrt{\frac{3}{2}}$. \\ 
&  &  &  & source for $f^{\prime }\left(\hat{s}\right)<0, \hat{s}<\sqrt{%
\frac{3}{2}}$. \\ 
&  &  &  & saddle otherwise. \\ \hline
$P_2(\hat{s})$ & $(-1, 0,\hat{s})$ & $f(\hat{s})=0$ & $6+2 \sqrt{6} \hat{s},
1, 2 \sqrt{6} f^{\prime }\left(\hat{s}\right)$ & nonhyperbolic for $%
f^{\prime }\left(\hat{s}\right)=0$, \\ 
&  &  &  & or $\hat{s}=-\sqrt{\frac{3}{2}}$. \\ 
&  &  &  & source for $f^{\prime }\left(\hat{s}\right)>0, \hat{s}>-\sqrt{%
\frac{3}{2}}$ \\ 
&  &  &  & saddle otherwise. \\ \hline
$P_3(\hat{s})$ & $\left(\sqrt{\frac{2}{3}} \hat{s}, 0, \hat{s}\right)$, & $f(%
\hat{s})=0$ & $2 \hat{s} ^2-3, 2 \left(\hat{s}^2-1\right), -4 \hat{s}
f^{\prime }\left(\hat{s}\right)$ & nonhyperbolic for $f^{\prime }\left(\hat{s%
}\right)=0$, \\ 
&  & $-\sqrt{\frac{3}{2}}\leq \hat{s}\leq \sqrt{\frac{3}{2}}$ &  & or $\hat{s%
}\in \left\{-\sqrt{\frac{3}{2}},-1, 0, 1, \sqrt{\frac{3}{2}}\right\}$. \\ 
&  &  &  & sink for $f^{\prime }\left(\hat{s}\right)<0, -1<\hat{s}<0$, \\ 
&  &  &  & or $f^{\prime }\left(\hat{s}\right)>0, 0<\hat{s}<1$. \\ 
&  &  &  & saddle otherwise. \\ \hline
$P_3^{0}$ & $(0, 0, 0)$ & always & $-2, -\frac{3}{2}\pm \frac{1}{2} \sqrt{%
9-48 f(0)}$ & nonhyperbolic for $f(0)=0$. \\ 
&  &  &  & sink for $f(0)>0$. \\ 
&  &  &  & saddle otherwise. \\ \hline
$P_4^{+}$ & $\left(\sqrt{\frac{2}{3}}, z_c, 1\right)$ & $z_c\in\mathbb{R},
f(1)=0$ & $-1, 0, -4 f^{\prime }(1)$ & nonhyperbolic. \\ \hline
$P_4^{-}$ & $\left(-\sqrt{\frac{2}{3}}, z_c, -1\right)$ & $z_c\in\mathbb{R}%
,f(-1)=0$ & $-1, 0, 4 f^{\prime }(-1)$ & nonhyperbolic. \\ \hline
\end{tabular*}%
\end{table*}
\begin{table*}[ht]
\caption{Case 1: Equilibrium points at the infinity region of the system 
\eqref{GenHLeqxcase1}, \eqref{GenHLeqzcase1}, \eqref{GenHLeqscase1}.}
\label{HLcritinfinity}\centering
\begin{tabular*}{\textwidth}{@{\extracolsep{\fill}}lrrrl}
\hline
Equil. & \multicolumn{1}{c}{$(x,Z,S)$} & \multicolumn{1}{c}{Existence} & 
\multicolumn{1}{c}{Eigenvalues} & \multicolumn{1}{c}{Stability} \\ 
Points &  &  &  &  \\ \hline
$Q_{1,2}(\hat{s})$ & $\left(-1, \pm 1, \frac{2}{\pi}\arctan(\hat{s})\right)$
& $f(\hat{s})=0$ & $-2, 2 \sqrt{6} \hat{s}+6, 2 \sqrt{6} f^{\prime }\left(%
\hat{s}\right)$ & nonhyperbolic for \\ 
&  &  &  & $f^{\prime }\left(\hat{s}\right)=0$, or \\ 
&  &  &  & $\hat{s}=-\sqrt{\frac{3}{2}}$. \\ 
&  &  &  & sink for $f^{\prime }\left(\hat{s}\right)<0, \hat{s}<-\sqrt{\frac{%
3}{2}}$. \\ 
&  &  &  & It is a saddle otherwise. \\ \hline
$Q_{3,4}(\hat{s})$ & $\left(1, \pm 1, \frac{2}{\pi}\arctan(\hat{s})\right)$
& $f(\hat{s})=0$ & $-2, 6- 2 \sqrt{6} \hat{s}, -2 \sqrt{6} f^{\prime }\left(%
\hat{s}\right)$ & nonhyperbolic for \\ 
&  &  &  & $f^{\prime }\left(\hat{s}\right)=0$, or $\hat{s}=\sqrt{\frac{3}{2}%
}$. \\ 
&  &  &  & It is a sink for $f^{\prime }\left(\hat{s}\right)>0, \hat{s}>%
\sqrt{\frac{3}{2}}$. \\ 
&  &  &  & It is a saddle otherwise. \\ \hline
$Q_{5,6}(\hat{s})$ & $\left(\sqrt{\frac{2}{3}} \hat{s},\pm 1,\frac{2}{\pi}%
\arctan(\hat{s})\right)$ & $f(\hat{s})=0,$ & $4(1- \hat{s}^2), 2 \hat{s}%
^2-3, -4 \hat{s} f^{\prime }\left(\hat{s}\right)$ & nonhyperbolic for $%
f^{\prime }\left(\hat{s}\right)=0$, \\ 
&  & $-\sqrt{\frac{3}{2}}\leq \hat{s}\leq \sqrt{\frac{3}{2}}$ &  & or $\hat{s%
}\in\left\{-\sqrt{\frac{3}{2}}, -1, 1, \sqrt{\frac{3}{2}}\right\}$. \\ 
&  &  &  & sink for $-\sqrt{\frac{3}{2}}<\hat{s}<-1, f^{\prime }\left(\hat{s}%
\right)<0$, \\ 
&  &  &  & or $1<\hat{s}<\sqrt{\frac{3}{2}}, f^{\prime }\left(\hat{s}%
\right)>0$. \\ 
&  &  &  & saddle otherwise. \\ \hline
$Q_{7,8}$ & $\left(0,\pm 1,0\right)$ & always & $4,-\frac{1}{2} \left(3 \pm 
\sqrt{9-48 f(0)}\right)$ & saddle. \\ \hline
\end{tabular*}%
\end{table*}

The field equations in this example become:

\begin{align}
& H^2= \frac{1}{6(3\lambda-1)}\left[\frac{3 \lambda -1}{4}\dot\phi^2+V(\phi)%
\right], 
\label{Cas61syst7a}\\
& \dot{H}=-\frac{1}{8}\dot\phi^2, \label{Cas61syst7b}\\
&\ddot\phi +3 H \dot \phi+\frac{2 V^{\prime }(\phi)}{3 \lambda -1}=0. \label{Cas61syst7c}
\end{align}

Now, we discuss the phase space of this model for arbitrary potentials and
next we specify for some potentials.

\subsection{Arbitrary Potentials}

\label{Case1A}

For an spatially flat spacetime without a cosmological constant whose evolution is given by \eqref{Cas61syst7a}, \eqref{Cas61syst7b}, \eqref{Cas61syst7c},  the
corresponding autonomous system writes: 
\begin{align}
&\frac{d x}{d N}=\left(3 x-\sqrt{6} s\right) \left(x^2-1\right),
\label{GenHLeqxcase1} \\
&\frac{d z}{d N}= \left(3 x^2-2\right) z,  \label{GenHLeqzcase1} \\
&\frac{d s}{d N}=-2 \sqrt{6} x f(s).  \label{GenHLeqscase1}
\end{align}
defined on the phase space $\{(x,z,s)\in\mathbb{R}^3: -1\leq x\leq 1\},$
where $N=\ln(a/a_0)$ denotes the new time variable.

The equilibrium points/curves of the dynamical system \eqref{GenHLeqxcase1}, \eqref{GenHLeqzcase1}, \eqref{GenHLeqscase1} at the finite region of the phase space are
presented in Table \ref{HLcrit}, where is shown the existence and stability
conditions. We proceed to the discussion of the more relevant features of
them.

\begin{itemize}
\item $P_1(\hat{s}): (x,z,s)=\left(1, 0, \hat{s}\right)$. Where $\hat{s}$
denotes a value of $s$, such that $f(\hat{s})=0$. This point is reduced to $%
P_1$ studied in \cite{Leon:2009rc}. It is a source for $f^{\prime }\left(%
\hat{s}\right)<0, \hat{s}<\sqrt{\frac{3}{2}}$.

\item $P_2(\hat{s}): (x,z,s)=\left(-1, 0, \hat{s}\right)$. Where $\hat{s}$
denotes a value of $s$,  such that $f(\hat{s})=0$. This point is reduced to $%
P_2$ studied in \cite{Leon:2009rc}. It is a source for $f^{\prime }\left(%
\hat{s}\right)>0, \hat{s}>-\sqrt{\frac{3}{2}}$.

\item $P_3(\hat{s}): (x,z,s)=\left(\sqrt{\frac{2}{3}} \hat{s}, 0, \hat{s}%
\right)$, where $\hat{s}$ denotes a value of $s$, such that $f(\hat{s})=0$
and $-\sqrt{\frac{3}{2}}\leq \hat{s}\leq \sqrt{\frac{3}{2}}$. This point is
reduced to $P_3$ studied in \cite{Leon:2009rc}. It is a sink for $f^{\prime
}\left(\hat{s}\right)<0, -1<\hat{s}<0$, or $f^{\prime }\left(\hat{s}%
\right)>0, 0<\hat{s}<1$.

\item $P_3^{0}: (x,z,s)=(0, 0, 0)$. This point is new, and it was not found
in \cite{Leon:2009rc}. It is a sink for $f(0)>0$.

\item $P_4^{+}: (x,z,s)=\left(\sqrt{\frac{2}{3}}, z_c, 1\right)$, where $z_c$
denotes an arbitrary number. Exists for $f(1)=0$. It is nonhyperbolic.

\item $P_4^{-}: (x,z,s)=\left(-\sqrt{\frac{2}{3}}, z_c, -1\right)$, where $%
z_c$ denotes an arbitrary number. Exists for $f(-1)=0$. It is nonhyperbolic.
The above lines are reduced to the line $P_4$ studied in \cite{Leon:2009rc}
when the analysis is restricted to the exponential potential (i.e., $s$
constant, $f\equiv 0$).
\end{itemize}

Owing to the fact that the dynamical system \eqref{GenHLeqxcase1}, %
\eqref{GenHLeqzcase1}, \eqref{GenHLeqscase1} is non-compact, there could be
features in the asymptotic regime which are non-trivial for the global
dynamics. Introducing the new  variables 
\begin{equation}
Z=\frac{z}{\sqrt{1+z^2}}, \quad S=\frac{2}{\pi}\arctan(s).
\end{equation}

The system \eqref{GenHLeqxcase1}, \eqref{GenHLeqzcase1}, %
\eqref{GenHLeqscase1} therefore becomes 
\begin{align}
&\frac{dx}{dN}=\left(x^2-1\right) \left(3 x-\sqrt{6} \,\tan\left(\frac{\pi S%
}{2}\right)\right), \label{Example1Sa}\\
&\frac{dZ}{dN}=\left(2-3 x^2\right) Z \left(Z^2-1\right), \label{Example1Sb}\\
&\frac{dS}{dN}=-\frac{2 \sqrt{6} x (\cos (\pi S)+1) f\left(\tan \left(\frac{%
\pi S}{2}\right)\right)}{\pi }, \label{Example1Sc}
\end{align}
defined on the compact phase space 
\begin{equation*}
\left\{(x,Z,S)\in\mathbb{R}^3: -1\leq x\leq 1, -1\leq Z \leq 1, -1\leq S\leq
1\right\}.
\end{equation*}

The points at the finite region of the phase space and their stability
remains the same (under the rescaling $s= \tan\left(\frac{\pi S}{2}\right)$%
). The points at the infinite region of the phase space are summarized in
table \ref{HLcritinfinity}. Now we discuss the relevant features of them.

\begin{itemize}
\item $Q_{1,2}(\hat{s}): (x,Z,S)=\left(-1, \pm 1, \frac{2}{\pi}\arctan(\hat{s%
})\right)$ , where $f(\hat{s})=0$. It is a sink for $f^{\prime }\left(\hat{s}%
\right)<0, \hat{s}<-\sqrt{\frac{3}{2}}$.

\item $Q_{3,4}(\hat{s}): (x,Z,S)=\left(1, \pm 1, \frac{2}{\pi}\arctan(\hat{s}%
)\right)$ where $f(\hat{s})=0$. It is a sink for $f^{\prime }\left(\hat{s}%
\right)>0, \hat{s}>\sqrt{\frac{3}{2}}$.

\item $Q_{5,6}(\hat{s}): (x,Z,S)=\left(\sqrt{\frac{2}{3}} \hat{s},\pm 1,%
\frac{2}{\pi}\arctan(\hat{s})\right)$, where $f(\hat{s})=0$, $-\sqrt{\frac{3%
}{2}}\leq \hat{s}\leq \sqrt{\frac{3}{2}}$. It is a sink for
$-\sqrt{\frac{3}{2}}<\hat{s}<-1, f^{\prime }\left(\hat{s}\right)<0$, or $1<%
\hat{s}<\sqrt{\frac{3}{2}}, f^{\prime }\left(\hat{s}\right)>0$.

\item $Q_{7,8}: (x,Z,S)=\left(0,\pm 1,0\right)$. These points are saddle
points.
\end{itemize}

These points at infinity satisfies $z\rightarrow \text{sgn}(\mu)\infty$,
which means $a^2 H\rightarrow 0$. This includes Minkowski and static
solutions.

\subsection{Exponential Potential}

\label{Case1B}

As Example 1, we implement the aforementioned procedure for the exponential
potential $V(\phi)=V_0 e^{-s \phi}$ \cite{Leon:2009rc}, where $s$ is
constant and $f$ is identically zero.

In this case the system \eqref{GenHLeqxcase1}-\eqref{GenHLeqzcase1} becomes 
\begin{align}
&\frac{d x}{d N}=\left(x^2-1\right) \left(3 x-\sqrt{6} s\right), \label{syst17a}\\
&\frac{dZ}{d N}=\left(2-3 x^2\right) Z \left(Z^2-1\right), \label{syst17b}
\end{align}
defined on the compact phase space \\
$\left\{(x,z)\in\mathbb{R}^2: -1\leq x\leq 1, -1\leq Z \leq 1\right\}.$

The equilibrium points $P_{1,2}$ are not relevant from a cosmological point
of view, since apart from being unstable they correspond to complete dark
matter domination, with the matter equation-of-state parameter being stiff.

The equilibrium point $P_3$ can be the late-time state of the universe. If
additionally we desire to keep the dark-matter equation-of-state parameter
in the physical range $0<w_M<1$ then we have to restrict the parameter $s$
in the range $\frac{\sqrt{3}}{2}<s<\sqrt{\frac{3}{2}}$. However, even in
this case the universe is finally completely dominated by dark matter. The
fact that $z_c=0$ means that in general this sub-class of universes will be
expand forever. The equilibrium points $P_4$ consist a stable late-time
solution, with a physical dark-matter equation-of-state parameter $w_M=1/3$,
but with zero dark energy density. We mention that the dark-matter
domination of the case at hand was expected, since in the absent of
curvature and of a cosmological constant the corresponding Ho\v{r}%
ava-Lifshitz universe is comprised only by dark matter. Note however that
the dark-energy equation-of-state parameter can be arbitrary.

\begin{figure*}[t!]
\centering
\includegraphics[scale=1.5]{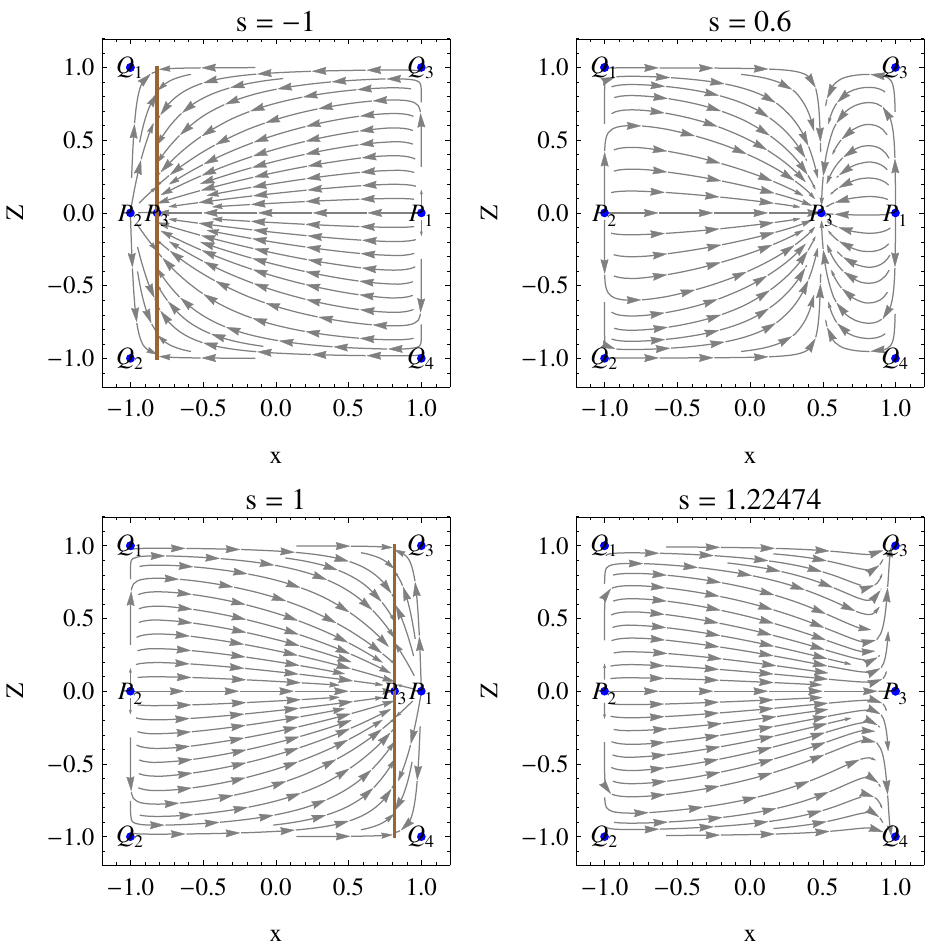} 
\caption{\textit{{(Color online) Compact phase portrait of \eqref{syst17a}-\eqref{syst17b}
for different choices of the parameter $s$. }}}
\label{fig:Case12HL}
\end{figure*}
Figure \ref{fig:Case12HL} a) and c) illustrates when $P_3$ becomes an stable
line of equilibrium points for the specific values $s=\pm 1$. Figure \ref%
{fig:Case12HL} b) illustrates when $P_3$ is an isolated late-time attractor.
Figure \ref{fig:Case12HL} d) illustrates when $Q_{5,6}: (x,Z)=\left(\sqrt{%
\frac{2}{3}} s,\pm 1\right)$ is an attractor at the region at infinity for $-%
\sqrt{\frac{3}{2}}<s<-1<0$, or $1<s<\sqrt{\frac{3}{2}}$.

\subsection{Powerlaw Potential}

\label{Case1C} As the Example 2, we consider the potential \cite%
{Alho:2015cza}: 

\begin{equation}
V(\phi)=\frac{1}{2n}(\mu \phi)^{2n}, \mu >0, n=1,2,\ldots.
\end{equation}
such that 
\begin{align}
f(s)=-\frac{s^2}{2 n}, \quad f(s)=0 \Leftrightarrow s=0, \quad f^{\prime
}(0)=0.
\end{align}
The system \eqref{Example1Sa}, \eqref{Example1Sb}, \eqref{Example1Sc} becomes 
\begin{align}
&\frac{dx}{dN}=\left(x^2-1\right) \left(3 x-\sqrt{6} \,\tan\left(\frac{\pi S%
}{2}\right)\right), \label{SYSTEM80a}\\
&\frac{dZ}{dN}=\left(2-3 x^2\right) Z \left(Z^2-1\right),  \label{SYSTEM80b}\\
&\frac{dS}{dN}=-\frac{\sqrt{6} x (\cos (\pi S)-1)}{\pi n},  \label{SYSTEM80c}
\end{align}
defined on the compact phase space\newline $\Big\{(x,S,Z)\in\mathbb{R}^3: -1\leq
x\leq 1, -1\leq S\leq 1, -1\leq Z \leq 1\Big\}.$ The equilibrium points of %
\eqref{SYSTEM80a}, \eqref{SYSTEM80b}, \eqref{SYSTEM80c}, and their stability conditions are summarized as follows.

\begin{itemize}
\item $P_{1}(0)$: $(x, Z, S):=\left(1, 0, 0\right)$. Eigenvalues $\{6,1,0\}$%
. Unstable.

\item $P_{2}(0)$: $(x, Z, S):=\left(-1, 0, 0\right)$. Eigenvalues $\{6,-5,0\}
$. Saddle.

\item $P_{3}^{0}$: $(x, Z, S):=\left(0, 0, 0\right)$. Eigenvalues $%
\{-3,-2,0\}$. Nonhyperbolic.

\item $Q_{1,2}(0)$: $(x, Z, S):=\left(-1, \pm 1, 0\right)$. Eigenvalues $%
\{10,6,0\}$. Unstable.

\item $Q_{3,4}(0)$: $(x, Z, S):=\left(1, \pm 1, 0\right)$. Eigenvalues $%
\{6,-2,0\}$. Saddle.

\item $Q_{5,6}(0)$: $(x, Z, S):=\left(0, \pm 1, 0\right)$. Eigenvalues $%
\{4,-3,0\}$. Saddle.
\end{itemize}

\subsubsection{Stability Analysis of the solution $P_{3}^{0}$ in Ho\v{r}%
ava-Lifshitz cosmology for the Flat universe with $\Lambda=0$ under the
detailed-balance condition for the powerlaw- potential.}

\label{PROPOSITION1}

\begin{prop}
\label{centerP30} The origin for the system \eqref{SYSTEM80a}, \eqref{SYSTEM80b}, \eqref{SYSTEM80c} is unstable
(saddle point).
\end{prop}

\textbf{Proof.} Taking the linear transformation 

\begin{equation}
(u,v_1,v_2)=\left(S, \frac{1}{6} \left(6 x-\sqrt{6} \pi S\right), Z\right),
\end{equation}
and taking Taylor series near $(u,v_1,v_2)=(0,0,0)$ up to fifth order we
obtain the system \eqref{SYSTEM80a} can be written into its Jordan canonical
form: 
\begin{equation}
\left(%
\begin{array}{c}
\frac{du}{dN} \\ 
\frac{dv_1}{dN} \\ 
\frac{dv_2}{dN}%
\end{array}%
\right)=\left(%
\begin{array}{ccc}
0 & 0 & 0 \\ 
0 & -3 & 0 \\ 
0 & 0 & -2%
\end{array}%
\right)\left(%
\begin{array}{c}
u \\ 
v_1 \\ 
v_2%
\end{array}%
\right)+\left(%
\begin{array}{c}
f(u,\mathbf{v}) \\ 
g_1(u,\mathbf{v}) \\ 
g_2(u,\mathbf{v})%
\end{array}%
\right),  \label{HLcentercenterPowerLawPa3}
\end{equation}%
where\newline
 $f(u,\mathbf{v})=-\frac{\pi u^2 \left(\pi ^2 u^2-12\right) \left(\pi u+%
\sqrt{6} v_1\right)}{24 n},$\newline 
$g_1(u,\mathbf{v})=\frac{\pi ^2 u^2 v_1 \left(\pi ^2 (1-2 n) u^2+12
(n-1)\right)}{24 n} -\frac{\pi u \left(\pi ^2 u^2-24\right) v_1^2}{4 \sqrt{6}%
}+3  v_1^3 +\frac{\pi ^5 (5-2 n) u^5+30 \pi ^3 (n-2) u^3}{120 \sqrt{6} n},$ 
\newline
$g_2(u,\mathbf{v})=2 v_2^3-\frac{1}{2} v_2  \left(v_2^2-1\right) \left(\pi
^2 u^2+2 \sqrt{6} \pi u  v_1+6 v_1^2\right).$ 

According to Theorem \ref{existenceCM}, there exists a 1-dimensional
invariant local center manifold $W^{c}\left( \mathbf{0}\right) $ of %
\eqref{HLcentercenterPowerLawPa3}, \newline
$W^{c}\left( \mathbf{0}\right) =\left\{ \left( u,\mathbf{v}\right) \in%
\mathbb{R}\times\mathbb{R}^{2}:\mathbf{v}=\mathbf{h}\left( u\right)\right\}$%
, satisfying $\mathbf{h}\left( 0\right) =\mathbf{0},\;D\mathbf{h}\left(
0\right) =\mathbf{0} ,\;\left\vert u\right\vert <\delta$ for $\delta$
sufficiently small. The restriction of (\ref{HLcentercenterPowerLawPa3}) to
the center manifold is $\frac{du}{dN}=f\left( u,\mathbf{h}\left( u\right)
\right)$, where the function $\mathbf{h}\left( u\right)$ satisfies %
\eqref{MaineqcM}:%
\begin{equation}
D\mathbf{h}\left( u\right) \left[ f\left( u,\mathbf{h}\left( u\right)
\right) \right] -P\mathbf{h}\left( u\right) -\mathbf{g}\left( u,\mathbf{h}%
\left( u\right) \right) =0,  \label{HLPowerLawP3h2}
\end{equation}
where 
\begin{equation*}
P=\left(%
\begin{array}{cc}
-3 & 0 \\ 
0 & -2%
\end{array}%
\right).
\end{equation*}
According to Theorem \ref{approximationCM}, the system \eqref{HLPowerLawP3h2}
can be solved approximately by expanding $\mathbf{h}\left( u\right) $ in
Taylor series at $u=0.$ Since $\mathbf{h}\left( 0\right) =\mathbf{0\ } $and $%
D\mathbf{h}\left( 0\right) =\mathbf{0},$ we propose the ansatsz {{\ 
\begin{equation}
\mathbf{h}\left( u\right) :=\left(%
\begin{array}{c}
h_{1}\left(u\right) \\ 
h_{2}\left(u\right)%
\end{array}%
\right)=\left(%
\begin{array}{c}
\sum_{j=1}^{4} a_j u^{j+1} +O\left( u^{6}\right) \\ 
\sum_{j=1}^{4} b_j u^{j+1} +O\left( u^{6}\right)%
\end{array}%
\right),
\end{equation}%
}} to substitute into (\ref{HLPowerLawP3h2}). By comparing the coefficients
of the equal powers of $u$ we find the non-null coefficients $a_2=\frac{\pi
^3 (n-2)}{12 \sqrt{6} n}, a_4= \frac{\pi ^5 (n (3 n-25)+40)}{360 \sqrt{6} n^2%
}.$ 

Therefore, the center manifold can be represented locally by the graph 
\begin{equation}
v_1=\frac{(n-2) \pi ^3 u^3}{12 \sqrt{6} n}+\frac{(n (3 n-25)+40) \pi ^5 u^5}{%
360 \sqrt{6} n^2}, v_2=0.
\end{equation}
That is, 
\begin{equation}
x=\frac{\pi S}{\sqrt{6}}+\frac{\pi ^3 (n-2) S^3}{12 \sqrt{6} n}+\frac{\pi ^5
(n (3 n-25)+40) S^5}{360 \sqrt{6} n^2}, Z=0.
\end{equation}
The dynamics on the center manifold is given by the gradient-like equation 
\begin{align}
&\frac{du}{dN}=-\nabla \Pi(u),  \notag \\
& \Pi(u)=-\frac{\pi ^2 u^4}{8 n}+\frac{\pi ^4 u^6}{72 n^2}-\frac{\left(\pi
^6 ((n-40) n+80)\right) u^8}{11520 n^3}.
\end{align}
We have $\Pi^{\prime }(0)=\Pi^{\prime \prime }(0)=\Pi^{\prime \prime \prime
(4)}(0)=-\frac{3 \pi ^2}{n}<0$. It follows that $u=0$ is a degenerated
maximum of the potential. Using the Theorem \ref{stabilityCM}, we conclude
that the center manifold of origin for the system %
\eqref{HLcentercenterPowerLawPa3}, and the origin itself are unstable
(saddle point). $\blacksquare$

\subsubsection{Alternative compactification}

In this example we can alternatively introduce the following
compactification inspired in the reference \cite{Alho:2015cza}. 
\begin{align}
& \Sigma =\frac{\dot\phi}{2 \sqrt{6} H}, \\
& Y=\frac{\mu \phi}{2^{\frac{1}{n}} 3^{\frac{1}{2 n}} ((3 \lambda -1) n)^{%
\frac{1}{2 n}} H^{\frac{1}{n}}}, \\
& T= \frac{c}{c+H^{\frac{1}{n}}},
\end{align}
where $c=2^{\frac{3}{2}-\frac{1}{n}} 3^{\frac{n-1}{2 n}} n^{-\frac{1}{2 n}}
(3 \lambda -1)^{-\frac{1}{2 n}}\mu$, such that 
\begin{align}
& \dot\phi= 2 \sqrt{6} \Sigma c^n \left(\frac{1}{T}-1\right)^n, \\
& \phi=\frac{2^{\frac{1}{n}} 3^{\frac{1}{2 n} } n^{\frac{1}{2 n}} c \left(%
\frac{1}{T}-1\right) Y (3 \lambda -1)^{\frac{1}{2 n}}}{\mu }, \\
& H= c^n\left(\frac{1}{T}-1\right)^n,
\end{align}
and 
\begin{align}
&x=\Sigma , \\
&y=Y^n, \\
&z=\frac{\mu c^{-n} \left(\frac{1}{T}-1\right)^{-n}}{4 (3 \lambda -1) a^2},
\\
&s=\frac{n T}{\sqrt{6} (T-1) Y}, \\
&Z=\frac{\mu c^{-n} \left(\frac{1}{T}-1\right)^{-n}}{\sqrt{16 a^2 (1-3
\lambda )^2+\mu ^2 c^{-2 n} \left(\frac{1}{T}-1\right)^{-2 n}}}, \\
& S=\frac{2 \;\text{arctan}\left(\frac{n T}{\sqrt{6} (T-1) Y}\right)}{\pi }.
\end{align}
\begin{figure*}[ht!]
\centering
\subfigure[]{\includegraphics[width=2.5in,
height=2.5in,angle=-90]{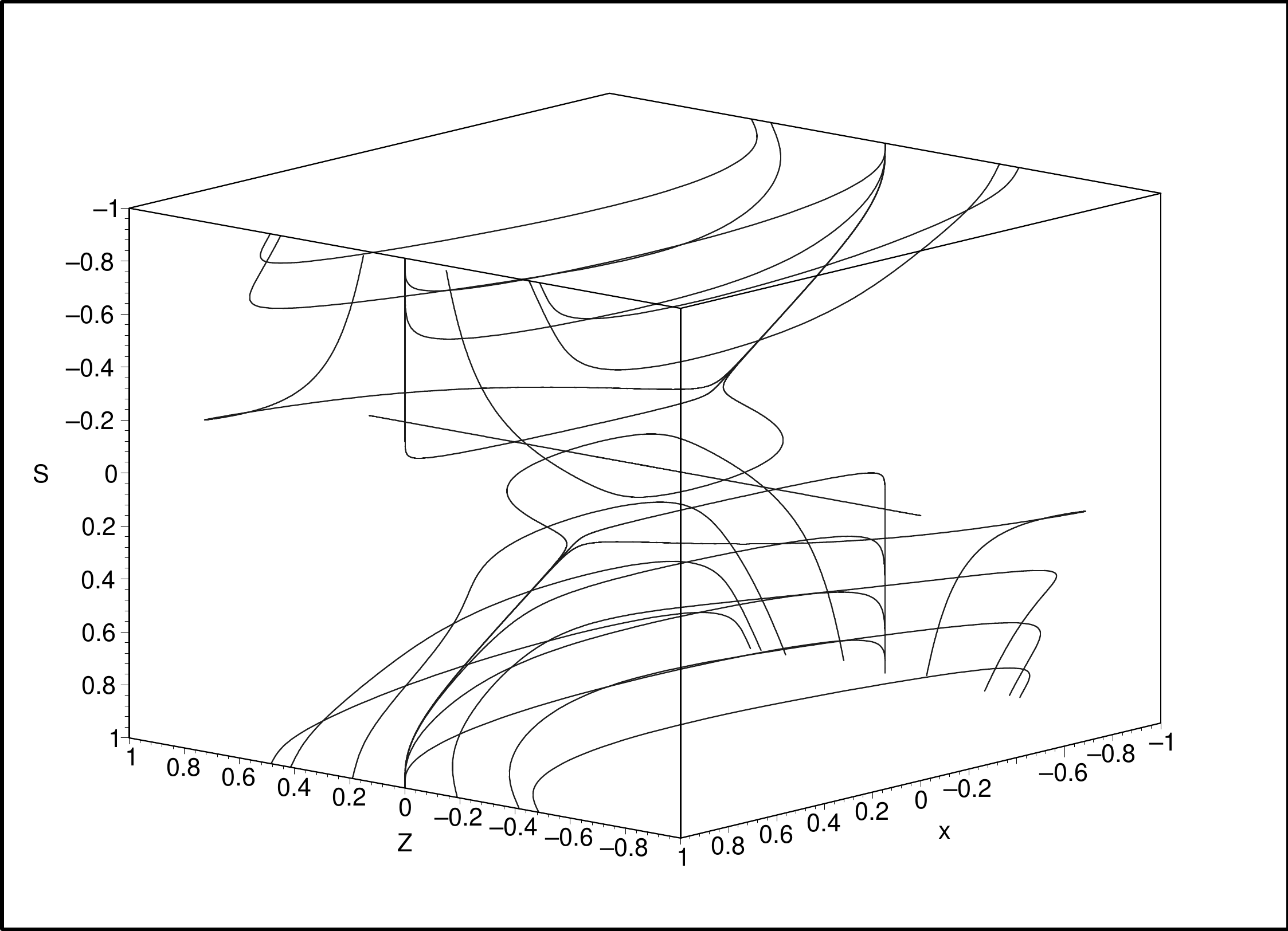}} \hspace{2cm}  \subfigure[]{%
\includegraphics[width=2.5in, height=2.5in,angle=-90]{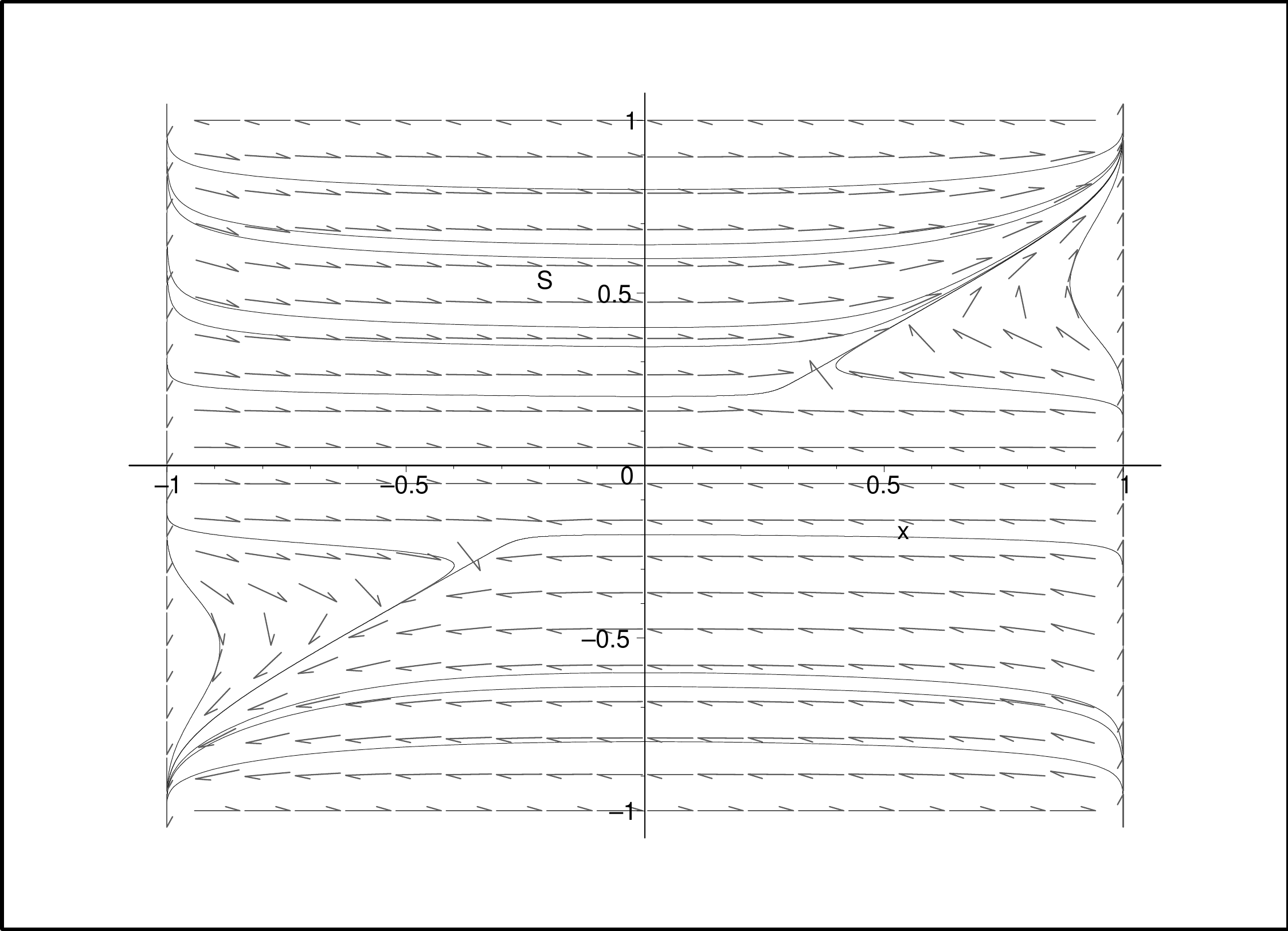}}  %
\subfigure[]{\includegraphics[width=2.5in, height=5in]{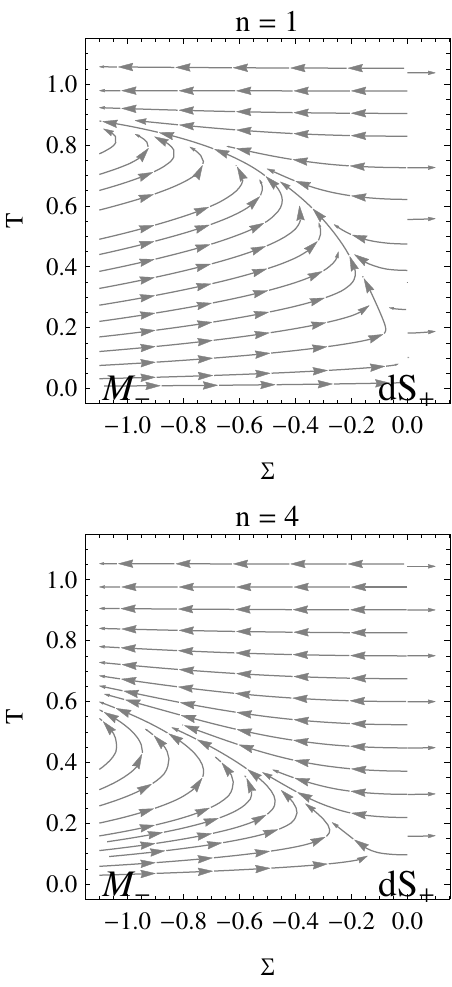}}\hspace{2cm%
}  \subfigure[]{\includegraphics[width=2.5in, height=5in]{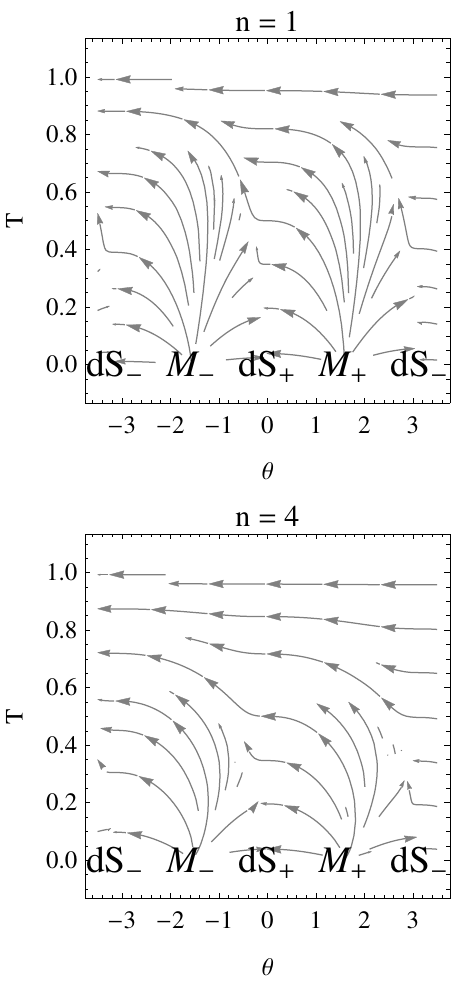}} 
\caption{\textit{{(a) Compact phase portrait of \eqref{SYSTEM80a}, \eqref{SYSTEM80b}, \eqref{SYSTEM80c} for the
choice $n=2$. (b) Dynamics in the invariant set $Z=0$. (c) Compact phase
portrait of \eqref{Eq:313} and \eqref{Eq:315} for $-1<Y<0$. (d) Unwrapped
solution space for the system \eqref{HLPL-unwrappeda}, \eqref{HLPL-unwrappedb} for $n=1, 4$.}}}
\label{fig:Case1HLPL}
\end{figure*}
and the new time variable $\frac{d\bar{\tau}}{d t}=H(1-T)^{-1}.$ Therefore,
we obtain the dynamical system 
\begin{align}
&\frac{d\Sigma }{d\bar{\tau}}=-n T Y^{2 n-1}-(2-q) \Sigma (1-T),
\label{Eq:313} \\
&\frac{d Y}{d\bar{\tau}}=\Sigma T-\frac{(q+1) (T-1) Y}{n}, \\
&\frac{d T}{d\bar{\tau}}=\frac{(q+1) (T-1)^2 T}{n},  \label{Eq:315}
\end{align}
where the fractional energy density of the scalar field energy density, $%
\Omega_\phi$, and the deceleration parameter, $q$ are: 
\begin{align}
& \Omega_\phi:=\Sigma + Y^{2 n}=1, \\
& q= \frac{1}{2} \left(-3 Y^{2 n}+3 \Sigma ^2+1\right)=-1+3\Sigma ^2.
\end{align}
Introducing the complementary global transformation 
\begin{align}
&\Sigma =F(\theta ) \sin (\theta ), \quad Y=\cos (\theta ) ,  \notag \\
&F(\theta )=\sqrt{\frac{1-\cos ^{2 n}(\theta )}{1-\cos ^2(\theta )}}={%
\sum_{k=0}^{n-1} \cos^{2k}(\theta)},
\end{align}
we obtain the following regular unconstrained 2D dynamical system 
\begin{align}
&\frac{d\theta}{d \bar{\tau}}=-T F(\theta )-\frac{3 (1-T) F(\theta )^2 \sin
(2\theta )}{2n}, \label{HLPL-unwrappeda}\\
&\frac{dT}{d \bar{\tau}}=\frac{3 (T-1)^2 T \left(1-\cos ^{2 n}(\theta
)\right)}{n}, \label{HLPL-unwrappedb}
\end{align}
and the deceleration parameter satisfies
\begin{equation}
q=2-3 \cos ^{2 n}(\theta ). 
\end{equation} 
We see that $F(\theta)\rightarrow \sqrt{n}$, as $\theta\rightarrow 0$. The
dynamics on the invariant set $T=0$ is restricted to the set $%
\Sigma^2+Y^{2n}=1$. The equilibrium points of the system %
\eqref{HLPL-unwrappeda}, \eqref{HLPL-unwrappedb} are $M_\pm=(\Sigma, Y)=(\pm 1, 0)$, corresponding to $%
\theta=\pm \frac{\pi}{2}$, representing massless scalar field cosmologies,
and $dS_\pm=(\Sigma, Y)=(0, \pm 1)$, representing de Sitter solutions,
corresponding to $\theta=0, \pi$ respectively. The following analysis is an
specific application of the results discussed in \cite{Alho:2015cza} for the
vacuum case. The attractors at this boundary set (the scalar field boundary)
are $dS_\pm$ and the sources are $M_\pm$. The solutions starting near $M_+$
approach the invariant set $T=1$ and tends to the limit-cycle given by the
circle $\Sigma^2+Y^{2n}, T=1$, that represents the Minkowski solution. In
the figure \ref{fig:Case1HLPL} it is shown (a) a compact phase portrait of %
\eqref{SYSTEM80a}, \eqref{SYSTEM80b}, \eqref{SYSTEM80c} for the choice $n=2$; (b) the dynamics in the invariant set $Z=0
$; (c) a compact phase portrait of \eqref{Eq:313} and \eqref{Eq:315} for $%
-1<Y<0$, and (d) the unwrapped solution space of  \eqref{HLPL-unwrappeda}, \eqref{HLPL-unwrappedb} for $n=1, 4$.

\subsubsection{Stability Analysis of the \emph{de Sitter} Solution in Ho\v{r}%
ava-Lifshitz cosmology for the Flat universe with $\Lambda=0$ under the
detailed-balance condition for the powerlaw- potential.}

\label{PROPOSITION2}

To find the center manifold of the de Sitter solution $dS_+$ it is
convenient to define $\tilde{T}=\frac{T}{1-T}$, and consider $N=\ln(a/a_0)$
as the time variable, therefore, we find 
\begin{align}
&\frac{d\theta}{d N}=\frac{3 \tilde{T} \left(1-\cos ^{2 n}(\theta )\right)}{n%
}, \label{HLcenterPowerLawa}\\
&\frac{d\tilde{T}}{d N}=-\tilde{T}F(\theta ) -\frac{3 F^2(\theta ) \sin
(2\theta )}{2n}.  \label{HLcenterPowerLawb}
\end{align}

\begin{prop}
\label{centerdS} The origin for the system \eqref{HLcenterPowerLawa}, \eqref{HLcenterPowerLawb} is
unstable (saddle point).
\end{prop}

\textbf{Proof.} Taking the linear transformation 

\begin{equation}
(u,v)=\left(\tilde{T}, \frac{1}{3} \left(3 \theta +\sqrt{n} \tilde{T}%
\right)\right),
\end{equation}
and taking Taylor series near $(\theta,\tilde{T})=(0,0)$ up to fifth order
we obtain the system \eqref{HLcenterPowerLawa}, \eqref{HLcenterPowerLawb} can be written in diagonal
form 
\begin{equation}
\left(%
\begin{array}{c}
\frac{du}{dN} \\ 
\frac{dv}{dN}%
\end{array}%
\right)=\left(%
\begin{array}{ccc}
0 & 0 &  \\ 
0 & -\sqrt{n} & 
\end{array}%
\right)\left(%
\begin{array}{c}
u \\ 
v%
\end{array}%
\right)+\left(%
\begin{array}{c}
f(u,v) \\ 
g(u,v)%
\end{array}%
\right),  \label{HLcenterPowerLaw3}
\end{equation}
\newline
\bigskip
$f(u,v)=-\frac{1}{162} u \left((3 n-1) \left(\sqrt{n} u-3
v\right)^2-54\right)  \left(\sqrt{n} u-3 v\right)^2$,\newline and 
\newline
\begin{small}
$g(u,v)=-\frac{n^{3/2} (n (5 n (3 n+2)-441)+144) u^4 v}{1944}+\frac{n^2
\left(n \left(55 n^2+50 n-2201\right)+720\right) u^5}{116640}$\newline
$+\frac{1}{216} \sqrt{n} u^2 v \left(24  \left(n^2+n-18\right)-(5 n (n (7
n+2)-89)+144) v^2\right)$\newline
$+\frac{u^3 \left(n (n (5 n (13 n+6)-1327)+432) v^2-4 n (n (3
n+5)-108)\right)}{1296}$\newline
$+\frac{1}{288} u \left((n (5  n (15 n+2)-453)+144) v^4-24 (n+3) (5 n-12)
v^2\right)$\newline
$+\frac{1}{30} \sqrt{n} v^3  \left(-5 n \left(n v^2-3\right)+v^2+5\right)$,
\end{small}
\newline
are nonlinear, vanish at $0$ and have vanishing derivatives at $0$. By the
theorem \ref{existenceCM}, there exists a 1-dimensional invariant local
center manifold $W^{c}\left(0\right) $ of \eqref{HLcenterPowerLaw3}, \newline$%
W^{c}\left(0\right) =\left\{ \left( u, v\right) \in\mathbb{R}\times\mathbb{R}%
:v=h\left(u\right)\right\}$, satisfying 
${h}\left(0\right) =0$, $h^{\prime }\left( 0\right)=0,\;\left\vert
u\right\vert <\delta$ for $\delta$ sufficiently small. According to Theorem %
\ref{stabilityCM}, the restriction of (\ref{HLcenterPowerLaw3}) to the
center manifold is $\frac{du}{dN}=f\left( u, h\left( u\right) \right)$,
where the function ${h}\left( u\right) $ that defines the local center
manifold satisfies%
\begin{equation}
h^{\prime }\left( u\right) \left[ f\left( u, {h}\left( u\right) \right) %
\right] +\sqrt{n} {h}\left( u\right) -{g}\left( u, {h}\left( u\right)
\right) =0.  \label{HLcenterPowerLawh}
\end{equation}
According to Theorem \ref{approximationCM}, we can use Taylor series as
follows $h(u)=a_1 u^2+a_2 u^3+a_3 u^4+a_4 u^5+ \mathcal{O}(u)^6,$ to obtain
the nonzero coefficients \newline $a_2= -\frac{1}{324} \sqrt{n} (n (3 n+5)-108)$, \newline 
\begin{small}
$%
a_4= \frac{\sqrt{n} \left(1080 n^{5/2}+1800 n^{3/2}-65 n^4-270 n^3+4079
n^2+8640 n-38880 \sqrt{n}-77760\right)}{116640}$.  
\end{small}\newline
The center manifold of $%
dS_+$ can expressed as
\begin{widetext}
	\begin{align}
&\theta=-\frac{\sqrt{n} \tilde{T}}{3}-\frac{1}{324} \left(\sqrt{n} (n (3 n+5)-108)\right) \tilde{T}^3 \nonumber \\
& +\frac{\sqrt{n} \left(1080 n^{5/2}+1800 n^{3/2}-65 n^4-270  n^3+4079 n^2+8640 n-38880 \sqrt{n}-77760\right) \tilde{T}^5}{116640} +O\left(\tilde{T}^6\right).
\end{align}
\end{widetext}
The dynamics on the center manifold can be approximated by the gradient-like
equation 
\begin{small} 
\begin{align}
&\frac{d u}{dN}=-\nabla \Pi(u),  \notag \\
& \Pi(u)=\frac{n \left(9 n^{5/2}+15 n^{3/2}-n^3+6 n^2+99 n-324 \sqrt{n}%
-810\right) u^8}{3888}  \notag \\
& -\frac{1}{162} (n-18) n u^6-\frac{n u^4}{12}.
\end{align}
\end{small}

Due to the first nonzero derivative of $\Pi$ evaluated a $u=0$ is $%
\Pi^{(4)}(0)=-2n<0$, it follows that $u=0$ is a degenerated maximum of the
potential. Using the Theorem \ref{stabilityCM}, we conclude that the center
manifold of origin for the system \eqref{HLcenterPowerLawa}, \eqref{HLcenterPowerLawb}, and the origin
itself are unstable (saddle point). Therefore, the center manifold of $dS_+$
is a good approximation for the early time attractor (see the reference \cite%
{Alho:2015cza}, Section 4). $\blacksquare$

\subsection{E-models}

\label{Case1D}

\begin{figure*}[ht!]
\centering
\subfigure[]{\includegraphics[width=2.5in, height=2.5in,angle=-90]{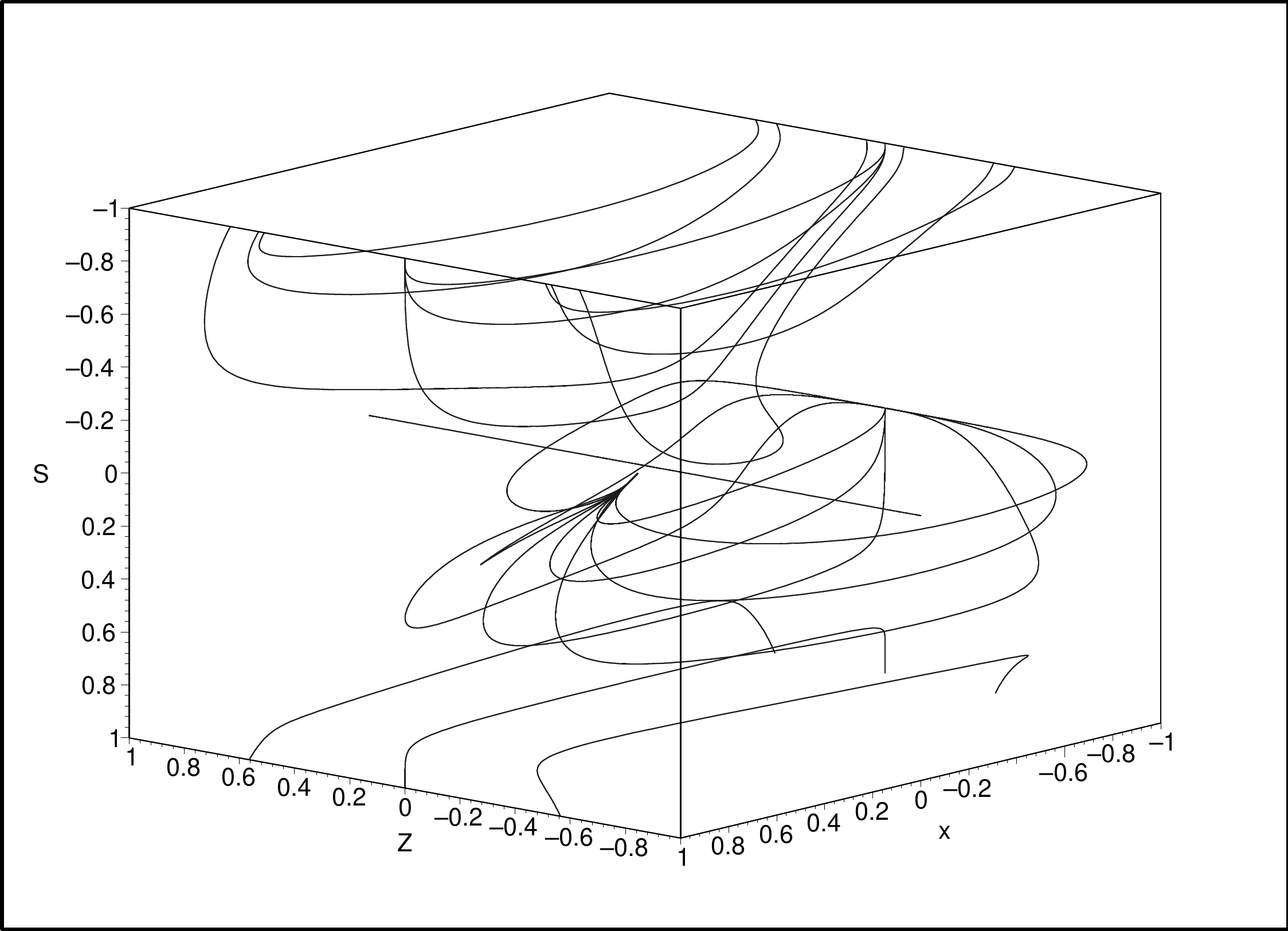}	} 
\hspace{2cm}  \subfigure[]{\includegraphics[width=2.5in,
height=2.5in,angle=-90]{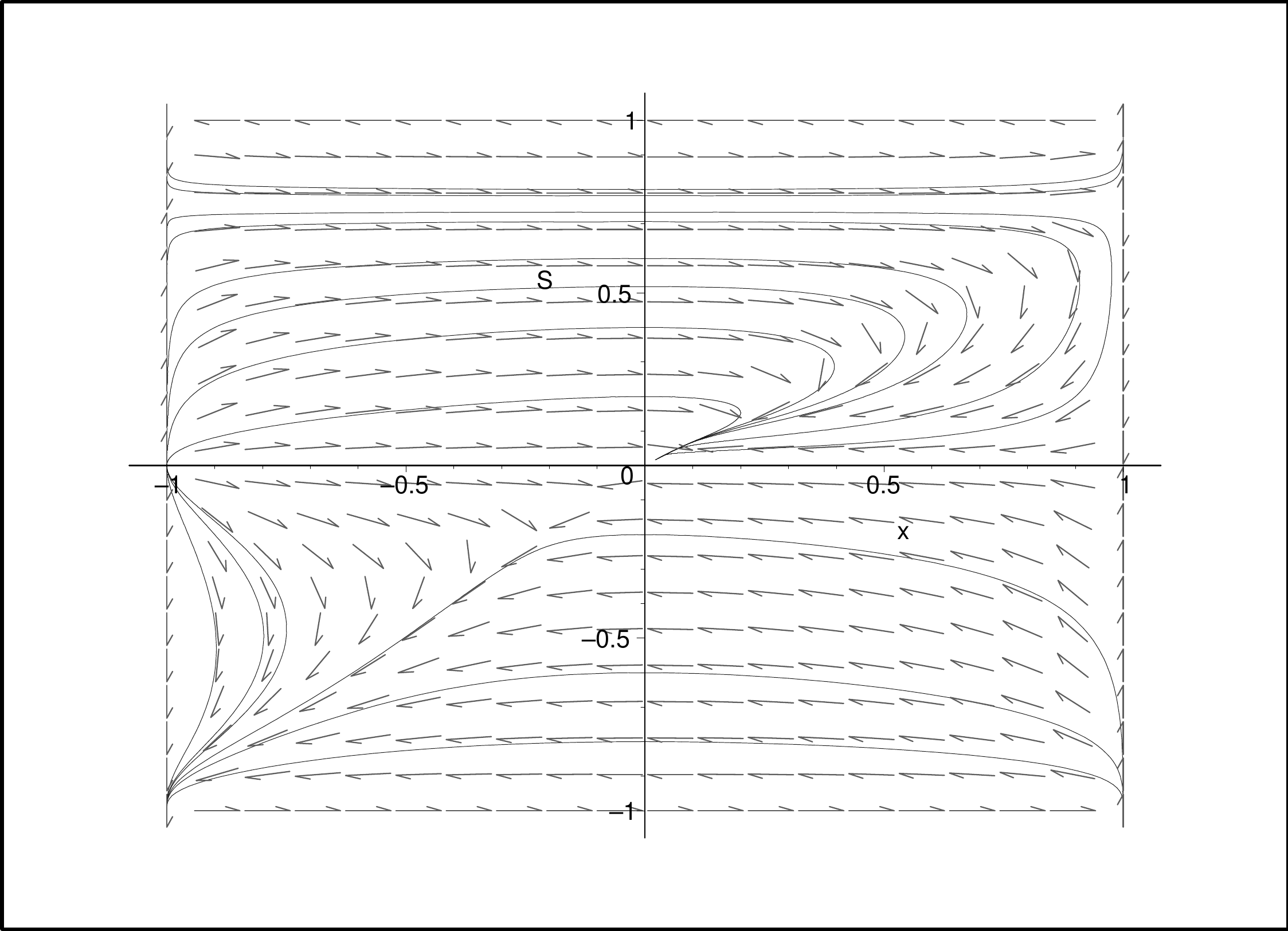}} \hspace{2cm}  \subfigure[]{%
\includegraphics[width=5in, height=5in]{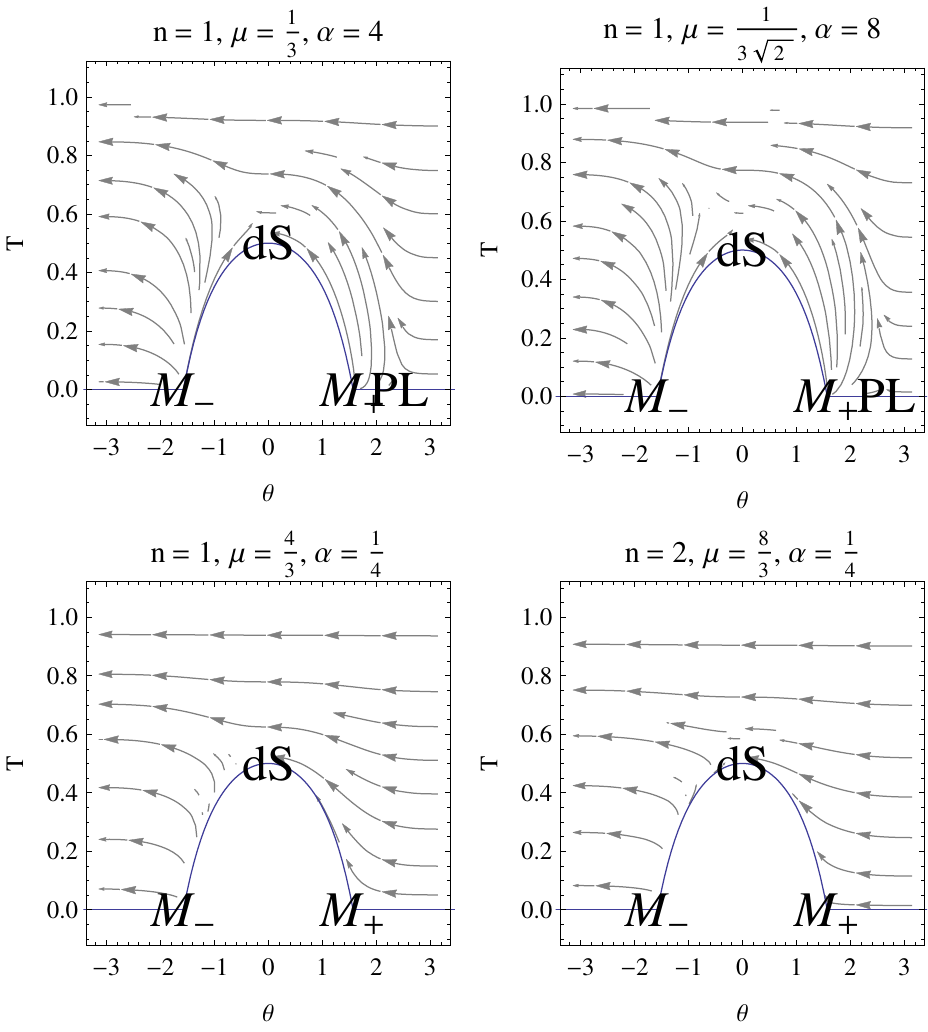}} 
\caption{\textit{{(a) Compact phase portrait of \eqref{SYSTEM92a}, \eqref{SYSTEM92b}, \eqref{SYSTEM92c} for the
choice $n=2, \protect\mu=1$. (b) Dynamics in the invariant set $Z=0$. (c)
Unwrapped solution space \eqref{HLE-unwrappeda}, \eqref{HLE-unwrappedb} for some values of $n,\protect%
\mu,\protect\alpha$.}}}
\label{fig:Case1HLE}
\end{figure*}

In this subsection we consider the E-model with potential
$V(\phi)=V_0\left(1-e^{-\sqrt{\frac{2}{3 \alpha}} \phi}\right)^{2 n}$. This is a non-negative potential with a single
minimum located at $(\phi,V(\phi))=(0,0)$. Therefore, the model admits an
Minkowski solution represented by the equilibrium point $(H,\dot\phi,
\phi)=(0,0,0)$. The potential has a plateau $V=V_0$, when $\phi\rightarrow
+\infty$, while $V\sim V_0 e^{-2 n \sqrt{\frac{2}{3 \alpha}} \phi}$ as $%
\phi\rightarrow -\infty$ \cite{Alho:2017opd}. At small $\phi$ the
E-potential behaves as $\phi^{2 n}$. 

For this choice we have 
\begin{align}
& f(s)=-\frac{s \left(s-\sqrt{6} \mu \right)}{2 n}, \quad \mu=\frac{2 n}{3 
\sqrt{\alpha}},  \notag \\
& f(s)=0 \Leftrightarrow s=0 \;\text{or}\; s=\sqrt{6} \mu,  \notag \\
& f^{\prime }(0)=\frac{\sqrt{\frac{3}{2}} \mu }{n}>0, \quad f^{\prime }(%
\sqrt{6} \mu)= -\frac{\sqrt{\frac{3}{2}} \mu }{n}<0.
\end{align}

The equations become 
\begin{align}
&\frac{dx}{dN}=\left(x^2-1\right) \left(3 x-\sqrt{6} \,\tan\left(\frac{\pi S%
}{2}\right)\right), 
\label{SYSTEM92a}\\
&\frac{dZ}{dN}=\left(2-3 x^2\right) Z \left(Z^2-1\right), \label{SYSTEM92b}\\
&\frac{dS}{dN}=\frac{x \left(-6 \mu \sin (\pi S)-\sqrt{6} \cos (\pi S)+\sqrt{%
6}\right)}{\pi n}, \label{SYSTEM92c}
\end{align}
defined on the compact phase space 
\begin{equation*}
\Big\{(x,Z,S)\in\mathbb{R}^3: -1\leq x\leq 1, -1\leq Z \leq 1, -1\leq S\leq 1%
\Big\}.
\end{equation*}

The equilibrium points of \eqref{SYSTEM92a}, \eqref{SYSTEM92b}, \eqref{SYSTEM92c}, and their stability conditions
are summarized as follows.

\begin{itemize}
\item $\left(x,Z,S\right)=(-1, -1, 0)$. The eigenvalues are $\{10,6, \frac{6
\mu}{n} \}$. It is a source.

\item $\left(x,Z,S\right)=\left(-1, -1, \frac{2 }{\pi}\text{arctan}\left(%
\sqrt{6 }\mu\right)\right)$. The eigenvalues are $\left\{10,12 \mu +6,-\frac{%
6 \mu }{n}\right\}$. It is a saddle.

\item $\left(x,Z,S\right)=(-1, 0, 0)$. The eigenvalues are $\left\{-5,6,%
\frac{6 \mu }{n}\right\}$. It is a saddle.

\item $\left(x,Z,S\right)=\left(-1, 0, \frac{2 }{\pi}\text{arctan}\left(%
\sqrt{6 }\mu\right)\right)$. The eigenvalues are
$\left\{-5,12 \mu +6,-\frac{6 \mu }{n}\right\}$. It is a saddle.

\item $\left(x,Z,S\right)=(-1, 1, 0)$. The eigenvalues are $\left\{6,10,%
\frac{6 \mu }{n}\right\}$. It is a source.

\item $\left(x,Z,S\right)=\left(-1, 1, \frac{2 }{\pi}\text{arctan}\left(%
\sqrt{6 }\mu\right)\right)$. The eigenvalues are $\left\{10,12 \mu +6,-\frac{%
6 \mu }{n}\right\}$. Saddle.

\item $\left(x,Z,S\right)=(0, -1, 0)$. The eigenvalues are $\{4,-3,0\}$.
Nonhyperbolic. Behaves as Saddle.

\item $\left(x,Z,S\right)= (0, 0, 0)$. The eigenvalues are $\{-3,-2,0\}$.
Nonhyperbolic.

\item $\left(x,Z,S\right)=(0, 1, 0)$. The eigenvalues are $\{4,-3,0\}$.
Nonhyperbolic. Behaves as a saddle.

\item $\left(x,Z,S\right)=(1, -1, 0)$. The eigenvalues are $\left\{-2,6,-%
\frac{6 \mu }{n}\right\}$. Saddle.

\item $\left(x,Z,S\right)=\left(1, -1, \frac{2 }{\pi}\text{arctan}\left(%
\sqrt{6 }\mu\right)\right)$. The eigenvalues are $\left\{-2,6-12 \mu ,\frac{%
6 \mu }{n}\right\}$. Saddle.

\item $\left(x,Z,S\right)=(1, 0, 0)$. The eigenvalues are $\left\{1,6,-\frac{%
6 \mu }{n}\right\} $. Saddle.

\item $\left(x,Z,S\right)=\left(1, 0, \frac{2 }{\pi}\text{arctan}\left(\sqrt{%
6 }\mu\right)\right)$. The eigenvalues are $\left\{1,6-12 \mu ,\frac{6 \mu }{%
n}\right\}$. Source for $\mu<\frac{1}{2}$.

\item $\left(x,Z,S\right)=(1, 1, 0)$. The eigenvalues are $\left\{-2,6,-%
\frac{6 \mu }{n}\right\}$. Saddle.

\item $\left(x,Z,S\right)=\left(1, 1, \frac{2 }{\pi}\text{arctan}\left(\sqrt{%
6 }\mu\right)\right)$. The eigenvalues are $\left\{-2,6-12 \mu ,\frac{6 \mu 
}{n}\right\}$. Saddle.

\item $\left(x,Z,S\right)=\left(2 \mu, -1, \frac{2 }{\pi}\text{arctan}\left(%
\sqrt{6 }\mu\right)\right)$. The eigenvalues are $\left\{4-48 \mu ^3,12 \mu
^2-3,\frac{12 \mu ^2}{n}\right\}$. Saddle.

\item $\left(x,Z,S\right)=\left(2 \mu , 0, \frac{2 }{\pi}\text{arctan}\left(%
\sqrt{6 }\mu\right)\right)$. The eigenvalues are $\left\{24 \mu ^3-2,12 \mu
^2-3,\frac{12 \mu ^2}{n}\right\}$. Source for $n>0, \mu >\frac{1}{2}$.
Saddle otherwise.

\item $\left(x,Z,S\right)=\left(2 \mu, 1, \frac{2 }{\pi}\text{arctan}\left(%
\sqrt{6 }\mu\right)\right)$. The eigenvalues are $\left\{4-48 \mu ^3,12 \mu
^2-3,\frac{12 \mu ^2}{n}\right\}$. Saddle.
\end{itemize}

\subsubsection{Stability Analysis of the solution $P_{3}^{0}:\left(x,Z,S%
\right)= (0, 0, 0)$ in Ho\v{r}ava-Lifshitz cosmology for the Flat universe
with $\Lambda=0$ under the detailed-balance condition for the E-model.}

\label{PROPOSITION3}

\begin{prop}
\label{centerP3Emodel} The origin for the system \eqref{SYSTEM92a}, \eqref{SYSTEM92b}, \eqref{SYSTEM92c} is
unstable (center-saddle).
\end{prop}

\textbf{Proof.} Taking the linear transformation 
\begin{equation}
(u,v_1,v_2)=\left(S, \frac{1}{6} \left(6 x-\sqrt{6} \pi S\right), Z\right),
\end{equation}
and taking Taylor series near $(u,v_1,v_2)=(0,0,0)$ up to fifth order we
obtain the system \eqref{SYSTEM92a}, \eqref{SYSTEM92b}, \eqref{SYSTEM92c} can be written into its Jordan canonical
form: %
\begin{equation}
\left(%
\begin{array}{c}
\frac{du}{dN} \\ 
\frac{dv_1}{dN} \\ 
\frac{dv_2}{dN}%
\end{array}%
\right)=\left(%
\begin{array}{ccc}
0 & 0 & 0 \\ 
0 & -3 & 0 \\ 
0 & 0 & -2%
\end{array}%
\right)\left(%
\begin{array}{c}
u \\ 
v_1 \\ 
v_2%
\end{array}%
\right)+\left(%
\begin{array}{c}
f(u,\mathbf{v}) \\ 
g_1(u,\mathbf{v}) \\ 
g_2(u,\mathbf{v})%
\end{array}%
\right),  \label{HLcenterS92}
\end{equation}%
\begin{small}
 $f(u,\mathbf{v})=-\frac{\pi ^4 u^5}{24 n}-\frac{\pi ^3 u^4 (v_1-4 \mu )}{%
4 \sqrt{6} n}+\frac{\pi ^2 u^3 (2 \mu v_1+1)}{2 n}+\frac{\sqrt{\frac{3}{2}}
\pi u^2 (v_1-2 \mu )}{n}-\frac{6 \mu u v_1}{n},$  $g_1(u,\mathbf{v})=\frac{%
\pi ^2 u^2 \left(720 \mu +\pi u \left(\sqrt{6} \pi ^2 (5-2 n) u^2+30 \sqrt{6}
(n-2)-120 \pi \mu u\right)\right)}{720 n}$\newline
$+\frac{\pi u v_1 \left(24 \sqrt{6} \mu +\pi u \left(\pi ^2 (1-2 n) u^2+12
(n-1)-4 \sqrt{6} \pi \mu u\right)\right)}{24 n}-\frac{\pi u \left(\pi ^2
u^2-24\right) v_1^2}{4 \sqrt{6}}+3 v_1^3,$\newline
$g_2(u,\mathbf{v})=\frac{1}{2} v_2 \left(4 v_2^2-\left(v_2^2-1\right)
\left(\pi ^2 u^2+2 \sqrt{6} \pi u v_1+6 v_1^2\right)\right).$
\end{small}\newline
According to Theorem \ref{existenceCM}, there exists a 1-dimensional
invariant local center manifold $W^{c}\left( \mathbf{0}\right) $ of %
\eqref{HLcenterS92}, \newline
$W^{c}\left( \mathbf{0}\right) =\left\{ \left( u,\mathbf{v}\right) \in%
\mathbb{R}\times\mathbb{R}^{2}:\mathbf{v}=\mathbf{h}\left( u\right)\right\}$%
, satisfying $\mathbf{h}\left( 0\right) =\mathbf{0},\;D\mathbf{h}\left(
0\right) =\mathbf{0} ,\;\left\vert u\right\vert <\delta$ for $\delta$
sufficiently small. The restriction of (\ref{HLcenterS92}) to the center
manifold is $\frac{du}{dN}=f\left( u,\mathbf{h}\left( u\right) \right)$,
where the function $\mathbf{h}\left( u\right)$ satisfies \eqref{MaineqcM}:%
\begin{equation}
D\mathbf{h}\left( u\right) \left[ f\left( u,\mathbf{h}\left( u\right)
\right) \right] -P\mathbf{h}\left( u\right) -\mathbf{g}\left( u,\mathbf{h}%
\left( u\right) \right) =0,  \label{HLS92P3h2}
\end{equation}
where 
\begin{equation*}
P=\left(%
\begin{array}{cc}
-3 & 0 \\ 
0 & -2%
\end{array}%
\right).
\end{equation*}
According to Theorem \ref{approximationCM}, the system \eqref{HLS92P3h2} can
be solved approximately by expanding $\mathbf{h}\left( u\right) $ in Taylor
series at $u=0.$ Using the ansatsz 
\begin{equation}
\mathbf{h}\left( u\right) :=\left(%
\begin{array}{c}
h_{1}\left(u\right) \\ 
h_{2}\left(u\right)%
\end{array}%
\right)=\left(%
\begin{array}{c}
\sum_{j=1}^{4} a_j u^{j+1} +O\left( u^{6}\right) \\ 
\sum_{j=1}^{4} b_j u^{j+1} +O\left( u^{6}\right)%
\end{array}%
\right),
\end{equation}%
 we find the non-null coefficients\newline 
$a_1=\frac{\pi ^2 \mu }{3 n}, a_2= \frac{\pi ^3 \left(24 \mu ^2+(n-2)
n\right)}{12 \sqrt{6} n^2}, a_3= \frac{\pi ^4 \mu \left(56 \mu ^2+n (2
n-7)\right)}{18 n^3}$,\newline $a_4= \frac{\pi ^5 \left(13600 \mu ^4+n^2 (n (3
n-25)+40)+20 \mu ^2 n (29 n-108)\right)}{360 \sqrt{6} n^4}$. \newline
Therefore, the center manifold can be represented locally by the graph 
\begin{small}
\begin{align}
&v_1= \frac{\pi ^2 \mu u^2}{3 n}+\frac{\pi ^3 \left(24 \mu ^2+(n-2) n\right)
u^3}{12 \sqrt{6} n^2}  \notag \\
& +\frac{\pi ^4 \mu \left(56 \mu ^2+n (2 n-7)\right) u^4}{18 n^3}  \notag \\
& +\frac{\pi ^5 \left(13600 \mu ^4+20 n (29 n-108) \mu ^2+n^2 (n (3
n-25)+40)\right) u^5}{360 \sqrt{6} n^4},  \notag \\
&v_2=0.
\end{align}%
\end{small}
 That is, 
\begin{small}
\begin{align}
&x=\frac{\pi S}{\sqrt{6}}+\frac{\pi ^2 \mu S^2}{3 n}+\frac{\pi ^3 \left(24
\mu ^2+(n-2) n\right) S^3}{12 \sqrt{6} n^2}  \notag \\
& +\frac{\pi ^4 \mu \left(56 \mu ^2+n (2 n-7)\right) S^4}{18 n^3}  \notag \\
& +\frac{\pi ^5 \left(13600 \mu ^4+20 n (29 n-108) \mu ^2+n^2 (n (3
n-25)+40)\right) S^5}{360 \sqrt{6} n^4},  \notag \\
&Z=0.
\end{align}%
\end{small}
The dynamics on the center manifold is given by the gradient-like
equation
\begin{small}
\begin{align}
&\frac{du}{dN}=-\nabla \Pi(u), \notag \\ & \Pi(u)=\frac{\sqrt{\frac{2}{3}} \pi \mu u^3}{n%
} -\frac{\pi ^2 u^4 \left(n-4 \mu ^2\right)}{8 n^2}  \notag \\
& -\frac{\pi ^3 \mu u^5 \left(n (n+4)-24 \mu ^2\right)}{10 \sqrt{6} n^3} +%
\frac{\pi ^4 u^6 \left(224 \mu ^4+n^2+4 \mu ^2 (n-10) n\right)}{72 n^4} 
\notag \\
& -\frac{\pi ^5 \mu u^7 \left(-6800 \mu ^4+n^2 (n (n+15)-55)+10 \mu ^2
(136-23 n) n\right)}{210 \sqrt{6} n^5}.
\end{align}%
\end{small}
 We have $\Pi^{\prime }(0)=\Pi^{\prime \prime }(0)=0, \Pi^{\prime \prime
\prime }(0)=\frac{2 \sqrt{6} \pi \mu }{n}$. It follows that $u=0$ is an
inflection of the potential. Using the Theorem \ref{stabilityCM}, we
conclude that the center manifold of origin for the system %
\eqref{HLcenterS92}, and the origin are unstable (center-saddle). $%
\blacksquare$

\subsubsection{Alternative compactification}
As we have commented before, $V(\phi)=V_0\left(1-e^{-\sqrt{\frac{2}{3
\alpha}} \phi}\right)^{2 n}$ is a non-negative potential with a single
minimum located at $(\phi,V(\phi))=(0,0)$, with $(H,\dot\phi,
\phi)=(0,0,0)$, corresponding to the 
Minkowski solution. The potential has a plateau $V=V_0$, when $\phi\rightarrow
+\infty$, while $V\sim V_0 e^{-2 n \sqrt{\frac{2}{3 \alpha}} \phi}$ as $%
\phi\rightarrow -\infty$ \cite{Alho:2017opd}. At small $\phi$ the
E-potential behaves as $\phi^{2 n}$, so, it can be implemented a similar
approach as the one used for $\sim \phi^{2n}$ potentials in section \ref%
{Case1C}. That is, we define new variables given by 
\begin{align}
&\Sigma =\frac{\dot\phi}{2 \sqrt{6} H}, \\
&Y=\left(\frac{V(\phi)}{6(3\lambda-1)H^2}\right)^{\frac{1}{2 n}}=\tilde{T}%
\left(1-e^{-\sqrt{\frac{2}{3 \alpha}} \phi}\right), \\
&\tilde{T}=\left[\frac{V_0}{6(3\lambda-1)H^2}\right]^{\frac{1}{2n}},
\end{align}
such that 
\begin{align}
&\dot\phi=\frac{2 \Sigma \sqrt{V_0} \tilde{T}^{-n}}{\sqrt{3 \lambda -1}}, \\
&\phi= -\frac{\sqrt{\frac{2}{3}} n \ln \left(1-\frac{Y}{\tilde{T}}\right)}{%
\mu}, \\
&H= \frac{\sqrt{V_0} \tilde{T}^{-n}}{\sqrt{3(6\lambda-2)}},
\end{align}
and 
\begin{align}
&x=\Sigma, \\
&y=Y^n, \\
&z=\frac{\sqrt{\frac{3}{2}} \mu \tilde{T}^n}{2 \sqrt{3 \lambda -1} \sqrt{V_0}
a^2}, \\
&u=\frac{\sqrt{\frac{3}{2}} \Lambda \mu \tilde{T}^n}{2 \sqrt{3 \lambda -1} 
\sqrt{V_0}}, \\
&s=\sqrt{6} \mu \left(1-\frac{\tilde{T}}{Y}\right), \\
&Z=\frac{\sqrt{3} \mu \tilde{T}^n}{\sqrt{24 \lambda V_0 a^4-8 V_0 a^4+3 \mu
^2 \tilde{T}^{2 n}}}, \\
&S=\frac{2 \;\text{arctan}\left(\sqrt{6} \mu \left(1-\frac{\tilde{T}}{Y}%
\right)\right)}{\pi },
\end{align}
where the fractional energy density of the scalar field energy density, $%
\Omega_\phi$, and the deceleration parameter, $q$ are: 
\begin{align}
&\Omega_\phi:=\Sigma ^2 + Y^{2 n}=1, \\
& q= \frac{1}{2} \left(-3 Y^{2 n}+3 \Sigma ^2+1\right)=-1+3\Sigma ^2.
\end{align}

We obtain the dynamical system: 
\begin{align}
&\frac{d\Sigma }{dN}= 6 \mu Y^{2 n-1} (Y-\tilde{T})+(q-2) \Sigma , \\
&\frac{dY}{dN}=\frac{Y (-6 \mu \Sigma +q+1)+6 \mu \Sigma \tilde{T}}{n}, \\
&\frac{d\tilde{T}}{dN}=\frac{(q+1) \tilde{T}}{n}.
\end{align}

This system have been extensively studied in \cite{Alho:2017opd} in the
context of a canonical scalar field cosmology. Now, we will discuss the more
relevant features of solutions space. It can be easily proven that $\tilde{T}
$ is monotonically increasing toward the future and decreasing towards the
past. The phase space is limited to the past by the invariant subset $\tilde{%
T}=0$, for $Y\leq 0$, and by $\tilde{T}-Y=0$, for $Y\geq 0$. The state space
is bounded when $\tilde{T}>0, \tilde{T}-Y>0$. As for the monomial potential
with odd exponent, the two past boundaries are intersected at the two
massless scalar field points $M_\pm=(\Sigma,Y)=(\pm 1, 0)$. The subset $%
\tilde{T}-Y=0$, on the other hand, it is divided in two disconnected regions
separated by the de Sitter equilibrium point $dS=(\tilde{T},Y)=(1,1)$. This
feature it is illustrated in the Figure \ref{fig:Case1HLE}, where it is seen
the phase-space has two disconnected regions, in different coordinates.

Introducing the complementary global transformation 
\begin{align}
&\Sigma =F(\theta ) \sin (\theta ), \quad Y=\cos (\theta ),  \notag \\
&F(\theta )=\sqrt{\frac{1-\cos ^{2 n}(\theta )}{1-\cos ^2(\theta )}},
\end{align}
we obtain the following regular unconstrained 2D dynamical system 
\label{HLE-unwrapped0}
\begin{align}
&\frac{d\theta}{dN}=-\frac{6 \mu F(\theta ) (\tilde{T}-\cos (\theta ))}{n}-%
\frac{3 F(\theta )^2 \sin (2\theta )}{2 n}, \label{HLE-unwrapped0a}\\
&\frac{d\tilde{T}}{dN}=\frac{3 \tilde{T} \left(1-\cos ^{2 n}(\theta )\right)%
}{n}, \label{HLE-unwrapped0b}
\end{align}
and the deceleration parameter becomes
\begin{equation}
q=2-3 \cos ^{2 n}(\theta ).
\end{equation}
Introducing the new compact variable $T=\frac{\tilde{T}}{1+\tilde{T}}$ and
the new time derivative $\frac{d{\bar{\tau}}}{d \ln a}=1+\tilde{T}=(1-T)^{-1}
$, we obtain the regular system:

\begin{align}
&\frac{d\theta}{d\bar{\tau}}=\frac{3 (T-1) F(\theta )^2 \sin (\theta ) \cos
(\theta )}{n}  \notag \\
& -\frac{6 \mu F(\theta ) ((T-1) \cos (\theta )+T)}{n}, \label{HLE-unwrappeda} \\
&\frac{d T}{d\bar{\tau}}=-\frac{3 (T-1)^2 T \left(\cos ^{2 n}(\theta
)-1\right)}{n}. \label{HLE-unwrappedb}
\end{align}
The past boundary is attached to the phase-space, and in the new variables $%
(\theta, T)$ it is defined by $\{T=0, \cos(\theta)\leq
0\}\cup\left\{T-(1-T)\cos\theta=0, \cos(\theta)>0\right\}$. It is also
included the future boundary $T=1$, which corresponds to $H=0$ and the final
state is the Minkowski point. 
The region \newline $\left\{T-(1-T)\cos\theta<0, \cos(\theta)>0\right\}$ is forbidden.

The equilibrium points of \eqref{HLE-unwrappeda}, \eqref{HLE-unwrappedb} are given by

\begin{itemize}
\item $M_{\pm}$: $\tilde{T}=T=0; \Sigma=\pm 1, Y=0; \theta=\pm\frac{\pi}{2}%
+2 k \pi, k=0,1,2 \ldots$. They are massless scalar field solutions. In the
original variables these solutions corresponds to $P_{1,2}(\sqrt{6}\mu):$%

$\left(x=\pm 1,y=0,z=0,s=\sqrt{6} \mu ; S= \frac{2 \;\text{arctan}\left(%
\sqrt{6} \mu \right)}{\pi }\right)$. They are saddle and source,
respectively, as it is confirmed in Figure \ref{fig:Case1HLE} (c) where its
presented the unwrapped solution space of \eqref{HLE-unwrappeda}, \eqref{HLE-unwrappedb} for some values
of $n,\mu,\alpha$.

\item $dS$: $\tilde{T}=1, T=\frac{1}{2}; \Sigma=0, Y=1; \theta=2 k \pi,
k=0,1,2 \ldots$. It is a de Sitter solution. In the original variables the
solution $dS$ corresponds to the point $(x=0,y=1,z=0; s=0,S=0)$, i.e, it is
represented by $P_3^{0}:(x,Z,S)=(0, 0, 0)$, which is is nonhyperbolic since $%
f(0)=0$.

\item $PL$: $\tilde{T}=T=0; \Sigma=2\mu; Y=-(1-4\mu^2)^{\frac{1}{2n}};
\theta=\pm\arccos Y$. Exists for $\mu<1/2$. It corresponds to a Powerlaw
selfsimilar solution for the exponential potential. It is equivalent to $P_3(%
\sqrt{6}\mu)$.
\end{itemize}

In Figure \ref{fig:Case1HLE} (a) it is shown a compact phase portrait of the
system \eqref{SYSTEM92a}, \eqref{SYSTEM92b}, \eqref{SYSTEM92c} for the choice $n=2, \mu=1$. In (b) it is shown the
dynamics in the invariant set $Z=0$. In (c) it is represented the unwrapped
solution space of the system \eqref{HLE-unwrappeda}, \eqref{HLE-unwrappedb} for some values of $n,\mu,\alpha$. This
plot clearly shows that the future boundary is $T=1$, which corresponds to $%
H=0$ and the final state is the Minkowski point given by a limit cycle.

\subsubsection{Stability Analysis of the \emph{de Sitter} Solution in Ho\v{r}%
ava-Lifshitz cosmology for the Flat universe with $\Lambda=0$ under the
detailed-balance condition for the E-potential.}

\label{PROPOSITION4} In order to analyze the stability of \emph{de Sitter}
solution we can use center manifold theorem. It is more convenient to use
the variables $(\theta, \tilde{T})$ whose evolution is given by the regular
unconstrained 2D dynamical system \eqref{HLE-unwrapped0a}, \eqref{HLE-unwrapped0b}.

\begin{prop}
\label{centerdSEmodel1} The \emph{de Sitter} solution $dS_+$, with $\theta=0$%
, for the system \eqref{HLE-unwrapped0a}, \eqref{HLE-unwrapped0b} is unstable (center-saddle).
\end{prop}

\textbf{Proof.} Taking the linear transformation 
\begin{equation}
(u,v)=\left(\tilde{T}-1,\theta +\frac{2 \mu (\tilde{T}-1)}{\sqrt{n}}\right),
\end{equation}
and truncating the Taylor series at fifth order we obtain that the system %
\eqref{HLE-unwrapped0a}, \eqref{HLE-unwrapped0b} can be written in diagonal form 
\begin{equation}
\left(%
\begin{array}{c}
\frac{du}{dN} \\ 
\frac{dv}{dN}%
\end{array}%
\right)=\left(%
\begin{array}{ccc}
0 & 0 &  \\ 
0 & -3 & 
\end{array}%
\right)\left(%
\begin{array}{c}
u \\ 
v%
\end{array}%
\right)+\left(%
\begin{array}{c}
f(u,v) \\ 
g(u,v)%
\end{array}%
\right),  \label{HLcenterEmodel13}
\end{equation}
$f(u,v)=\frac{4 \mu ^2 u^3 \left(3 \sqrt{n} \left((1-3 n) v^2+1\right)+4 \mu
(3 n-1) v\right)}{n^{3/2}}$\newline
$+\frac{8 \mu ^4 (1-3 n) u^5}{n^2}+\frac{8 \mu ^3 (3 n-1) u^4 \left(2 \sqrt{n%
} v-\mu \right)}{n^2}$\newline
$+\frac{4 \mu u^2 \left(3 \mu \left((1-3 n) v^2+1\right)+\sqrt{n} v \left((3
n-1) v^2-3\right)\right)}{n}$\newline
$+\frac{u v \left(5 \sqrt{n} v \left((1-3 n) v^2+6\right)-8 \mu \left((5-15
n) v^2+2 v^4+15\right)\right)}{10 \sqrt{n}}$\newline
$+\frac{1}{2} \left((1-3 n) v^4+6 v^2\right)$ and
$g(u,v)=\frac{\mu ^5 (5 n (11 n-38)+39) u^5}{5 n^{5/2}}$ 
\newline$ -\frac{2 \mu ^4 u^4
\left(-4 \mu +18 \mu n+\sqrt{n} (3 n-7) (5 n-1) v\right)}{n^{5/2}}$\newline
$+\frac{\mu ^2 u^2 \left(12 \mu \left((2-9 n) v^2+2\right)+(1-n) \sqrt{n} v
\left((35 n-3) v^2-24\right)\right)}{2 n^{3/2}}$\newline
$+\frac{\mu ^3 u^3 \left(\sqrt{n} \left((n (65 n-114)+17) v^2-12
n+28\right)+16 \mu (9 n-2) v\right)}{2 n^2}$\newline
$+\frac{\mu u v \left(32 \mu \left((9 n-2) v^2-6\right)+\sqrt{n} v \left((n
(75 n-38)-5) v^2-120 n+24\right)\right)}{16 n}$\newline
$+\frac{v^2 \left(5 \mu \left((2-9 n) v^2+12\right)+2 \sqrt{n} v \left(-5 n
\left(n v^2-3\right)+v^2+5\right)\right)}{20 \sqrt{n}}$.
\newline
According to Theorem \ref{existenceCM}, there exists a 1-dimensional
invariant local center manifold of \eqref{HLcenterEmodel13}, \newline
$W^{c}\left( \mathbf{0}\right) =\left\{ \left( u,v\right) \in\mathbb{R}%
^2:v=h\left( u\right)\right\}$, where $h\left(0\right) =0,\;h^{\prime
}\left( 0\right) =0,\;\left\vert u\right\vert <\delta$ for $\delta$
sufficiently small. The restriction of (\ref{HLcenterEmodel13}) to the
center manifold is $\frac{du}{dN}=f\left( u, h\left( u\right) \right)$, such
that the function that defines the center manifold satisfies 
\begin{align}
h^{\prime }\left( u\right) \left[ f\left( u, {h}\left( u\right) \right) %
\right] +3 {h}\left( u\right) -{g}\left( u, {h}\left( u\right) \right) =0.
\label{HLcenterEmodel1h}
\end{align}
According to Theorem \ref{approximationCM}, we can use Taylor series as
follows $h(u)=a_1 u^2+a_2 u^3+a_3 u^4+a_4 u^3+ \mathcal{O}(u)^6,$ to obtain  
		$a_1= \frac{4 \mu ^3}{n^{3/2}}, a_2= -\frac{2 \mu ^3 \left(72 \mu ^2+n (3 n-7)\right)}{3 n^{5/2}}, a_3= \frac{12
   \mu ^5 \left(76 \mu ^2+n (3 n-10)\right)}{n^{7/2}}$,\\
	$a_4= \frac{-338880 \mu ^9+\mu ^5 n^2 \left(-65 n^2+570 n-1081\right)+120 \mu ^7
   (459-121 n) n}{15 n^{9/2}}$. 	\newline
		The center manifold can be expressed, in terms of $\tilde{T}$,  by
		\begin{widetext}
\begin{align}
& \theta=-\frac{2 \mu  (\tilde{T}-1)}{\sqrt{n}}+\frac{4 \mu ^3 (\tilde{T}-1)^2}{n^{3/2}}  +\frac{12 \mu ^5 (\tilde{T}-1)^4 \left(76 \mu ^2+n (3 n-10)\right)}{n^{7/2}}  -\frac{2 \mu ^3 (\tilde{T}-1)^3 \left(72 \mu ^2+n
   (3 n-7)\right)}{3 n^{5/2}}\nonumber\\
	&+\frac{(\tilde{T}-1)^5 \left(-338880 \mu ^9+\mu ^5 n^2 \left(-65 n^2+570 n-1081\right)+120 \mu ^7 (459-121 n) n\right)}{15
   n^{9/2}}.
	\end{align}	
	\end{widetext}
The dynamics on the center manifold can be approximated by the gradient-like equation 
\begin{small}
	\begin{align}
&\frac{d u}{dN}=-\nabla \Pi(u), \notag \\
& \Pi(u)=-\frac{4 \mu
   ^2 u^3}{n}-\frac{3 \mu ^2 u^4 \left(n-4 \mu ^2\right)}{n^2}+\frac{48 \mu ^4 u^5 \left(2 n-13 \mu ^2\right)}{5 n^3} \notag \\
& +\frac{8 \mu ^4 u^6 \left(252 \mu ^4+n^2+2 \mu ^2 n (3 n-22)\right)}{n^4} \nonumber\\
&	+\frac{48 \mu ^6 u^7 \left(-6248 \mu
   ^4+n^2 (12 n-49)+2 \mu ^2 (619-106 n) n\right)}{7 n^5},
		\end{align}
\end{small}
with $\Pi^{\prime \prime \prime }(u)=-\frac{24 \mu ^2}{n}<0$, that is, the
origin is an inflection point. Using the Theorem \ref{stabilityCM}, we
conclude that the center manifold of origin for the system %
\eqref{HLcenterEmodel13}, and therefore $dS_+$ is unstable (center-saddle)
(see \cite{Alho:2015cza}, Section 4). $\blacksquare$

\section{Case 2: Non-flat universe with $\Lambda=0$ under the
detailed-balance condition}

\label{Case2} 

\begin{table*}[ht]
\caption{Case 2: Equilibrium points at the finite region of the system 
\eqref{GenHLeqxcase2}, \eqref{GenHLeqzcase2}, \eqref{GenHLeqscase2}.}
\label{HLcrit2}\centering
\begin{tabular*}{\textwidth}{@{\extracolsep{\fill}}lrrrl}
\hline
Equil. & \multicolumn{1}{c}{$(x,z,s)$} & \multicolumn{1}{c}{Existence} & 
\multicolumn{1}{c}{Eigenvalues} & \multicolumn{1}{c}{Stability} \\ 
Points &  &  &  &  \\ \hline
$P_1(\hat{s})$ & $\left(1, 0, \hat{s}\right)$ & $f(\hat{s})=0$ & $6-2 \sqrt{6%
} \hat{s}, 1, -2 \sqrt{6} f^{\prime }\left(\hat{s}\right)$ & nonhyperbolic
for $f^{\prime }\left(\hat{s}\right)=0$, or $\hat{s}=\sqrt{\frac{3}{2}}$. \\ 
&  &  &  & source for $f^{\prime }\left(\hat{s}\right)<0, \hat{s}<\sqrt{%
\frac{3}{2}}$. \\ 
&  &  &  & saddle otherwise. \\ \hline
$P_2(\hat{s})$ & $\left(-1, 0, \hat{s}\right)$ & $f(\hat{s})=0$ & $6+2 \sqrt{%
6} \hat{s}, 1, 2 \sqrt{6} f^{\prime }\left(\hat{s}\right)$. & nonhyperbolic
for $f^{\prime }\left(\hat{s}\right)=0$, or $\hat{s}=-\sqrt{\frac{3}{2}}$.
\\ 
&  &  &  & source for $f^{\prime }\left(\hat{s}\right)>0, \hat{s}>-\sqrt{%
\frac{3}{2}}$ \\ 
&  &  &  & saddle otherwise. \\ \hline
$P_3(\hat{s})$ & $\left(\sqrt{\frac{2}{3}} \hat{s}, 0, \hat{s}\right)$ & $f(%
\hat{s})=0$ & $2 \hat{s} ^2-3, 2 \left(\hat{s}^2-1\right), -4 \hat{s}
f^{\prime }\left(\hat{s}\right)$ & nonhyperbolic for$f^{\prime }\left(\hat{s}%
\right)=0$, or $\hat{s}\in \left\{-\sqrt{\frac{3}{2}},-1, 0, 1, \sqrt{\frac{3%
}{2}}\right\}$. \\ 
&  & $-\sqrt{\frac{3}{2}}\leq \hat{s}\leq \sqrt{\frac{3}{2}}$ &  & sink for $%
f^{\prime }\left(\hat{s}\right)<0, -1<\hat{s}<0$, or $f^{\prime }\left(\hat{s%
}\right)>0, 0<\hat{s}<1$. \\ 
&  &  &  & saddle otherwise. \\ \hline
$P_3^{0}:$ & $(0, 0, 0)$ & always & $-2, -\frac{3}{2}\pm \frac{1}{2} \sqrt{%
9-48 f(0)}$ & nonhyperbolic for $f(0)=0$. \\ 
&  &  &  & sink for $f(0)>0$. \\ 
&  &  &  & saddle otherwise. \\ \hline
$P_{5,6}(\hat{s})$ & $\left(\frac{\sqrt{\frac{2}{3}}}{\hat{s}}, \pm \frac{%
\sqrt{1-\hat{s}^2}}{\hat{ s}}, \hat{s}\right)$ & $f(\hat{s})=0$ & $-\frac{1}{%
2}\pm \frac{\sqrt{16 \hat{s}^2-15 \hat{s}^4}}{2 \hat{s}^2}, -\frac{4
f^{\prime }\left(\hat{s}\right)}{\hat{s}}$ & nonhyperbolic for $f^{\prime
}\left(\hat{s}\right)=0$, or $\hat{s}\in \left\{-1, 1\right\}$. \\ 
&  & $-1\leq \hat{s}\leq 1, \hat{s}\neq 0$ &  & sink for $f^{\prime }\left(%
\hat{s}\right)<0, \hat{s}<-1$ or $f^{\prime }\left(\hat{s}\right)>0, \hat{s}%
>1$. \\ 
&  &  &  & saddle otherwise. \\ \hline
\end{tabular*}
\caption{Case 2: Equilibrium points at the infinity region of the system 
\eqref{GenHLeqxcase2}, \eqref{GenHLeqzcase2}, \eqref{GenHLeqscase2}.}
\label{HLcrit2infinity}
\begin{tabular*}{\textwidth}{@{\extracolsep{\fill}}lrrrl}
\hline
Equil. & \multicolumn{1}{c}{$(x,Z,S)$} & \multicolumn{1}{c}{Existence} & 
\multicolumn{1}{c}{Eigenvalues} & \multicolumn{1}{c}{Stability} \\ 
Points &  &  &  &  \\ \hline
$Q_{9,10}(\hat{s})$ & $\left(-1, \pm 1, \frac{2}{\pi}\arctan(\hat{s})\right)$
& $f(\hat{s})=0$ & $-\infty, 2 \sqrt{6} \hat{s}, 2 \sqrt{6} f^{\prime }\left(%
\hat{s}\right)$ & sinks for $\hat{s}<0, f^{\prime }\left(\hat{ s}\right)<0$
\\ \hline
$Q_{11,12}(\hat{s})$ & $\left(1, \pm 1, \frac{2}{\pi}\arctan(\hat{s})\right)$
& $f(\hat{s})=0$ & $-\infty, -2 \sqrt{6} \hat{s}, -2 \sqrt{6} f^{\prime
}\left(\hat{s}\right)$ & sinks for $\hat{s}>0, f^{\prime }\left(\hat{s}%
\right)>0$. \\ \hline
$Q_{13,14}$ & $\left(0, \pm 1, 0\right)$ & always & $\infty ,-2 \sqrt{3} 
\sqrt{-f(0)},2 \sqrt{3} \sqrt{-f(0)}$ & Saddle for $f(0)<0$. Center for $%
f(0)>0$. \\ \hline
$Q_{15,16}$ & $\left(-\sqrt{\frac{3}{2}}, \pm 1, 0\right)$ & $f(0)=0$ & $0,
0, 4 f^{\prime }(0)$ & nonhyperbolic. \\ \hline
$Q_{17,18}$ & $\left(\sqrt{\frac{3}{2}}, \pm 1, 0\right)$ & $f(0)=0$ & $0,
0, -4 f^{\prime }(0)$ & nonhyperbolic. \\ \hline
\end{tabular*}%
\end{table*}

For this case the equations are: 
\begin{align}  
& H^2= \frac{1}{6(3\lambda-1)}\left[\frac{3 \lambda -1}{4}\dot\phi^2+V(\phi)%
\right]  -\frac{\mu^2 k^2}{16(3\lambda-1)^2a^4}, \label{Case2syst7a}\\
&8 \dot H+ \dot\phi^2=\frac{k^2 \mu ^2}{(3 \lambda -1)^2 a^4}, \label{Case2systb}\\
& \ddot\phi +3 H \dot \phi+\frac{2 V^{\prime }(\phi)}{3 \lambda -1}=0. \label{Case2systc}
\end{align}

\subsection{Arbitrary Potential}
\label{Case2A}
In this example the system  \eqref{Case2syst7a}, \eqref{Case2systb}, \eqref{Case2systc}, is reduced to the autonomous form:  
\begin{align}
&\frac{dx}{dN}= x \left(3 x^2-2 z^2-3\right)+\sqrt{6} s
\left(1-x^2+z^2\right),  \label{GenHLeqxcase2} \\
&\frac{dz}{dN}= z \left[3 x^2-2 \left(z^2+1\right)\right],
\label{GenHLeqzcase2} \\
&\frac{ds}{dN}= -2 \sqrt{6} x f(s).  \label{GenHLeqscase2}
\end{align}
defined on the phase space $\{(x,z,s)\in\mathbb{R}^3: x^2-z^2 \leq 1\}.$

The equilibrium points/curves of the system \eqref{GenHLeqxcase2}, \eqref{GenHLeqzcase2}, \eqref{GenHLeqscase2}, at the finite region of the phase space is
presented in Table \ref{HLcrit2}. Now we discuss the more relevant features
of them.

\begin{itemize}
\item $P_1(\hat{s}): (x,z,s)=\left(1, 0, \hat{s}\right)$. Always exists. It
is a source for $f^{\prime }\left(\hat{s}\right)<0, \hat{s}<\sqrt{\frac{3}{2}%
}$.

\item $P_2(\hat{s}): (x,z,s)=\left(-1, 0, \hat{s}\right)$. Always exists. It
is a source for $f^{\prime }\left(\hat{s}\right)>0, \hat{s}>-\sqrt{\frac{3}{2%
}}$.

\item $P_3(\hat{s}): (x,z,s)=\left(\sqrt{\frac{2}{3}} \hat{s}, 0, \hat{s}%
\right)$. Exists for $-\sqrt{\frac{3}{2}}\leq \hat{s}\leq \sqrt{\frac{3}{2}}$%
. It is a sink for  $f^{\prime }\left(\hat{s}\right)<0, -1<\hat{s}<0$, or  $%
f^{\prime }\left(\hat{s}\right)>0, 0<\hat{s}<1$.

\item $P_3^{0}: (x,z,s)=(0, 0, 0)$. It is a sink for $f(0)>0$.

\item $P_{5,6}(\hat{s}): \left(\frac{\sqrt{\frac{2}{3}}}{\hat{s}}, \pm \frac{%
\sqrt{1-\hat{s}^2}}{\hat{s}}, \hat{s}\right)$. Exists for $-1\leq \hat{s}%
\leq 1, \hat{s}\neq 0$. It is a sink for $f^{\prime }\left(\hat{s}\right)<0, 
\hat{s}<-1$ or $f^{\prime }\left(\hat{s}\right)>0, \hat{s}>1$.

\item There are two lines of equilibrium points $P_{7,8}(s_c):
(x,z,s)=(0,\pm i, s_c), s_c\in \mathbb{R}$ which are not considered since
they are complex valued.
\end{itemize}

Owing to the fact that the dynamical system \eqref{GenHLeqxcase2}, %
\eqref{GenHLeqzcase2}, \eqref{GenHLeqscase2} is unbounded, we introduce the
new variables 
\begin{equation}
X= \frac{x}{\sqrt{1+z^2}}, \quad Z=\frac{z}{\sqrt{1+z^2}}, \quad S=\frac{2}{%
\pi}\arctan(s),
\end{equation}
and the time rescaling 
\begin{equation}
\frac{d f}{d\tau}= \sqrt{1-Z^2} \frac{d f}{dN},
\end{equation}
to obtain the dynamical system 
\begin{align}
&\frac{dX}{d\tau}=\left(X^2-1\right) \left(3 X \sqrt{1-Z^2}-\sqrt{6}\,
\tan\left(\frac{\pi S}{2}\right)\right), \label{Example2Sa}\\
&\frac{dZ}{d\tau}=\left(3 X^2-2\right) Z \sqrt{1-Z^2}, \label{Example2Sb}\\
&\frac{dS}{d\tau}=-\frac{2 \sqrt{6} X (\cos (\pi S)+1) f\left(\tan \left(%
\frac{\pi S}{2}\right)\right)}{\pi },  \label{Example2Sc}
\end{align}
defined on the compacted phase space 
\begin{equation*}
\{(X,Z,S)\in\mathbb{R}^3: -1\leq X \leq 1, -1\leq Z \leq 1, -1\leq S \leq
1\}. 
\end{equation*}
The equilibrium points at the infinity region of the system %
\eqref{GenHLeqxcase2}, \eqref{GenHLeqzcase2}, \eqref{GenHLeqscase2} are
summarized in Table \ref{HLcrit2infinity}. Here we discuss the main features
of them.

\begin{itemize}
\item $Q_{9,10}(\hat{s}): (X,Z,S)=\left(-1, \pm 1, \frac{2}{\pi}\arctan(\hat{%
s})\right)$, where $f(\hat{s})=0$. They are sinks for $\hat{s}<0, f^{\prime
}\left(\hat{s}\right)<0$

\item $Q_{11,12}(\hat{s}): (X,Z,S)=\left(1, \pm 1, \frac{2}{\pi}\arctan(\hat{%
s})\right)$, where $f(\hat{s})=0$. They are sinks for $\hat{s}>0, f^{\prime
}\left(\hat{s}\right)>0$.

\item $Q_{13,14}: (X,Z,S)=\left(0, \pm 1, 0\right)$. Saddle for $f(0)<0$.
Center if $f(0)>0$.

\item $Q_{15,16}: (X,Z,S)=\left(-\sqrt{\frac{3}{2}}, \pm 1, 0\right)$ ,
where $f(0)=0$. It is nonhyperbolic.

\item $Q_{17,18}: (X,Z,S)=\left(\sqrt{\frac{3}{2}}, \pm 1, 0\right)$ , where 
$f(0)=0$. It is nonhyperbolic.
\end{itemize}

\subsection{Exponential Potential}
\label{Case2B}

\begin{figure*}[t!]
\centering
\includegraphics[scale=1.5]{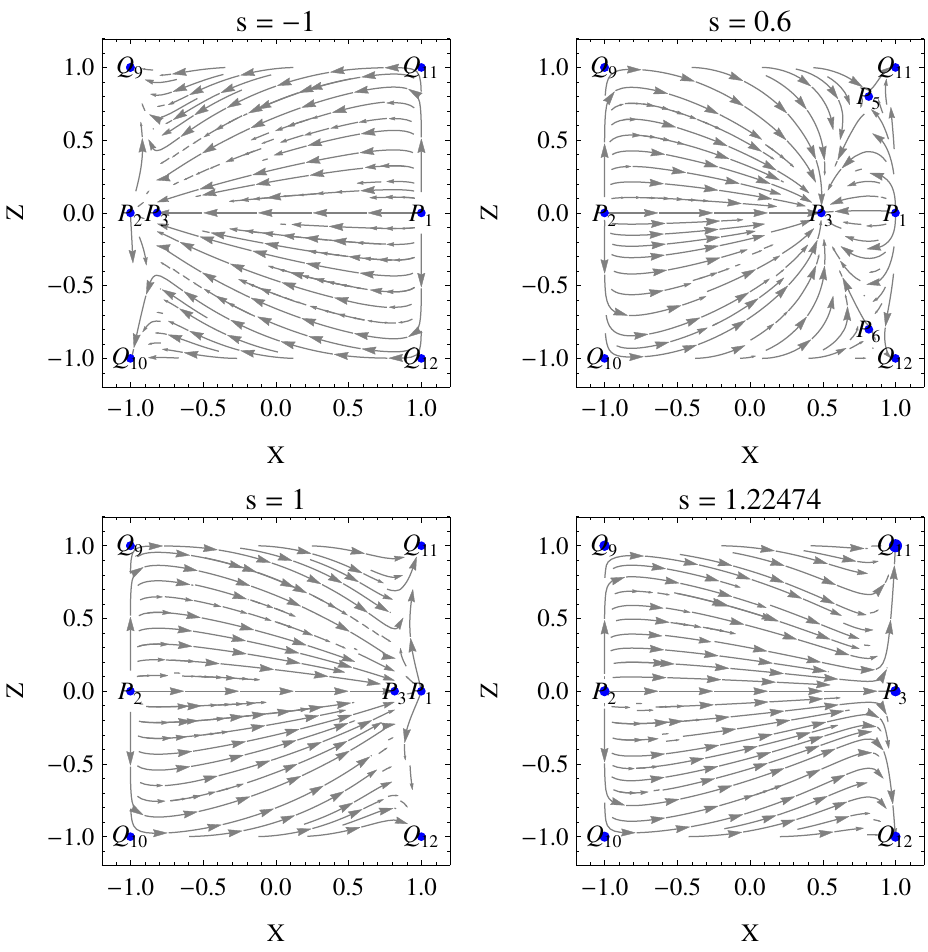} 
\caption{\textit{{(Color online) Compact phase portrait of \eqref{HLinfty2a}, \eqref{HLinfty2b}
for different choices of the parameter $s$.}}}
\label{fig:Case2HL}
\end{figure*}

In this case the system \eqref{Example2Sa}, \eqref{Example2Sb}, \eqref{Example2Sc}, becomes 
\begin{align}
&\frac{dX}{d\tau}=\left(X^2-1\right) \left(3 X \sqrt{1-Z^2}-\sqrt{6}
s\right),  \label{HLinfty2a} \\
&\frac{dZ}{d\tau}=\left(3 X^2-2\right) Z \sqrt{1-Z^2},  \label{HLinfty2b}
\end{align}
defined on the compact phase space\newline
$\{(X,Z)\in\mathbb{R}^2: -1\leq X \leq 1, -1\leq Z \leq 1\}.$

The equilibrium points $P_{1,2,3}$ are exactly the same as in case 1, and
thus the physical implications are the same. The equilibrium points $P_{5,6}$
are unstable, corresponding to a dark-matter dominated universe. This was
expected since in the absence of the cosmological constant $\Lambda$, the
curvature role is downgrading as the scale factor increases and thus in the
end this case tends to the case 1 above. Note however that at early times,
where the scale factor is small, the behavior of the system will be
significantly different than case 1, with the dark energy playing an
important role. Figure \ref{fig:Case2HL} a) illustrates when $Q_{9,10}:
(X,Z)=\left(-1, \pm 1\right)$ are sinks for $s<0$. Figures \ref{fig:Case2HL}
b), c) and d) illustrates when $Q_{11,12}: (X,Z)=\left(1, \pm 1\right)$ are
sinks for $s>0$.

\subsection{Powerlaw potential}

\label{Case2C}

\begin{figure*}[ht!]
\centering
\subfigure[]{\includegraphics[width=2.5in, height=2.5in,angle=-90]{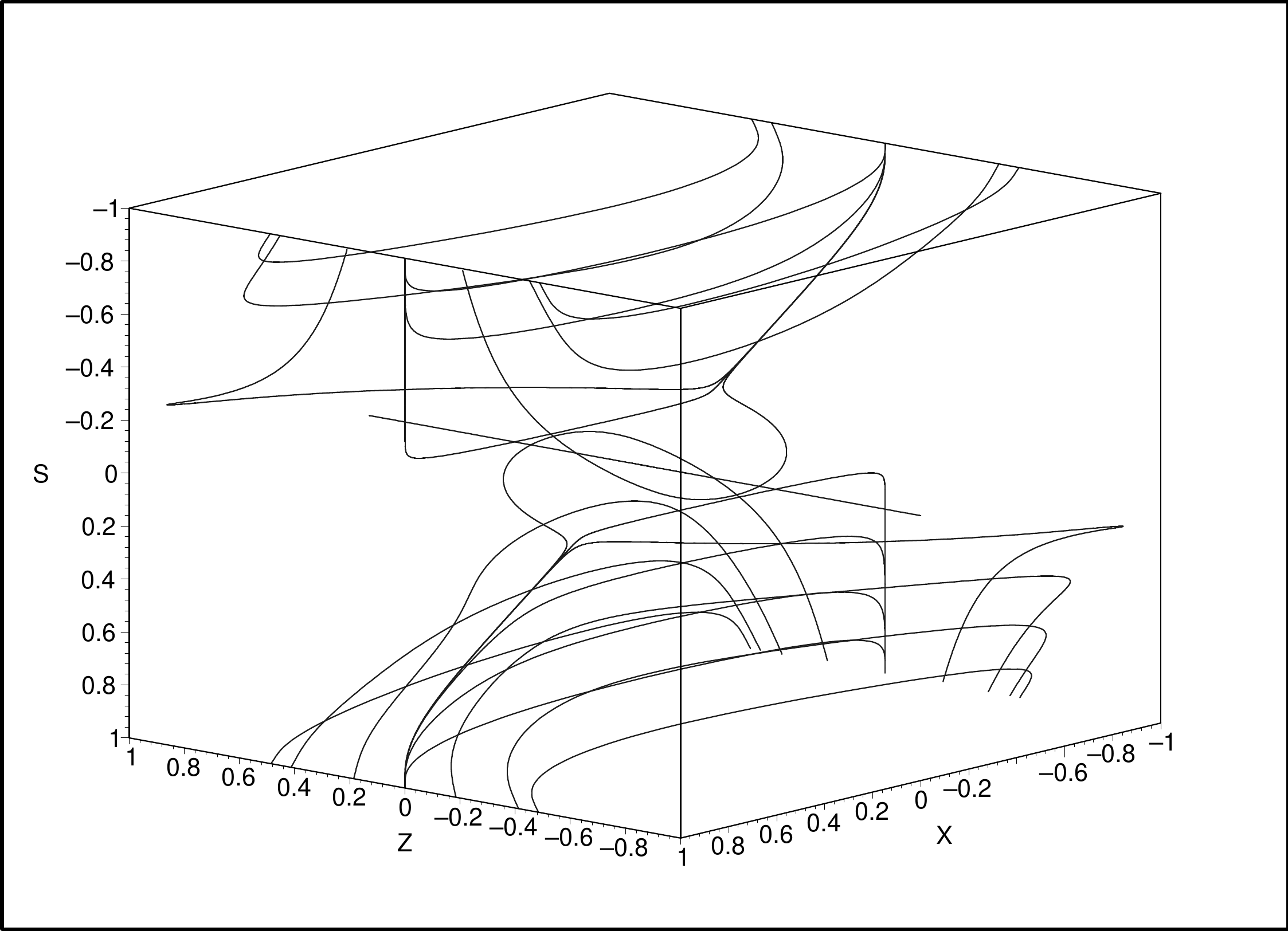}} 
\hspace{2cm}  
\subfigure[]{
					\includegraphics[width=2.5in, height=2.5in,angle=-90]{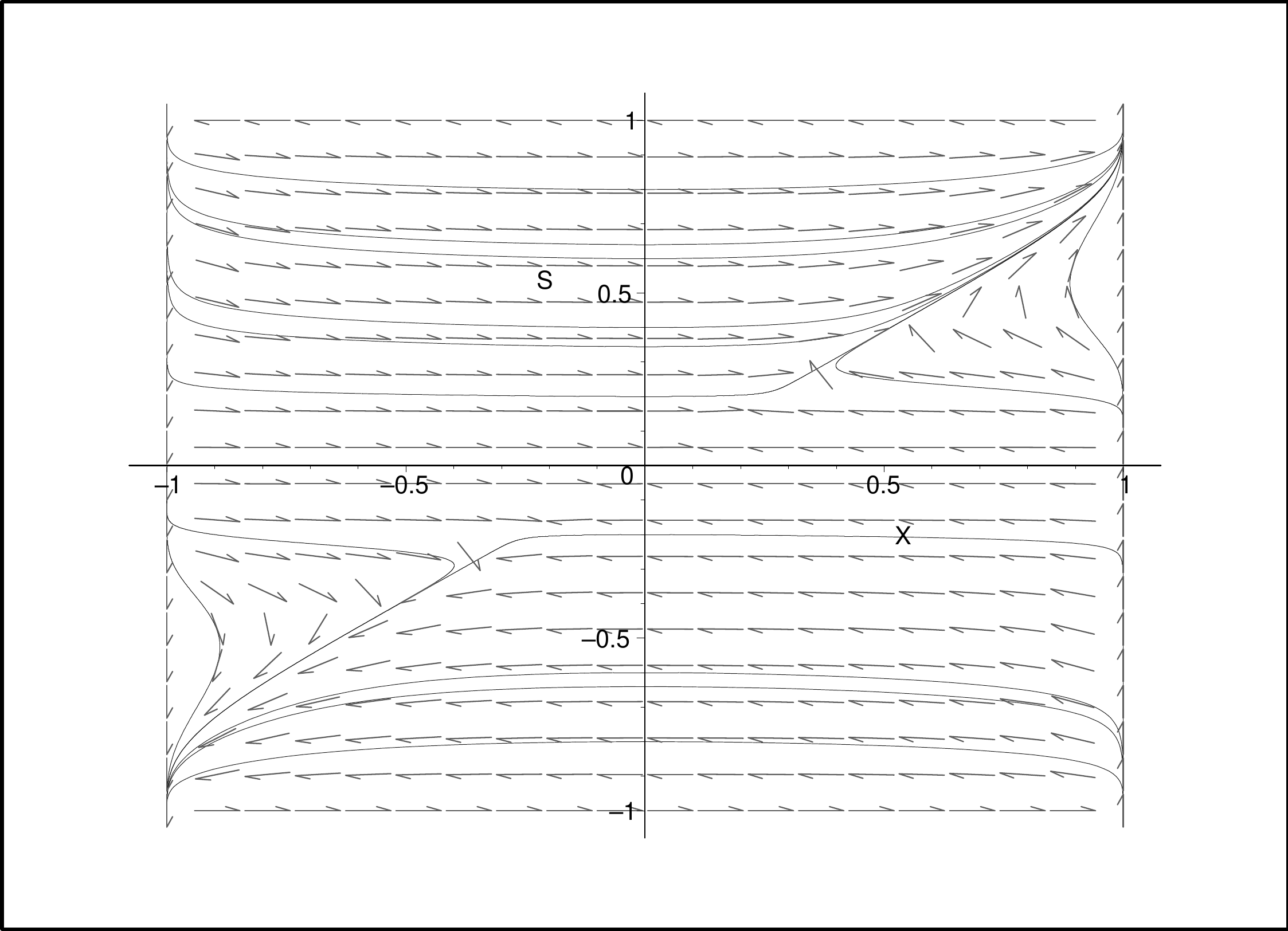}}  
\subfigure[]{
					\includegraphics[width=2.5in, height=2.5in,angle=-90]{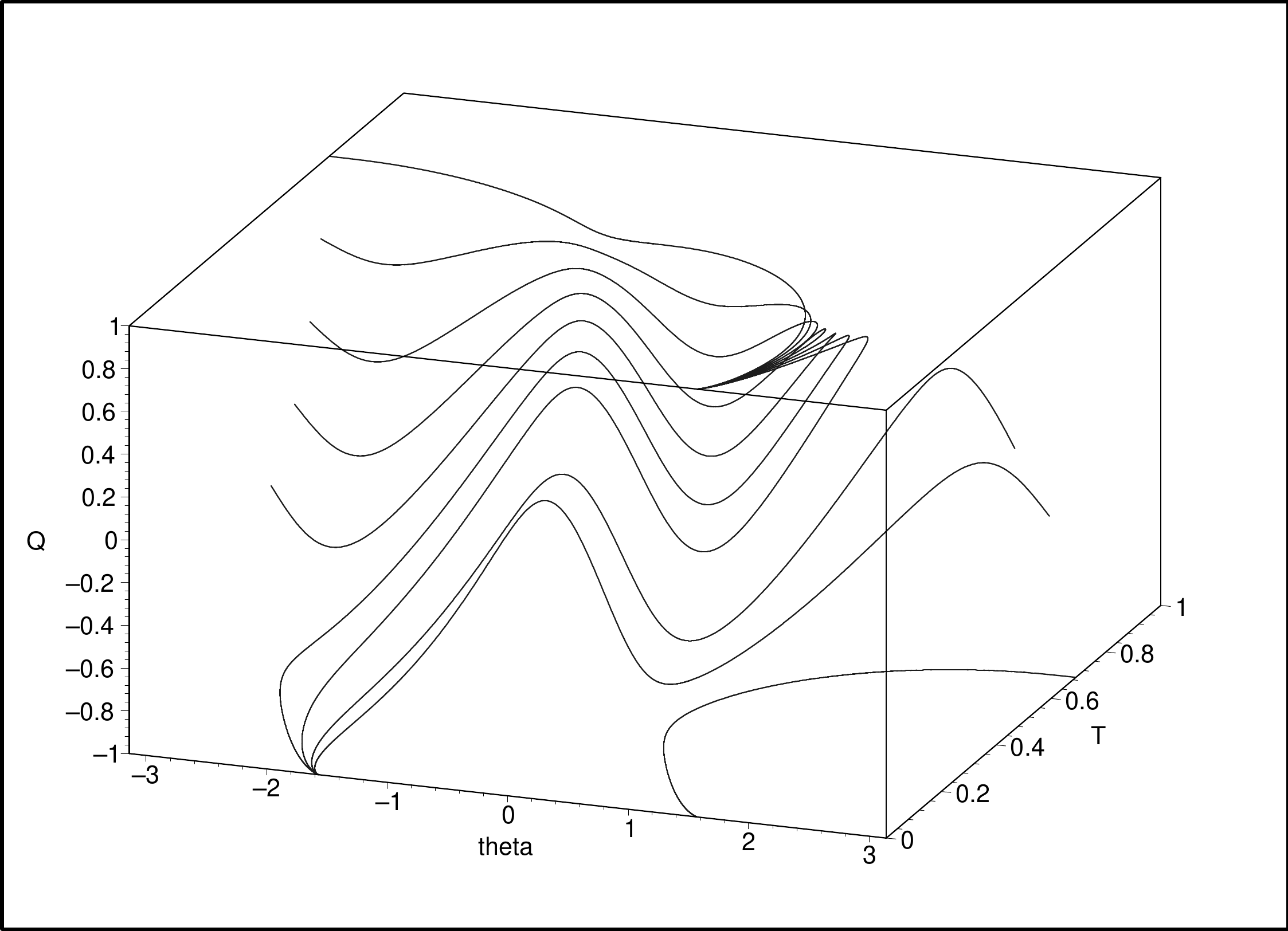}} \hspace{%
2cm}  
\subfigure[]{
					\includegraphics[width=2.5in, height=2.5in,angle=-90]{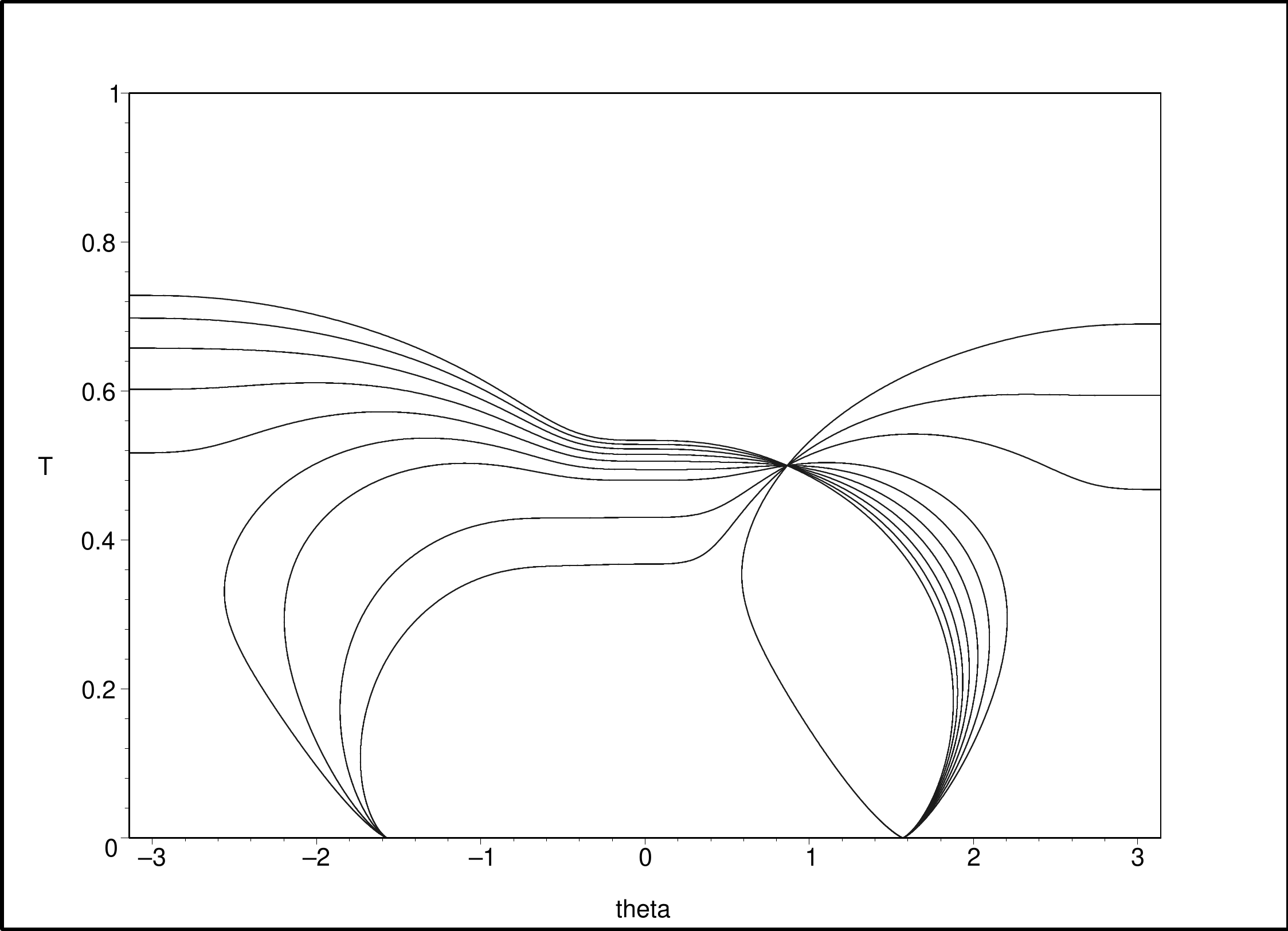}} \hspace{%
2cm}  
\subfigure[]{
					\includegraphics[width=2.5in, height=2.5in,angle=-90]{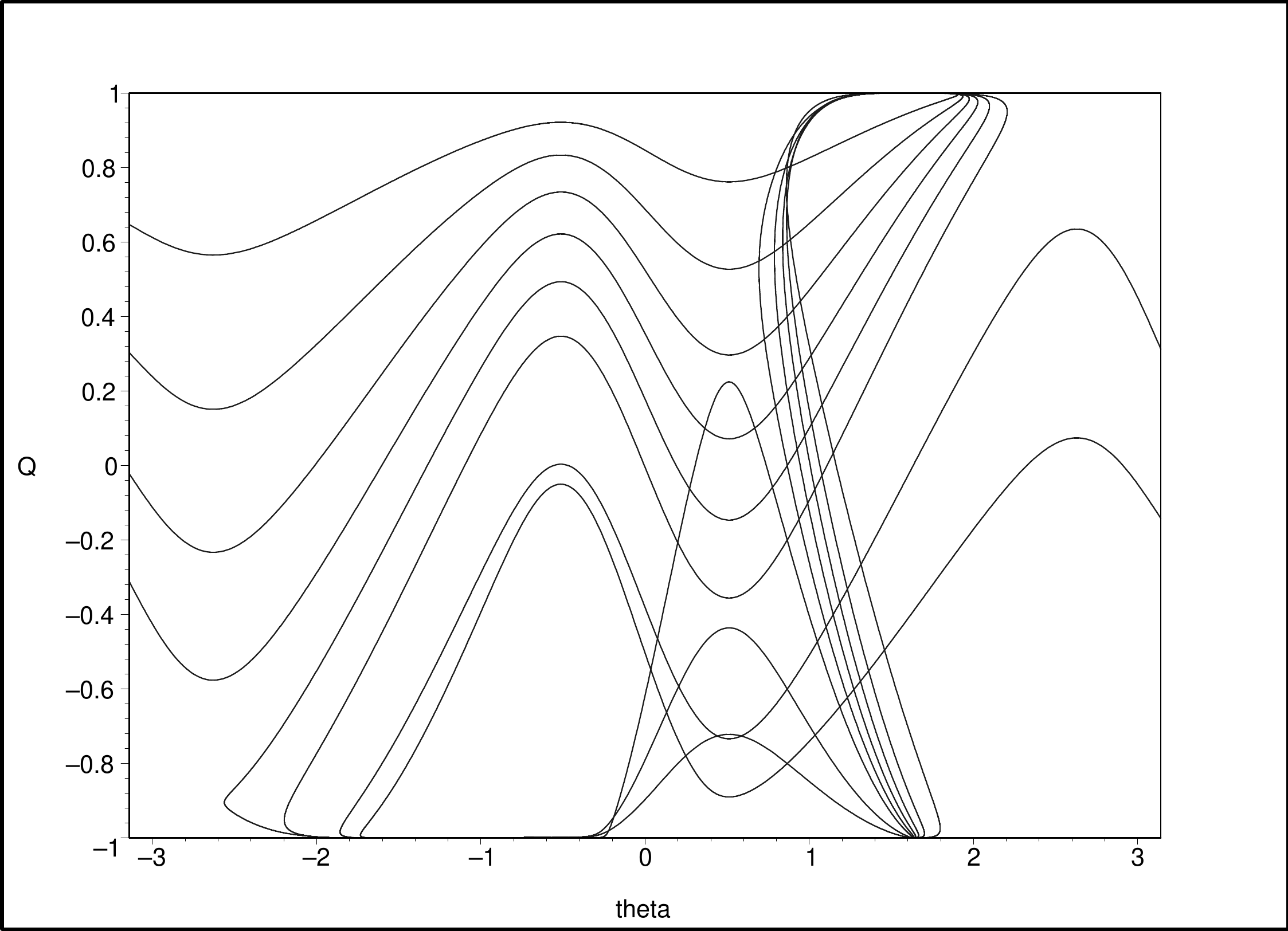}} \hspace{%
2cm}  
\subfigure[]{
					\includegraphics[width=2.5in, height=2.5in,angle=-90]{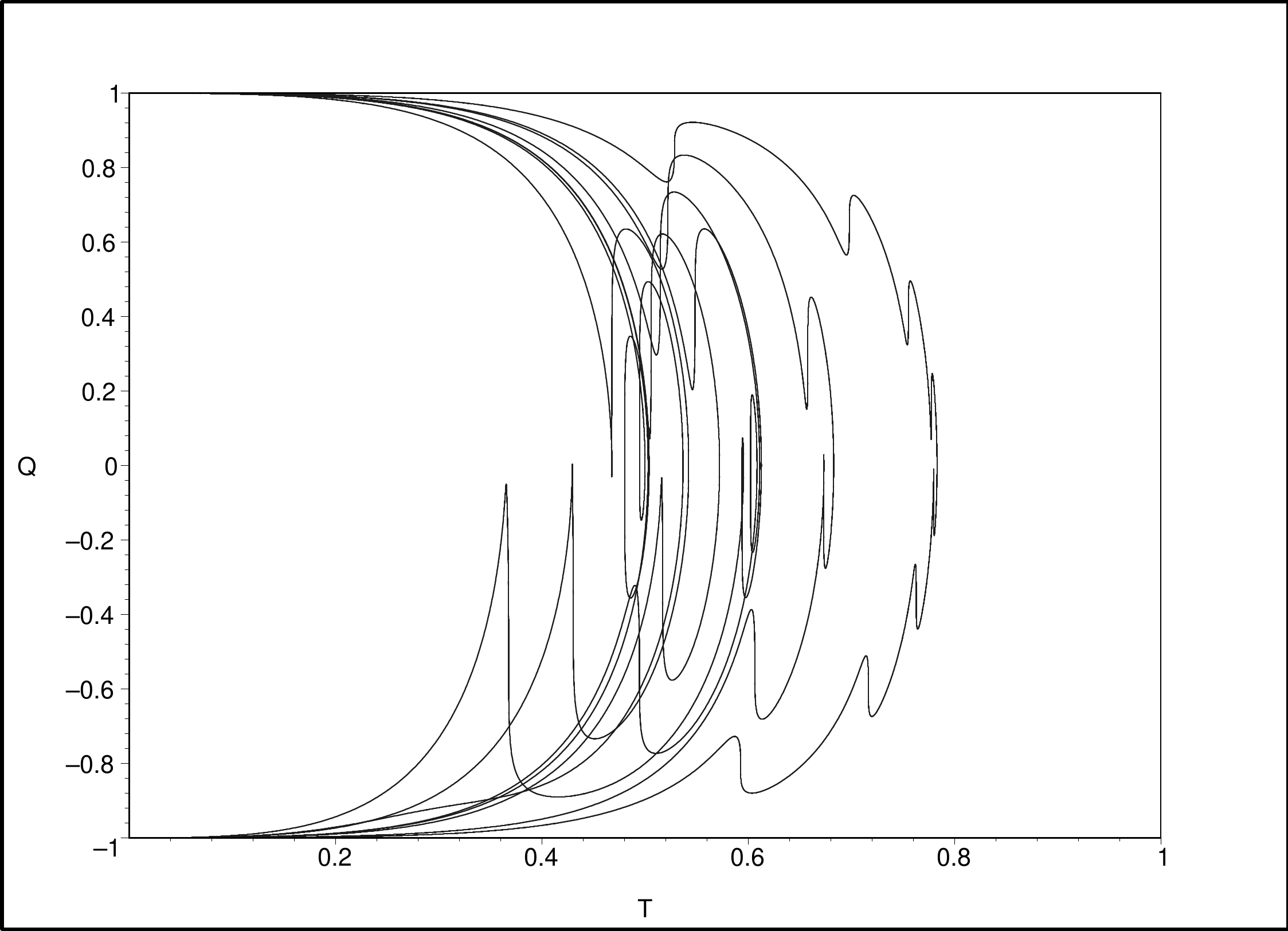}} 
\caption{\textit{{(a) Compact phase portrait of \eqref{SYSTEM81a}, \eqref{SYSTEM81b}, \eqref{SYSTEM81c} for the
choice $n=2$. (b) Dynamics in the invariant set $Z=0$. (c)-(f). Dynamics of the system 
\eqref{Case2_uggla-varsa}, \eqref{Case2_uggla-varsb}, \eqref{Case2_uggla-varsc} and some 2D projections for the choice $n=4$.}}}
\label{fig:Case2HLPL}
\end{figure*}

In this example, the system \eqref{Example2Sa}, \eqref{Example2Sb}, \eqref{Example2Sc}, becomes 
\begin{align}
&\frac{dX}{d\tau}=\left(X^2-1\right) \left(3 X \sqrt{1-Z^2}-\sqrt{6}\,
\tan\left(\frac{\pi S}{2}\right)\right), \label{SYSTEM81a}\\
&\frac{dZ}{d\tau}=\left(3 X^2-2\right) Z \sqrt{1-Z^2}, \label{SYSTEM81b}\\
&\frac{dS}{d\tau}=-\frac{\sqrt{6} X (\cos (\pi S)-1)}{\pi n}, \label{SYSTEM81c}
\end{align}
defined on the compact phase space 

\begin{equation*}
\Big\{(X,Z,S)\in\mathbb{R}^3: -1\leq X \leq 1, -1\leq Z \leq 1, -1\leq S
\leq 1\Big\}.
\end{equation*}

The equilibrium points of \eqref{SYSTEM81a}, \eqref{SYSTEM81b}, \eqref{SYSTEM81c} and their stability conditions
are summarized as follows.

\begin{itemize}
\item $(X,Z,S)=(-1,-1, 0)$. The eigenvalues are $\{0,-\infty ,0\}$.
Nonhyperbolic.

\item $(X,Z,S)=(-1, 1, 0)$. The eigenvalues are $\{0,-\infty ,0\}$.
Nonhyperbolic.

\item $(X,Z,S)=(0, -1, 0)$. The eigenvalues are $\{0,0,\infty \}$.
Nonhyperbolic.

\item $(X,Z,S)=(0, 0, 0)$. The eigenvalues are $\{0,-3,-2\}$. Nonhyperbolic.

\item $(X,Z,S)=(0, 1, 0)$. The eigenvalues are $\{0,0,\infty \}$.
Nonhyperbolic.

\item $(X,Z,S)=(1, -1, 0)$. The eigenvalues are $\{0,-\infty ,0\}$.
Nonhyperbolic.

\item $(X,Z,S)=(1, 1, 0)$. The eigenvalues are $\{0,-\infty ,0\}$.
Nonhyperbolic.

\item $(X,Z,S)=\left(-\sqrt{\frac{2}{3}}, -1, 0\right)$. The eigenvalues are 
$\{0,-2,2\}$. Nonhyperbolic. Behaves as saddle.

\item $(X,Z,S)=\left(\sqrt{\frac{2}{3}}, -1, 0\right)$. The eigenvalues are $%
\{0,-2,2\}$. Nonhyperbolic. Behaves as saddle.

\item $(X,Z,S)=(-1, 0, 0)$. The eigenvalues are $\{0,1,6\}$. Nonhyperbolic.
Unstable.

\item $(X,Z,S)=(1, 0, 0)$. The eigenvalues are $\{0,1,6\}$. Nonhyperbolic.
Unstable.

\item $(X,Z,S)=\left(-\sqrt{\frac{2}{3}}, 1, 0\right)$. The eigenvalues are $%
\{0,-2,2\}$. Nonhyperbolic. Behaves as saddle.

\item $(X,Z,S)=\left(\sqrt{\frac{2}{3}}, 1, 0\right)$. The eigenvalues are $%
\{0,-2,2\}$. Nonhyperbolic. Behaves as saddle.
\end{itemize}

\subsubsection{Stability Analysis of the solution $P_{3}^{0}:(X,Z,S)=(0,0,0)$%
, in Ho\v{r}ava-Lifshitz cosmology for the Non-flat universe with $\Lambda =
0$ under the detailed-balance condition the powerlaw- potential.}

\label{PROPOSITION5}

\begin{prop}
\label{centerP3PLCase2} The origin for the system \eqref{SYSTEM81a}, \eqref{SYSTEM81b}, \eqref{SYSTEM81c} is
unstable (saddle point).
\end{prop}

\textbf{Proof.} Taking the linear transformation 
\begin{equation}
(u,v_1,v_2)=\left(S, \frac{1}{6} \left(6 X-\sqrt{6} \pi S\right), Z\right),
\end{equation}
and taking Taylor series near $(u,v_1,v_2)=(0,0,0)$ up to fifth order we
obtain the system \eqref{SYSTEM81a}, \eqref{SYSTEM81b}, \eqref{SYSTEM81c} can be written into its Jordan canonical
form: %
\begin{equation}
\left(%
\begin{array}{c}
\frac{du}{dN} \\ 
\frac{dv_1}{dN} \\ 
\frac{dv_2}{dN}%
\end{array}%
\right)=\left(%
\begin{array}{ccc}
0 & 0 & 0 \\ 
0 & -3 & 0 \\ 
0 & 0 & -2%
\end{array}%
\right)\left(%
\begin{array}{c}
u \\ 
v_1 \\ 
v_2%
\end{array}%
\right)+\left(%
\begin{array}{c}
f(u,\mathbf{v}) \\ 
g_1(u,\mathbf{v}) \\ 
g_2(u,\mathbf{v})%
\end{array}%
\right),  \label{HLPLS81P3}
\end{equation}%
 $f(u,\mathbf{v})=-\frac{\pi u^2 \left(\pi ^2 u^2-12\right) \left(\pi u+%
\sqrt{6} v_1\right)}{24 n},$\newline 
$g_1(u,\mathbf{v})=\frac{\pi ^5 (5-2 n) u^5}{%
120 \sqrt{6} n}+\frac{\pi ^4 (1-2 n) u^4 v_1}{24 n}-\frac{\pi ^3 u^3 \left(n
\left(v_1^2+v_2^2-1\right)+2\right)}{4 \sqrt{6} n}-\frac{\pi ^2 u^2 v_1
\left(n \left(3 v_2^2-2\right)+2\right)}{4 n}$
$+\frac{3}{8} v_1 \left(-4  \left(v_1^2-1\right) v_2^2+8 v_1^2+v_2^4\right)$%
\newline
$+\frac{1}{8} \sqrt{\frac{3}{2}} \pi u \left(v_1^2 \left(16-3 v_2^2
\left(v_2^2+4\right)\right)+v_2^4+4 v_2^2\right),$\newline  
$g_2(u,\mathbf{v})=-\frac{1}{4} \pi ^2 u^2 \left(v_2^2-2\right) v_2-\sqrt{%
\frac{3}{2}} \pi u v_1 \left(v_2^2-2\right) v_2$\newline
$-\frac{3}{2}  v_1^2 \left(v_2^2-2\right) v_2+\frac{v_2^5}{4}+v_2^3.$ 

According to Theorem \ref{existenceCM}, there exists a 1-dimensional
invariant local center manifold of \eqref{HLPLS81P3}, \newline
$W^{c}\left( \mathbf{0}\right) =\left\{ \left( u,\mathbf{v}\right) \in%
\mathbb{R}\times\mathbb{R}^{2}:\mathbf{v}=\mathbf{h}\left( u\right)\right\}$%
, satisfying $\mathbf{h}\left( 0\right) =\mathbf{0},\;D\mathbf{h}\left(
0\right) =\mathbf{0} ,\;\left\vert u\right\vert <\delta$ for $\delta$
sufficiently small. The restriction of (\ref{HLPLS81P3}) to the center
manifold is $\frac{du}{dN}=f\left( u,\mathbf{h}\left( u\right) \right)$,
where the function $\mathbf{h}\left( u\right)$ satisfies \eqref{MaineqcM}:%
\begin{equation}
D\mathbf{h}\left( u\right) \left[ f\left( u,\mathbf{h}\left( u\right)
\right) \right] -P\mathbf{h}\left( u\right) -\mathbf{g}\left( u,\mathbf{h}%
\left( u\right) \right) =0,  \label{HLPLS81P3h}
\end{equation}
where 
\begin{equation*}
P=\left(%
\begin{array}{cc}
-3 & 0 \\ 
0 & -2%
\end{array}%
\right).
\end{equation*}
According to Theorem \ref{approximationCM}, the system \eqref{HLPLS81P3h}
can be solved approximately by expanding $\mathbf{h}\left( u\right) $ in
Taylor series at $u=0.$ Since $\mathbf{h}\left( 0\right) =\mathbf{0\ } $and $%
D\mathbf{h}\left( 0\right) =\mathbf{0},$ we propose the ansatsz 
\begin{equation}
\mathbf{h}\left( u\right) :=\left(%
\begin{array}{c}
h_{1}\left(u\right) \\ 
h_{2}\left(u\right)%
\end{array}%
\right)=\left(%
\begin{array}{c}
\sum_{j=1}^{4} a_j u^{j+1} +O\left( u^{6}\right) \\ 
\sum_{j=1}^{4} b_j u^{j+1} +O\left( u^{6}\right)%
\end{array}%
\right),
\end{equation}%
to find the non-null coefficients\newline $a_2= \frac{\pi ^3 (n-2)}{12 \sqrt{6} n}
$, $a_4= \frac{\pi ^5 (n (3 n-25)+40)}{360 \sqrt{6} n^2}.$\newline Therefore, the
center manifold can be represented locally by 
\begin{equation}
v_1=\frac{(n-2) \pi ^3 u^3}{12 \sqrt{6} n}+\frac{(n (3 n-25)+40) \pi ^5 u^5}{%
360 \sqrt{6} n^2}, v_2=0.
\end{equation}
That is, 
\begin{equation}
X=\frac{\pi S}{\sqrt{6}}+\frac{\pi ^3 (n-2) S^3}{12 \sqrt{6} n}+\frac{\pi ^5
(n (3 n-25)+40) S^5}{360 \sqrt{6} n^2}, Z=0.
\end{equation}
The dynamics on the center manifold is given by the gradient-like equation 
\begin{align}
&\frac{du}{dN}=-\nabla \Pi(u),  \notag \\
& \Pi(u)=-\frac{\pi ^2 u^4}{8 n}+\frac{\pi ^4 u^6}{72 n^2}-\frac{\left(\pi
^6 ((n-40) n+80)\right) u^8}{11520 n^3}.
\end{align}
We have $\Pi^{\prime }(0)=\Pi^{\prime \prime }(0)=\Pi^{\prime \prime \prime
(4)}(0)=-\frac{3 \pi ^2}{n}<0$. It follows that $u=0$ is a degenerated
maximum of the potential. Using the Theorem \ref{stabilityCM}, we conclude
that the center manifold of origin for the system \eqref{HLPLS81P3}, and the
origin itself are unstable (saddle point). $\blacksquare$

\subsubsection{Alternative compactification}

In this example we can alternatively introduce the following
compactification 

\begin{align}
& \bar{H}=\sqrt{H^2+\frac{\mu^2 k^2}{16(3\lambda-1)^2a^4}}, \\
& Q=\frac{H}{\bar{H}}, \\
& \Sigma=\frac{\dot\phi}{2 \sqrt{6} \bar{H}}, \\
& Y=\frac{\mu \phi}{2^{\frac{1}{n}} 3^{\frac{1}{2 n}} ((3 \lambda -1) n)^{%
\frac{1}{2 n}} \bar{H}^{\frac{1}{n}}}, \\
& T= \frac{c}{c+\bar{H}^{\frac{1}{n}}}, \quad c=2^{\frac{3}{2}-\frac{1}{n}}
3^{\frac{n-1}{2 n}} n^{-\frac{1}{2 n}} (3 \lambda -1)^{-\frac{1}{2 n}}\mu, \\
& \Sigma+ Y^{2n}=1, \\
& \frac{d\bar{\tau}}{d t}=\bar{H}(1-T)^{-1},
\end{align}
such that 
\begin{align}
&\dot\phi = \frac{2 \sqrt{6} \Sigma c^n \left(\frac{1}{ {T}}-1\right)^n}{Q},
\\
&\phi=\frac{2^{\frac{1}{n}} 3^{\frac{1}{2 n}} n^{\frac{1}{2 n}} \left(\frac{1%
}{ {T}}-1\right) Y (3 \lambda -1)^{\frac{1}{2 n}} Q^{-1/n} c}{\mu }, \\
& a= \frac{\sqrt{\mu } \sqrt{Q} c^{-n/2} \left(\frac{1}{ {T}}-1\right)^{-n/2}%
}{2 \sqrt{3 \lambda -1} \sqrt[4]{1-Q^2}}, \\
&\bar{H}=c^n \left(\frac{1}{ {T}}-1\right)^n.
\end{align}
Therefore, we obtain the dynamical system 
\begin{align}
&\frac{d\Sigma}{d\bar{\tau}}=-n {T} Y^{2 n-1}-Q \Sigma( {T}-1) \left((q-1)
Q^2-1\right), \\
&\frac{dY}{d\bar{\tau}}=\Sigma {T}-\frac{Q ( {T}-1) Y \left((q-1)
Q^2+2\right)}{n}, \\
&\frac{d {T}}{d\bar{\tau}}=\frac{Q ( {T}-1)^2 {T} \left((q-1) Q^2+2\right)}{n%
}, \\
&\frac{dQ}{d\bar{\tau}}= (1-q) Q^2 \left(Q^2-1\right) ( {T}-1).
\end{align}
where we have the relation 
\begin{equation}
Q^2 q=\frac{1}{2} \left(-3 Y^{2 n}+2 Q^2+3 \Sigma ^2-1\right)=Q^2+3
\Sigma^2-2.
\end{equation}

Introducing the complementary global transformation 
\begin{align}
&\Sigma=F(\theta ) \sin (\theta ), \quad Y=\cos (\theta ),  \notag \\
& F(\theta )=\sqrt{\frac{1-\cos ^{2 n}(\theta )}{1-\cos ^2(\theta )}},
\end{align}
we obtain the following unconstrained 3D dynamical system 
\begin{align}
&\frac{d\theta}{d\bar{\tau}}=- {T} F(\theta )+\frac{3 Q ( {T}-1) F(\theta
)^2 \sin (2 \theta )}{2 n}, \label{Case2_uggla-varsa}\\
&\frac{d {T}}{d\bar{\tau}}= -\frac{3 Q ( {T}-1)^2 {T} \left(\cos ^{2
n}(\theta )-1\right)}{n}, \label{Case2_uggla-varsb}\\
&\dfrac{dQ}{d\bar{\tau}}=\left(Q^2-1\right) ( {T}-1) \left(3 \cos ^{2
n}(\theta )-1\right), \label{Case2_uggla-varsc}
\end{align}
and the deceleration parameter satisfies 
\begin{equation}
q Q^2=1-3 \cos ^{2 n}(\theta )+Q^2.
\end{equation}

The system \eqref{Case2_uggla-varsa}, \eqref{Case2_uggla-varsb}, \eqref{Case2_uggla-varsc} has the equilibrium points:

\begin{itemize}
\item $(\theta, {T},Q)=\left(\frac{\pi}{2}+2k\pi,0, \pm 1\right)$; $%
\left(\Sigma, Y\right)=(1,0)$, with eigenvalues $\left\{\pm \frac{3}{n},\pm%
\frac{3}{n},\pm 2\right\}$. For $Q=+1$ it is a source. For $Q=-1$ it is a
sink.

\item $(\theta, {T},Q)=\left(-\frac{\pi}{2}+2k\pi,0, \pm 1\right)$; $%
\left(\Sigma, Y\right)=(-1,0)$, with eigenvalues $\left\{\pm \frac{3}{n},\pm%
\frac{3}{n},\pm 2\right\}$. For $Q=+1$ it is a source. For $Q=-1$ it is a
sink.

\item $(\theta, {T},Q)=(2k\pi,0, \pm 1)$; $\left(\Sigma, Y\right)=(0,1)$ and 
$(\theta, {T},Q)=((2k+1)\pi,0, \pm 1)$; $\left(\Sigma, Y\right)=(0,-1)$. The
eigenvalues are $\{0,\mp 4 ,\mp 3\}$. For $Q=+1$ it is nonhyperbolic with a
2D stable manifold. For $Q=-1$ it is nonhyperbolic with a 2D unstable
manifold.

\item $(\theta, {T},Q)=\left(\pm\text{arccos}\left(3^{-\frac{1}{2 n}%
}\right), 0, 0\right)$; \newline
$\left(\Sigma, Y\right)=\left(\pm\sqrt{\frac{2}{3}},3^{-\frac{1}{2 n}}\right)
$, with eigenvalues $0,\pm 2, \mp 2$. These solutions corresponds to static
universe $H=0$. Nonhyperbolic. Behaves as saddle.
\end{itemize}

Substituting ${T}=1$ in the above equations we obtain $\frac{d\theta}{d\bar{%
\tau}}=-F(\theta )$ that can be integrated in quadratures as 
\begin{equation}
\cos (\theta ) \, _2F_1\left(\frac{1}{2},\frac{1}{2 n};1+\frac{1}{2 n};\cos
^{2 n}(\theta )\right)=\bar{\tau} -\bar{\tau}_{0}.
\end{equation}

In figure \ref{fig:Case2HLPL} it is shown (a) a compact phase portrait of %
\eqref{SYSTEM81a}, \eqref{SYSTEM81b}, \eqref{SYSTEM81c} for the choice $n=2$. (b) Dynamics in the invariant set $Z=0
$. Additionally it is presented in (c)-(f) the dynamics of the system %
\eqref{Case2_uggla-varsa}, \eqref{Case2_uggla-varsb}, \eqref{Case2_uggla-varsc} and some 2D projections for the choice $n=4$.

\subsubsection{Center manifold of the de Sitter solution $(\protect\theta,{T}%
,Q)=(0,0,1)$ for Ho\v{r}ava-Lifshitz with non-flat universe with $\Lambda=0$
and powerlaw potential under the detailed-balance condition.}
\label{PROPOSITION6}

\begin{prop}
\label{centerP3PLCase2DS} The point $(\theta,{T},Q)=(0,0,1)$ of the system %
\eqref{Case2_uggla-varsa}, \eqref{Case2_uggla-varsb}, \eqref{Case2_uggla-varsc} is unstable (saddle point).
\end{prop}

\textbf{Proof.} Taking the linear transformation 
\begin{equation}
(u,v_1,v_2)=\left(T, 1-Q, \frac{1}{3} \left(3 \theta +\sqrt{n}
T\right)\right),
\end{equation}
and truncating the Taylor series at fifth order we obtain that the system %
\eqref{Case2_uggla-varsa}, \eqref{Case2_uggla-varsb}, \eqref{Case2_uggla-varsc} can be written in diagonal form 
\begin{equation}
\left(%
\begin{array}{c}
\frac{du}{d\bar{\tau}} \\ 
\frac{dv_1}{d\bar{\tau}} \\ 
\frac{dv_2}{d\bar{\tau}}%
\end{array}%
\right)=\left(%
\begin{array}{ccc}
0 & 0 & 0 \\ 
0 & -4 & 0 \\ 
0 & 0 & -3%
\end{array}%
\right)\left(%
\begin{array}{c}
u \\ 
v_1 \\ 
v_2%
\end{array}%
\right)+\left(%
\begin{array}{c}
f(u,\mathbf{v}) \\ 
g_1(u,\mathbf{v}) \\ 
g_2(u,\mathbf{v})%
\end{array}%
\right),  \label{HLcenterPLmodel2DS}
\end{equation}
\bigskip
where
\begin{widetext}
\begin{small}
$f(u,\mathbf{v})=\frac{1}{162} n \left(-3 n^2+n+54\right) u^5+\frac{1}{3} u^3 \left(\left(-3 n^2+n+9\right) v_2^2-12 \sqrt{n} (v_1-1) v_2-n v_1+n\right)+\frac{1}{27} u^4 \left(18 n (v_1-1)+2 \sqrt{n} (n
   (3 n-1)-27) v_2\right)+\frac{2}{3} u^2 v_2 \left(3 \sqrt{n} (v_1-1)+\sqrt{n} (3 n-1) v_2^2+9 (v_1-1) v_2\right)+\frac{1}{2} u v_2^2 \left((1-3 n) v_2^2-6
   v_1+6\right),$
\\
$g_1(u,\mathbf{v})=\frac{1}{3} u v_1 \left(6 n^{3/2} (v_1-2) v_2+4 n^{3/2} (3 n-1) v_2^3+9 n (v_1-2) v_2^2-6 (v_1-2)\right)+\frac{1}{81} (1-3 n) n^3 u^4 v_1+\frac{1}{27}
   u^3 \left(4 (3 n-1) n^{5/2} v_1 v_2+9 n^2 (v_1-2) v_1\right)-\frac{1}{3} u^2 v_1 \left(6 n^{3/2} (v_1-2) v_2+6 n^3 v_2^2+n^2 \left(v_1-2
   \left(v_2^2+1\right)\right)\right)-v_1 \left(3 n (v_1-2) v_2^2+n (3 n-1) v_2^4-2 v_1\right),$
\\
$g_2(u,\mathbf{v})=\frac{n^{3/2} (n (5 n (11 n-38)+39)+4320) u^5}{38880}+\frac{1}{648} u^4 \left(-12 (3 n-11)
   n^{3/2} (v_1-1)-(n (3 n-7) (5 n-1)+432) n v_2\right)+\frac{1}{432} u^3 \left(4 n^{3/2} (n (6 v_1-3)-10 v_1+7)+72 n (3 n-7) (v_1-1) v_2+\sqrt{n} (n (n (65 n-114)+17)+432)
   v_2^2\right)$\\$+\frac{1}{10} \left(\left(1-5 n^2\right) v_2^5-5 (3 n+1) (v_1-1) v_2^3+30 v_1 v_2\right)+u^2 \left(-\frac{3}{2} (n-1) \sqrt{n} (v_1-1) v_2^2-\frac{1}{6} (n-1) n
   (3 v_1-2) v_2+\sqrt{n} (v_1-1)-\frac{1}{72} (n-1) n (35 n-3) v_2^3\right)+u \left(\frac{1}{2} (3 n+1) (v_1-1) v_2^3+\frac{1}{4} \sqrt{n} v_2^2 (n (6 v_1-5)-2
   v_1+1)-\sqrt{n} v_1+\frac{1}{96} \sqrt{n} (n (75 n-38)-5) v_2^4-3 (v_1-1) v_2\right).$
	\end{small}
\end{widetext}
According to Theorem \ref{existenceCM}, there exists a 1-dimensional
invariant local center manifold
$W^{c}\left( \mathbf{0}\right) $ of \eqref{HLcenterPLmodel2DS}, \newline $W^{c}\left( 
\mathbf{0}\right) =\left\{ \left( u,\mathbf{v}\right) \in\mathbb{R}\times%
\mathbb{R}^{2}:\mathbf{v}=\mathbf{h}\left( u\right)\right\}$, satisfying $%
\mathbf{h}\left( 0\right) =\mathbf{0},\;D\mathbf{h}\left( 0\right) =\mathbf{0%
} ,\;\left\vert u\right\vert <\delta$ for $\delta$ sufficiently small.

The restriction of (\ref{HLcenterPLmodel2DS}) to the center manifold is $%
\frac{du}{dN}=f\left( u,\mathbf{h}\left( u\right) \right)$, where the
function $\mathbf{h}\left( u\right)$ satisfies \eqref{MaineqcM}:%
\begin{equation}
D\mathbf{h}\left( u\right) \left[ f\left( u,\mathbf{h}\left( u\right)
\right) \right] -P\mathbf{h}\left( u\right) -\mathbf{g}\left( u,\mathbf{h}%
\left( u\right) \right) =0,  \label{HLcenterPLmodel2DSP3h}
\end{equation}
where 
\begin{equation*}
P=\left(%
\begin{array}{cc}
-4 & 0 \\ 
0 & -3%
\end{array}%
\right).
\end{equation*}
According to Theorem \ref{approximationCM}, the system %
\eqref{HLcenterPLmodel2DSP3h} can be solved approximately by expanding $%
\mathbf{h}\left( u\right) $ in Taylor series at $u=0.$ Since $\mathbf{h}%
\left( 0\right) =\mathbf{0\ } $and $D\mathbf{h}\left( 0\right) =\mathbf{0},$
we propose the ansatsz
\begin{equation}
\mathbf{h}\left( u\right) :=\left(%
\begin{array}{c}
h_{1}\left(u\right) \\ 
h_{2}\left(u\right)%
\end{array}%
\right)=\left(%
\begin{array}{c}
\sum_{j=1}^{4} a_j u^{j+1} +O\left( u^{6}\right) \\ 
\sum_{j=1}^{4} b_j u^{j+1} +O\left( u^{6}\right)%
\end{array}%
\right),
\end{equation}%
 to find the non-null coefficients\newline
$b_1=-\frac{\sqrt{n}}{3}, b_2= -\frac{1}{324} \sqrt{n} (n (3 n-7)+108)$,%
\newline
$b_3= -\frac{1}{108} \sqrt{n} (n (3 n-7)+36)$,\newline
$b_4=  -\frac{\sqrt{n} \left(n \left(n \left(65 n^2-570
n+7561\right)-15120\right)+38880\right)}{116640}$.\newline  Therefore, the center
manifold can be represented locally by the graph 
\begin{align}
& v_1=0, v_2= -\frac{\sqrt{n} u^2}{3}-\frac{1}{324} \left(\sqrt{n} (n (3
n-7)+108)\right) u^3  \notag \\
& -\frac{1}{108} \left(\sqrt{n} (n (3 n-7)+36)\right) u^4  \notag \\
& -\frac{\left(\sqrt{n} \left(n \left(n \left(65 n^2-570
n+7561\right)-15120\right)+38880\right)\right) u^5}{116640}.
\end{align}%
That is,  
\begin{align}
&\theta= -\frac{\sqrt{n} T}{3}-\frac{\sqrt{n} T^2}{3}-\frac{1}{324} \left(%
\sqrt{n} (n (3 n-7)+108)\right) T^3  \notag \\
& -\frac{1}{108} \left(\sqrt{n} (n (3 n-7)+36)\right) T^4  \notag \\
& -\frac{\left(\sqrt{n} \left(n \left(n \left(65 n^2-570
n+7561\right)-15120\right)+38880\right)\right) T^5}{116640},  \notag \\
& Q= 1.
\end{align}%
The dynamics on the center manifold is given by the gradient-like
equation 
\begin{align}
&\frac{du}{dN}=-\nabla \Pi(u), \Pi(u)=-\frac{n u^4}{12}+\frac{n^2 u^6}{162}+%
\frac{1}{567} n^2 (3 n+5) u^7  \notag \\
& +\frac{n^2 (n (15 n (n+8)+2434)+1485) u^8}{174960}.
\end{align}%
We have $\Pi^{\prime }(0)=\Pi^{\prime \prime }(0)=\Pi^{\prime \prime
\prime (4)}(0)=-2n<0$. It follows that $u=0$ is a degenerated maximum of the
potential. Using the Theorem \ref{stabilityCM}, we conclude that the center
manifold of the point $(\theta,{T},Q)=(0,0,1)$ of the system %
\eqref{Case2_uggla-varsa}, \eqref{Case2_uggla-varsb}, \eqref{Case2_uggla-varsc}, and the point itself, are unstable (saddle point). 
$\blacksquare$

\subsection{E-models}

\label{Case2D}

\begin{figure*}[ht!]
\centering
\subfigure[]{\includegraphics[width=2.5in, height=2.5in,angle=-90]{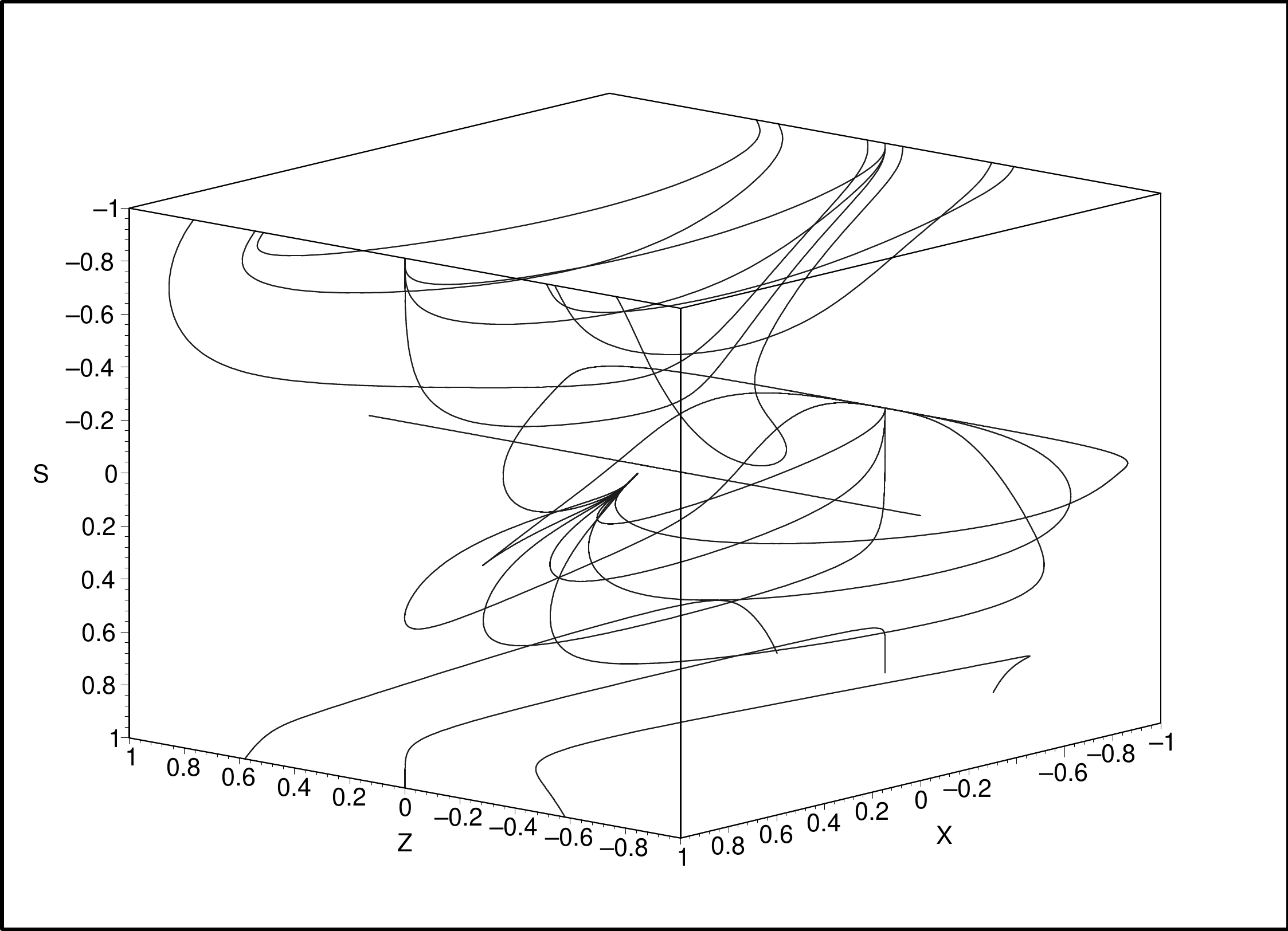}} 
\hspace{2cm}  \subfigure[]{\includegraphics[width=2.5in,
height=2.5in,angle=-90]{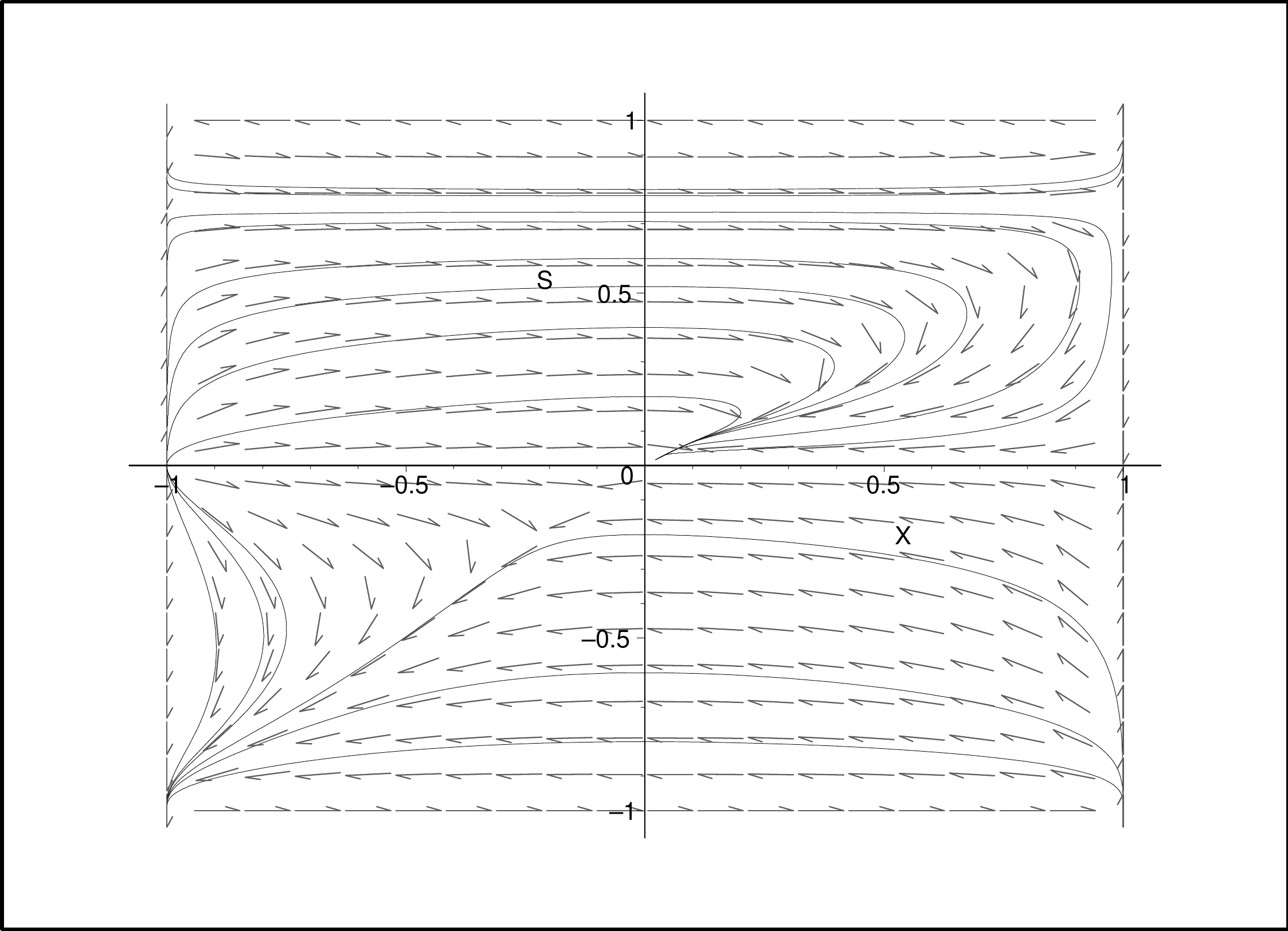}}  \subfigure[]{\includegraphics[width=2.5in,
height=2.5in,angle=-90]{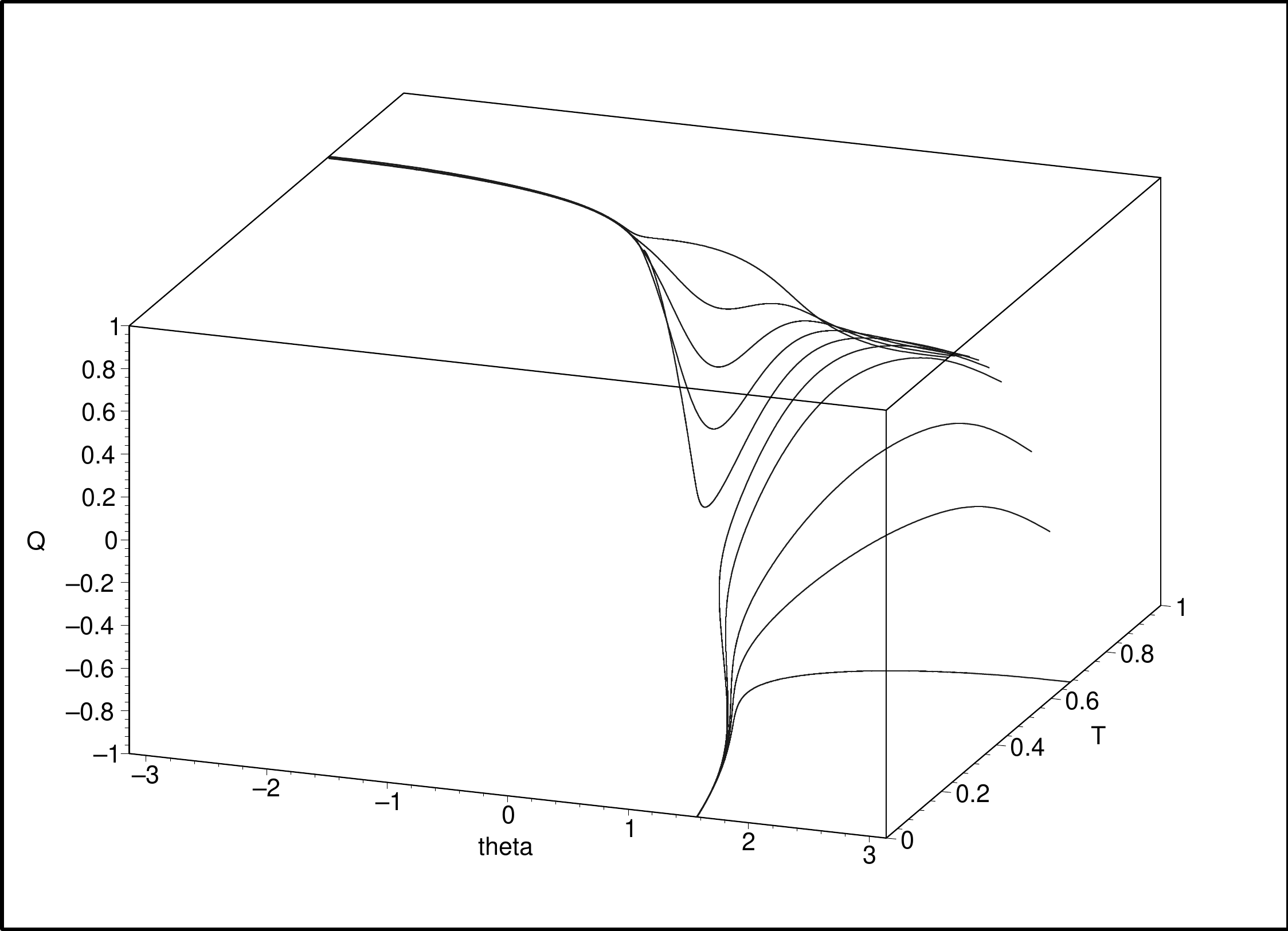}} \hspace{2cm}  \subfigure[]{%
\includegraphics[width=2.5in, height=2.5in,angle=-90]{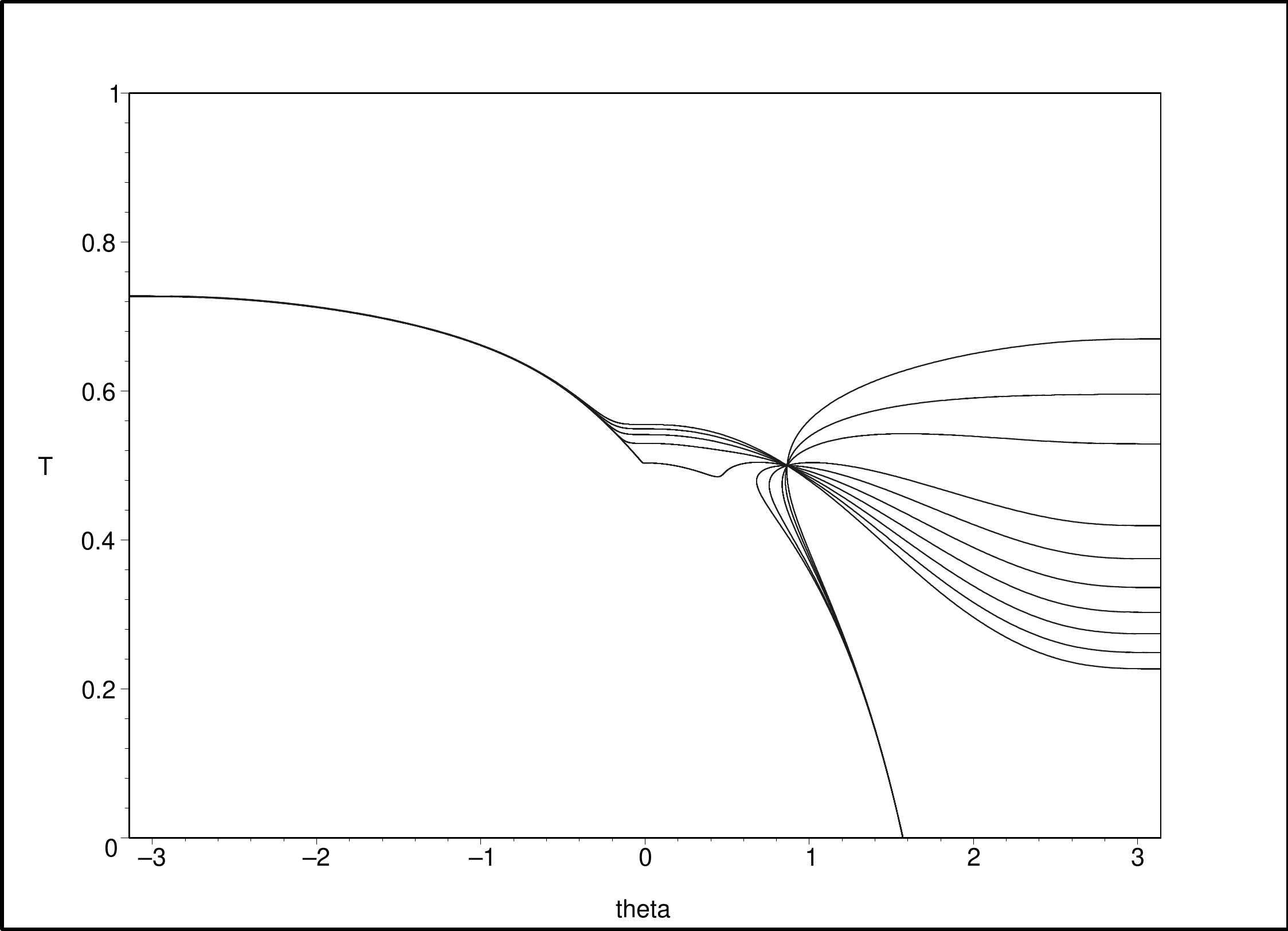}}  %
\subfigure[]{\includegraphics[width=2.5in, height=2.5in,angle=-90]{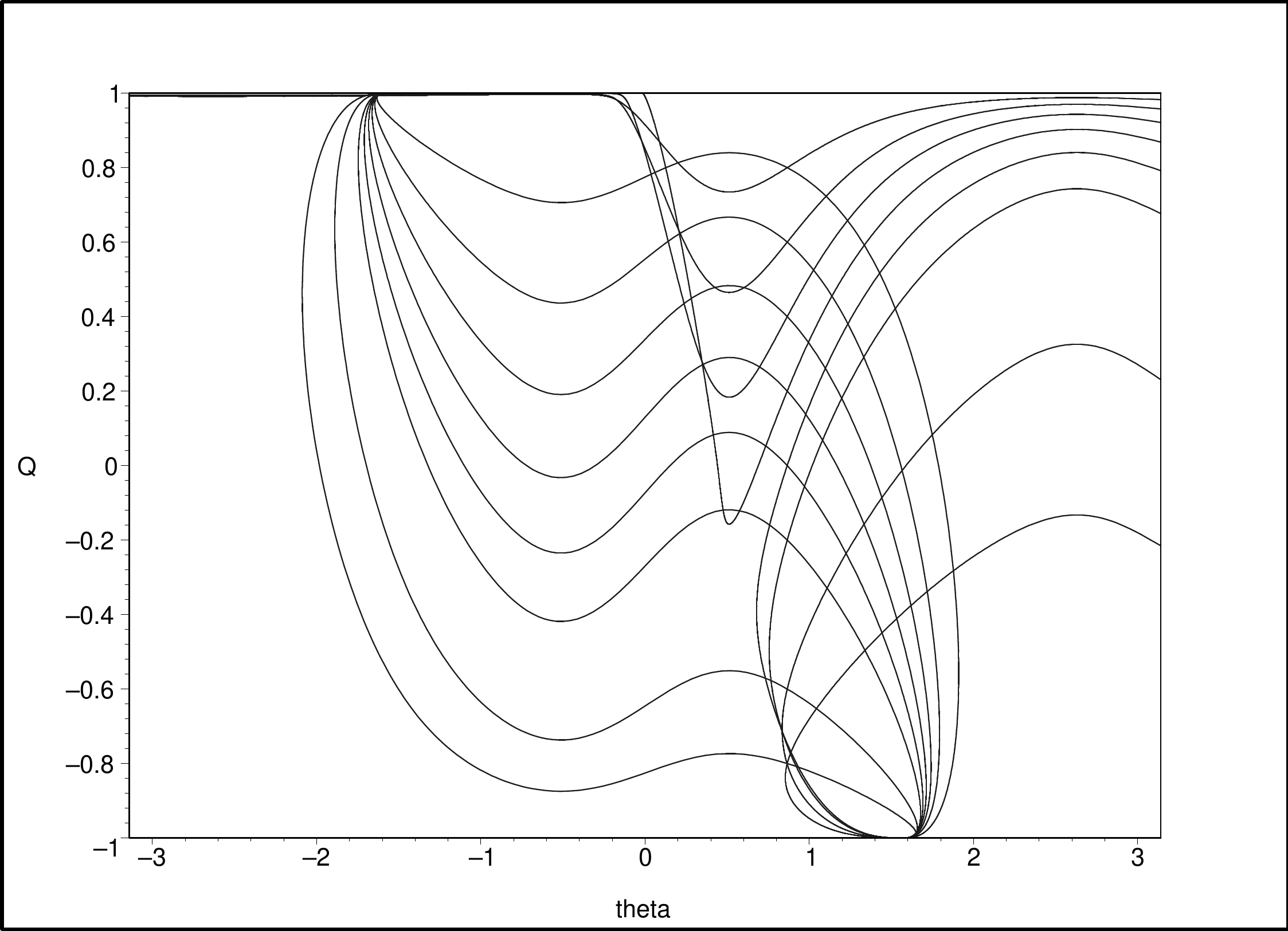}} 
\hspace{2cm}  \subfigure[]{\includegraphics[width=2.5in,
height=2.5in,angle=-90]{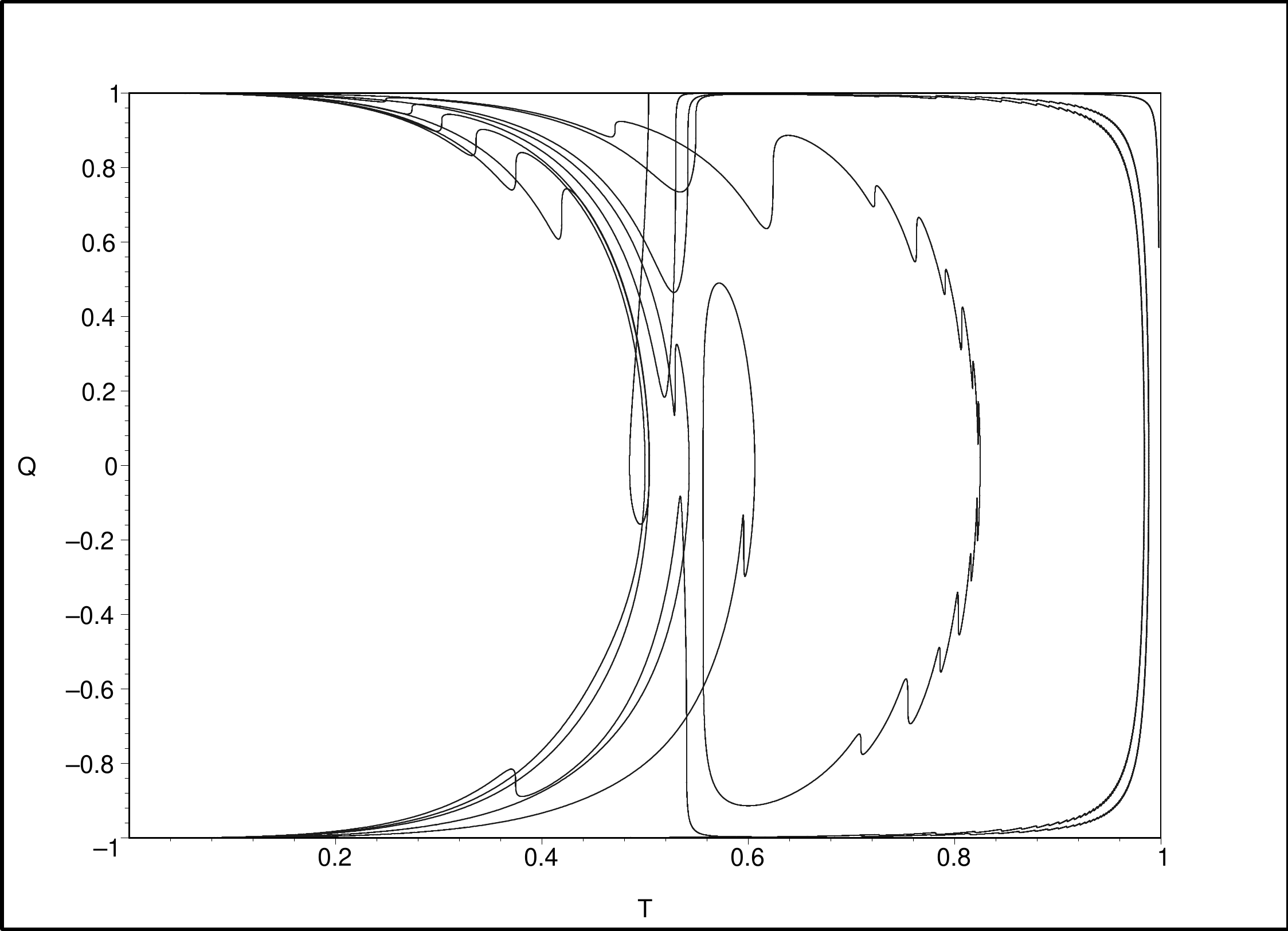}}
\caption{\textit{{(a) Compact phase portrait of \eqref{SYSTEM93a}, \eqref{SYSTEM93b}, \eqref{SYSTEM93c} for the
choice $n=2, \protect\mu=1$. (b) Dynamics in the invariant set $Z=0$.
(c)-(f): Dynamics of the system \eqref{Case2_uggla-varsX2a}, \eqref{Case2_uggla-varsX2b}, \eqref{Case2_uggla-varsX2c} and some 2D projections for $%
n=4, \protect\mu=1$.}}}
\label{fig:Case2HLE}
\end{figure*}

In this case the system  \eqref{Example2Sa}, \eqref{Example2Sb}, \eqref{Example2Sc}, becomes
\begin{align}
&\frac{dX}{d\tau}=\left(X^2-1\right) \left(3 X \sqrt{1-Z^2}-\sqrt{6}\,
\tan\left(\frac{\pi S}{2}\right)\right), \label{SYSTEM93a}\\
&\frac{dZ}{d\tau}=\left(3 X^2-2\right) Z \sqrt{1-Z^2}, \label{SYSTEM93b}\\
&\frac{dS}{d\tau}=\frac{X \left(-6 \mu \sin (\pi S)-\sqrt{6} \cos (\pi S)+%
\sqrt{6}\right)}{\pi n}, \label{SYSTEM93c}
\end{align}
defined on the compact phase space\newline
$\Big\{(X,Z,S)\in\mathbb{R}^3: -1\leq X \leq 1, -1\leq Z \leq 1, -1\leq S
\leq 1\Big\}.$ The equilibrium points of \eqref{SYSTEM93a}, \eqref{SYSTEM93b}, \eqref{SYSTEM93c} and their
stability conditions are summarized as follows:

\begin{itemize}
\item $(X,Z,S)=(0, -1, 0)$. The eigenvalues are $\{0,0,\infty \}$.
Nonhyperbolic.

\item $(X,Z,S)=(0, 0, 0)$. The eigenvalues are $\{0,-3,-2\}$. Nonhyperbolic.

\item $(X,Z,S)=(0, 1, 0)$. The eigenvalues are $\{0,0,\infty\}$.
Nonhyperbolic.

\item $(X,Z,S)=(-1, -1, 0)$. The eigenvalues $\left\{0,-\infty ,\frac{6 \mu 
}{n}\right\}$. Nonhyperbolic. Behaves as saddle.

\item $(X,Z,S)=\left(-1, -1, \frac{2 \,\text{arctan}\left(\sqrt{6} \mu
\right)}{\pi }\right)$. The eigenvalues are
$\left\{-\infty ,12 \mu ,-\frac{6 \mu }{n}\right\}$. Saddle.

\item $(X,Z,S)=(1, -1, 0)$. The eigenvalues are $\left\{0,-\infty ,-\frac{6
\mu }{n}\right\}$. Nonhyperbolic.

\item $(X,Z,S)=\left(1, -1, \frac{2 \,\text{arctan}\left(\sqrt{6} \mu \right)%
}{\pi }\right)$. The eigenvalues are
$\left\{-\infty ,-12 \mu ,\frac{6 \mu }{n}\right\}$. Saddle.

\item $(X,Z,S)=(-1, 0, 0)$. The eigenvalues are $\left\{1,6,\frac{6 \mu }{n}%
\right\}$. Source.

\item $(X,Z,S)=\left(-1, 0, \frac{2 \,\text{arctan}\left(\sqrt{6} \mu \right)%
}{\pi }\right)$. The eigenvalues are
$\left\{1,12 \mu +6,-\frac{6 \mu }{n}\right\}$. Saddle.

\item $(X,Z,S)=(1, 0, 0)$. The eigenvalues are $\left\{6,1,-\frac{6 \mu }{n}%
\right\}$. Saddle.

\item $(X,Z,S)=\left(1, 0, \frac{2 \,\text{arctan}\left(\sqrt{6} \mu \right)%
}{\pi }\right)$. The eigenvalues are\newline
$\left\{1,6-12 \mu ,\frac{6 \mu }{n}\right\}$. Source for $\mu<\frac{1}{2}$.

\item $(X,Z,S)=(-1, 1, 0)$. The eigenvalues are $\left\{0,-\infty ,\frac{6
\mu }{n}\right\}$. Nonhyperbolic. Behaves as saddle.

\item $(X,Z,S)=\left(-1, 1, \frac{2 \,\text{arctan}\left(\sqrt{6} \mu \right)%
}{\pi }\right)$. The eigenvalues are
$\left\{-\infty ,12 \mu ,-\frac{6\mu }{n}\right\}$. Saddle.

\item $(X,Z,S)=(1, 1, 0)$. The eigenvalues are $\left\{0,-\infty ,-\frac{6
\mu }{n}\right\}$. Nonhyperbolic.

\item $(X,Z,S)=\left(1, 1, \frac{2 \,\text{arctan}\left(\sqrt{6} \mu \right)%
}{\pi }\right)$. The eigenvalues are\newline
$\left\{-\infty ,-12 \mu ,\frac{6 \mu }{n}\right\}$. Saddle.
\end{itemize}

\subsubsection{Stability Analysis of the solution $P_{3}^{0}:(X,Z,S)=(0,0,0)$%
, in Ho\v{r}ava-Lifshitz cosmology for the Non-flat universe with $\Lambda =
0$ under the detailed-balance condition the E-model.}

\label{PROPOSITION7}

\begin{prop}
\label{centerP3EPCase2} The origin for the system \eqref{SYSTEM93a}, \eqref{SYSTEM93b}, \eqref{SYSTEM93c} is
unstable (center-saddle).
\end{prop}

\textbf{Proof.} Taking the linear transformation 

\begin{equation}
(u,v_1,v_2)=\left(S, \frac{1}{6} \left(6 X-\sqrt{6} \pi S\right), Z\right),
\end{equation}
and taking Taylor series near $(u,v_1,v_2)=(0,0,0)$ up to fifth order we
obtain the system \eqref{SYSTEM93a}, \eqref{SYSTEM93b}, \eqref{SYSTEM93c} can be written into its Jordan canonical
form:%
\begin{equation}
\left(%
\begin{array}{c}
\frac{du}{dN} \\ 
\frac{dv_1}{dN} \\ 
\frac{dv_2}{dN}%
\end{array}%
\right)=\left(%
\begin{array}{ccc}
0 & 0 & 0 \\ 
0 & -3 & 0 \\ 
0 & 0 & -2%
\end{array}%
\right)\left(%
\begin{array}{c}
u \\ 
v_1 \\ 
v_2%
\end{array}%
\right)+\left(%
\begin{array}{c}
f(u,\mathbf{v}) \\ 
g_1(u,\mathbf{v}) \\ 
g_2(u,\mathbf{v})%
\end{array}%
\right)  \label{HLcenterS93}
\end{equation}%
 where \newline
\begin{small}
$f(u,\mathbf{v})=-\frac{\pi ^4 u^5}{24 n}-\frac{\pi ^3 u^4 (v_1-4 \mu )}{4 
\sqrt{6} n}+\frac{\pi ^2 u^3 (2 \mu v_1+1)}{2 n}+\frac{\sqrt{\frac{3}{2}}
\pi u^2 (v_1-2 \mu )}{n}-\frac{6 \mu u v_1}{n},$\newline $g_1(u,\mathbf{v})=\frac{%
\pi ^5 (5-2 n) u^5}{120 \sqrt{6} n}+\frac{\pi ^4 u^4 (-4 \mu -2 n v_1+v_1)}{%
24 n}$\newline
$-\frac{\pi ^3 u^3 \left(n \left(v_1^2+v_2^2-1\right)+4 \mu v_1+2\right)}{4 
\sqrt{6} n}+\frac{\pi ^2 u^2 \left(4 \mu +v_1 \left(n \left(2-3
v_2^2\right)-2\right)\right)}{4 n}$\newline
$+\frac{\sqrt{\frac{3}{2}} \pi u \left(n \left(4 v_1^2 \left(4-3
v_2^2\right)+v_2^4+4 v_2^2\right)+16 \mu v_1\right)}{8 n}$
$+\frac{3}{8} v_1 \left(-4 \left(v_1^2-1\right) v_2^2+8 v_1^2+v_2^4\right),$
\newline
$g_2(u,\mathbf{v})=-\frac{1}{4} \pi ^2 u^2 v_2 \left(v_2^2-2\right)-\sqrt{%
\frac{3}{2}} \pi u v_1 v_2 \left(v_2^2-2\right)$\newline
$+\frac{1}{4} v_2 \left(-6 v_1^2 \left(v_2^2-2\right)+v_2^4+4  v_2^2\right).$
\end{small}\newline
According to Theorem \ref{existenceCM}, there exists a 1-dimensional
invariant local center manifold $W^{c}\left( \mathbf{0}\right) $ of %
\eqref{HLcenterS93}, \newline
$W^{c}\left( \mathbf{0}\right) =\left\{ \left( u,\mathbf{v}\right) \in%
\mathbb{R}\times\mathbb{R}^{2}:\mathbf{v}=\mathbf{h}\left( u\right)\right\}$%
, satisfying $\mathbf{h}\left( 0\right) =\mathbf{0},\;D\mathbf{h}\left(
0\right) =\mathbf{0} ,\;\left\vert u\right\vert <\delta$ for $\delta$
sufficiently small. The restriction of (\ref{HLcenterS93}) to the center
manifold is $\frac{du}{dN}=f\left( u,\mathbf{h}\left( u\right) \right)$,
where the function $\mathbf{h}\left( u\right)$ that defines the local center
manifold satisfies \eqref{MaineqcM}:%
\begin{equation}
D\mathbf{h}\left( u\right) \left[ f\left( u,\mathbf{h}\left( u\right)
\right) \right] -P\mathbf{h}\left( u\right) -\mathbf{g}\left( u,\mathbf{h}%
\left( u\right) \right) =0,  \label{HLS93P3h}
\end{equation}
where 
\begin{equation*}
P=\left(%
\begin{array}{cc}
-3 & 0 \\ 
0 & -2%
\end{array}%
\right).
\end{equation*}
According to Theorem \ref{approximationCM}, the system \eqref{HLS93P3h} can
be solved approximately by expanding $\mathbf{h}\left( u\right) $ in Taylor
series at $u=0.$ Assuming that 
\begin{equation}
\mathbf{h}\left( u\right) :=\left(%
\begin{array}{c}
h_{1}\left(u\right) \\ 
h_{2}\left(u\right)%
\end{array}%
\right)=\left(%
\begin{array}{c}
\sum_{j=1}^{4} a_j u^{j+1} +O\left( u^{6}\right) \\ 
\sum_{j=1}^{4} b_j u^{j+1} +O\left( u^{6}\right)%
\end{array}%
\right),
\end{equation}%
 we find the non-null coefficients\newline
$a_1=\frac{\pi ^2 \mu }{3 n}, a_2= \frac{\pi ^3 \left(24 \mu ^2+(n-2)
n\right)}{12 \sqrt{6} n^2}, a_3= \frac{\pi ^4 \mu \left(56 \mu ^2+n (2
n-7)\right)}{18 n^3}$,\newline
$a_4= \frac{\pi ^5 \left(13600 \mu ^4+n^2 (n (3 n-25)+40)+20 \mu ^2 n (29
n-108)\right)}{360 \sqrt{6} n^4}$. \newline Therefore, the center manifold can be
represented locally by the graph 
\begin{small}
\begin{align}
&v_1=\frac{\pi ^2 \mu u^2}{3 n}+\frac{\pi ^3 \left(24 \mu ^2+(n-2) n\right)
u^3}{12 \sqrt{6} n^2} +\frac{\pi ^4 \mu \left(56 \mu ^2+n (2 n-7)\right) u^4%
}{18 n^3}  \notag \\
& +\frac{\pi ^5 \left(13600 \mu ^4+20 n (29 n-108) \mu ^2+n^2 (n (3
n-25)+40)\right) u^5}{360 \sqrt{6} n^4},  \notag \\
& v_2=0.
\end{align}%
\end{small} That is, 
\begin{small}
\begin{align}
&X=\frac{\pi S}{\sqrt{6}}+\frac{\pi ^2 \mu S^2}{3 n}+\frac{\pi ^3 \left(24
\mu ^2+(n-2) n\right) S^3}{12 \sqrt{6} n^2}  \notag \\
&+\frac{\pi ^4 \mu \left(56 \mu ^2+n (2 n-7)\right) S^4}{18 n^3}  \notag \\
& +\frac{\pi ^5 \left(13600 \mu ^4+20 n (29 n-108) \mu ^2+n^2 (n (3
n-25)+40)\right) S^5}{360 \sqrt{6} n^4},  \notag \\
& Z=0.
\end{align}%
\end{small}
The dynamics on the center manifold is given by the gradient-like
equation 
\begin{small}
\begin{align}
&\frac{du}{dN}=-\nabla \Pi(u),  \notag \\
&\Pi(u)=\frac{\sqrt{\frac{2}{3}} \pi \mu u^3}{n}-\frac{\pi ^2 u^4 \left(n-4
\mu ^2\right)}{8 n^2}-\frac{\pi ^3 \mu u^5 \left(n (n+4)-24 \mu ^2\right)}{%
10 \sqrt{6} n^3}  \notag \\
& +\frac{\pi ^4 u^6 \left(224 \mu ^4+n^2+4 \mu ^2 (n-10) n\right)}{72 n^4} 
\notag \\
& -\frac{\pi ^5 \mu u^7 \left(-6800 \mu ^4+n^2 (n (n+15)-55)+10 \mu ^2
(136-23 n) n\right)}{210 \sqrt{6} n^5}.
\end{align}
\end{small}
We have $\Pi^{\prime }(0)=\Pi^{\prime \prime }(0)=0, \Pi^{\prime \prime
\prime }(0)=\frac{2 \sqrt{6} \pi \mu }{n}$. It follows that $u=0$ is an
inflection of the potential. Using the Theorem \ref{stabilityCM}, we
conclude that the center manifold of origin for the system %
\eqref{HLcenterS93}, and the origin are unstable (center-saddle). $%
\blacksquare$

\subsubsection{Alternative compactification}

In this example we can alternatively introduce the following
compactification 
\begin{align}
&\bar{H}=\sqrt{H^2+\frac{\mu^2 k^2}{16(3\lambda-1)^2a^4}}, \\
& Q=\frac{H}{\bar{H}}, \quad \Sigma =\frac{\dot\phi}{2 \sqrt{6} \bar{H}}, \\
&Y=\left(\frac{V(\phi)}{6(3\lambda-1)\bar{H}^2}\right)^{\frac{1}{2 n}}=\bar{%
T}\left(1-e^{-\sqrt{\frac{2}{3 \alpha}} \phi}\right), \\
& \bar{T}=\left[\frac{V_0}{6(3\lambda-1)\bar{H}^2}\right]^{\frac{1}{2n}},
\end{align}
such that 
\begin{align}
&\dot\phi =\frac{\sqrt{\frac{2}{3}} \Sigma V_0 \bar{T}^{-2 n}}{3 \lambda -1},
\\
&\phi= -\frac{\sqrt{\frac{2}{3}} n \ln \left(1-\frac{Y}{T}\right)}{\mu }, \\
& H= Q \sqrt{\frac{V_0 }{6 (3\lambda -1)}}\bar{T}^{-n}, \\
& a= \frac{\sqrt[4]{3} \sqrt{\mu } \bar{T}^{n/2}}{2^{3/4} \sqrt[4]{3 \lambda
-1} \sqrt[4]{1-Q^2} \sqrt[4]{V_0}},
\end{align}
and
\begin{align}
&\Sigma + Y^{2 n}=1, \\
&Q^2 q= \frac{1}{2} \left(-3 Y^{2 n}+2 Q^2+3 \Sigma ^2-1\right)=Q^2+3 \Sigma
^2-2.
\end{align}
Introducing the new time variable $\frac{dM}{d t}=\frac{H}{Q}$, the
dynamical system becomes 
\begin{align}
&\frac{d \Sigma }{dM}=6 \mu Y^{2 n-1} (Y-\bar{T})+Q \Sigma \left((q-1)
Q^2-1\right), \\
&\frac{d Y}{dM}=\frac{(q-1) Q^3 Y+2 Q Y+6 \mu \Sigma (\bar{T}-Y)}{n}, \\
&\frac{d \bar{T}}{dM}=\frac{Q \bar{T} \left((q-1) Q^2+2\right)}{n}, \\
&\frac{d Q}{dM}=(q-1) Q^2 \left(Q^2-1\right).
\end{align}
and the deceleration parameter satisfies 
\begin{equation}
q Q^2=1-3 \cos ^{2 n}(\theta )+Q^2.
\end{equation}
Introducing the complementary global transformation 
\begin{align}
&\Sigma =F(\theta ) \sin (\theta ), \quad Y=\cos (\theta ),  \notag \\
& F(\theta )=\sqrt{\frac{1-\cos ^{2 n}(\theta )}{1-\cos ^2(\theta )}},
\end{align}
we obtain the following regular unconstrained 3D dynamical system 
\begin{align}
&\frac{d\theta}{dM}=-\frac{6 \mu F(\theta ) (\bar{T}-\cos (\theta ))}{n}-%
\frac{3 Q F(\theta )^2 \sin (2\theta )}{2 n}, \\
&\frac{d\bar{T}}{dM}= -\frac{3 Q \bar{T} \left(\cos ^{2 n}(\theta )-1\right)%
}{n}, \\
&\frac{d {Q}}{dM}=\left(1-Q^2\right) \left(3 \cos ^{2 n}(\theta )-1\right).
\end{align}
Introducing the compact variable $T=\frac{\bar{T}}{1+\bar{T}}$ and the time
derivative $\frac{d{\bar{\tau}}}{d M}=1+\bar{T}=(1-T)^{-1}$, we obtain the
regular system: 
\begin{align}
&\frac{d\theta}{d\bar{\tau}}=-\frac{6 \mu F(\theta ) ((T-1) \cos (\theta )+T)%
}{n}  \notag \\
&+\frac{3 Q (T-1) F(\theta )^2 \sin (2\theta ) }{2n}, \label{Case2_uggla-varsX2a}\\
&\frac{d T}{d\bar{\tau}}=-\frac{3 Q (T-1)^2 T \left(\cos ^{2 n}(\theta
)-1\right)}{n}, \label{Case2_uggla-varsX2b}\\
&\frac{d Q}{d\bar{\tau}}=\left(Q^2-1\right) (T-1) \left(3 \cos ^{2 n}(\theta
)-1\right). \label{Case2_uggla-varsX2c}
\end{align}
The past boundary is attached to the phase-space, and it is $\{T=0,
\cos(\theta)\leq 0\}\cup\left\{T-(1-T)\cos\theta=0, \cos(\theta)>0\right\}$.
It is also included the future boundary $T=1$, which corresponds to $H=0$
and the final state is the Minkowski point. The region $\left\{T-(1-T)\cos%
\theta<0, \cos(\theta)>0\right\}$ is forbidden. We include the boundaries $%
Q=\pm 1$ too. The equilibrium points of \eqref{Case2_uggla-varsX2a}, \eqref{Case2_uggla-varsX2b}, \eqref{Case2_uggla-varsX2c} and their
stability is summarized as follows.

\begin{itemize}
\item ${}_\pm M_{\pm}$: $\bar{T}=T=0; \Sigma=\pm 1, Q=\pm 1, Y=0; \theta=\pm 
\frac{\pi}{2}+2 k \pi, k\in\mathbb{Z}$, were we have used the Kernel: ${}_{%
\text{sign}(Q)} M_{\text{sign}(\Sigma)}$. The eigenvalues of ${}_{-}M_{+}$
are $\left\{-\frac{3}{n},-2,-\frac{6 \mu +3}{n}\right\}$. It is a sink. The
eigenvalues of ${}_{-}M_{-}$ are $\left\{-\frac{3}{n},-2,-\frac{3-6 \mu}{n}%
\right\}$. It is a sink for $\mu<\frac{1}{2}$ or a saddle otherwise. The
eigenvalues of ${}_{+}M_{+}$ are $\left\{\frac{3}{n},2,\frac{3-6 \mu }{n}%
\right\}$. It is a source for $\mu<\frac{1}{2}$ or a saddle otherwise. The
eigenvalues of ${}_{+}M_{-}$ are $\left\{\frac{3}{n},2,\frac{6 \mu +3}{n}%
\right\}$. It is a source.

\item $dS^{\pm}$: $\bar{T}=1, T=\frac{1}{2}; \Sigma=0, Q=\pm 1 ,
Y=1;\theta=2 k \pi, k\in\mathbb{Z}$. The eigenvalues of $dS^{+}$ are $%
\left\{0,-2,-\frac{3}{2}\right\}$. Nonhyperbolic with 2D stable manifold.
The eigenvalues of $dS^{-}$ are $\left\{0,2,\frac{3}{2}\right\}$.
Nonhyperbolic with 2D unstable manifold.

\item $PL^{\pm}$: $\bar{T}=T=0; \Sigma=\pm 2\mu, Q=\pm 1; Y=-(1-4\mu^2)^{%
\frac{1}{2n}}$;
$\theta=\pm\arccos Y$. Exists for $\mu<1/2$. The eigenvalues are
$\left\{\pm\frac{12 \mu ^2}{n},\pm(24 \mu ^2-4),\pm(12 \mu ^2-3)\right\}$.
It is a saddle.

\item $L_{1}^{\pm}$: $\bar{T}=T=0; \Sigma=\pm\sqrt{\frac{2}{3}}, Q=\pm\sqrt{6%
}\mu$;
$Y=3^{-\frac{1}{2 n}}, \theta=\pm\text{arccos}\left(3^{-\frac{1}{2 n}}\right)
$. The eigenvalues are\newline
$\left\{\pm\frac{2 \sqrt{6} \mu }{n},\mp\frac{\sqrt{3} \mu +\sqrt{8-45 \mu ^2%
}}{\sqrt{2}},\mp\frac{\sqrt{3} \mu -\sqrt{8-45 \mu ^2}}{\sqrt{2}}\right\}$.
Saddle.

\item $S^{\pm}$: $\bar{T}=3^{-\frac{1}{2n}}, T=\frac{1}{1+3^{\frac{1}{2n}}};
\Sigma=\pm\sqrt{\frac{2}{3}}, Q=0; Y=3^{-\frac{1}{2 n}}$;
$\theta=\pm\text{arccos}\left(3^{-\frac{1}{2 n}}\right)$. The eigenvalues are%

$\left\{\pm 2 \frac{ 3^{\frac{1}{2 n}}}{3^{\frac{1}{2 n}}+1},\mp 2\frac{ 3^{%
\frac{1}{2 n}}}{3^{\frac{1}{2 n}}+1},\mp\frac{2 \sqrt{6} \mu}{n} \frac{ 3^{%
\frac{1}{2 n}}}{3^{\frac{1}{2 n}}+1}\right\}$. It is a saddle.
\end{itemize}

Substituting ${T}=1$ in the above equations we obtain $\frac{d\theta}{d\bar{%
\tau}}=-\frac{6\mu}{n}F(\theta )$. This equation can be integrated in
quadratures as 

\begin{equation}
\cos (\theta ) \, _2F_1\left(\frac{1}{2},\frac{1}{2 n};1+\frac{1}{2 n};\cos
^{2 n}(\theta )\right)=\frac{6\mu}{n}\left(\bar{\tau} -\bar{\tau}_{0}\right).
\end{equation}

In figure \ref{fig:Case2HLE} it is shown (a) a compact phase portrait of %
\eqref{SYSTEM93a}, \eqref{SYSTEM93b}, \eqref{SYSTEM93c} for the choice $n=2, \mu=1$. (b) Dynamics in the
invariant set $Z=0$. In (c)-(f) it is presented the dynamics of %
\eqref{Case2_uggla-varsX2a}, \eqref{Case2_uggla-varsX2b}, \eqref{Case2_uggla-varsX2c} and some 2D projections for $n=4, \mu=1$.

\subsubsection{Center manifold of the de Sitter solution $(\protect\theta, {T%
},Q)=\left(0,\frac{1}{2}, 1\right)$ for Ho\v{r}ava-Lifshitz with non-flat
universe with $\Lambda=0$ and E-potential under the detailed-balance
condition.}
\label{PROPOSITION8}

\begin{prop}
\label{centerP3EmodelCase2DS} The point $(\theta,{T},Q)=\left(0,\frac{1}{2},
1\right)$ of the system \eqref{Case2_uggla-varsX2a}, \eqref{Case2_uggla-varsX2b}, \eqref{Case2_uggla-varsX2c} is unstable
(center-saddle).
\end{prop}

\textbf{Proof.} Taking the linear transformation 
\begin{equation}
(u,v_1,v_2)=\left(\frac{1}{2} (2 T-1), 1-Q, \frac{-4 \mu +\theta \sqrt{n}+8
\mu T}{\sqrt{n}}\right),
\end{equation}
 and taking Taylor series near $(u,v_1,v_2)=(0,0,0)$ up to fifth order we
obtain the system \eqref{Case2_uggla-varsX2a}, \eqref{Case2_uggla-varsX2b}, \eqref{Case2_uggla-varsX2c} can be written into its Jordan
canonical form: %
\begin{equation}
\left(%
\begin{array}{c}
\frac{du}{dN} \\ 
\frac{dv_1}{dN} \\ 
\frac{dv_2}{dN}%
\end{array}%
\right)=\left(%
\begin{array}{ccc}
0 & 0 & 0 \\ 
0 & -2 & 0 \\ 
0 & 0 & -\frac{3}{2}%
\end{array}%
\right)\left(%
\begin{array}{c}
u \\ 
v_1 \\ 
v_2%
\end{array}%
\right)+\left(%
\begin{array}{c}
f(u,\mathbf{v}) \\ 
g_1(u,\mathbf{v}) \\ 
g_2(u,\mathbf{v})%
\end{array}%
\right),  \label{HLcenterCase2_uggla-varsX2}
\end{equation}%
 where 
\begin{widetext}
$f(u,\mathbf{v})=\frac{u^3 \left(-3 n^{3/2} (v_1-1) v_2^2+48 \mu ^2 \sqrt{n} \left((3 n-1) v_2^2+v_1-1\right)-128 \mu ^3 (3 n-1) (v_1-1) v_2-24 \mu  n (v_1-1) v_2\right)}{n^{3/2}}-\frac{64 \mu
   ^2 u^5 \left(8 \mu ^2+3 n \left(-8 \mu ^2+v_1-1\right)\right)}{n^2}+\frac{16 \mu  u^4 \left(-3 n^{3/2} v_2+16 \mu ^3 (3 n-1) (v_1-1)+6 \mu  n (v_1-1)+16 \mu ^2 (1-3 n) \sqrt{n}
   v_2\right)}{n^2}+\frac{u^2 \left(48 \mu ^2 (v_1-1) \left((3 n-1) v_2^2-1\right)+8 \mu  \sqrt{n} v_2 \left((1-3 n) v_2^2-3 v_1+3\right)+3 n (v_1-1) v_2^2\right)}{2
   n}+\frac{u v_2 \left(\sqrt{n} v_2 \left(v_2^2 (3 n+v_1-1)+6 (v_1-1)\right)-16 \mu  (v_1-1) \left((3 n-1) v_2^2-3\right)\right)}{8 \sqrt{n}}+\frac{1}{16} (v_1-1) v_2^2
   \left((3 n-1) v_2^2-6\right),$ \newline
$g_1(u,\mathbf{v})=\frac{2048 (1-3 n) u^4 v_1 \mu ^4}{n}+64 u^3 v_1 \left(3 v_1+\frac{16 (3 n-1) v_2 \mu }{\sqrt{n}}-6\right) \mu ^2-48 u^2 v_1 \left(\sqrt{n} (v_1-2)
   v_2+2 \left((6 n-2) v_2^2+v_1-2\right) \mu \right) \mu +\frac{1}{2} v_1 \left((1-3 n) n v_2^4-3 n (v_1-2) v_2^2+2 v_1\right)+u v_1 \left(3 n^2 v_2^4-n
   v_2^4-6 n v_2^2+3 n v_1 v_2^2+8 \sqrt{n} \left((6 n-2) v_2^2+3 v_1-6\right) \mu  v_2-2 v_1+4\right),$
	\newline
$g_2(u,\mathbf{v})=\frac{512 \mu ^3 \left((5 n (11 n+28)-61) \mu ^2-15 n
   (v_1-1)\right) u^5}{5 n^{5/2}}+\frac{128 \mu ^2 \left(4 (-4 v_1+3 n (4 v_1-3)+2) \mu ^3-2 \sqrt{n} (n (15 n+28)-13) v_2 \mu ^2-2 n (3 n-2) (v_1-1) \mu -3 n^{3/2} v_2\right)
   u^4}{n^{5/2}}+\frac{8 \mu  \left(-32 (-4 v_1+3 n (4 v_1-3)+2) v_2 \mu ^3+2 \sqrt{n} \left((n (65 n+84)-43) v_2^2+4 (6 v_1 n-3 n+8 v_1-8)\right) \mu ^2+12 n (3 n-1) (v_1-1)
   v_2 \mu +3 n^{3/2} v_2^2\right) u^3}{n^2}$\\$+\frac{4 \mu  \left(12 \left((-4 v_1+3 n (4 v_1-3)+2) v_2^2-4 v_1+2\right) \mu ^2+\sqrt{n} v_2 \left((17-7 n (5 n+4)) v_2^2+12 (2
   n-3 (n+1) v_1+3)\right) \mu -3 n (v_1-1) \left(3 n v_2^2-2\right)\right) u^2}{n^{3/2}}$\\$+\frac{\left(-160 v_2 \left((-4 v_1+3 n (4 v_1-3)+2) v_2^2-12 v_1+6\right) \mu ^2+5
   \sqrt{n} \left((n (75 n+28)-25) v_2^4+24 (6 v_1 n-5 n+4 v_1-4) v_2^2-96 v_1\right) \mu +4 n v_2 \left(-v_2^4+5 (3 n+1) (v_1-1) v_2^2-30 (v_1-1)\right)\right)
   u}{40 n}+\frac{v_2 \left(2 \sqrt{n} \left(\left(1-5 n^2\right) v_2^4-5 (3 n+1) (v_1-1) v_2^2+30 v_1\right)+5 v_2 \left((-4 v_1+3 n (4 v_1-3)+2) v_2^2-24
   v_1+12\right) \mu \right)}{40 \sqrt{n}}.$
\end{widetext}
\newline
According to Theorem \ref{existenceCM}, there exists a 1-dimensional
invariant local center manifold$W^{c}\left( \mathbf{0}\right) $ of %
\eqref{HLcenterCase2_uggla-varsX2}, \newline 
$W^{c}\left( \mathbf{0}\right) =\left\{ \left( u,\mathbf{v}\right) \in%
\mathbb{R}\times\mathbb{R}^{2}:\mathbf{v}=\mathbf{h}\left( u\right)\right\}$%
, satisfying $\mathbf{h}\left( 0\right) =\mathbf{0},\;D\mathbf{h}\left(
0\right) =\mathbf{0} ,\;\left\vert u\right\vert <\delta$ for $\delta$
sufficiently small. The restriction of (\ref{HLcenterCase2_uggla-varsX2}) to
the center manifold is $\frac{du}{dN}=f\left( u,\mathbf{h}\left( u\right)
\right), $ where the function $\mathbf{h}\left( u\right)$ satisfies %
\eqref{MaineqcM}: 
\begin{equation}
D\mathbf{h}\left( u\right) \left[ f\left( u,\mathbf{h}\left( u\right)
\right) \right] -P\mathbf{h}\left( u\right) -\mathbf{g}\left( u,\mathbf{h}%
\left( u\right) \right) =0,  \label{HLCase2_uggla-varsX2h}
\end{equation}
where 
\begin{equation*}
P=\left(%
\begin{array}{cc}
-2 & 0 \\ 
0 & -\frac{3}{2}%
\end{array}%
\right).
\end{equation*}

According to Theorem \ref{approximationCM}, the system %
\eqref{HLCase2_uggla-varsX2h} can be solved approximately by expanding $%
\mathbf{h}\left( u\right) $ in Taylor series at $u=0.$ Since $\mathbf{h}%
\left( 0\right) =\mathbf{0\ } $and $D\mathbf{h}\left( 0\right) =\mathbf{0},$
we propose the ansatsz 
\begin{equation}
\mathbf{h}\left( u\right) :=\left(%
\begin{array}{c}
h_{1}\left(u\right) \\ 
h_{2}\left(u\right)%
\end{array}%
\right)=\left(%
\begin{array}{c}
\sum_{j=1}^{4} a_j u^{j+1} +O\left( u^{6}\right) \\ 
\sum_{j=1}^{4} b_j u^{j+1} +O\left( u^{6}\right)%
\end{array}%
\right),
\end{equation}%
Comparing the coefficients of the equal powers of $u$
we find the non-null coefficients \\
$b_1= -\frac{16 \mu  \left(n-4 \mu ^2\right)}{n^{3/2}}, b_2= -\frac{32 \left(288 \mu ^5+3 \mu  n^2+4 \mu ^3 n (3 n-13)\right)}{3
   n^{5/2}}$, \newline
	$b_3= \frac{64 \mu  \left(3648 \mu ^6-n^3+4 \mu ^2 (10-3 n) n^2+48 \mu ^4 n (3 n-16)\right)}{n^{7/2}}$, \newline
	\begin{small}
	$b_4= \frac{128 \mu  \left(-2711040 \mu ^8-15 n^4-360 \mu ^2 (n-3) n^3-8 \mu ^4 n^2 (5 n (13
   n-330)+5761)-960 \mu ^6 n (121 n-687)\right)}{15 n^{9/2}}$.
	\end{small}
\newline		
Therefore, the center manifold can be represented locally by the graph
\begin{widetext}
\begin{align}
&v_1=0, \nonumber\\
&v_2= -\frac{16 \left(\mu  \left(n-4 \mu ^2\right)\right) u^2}{n^{3/2}}  -\frac{32 \left(288 \mu ^5+4 n (3 n-13) \mu ^3+3 n^2 \mu \right) u^3}{3 n^{5/2}} \notag\\
& +\frac{64 \mu  \left(3648
   \mu ^6+48 n (3 n-16) \mu ^4+4 (10-3 n) n^2 \mu ^2-n^3\right) u^4}{n^{7/2}}\nonumber\\
	& +\frac{128 \mu  \left(-2711040 \mu ^8-960 n (121 n-687) \mu ^6-8 n^2 (5 n (13 n-330)+5761) \mu ^4-360 (n-3) n^3 \mu ^2-15 n^4\right) u^5}{15
   n^{9/2}}.
\end{align}
That is, 
\begin{align}
&\theta =-\frac{4 \mu  (2
   T-1)}{\sqrt{n}} -\frac{4 \mu  (2 T-1)^2 \left(n-4 \mu ^2\right)}{n^{3/2}}-\frac{4 (2 T-1)^3 \left(288 \mu ^5+3 \mu  n^2+4 \mu ^3 n (3 n-13)\right)}{3 n^{5/2}}
\nonumber \\& +\frac{4 \mu  (2 T-1)^4 \left(3648 \mu ^6-n^3+4 \mu ^2 (10-3 n)
   n^2+48 \mu ^4 n (3 n-16)\right)}{n^{7/2}} \nonumber \\
	& +\frac{4 \mu  (2 T-1)^5 \left(-2711040 \mu ^8-15 n^4-360 \mu ^2 (n-3) n^3-8 \mu ^4 n^2 (5 n (13 n-330)+5761)-960 \mu ^6 n (121 n-687)\right)}{15 n^{9/2}}, \nonumber\\
&	Q=1.
\end{align}
The dynamics on the center manifold is given by the gradient-like equation 
 \begin{align}
&\frac{du}{dN}=-\nabla \Pi(u),  \notag \\
& \Pi(u)=-\frac{8 \mu ^2 u^3}{n}-\frac{1536 \mu ^4 u^5 \left(13 \mu ^2-2
n\right)}{5 n^3}-\frac{12 \mu ^2 u^4 \left(n-8 \mu ^2\right)}{n^2}  \notag \\
& +\frac{512 u^6 \left(504 \mu ^8+4 \mu ^4 n^2+\mu ^6 n (12 n-101)\right)}{%
n^4}  \notag \\
& -\frac{6144 \mu ^4 u^7 \left(24992 \mu ^6-6 n^3+\mu ^2 (385-72 n) n^2+8
\mu ^4 n (106 n-745)\right)}{7 n^5}.
\end{align}%
\end{widetext}

We have $\Pi^{\prime }(0)=\Pi^{\prime \prime }(0)=0, \Pi^{\prime \prime
\prime }(0)=-\frac{48 \mu ^2}{n}<0$. It follows that $u=0$ is an inflection
of the potential. Using the Theorem \ref{stabilityCM}, we conclude that the
center manifold of origin for the system \eqref{HLcenterCase2_uggla-varsX2},
and the origin are unstable (center-saddle). $\blacksquare$

\section{Case 3: Flat universe with $\Lambda \neq 0$ under the
detailed-balance condition}
\label{Case3}

\begin{table*}[ht]
\caption{Case 3: Equilibrium points at the finite region of the system 
\eqref{GenHLeqxcase3}, \eqref{GenHLequcase3}, \eqref{GenHLeqscase3}.}
\label{HLcrit3}\centering
\begin{tabular*}{\textwidth}{@{\extracolsep{\fill}}lrrrl}
\hline
Equil. & \multicolumn{1}{c}{$(x,u,s)$} & \multicolumn{1}{c}{Existence} & 
\multicolumn{1}{c}{Eigenvalues} & \multicolumn{1}{c}{Stability} \\ 
Points &  &  &  &  \\ \hline
$P_9(\hat{s})$ & $\left(1, 0, \hat{s}\right)$ & $f(\hat{s})=0$ & $6-2 \sqrt{6%
} \hat{s}, 3, -2 \sqrt{6} f^{\prime }\left(\hat{s}\right)$ & nonhyperbolic
for \\ 
&  &  &  & $f^{\prime }\left(\hat{s}\right)=0$, or $\hat{s}=\sqrt{\frac{3}{2}%
}$. \\ 
&  &  &  & source for $f^{\prime }\left(\hat{s}\right)<0, \hat{s}<\sqrt{%
\frac{3}{2}}$. \\ 
&  &  &  & saddle otherwise. \\ \hline
$P_{10}(\hat{s})$ & $\left(-1, 0, \hat{s}\right)$ & $f(\hat{s})=0$ & $6+2 
\sqrt{6} \hat{s}, 3, 2 \sqrt{6} f^{\prime }\left(\hat{s}\right)$ & 
nonhyperbolic for \\ 
&  &  &  & $f^{\prime }\left(\hat{s}\right)=0$, or $\hat{s}=-\sqrt{\frac{3}{2%
}}$. \\ 
&  &  &  & source for $f^{\prime }\left(\hat{s}\right)>0, \hat{s}>-\sqrt{%
\frac{3}{2}}$. \\ 
&  &  &  & saddle otherwise. \\ \hline
$P_{11}(\hat{s})$ & $\left(\sqrt{\frac{2}{3}} \hat{s}, 0, \hat{s}\right)$ & $%
f(\hat{s})=0$ & $2 \hat{s}^2,2 \hat{s}^2-3,-4 \hat{s} f^{\prime }\left(\hat{s%
}\right)$ & nonhyperbolic for \\ 
&  & $-\sqrt{\frac{3}{2}}\leq \hat{s}\leq \sqrt{\frac{3}{2}}$ &  & $%
f^{\prime }\left(\hat{s}\right)=0$, or $\hat{s}\in \left\{-\sqrt{\frac{3}{2}}%
, 0, \sqrt{\frac{3}{2}}\right\}$. \\ 
&  &  &  & saddle otherwise. \\ \hline
$P_{11}^{0}(u_c):$ & $(0, u_c, 0)$. & $u_c\in\mathbb{R}$ & $0, -\frac{3}{2}
\pm \frac{1}{2} \sqrt{9-48 f(0) \left(u_c^2+1\right)}$ & nonhyperbolic. \\ 
\hline
\end{tabular*}
\centering
\caption{Case 3: Equilibrium points at the infinity region of the system 
\eqref{GenHLeqxcase3}, \eqref{GenHLequcase3}, \eqref{GenHLeqscase3}.}
\label{HLcrit3infinity}
\begin{tabular*}{\textwidth}{@{\extracolsep{\fill}}lrrrl}
\hline
Equil. & \multicolumn{1}{c}{$(x,U,S)$} & \multicolumn{1}{c}{Existence} & 
\multicolumn{1}{c}{Eigenvalues} & \multicolumn{1}{c}{Stability} \\ 
Points &  &  &  &  \\ \hline
$Q_{19,20}(\hat{s})$ & $\left(-1, \pm 1, \frac{2}{\pi}\arctan(\hat{s})\right)
$ & $f(\hat{s})=0$ & $-\infty , 2 \sqrt{6} \hat{s}, 2 \sqrt{6} f^{\prime
}\left(\hat{s}\right)$ & sinks for $\hat{s}<0, f^{\prime }\left(\hat{ s}%
\right)<0$. \\ \hline
$Q_{21,22}(\hat{s})$ & $\left(1, \pm 1, \frac{2}{\pi}\arctan(\hat{s})\right)$
& $f(\hat{s})=0$ & $-\infty, -2 \sqrt{6} \hat{s}, -2 \sqrt{6} f^{\prime
}\left(\hat{s}\right)$ & sinks for $\hat{s}>0, f^{\prime }\left(\hat{s}%
\right)>0$. \\ \hline
$Q_{23,24}$ & $\left(0, \pm 1, 0\right)$ & always & $0,-2 i \sqrt{3} \sqrt{%
f(0)},2 i \sqrt{3} \sqrt{ f(0)}$ & Saddle for $f(0)<0$. Center for $f(0)>0$.
\\ \hline
\end{tabular*}%
\end{table*}

\begin{figure*}[ht!]
\centering
\includegraphics[scale=1.5]{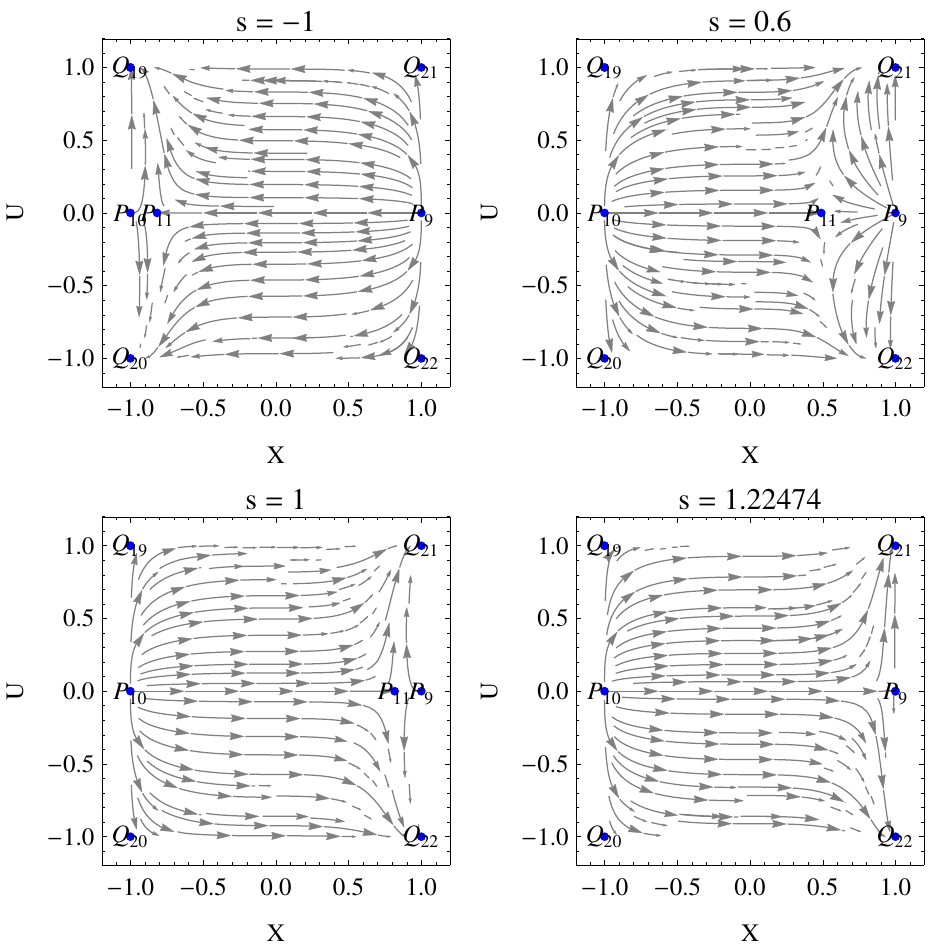} 
\caption{\textit{{(Color online) Compact phase portrait of \eqref{HLinfty3a}, \eqref{HLinfty3b}
for different choices of the parameter $s$.}}}
\label{fig:Case3HL}
\end{figure*}

The field equations of HL for a Flat universe with $\Lambda \neq 0$ under
the detailed-balance condition are: 
\label{Case3syst7}
\begin{align} 
& H^2= \frac{1}{6(3\lambda-1)}\left[\frac{3 \lambda -1}{4}\dot\phi^2+V(\phi)%
\right]  \notag \\
& -\frac{\mu^2\Lambda ^2}{16(3\lambda-1)^2}, \label{Case3syst7a}\\
&\dot{H}+\frac{3}{2}H^2= -\frac{1}{4(3\lambda-1)}\left[\frac{3 \lambda -1}{4}%
\dot\phi^2-V(\phi)\right]  \notag \\
& -\frac{3\mu^2\Lambda ^2}{32(3\lambda-1)^2}, \label{Case3syst7b}\\
& \ddot\phi +3 H \dot \phi+\frac{2 V^{\prime }(\phi)}{3 \lambda -1}=0. \label{Case3syst7c}
\end{align}

\subsection{Arbitrary Potential}
\label{Case3A} 
In this case the system \eqref{Case3syst7a}, \eqref{Case3syst7b}, \eqref{Case3syst7c} is given in its autonomous form: 
\begin{align}
&\frac{dx}{dN}= \sqrt{6} s \left(u^2-x^2+1\right)+3 x \left(x^2-1\right),
\label{GenHLeqxcase3} \\
&\frac{du}{dN}= 3 u x^2,  \label{GenHLequcase3} \\
&\frac{ds}{dN}= -2 \sqrt{6} x f(s).  \label{GenHLeqscase3}
\end{align}
defined on the phase space $\{(x,u,s)\in\mathbb{R}^3: x^2-u^2 \leq 1\}.$

The equilibrium points/curves at the finite region of the phase space of \eqref{GenHLeqxcase3}, \eqref{GenHLequcase3}, \eqref{GenHLeqscase3} are
presented in Table \ref{HLcrit3}, where is shown the existence and stability
conditions. We proceed to the discussion of the more relevant features of
them.

\begin{itemize}
\item $P_9(\hat{s}): (x,u,s)=\left(1, 0, \hat{s}\right)$, where $f(\hat{s})=0
$. It is a source for $f^{\prime }\left(\hat{s}\right)<0, \hat{s}<\sqrt{%
\frac{3}{2}}$.

\item $P_{10}(\hat{s}): (x,u,s)=\left(-1, 0, \hat{s}\right)$, where $f(\hat{s%
})=0$. It is a source for $f^{\prime }\left(\hat{s}\right)>0, \hat{s}>-\sqrt{%
\frac{3}{2}}$.

\item $P_{11}(\hat{s}): (x,u,s)=\left(\sqrt{\frac{2}{3}} \hat{s}, 0, \hat{s}%
\right)$, where $f(\hat{s})=0$, $-\sqrt{\frac{3}{2}}\leq \hat{s}\leq \sqrt{%
\frac{3}{2}}$. It is a saddle.

\item $P_{11}^{0}(u_c): (x,u,s)=(0, u_c, 0)$. This line of equilibrium
points is new, and it was not found in \cite{Leon:2009rc}. It is
nonhyperbolic.

\item There are two lines of equilibrium points
$P_{12,13}(s_c): (x,u,s)=(0,\pm i, s_c), s_c\in \mathbb{R} $ which are not
considered since they are complex valued.
\end{itemize}

Due to the dynamical system \eqref{GenHLeqxcase3}, \eqref{GenHLequcase3}, %
\eqref{GenHLeqscase3} is unbounded, we introduce the new variables 
\begin{equation}
X= \frac{x}{\sqrt{1+u^2}}, \quad U=\frac{u}{\sqrt{1+u^2}}, \quad S=\frac{2}{%
\pi}\arctan(s),
\end{equation}
and the time rescaling $\frac{df}{d\tau}= \sqrt{1-U^2} \frac{d f}{dN}$ to
obtain the dynamical system 
\begin{align}
&\frac{dX}{d\tau}=\left(X^2-1\right) \left(3 \sqrt{1-U^2} X-\sqrt{6} \,\tan
\left(\frac{\pi S}{2}\right)\right), \label{Example3Sa}\\
&\frac{dU}{d\tau}=3 U \sqrt{1-U^2} X^2, \label{Example3Sb}\\
&\frac{dS}{d\tau}=-\frac{2 \sqrt{6} X (\cos (\pi S)+1) f\left(\tan \left(%
\frac{\pi S}{2}\right)\right)}{\pi },  \label{Example3Sc}
\end{align}
defined on the compacted phase space 
\begin{equation*}
\{(X, U, S)\in\mathbb{R}^3: -1\leq X\leq 1, -1\leq U \leq 1, -1\leq S \leq
1\}.
\end{equation*}

The points at the infinite region of the phase space are summarized in table %
\ref{HLcrit3infinity}. Now we discuss the relevant features of them.
\begin{itemize}
\item $Q_{19,20}(\hat{s}): (X, U, S)=\left(-1, \pm 1, \frac{2}{\pi}\arctan(%
\hat{s})\right)$. It is a sink for $\hat{s}<0, f^{\prime }\left(\hat{s}%
\right)<0$

\item $Q_{21,22}(\hat{s}): (X, U, S)=\left(1, \pm 1, \frac{2}{\pi}\arctan(%
\hat{s})\right)$ , where $f(\hat{s}) =0$. It is a sink for $\hat{s}>0,
f^{\prime }\left(\hat{s}\right)>0$.

\item $Q_{23,24}: (X, U, S)= \left(0, \pm 1, 0\right)$. Always exists. It is
a Saddle for $f(0)<0$. It is a Center for $f(0)>0$.
\end{itemize}

\subsection{Exponential Potential}
\label{Cas3B}
In this example the system \eqref{Example3Sa}, \eqref{Example3Sb} becomes 
\begin{align}
&\frac{dX}{d\tau}=\left(X^2-1\right) \left(3 X\sqrt{1-U^2} -\sqrt{6}
s\right),  \label{HLinfty3a} \\
&\frac{dU}{d\tau}=3 X^2 U \sqrt{1-U^2} ,  \label{HLinfty3b}
\end{align}
defined on the compact phase space\newline
$\{(X,U)\in\mathbb{R}^2: -1\leq X \leq 1, -1\leq U \leq 1\}.$

Under this scenario, the Ho\v{r}ava-Lifshitz universe admits two unstable
equilibrium points ($P_{9,10}$), completely dominated by stiff dark matter.

Point $P_{11}$ exhibits a more physical dark matter equation-of-state
parameter, but still with negligible dark energy at late times.

Figure \ref{fig:Case3HL} a) illustrates when $Q_{19,20}: (X, U)=\left(-1,
\pm 1\right)$ is a sink for $s<0$. Figures \ref{fig:Case3HL} b), c) and d)
illustrates when $Q_{21,22}: (X, U)=\left(1, \pm 1\right)$ is a sink for $s>0
$.

\subsection{Powerlaw Potential}

\label{Case3C}

In this example the system \eqref{Example3Sa}, \eqref{Example3Sb}, \eqref{Example3Sc} becomes
\begin{align}
&\frac{dX}{d\tau}=\left(X^2-1\right) \left(3 \sqrt{1-U^2} X-\sqrt{6} \,\tan
\left(\frac{\pi S}{2}\right)\right), \label{SYSTEM82a}\\
&\frac{dU}{d\tau}=3 U \sqrt{1-U^2} X^2, \label{SYSTEM82b}\\
&\frac{dS}{d\tau}=-\frac{\sqrt{6} X (\cos (\pi S)-1)}{\pi n}, \label{SYSTEM82c}
\end{align}
defined on the compact phase space 

\begin{equation*}
\Big\{(X, U, S)\in\mathbb{R}^3: -1\leq X\leq 1, -1\leq U \leq 1, -1\leq S
\leq 1\Big\}.
\end{equation*}

The coordinates equilibrium points of \eqref{SYSTEM82a}, \eqref{SYSTEM82b}, \eqref{SYSTEM82c} and their stability
conditions are summarized as follows:

\begin{itemize}
\item $(x,U,S)=(0,U_c,0)$ with eigenvalues $\left\{0,0,-3 \sqrt{1-U_c^2}%
\right\}$.
Non-hyperbolic with a 1 dimensional stable manifold.

\item $(x,U,S)=(-1,0,0)$ with eigenvalues $\{6,3,0\}$. \newline 
Non-hyperbolic with a 2 dimensional unstable manifold.

\item $(x,U,S)=(1,0,0)$ with eigenvalues $\{6,3,0\}$. \newline 
Non-hyperbolic with a 2 dimensional unstable manifold.

\item $(x,U,S)=(-1, -1, 0)$ with eigenvalues $\{0,-\infty ,0\}$.
Non-hyperbolic with a 1 dimensional stable manifold.

\item $(x,U,S)=(-1, 1, 0)$ with eigenvalues $\{0,-\infty ,0\}$.
Non-hyperbolic with a 1 dimensional stable manifold.

\item $(x,U,S)=(1, -1, 0)$ with eigenvalues $\{0,-\infty ,0\}$.
Non-hyperbolic with a 1 dimensional stable manifold.

\item $(x,U,S)=(1, 1, 0)$ with eigenvalues $\{0,-\infty ,0\}$.
Non-hyperbolic with a 1 dimensional stable manifold.
\end{itemize}

\subsubsection{Alternative compactification}

\begin{figure*}[ht!]
\centering
\subfigure[]{\includegraphics[width=2.5in, height=2.5in,angle=-90]{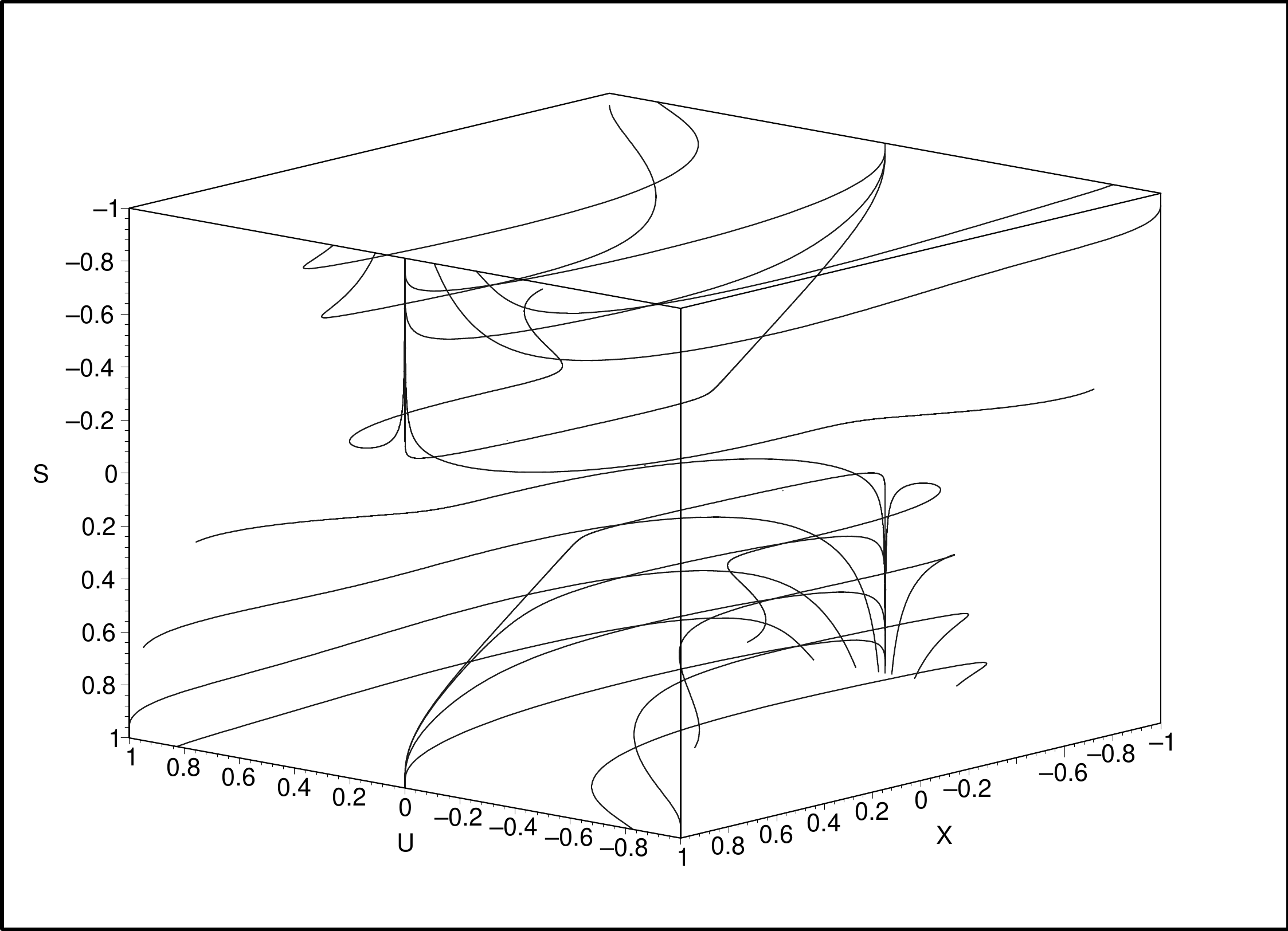}	} 
\hspace{2cm}  \subfigure[]{\includegraphics[width=2.5in,
height=2.5in,angle=-90]{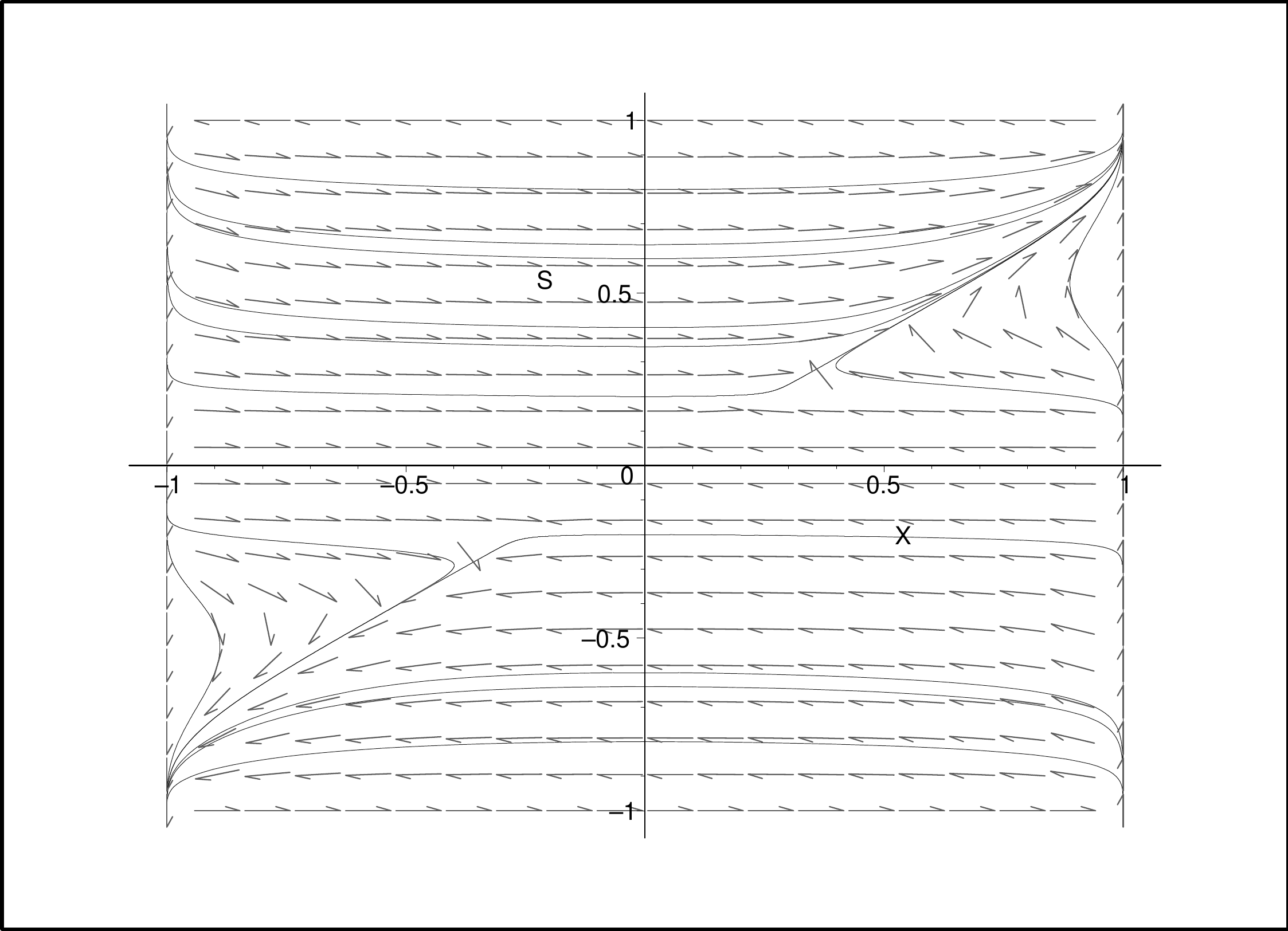}}  \subfigure[]{\includegraphics[width=2.5in,
height=2.5in,angle=-90]{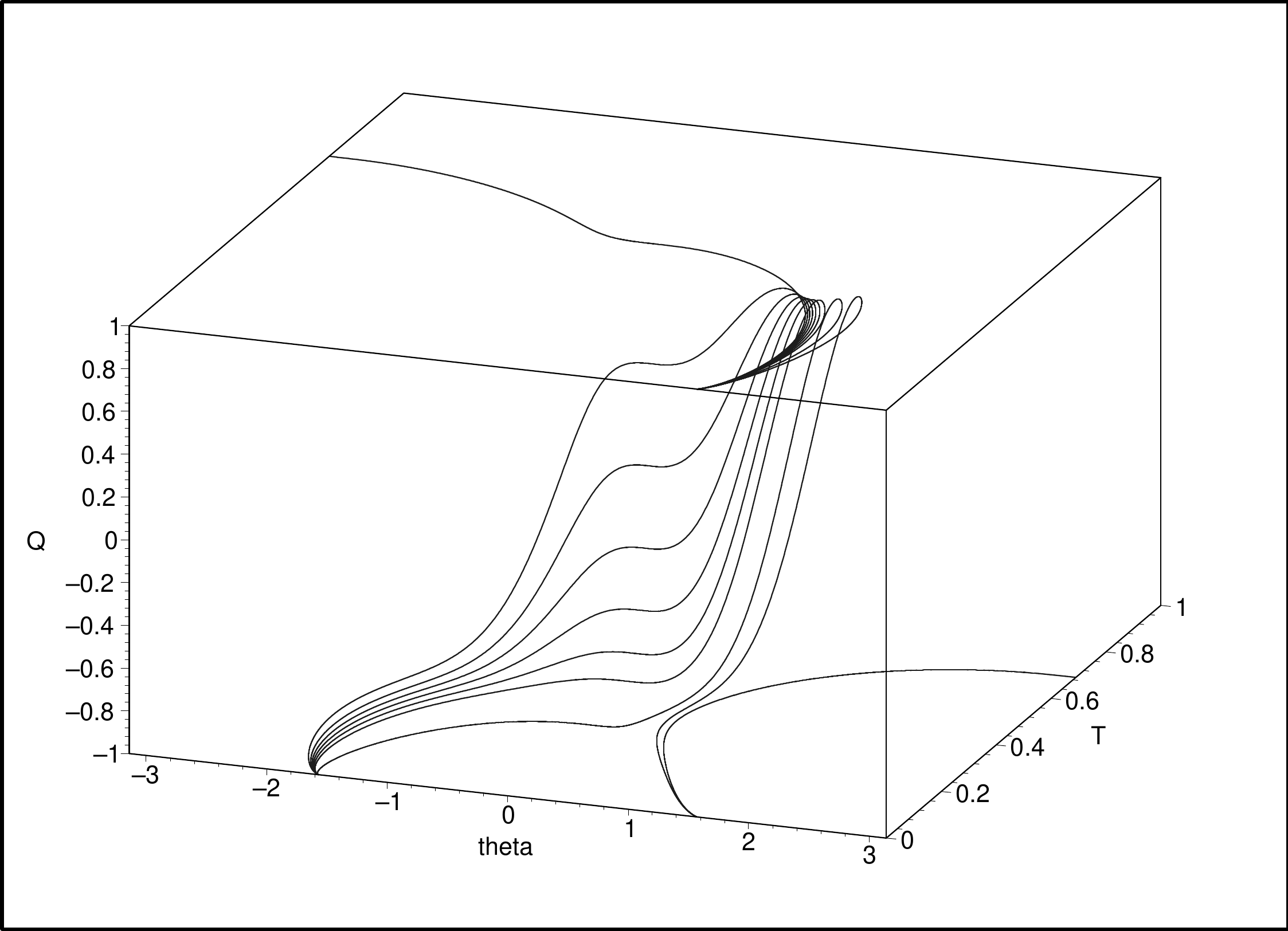}	} \hspace{2cm}  \subfigure[]{%
\includegraphics[width=2.5in, height=2.5in,angle=-90]{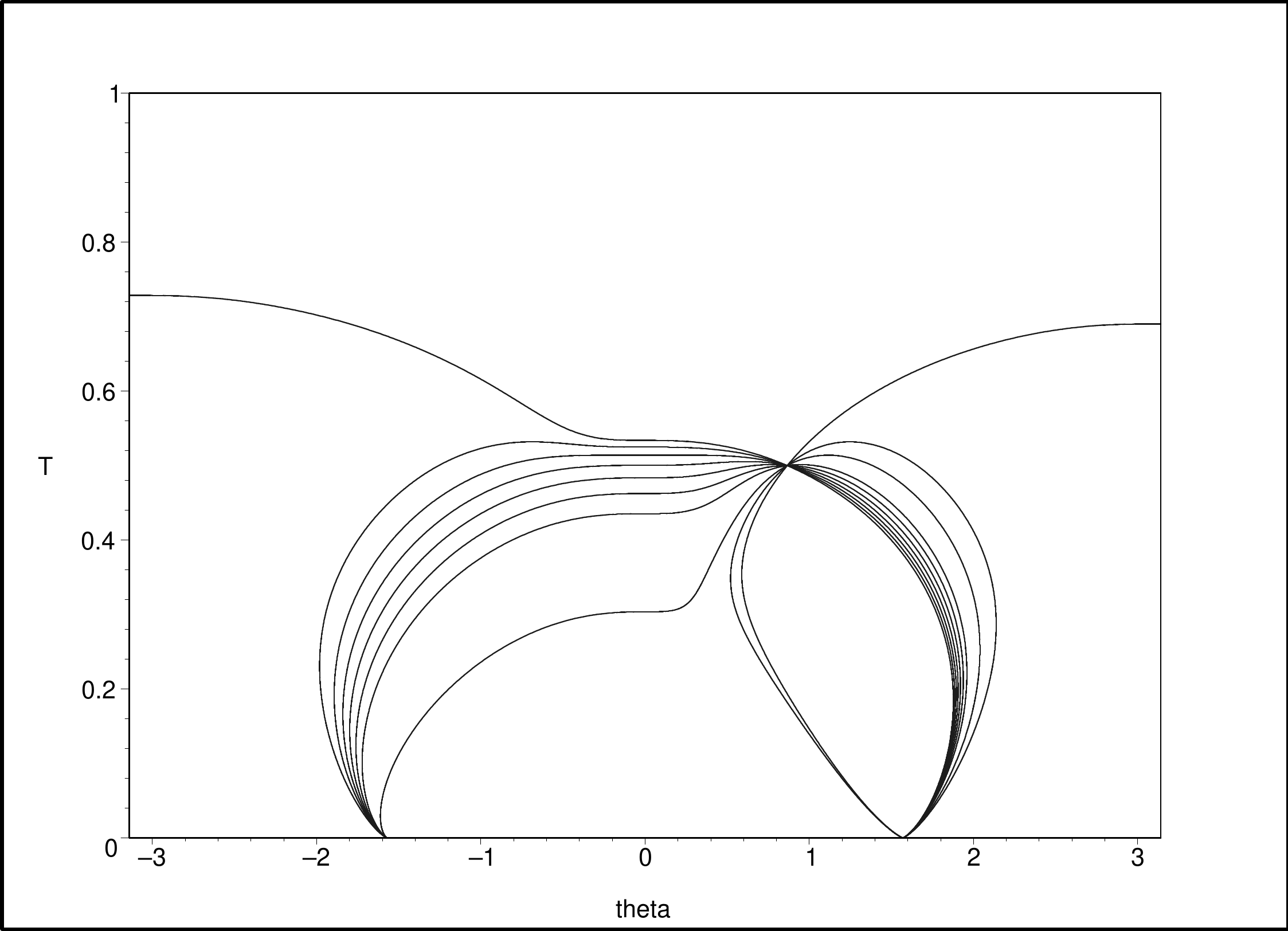}}  %
\subfigure[]{\includegraphics[width=2.5in, height=2.5in,angle=-90]{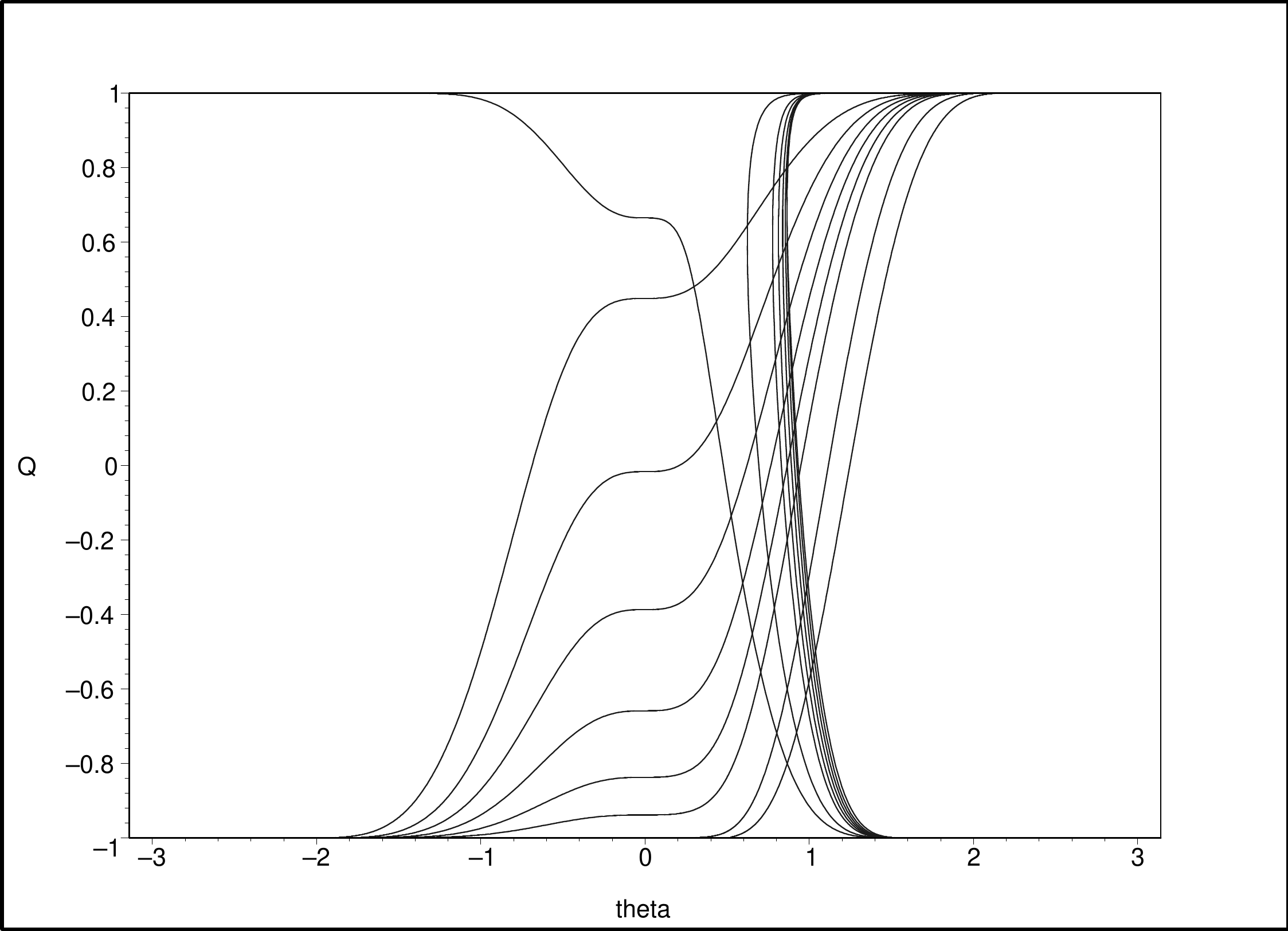}	} 
\hspace{2cm}  \subfigure[]{\includegraphics[width=2.5in,
height=2.5in,angle=-90]{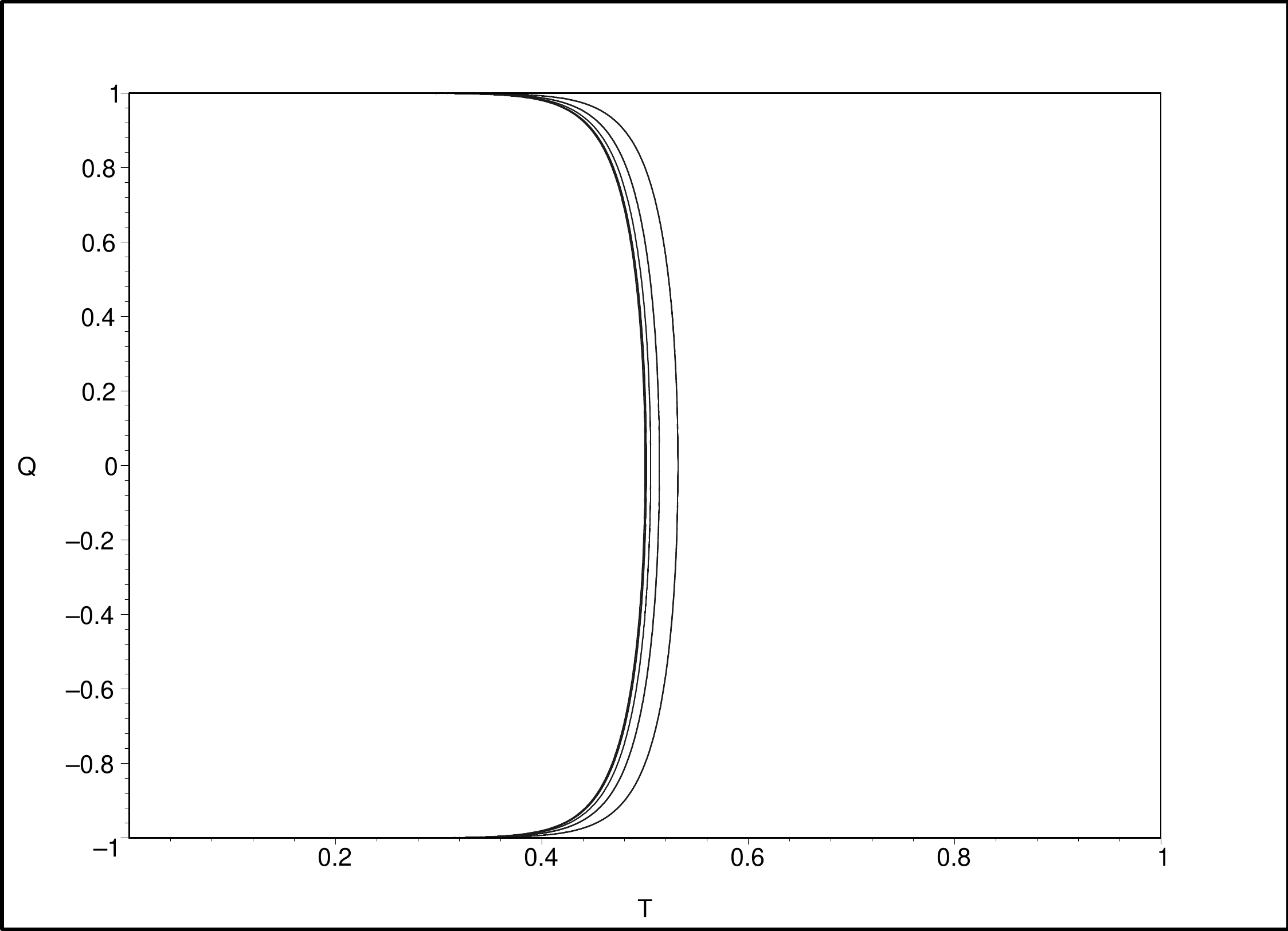}} 
\caption{\textit{{(a) Compact phase portrait of \eqref{SYSTEM82a}, \eqref{SYSTEM82b}, \eqref{SYSTEM82c} for the
choice $n=2$. (b) Dynamics in the invariant set $U=0$. (c)-(f): Dynamics of 
\eqref{system517a}, \eqref{system517b}, \eqref{system517c}, and some 2D projections of the solutions for $n=4, \protect%
\mu=1$.}}}
\label{fig:Case3HLPL}
\end{figure*}

Introducing the following compactification
 
\begin{align}
& \hat{H}=\sqrt{H^2+\frac{\mu^2 \Lambda^2}{16(3\lambda-1)^2}}, \\
&Q=\frac{H}{\hat{H}}, \\
&\Sigma=\frac{\dot\phi}{2 \sqrt{6} \hat{H}}, \\
&Y=\frac{\mu \phi}{2^{\frac{1}{n}} 3^{\frac{1}{2 n}} ((3 \lambda -1) n)^{%
\frac{1}{2 n}} \hat{H}^{\frac{1}{n}}}, \\
& {T}= \frac{c}{c+\hat{H}^{\frac{1}{n}}}, \quad c=2^{\frac{3}{2}-\frac{1}{n}%
} 3^{\frac{n-1}{2 n}} n^{-\frac{1}{2 n}} (3 \lambda -1)^{-\frac{1}{2 n}}\mu,
\\
&\Sigma+ Y^{2n}=1.
\end{align}
such that 

\begin{align}
&\dot\phi = 2 \sqrt{6} \Sigma c^n \left(\frac{1}{ {T}}-1\right)^n, \\
&\phi=\frac{2^{\frac{1}{n}} 3^{\frac{1}{2 n}} n^{\frac{1}{2 n}} \left(\frac{1%
}{ {T}}-1\right) Y (3 \lambda -1)^{\frac{1}{2 n}}c }{\mu }, \\
& \Lambda= \frac{4 (3 \lambda -1) \sqrt{1-Q^2} c^n \left(\frac{1}{ {T}}%
-1\right)^n}{\mu }, \\
&\hat{H}=c^n \left(\frac{1}{ {T}}-1\right)^n,
\end{align}
and the new time variable $\frac{d\hat{\tau}}{d t}=\hat{H}(1- {T})^{-1}$, we
obtain the dynamical system 

\begin{align}
&\frac{d\Sigma}{d\hat{\tau}}=-n {T} Y^{2 n-1}-Q\Sigma( {T}-1) \left((q+1)
Q^2-3\right), \\
&\frac{dY}{d\hat{\tau}}=\Sigma {T}-\frac{(q+1) Q^3 ( {T}-1) Y}{n}, \\
&\frac{d {T}}{d\hat{\tau}}=\frac{(q+1) Q^3 ( {T}-1)^2 {T}}{n}, \\
&\frac{dQ}{d\hat{\tau}}= -(q+1) Q^2 \left(Q^2-1\right) ( {T}-1).
\end{align}
where
\begin{equation}
Q^2 q= \frac{1}{2} \left(-3 Y^{2 n}-2 Q^2+3 \Sigma ^2+3\right)=3 \Sigma
^2-Q^2.
\end{equation}

Introducing the complementary global transformation 
\begin{align}
&\Sigma=F(\theta ) \sin (\theta ), \quad Y=\cos (\theta ),  \notag \\
& F(\theta )=\sqrt{\frac{1-\cos ^{2 n}(\theta )}{1-\cos ^2(\theta )}},
\end{align}
we obtain the following unconstrained 3D dynamical system

\begin{align}
&\frac{d\theta}{d\hat{\tau}}=\frac{3 Q ( {T}-1) F(\theta )^2 \sin (2\theta )%
}{2n}- {T} F(\theta ), \label{system517a}\\
&\frac{d {T}}{d\hat{\tau}}=\frac{3 Q ( {T}-1)^2 {T} \left(1-\cos ^{2
n}(\theta )\right)}{n}, \label{system517b}\\
&\frac{dQ}{d\hat{\tau}}=-3 \left(Q^2-1\right) ( {T}-1) \left(1-\cos ^{2
n}(\theta )\right). \label{system517c}
\end{align}
where 
\begin{equation}
Q^2 q= -3 \cos ^{2 n}(\theta )-Q^2+3.
\end{equation}
The equilibrium points of \eqref{system517a}, \eqref{system517b}, \eqref{system517c} and their stability conditions
are summarized as follows:

\begin{itemize}
\item ${}_{\pm} M_{+}$: $(\theta, T, Q)=(\frac{\pi}{2}+2 k \pi, 0, \pm 1)$. $%
(\Sigma, Y)=(1,0)$. The eigenvalues are $\left\{\pm 6, \pm \frac{3}{n}, \pm 
\frac{3}{n}\right\}$. ${}_{-} M_{+}$ is a sink whereas ${}_{+} M_{+}$ is a
source.

\item ${}_{\pm} M_{-}$: $(\theta, T, Q)=( -\frac{\pi}{2}+2 k\pi, 0, \pm 1)$. 
$(\Sigma, Y)=(-1,0)$. In the last two cases we used the Kernel: ${}_{\text{%
sign}(Q)} M_{\text{sign}(\Sigma)}$. The eigenvalues are $\left\{\pm 6, \pm 
\frac{3}{n}, \pm \frac{3}{n}\right\}$. ${}_{-} M_{-}$ is a sink whereas $%
{}_{+} M_{-}$ is a source.

\item $dS^{\pm}$: $(\theta, T, Q)=(2 k\pi, 0, \pm 1)$; $(\Sigma, Y)=(0,1)$
and $dS^{\pm}_{-}$: $(\theta, T, Q)=((2 k+1)\pi, 0, \pm 1)$; $(\Sigma,
Y)=(0,-1)$. The eigenvalues are $\mp 3,0,0$. Nonhyperbolic, 2 dimensional
center manifold. In general we have the line $dS_+(Q^*)$: $(\theta, T, Q)=(2
k\pi, 0, Q^{*})$; $(\Sigma, Y)=(0,1)$ and $dS_{-}(Q^{*})$: $(\theta, T,
Q)=((2 k+1)\pi, 0, Q^{*})$; $(\Sigma, Y)=(0,-1)$. The eigenvalues are $-3
Q^{*},0,0$. Nonhyperbolic, 2 dimensional center manifold.
\end{itemize}

Substituting ${T}=1$ in the above equations we obtain $\frac{d\theta}{d\hat{%
\tau}}=-F(\theta )$. This equation can be integrated in quadratures as 
\begin{eqnarray}
\cos (\theta ) \, _2F_1\left(\frac{1}{2},\frac{1}{2 n};1+\frac{1}{2 n};\cos
^{2 n}(\theta )\right)=\left(\hat{\tau} -\hat{\tau}_{0}\right).
\end{eqnarray}

In the figure \ref{fig:Case3HLPL} it is shown (a) a compact phase portrait
of the system \eqref{SYSTEM82a}, \eqref{SYSTEM82b}, \eqref{SYSTEM82c} for the choice $n=2$. (b) Dynamics in the invariant set $%
U=0$. In Fig. \ref{fig:Case3HLPL} (c)-(f) it is presented the dynamics of the system %
\eqref{system517a}, \eqref{system517b}, \eqref{system517c} and some 2D projections of the solutions for $n=4, \mu=1$.
\bigskip

\subsubsection{Center manifold of the de Sitter solution for: (i) $dS^{+}: (%
\protect\theta, T, Q)=(0, 0, 1)$, (ii) $dS^{+}(Q^{*}):(\protect\theta, T,
Q)=(0, 0, Q^{*}) , Q^{*}\neq \pm 1,0$, for Ho\v{r}ava-Lifshitz with flat
universe with $\Lambda \neq 0$ and powerlaw potential under the
detailed-balance.}

\label{PROPOSITION9}

\begin{prop}
\label{centerP3PLCase3DS} (i) The point $dS^+$: $(\theta, T, Q)=(0, 0, 1)$
of the system \eqref{system517a}, \eqref{system517b}, \eqref{system517c} is unstable (saddle point). (ii) The line of
fixed points $dS_{+}(Q^{*})$ :$(\theta, T, Q)=(0, 0, Q^{*}) , Q^{*}\neq \pm
1, 0$ of the system \eqref{system517a}, \eqref{system517b}, \eqref{system517c} is unstable (saddle).
\end{prop}

\textbf{Proof Part (i)}: Taking the linear transformation 
\begin{align}
& (u_1, u_2, v)=\left(T,1-Q,\frac{1}{3} \left(3 \theta +\sqrt{n}
T\right)\right),  \notag \\
& \;\;\;\;\;\;\; \;\;\;\;\;\;\; \;\;\;\;\;\;\; u_1\in[0,1], u_2\in[0,2],
\end{align}
and taking Taylor series near $(u_1,u_2,v)=(0,0,0)$ up to fifth order we
obtain the system \eqref{system517a}, \eqref{system517b}, \eqref{system517c} can be written into its Jordan canonical
form: %
\begin{equation}
\left(%
\begin{array}{c}
\frac{du_1}{dN} \\ 
\frac{du_2}{dN} \\ 
\frac{d v}{dN}%
\end{array}%
\right)=\left(%
\begin{array}{ccc}
0 & 0 & 0 \\ 
0 & 0 & 0 \\ 
0 & 0 & -3%
\end{array}%
\right)\left(%
\begin{array}{c}
u_1 \\ 
u_2 \\ 
v%
\end{array}%
\right)+\left(%
\begin{array}{c}
f_1(\mathbf{u},v) \\ 
f_2(\mathbf{u},v) \\ 
g(\mathbf{u},v)%
\end{array}%
\right),  \label{HLcenterCase3_system517}
\end{equation}%
where
\begin{widetext}
$f_1(\mathbf{u},v)=\frac{1}{162} n \left(-3 n^2+n+54\right) u_1^5+u_1^3 \left(\frac{1}{3} \left(\left(-3 n^2+n+9\right) v^2+12 \sqrt{n} v+n\right)+u_2 \left(-4 \sqrt{n} v-\frac{n}{3}\right)\right)+u_1^4
   \left(\frac{2}{3} u_2 \left(3 \sqrt{n} v+n\right)+\frac{2}{27} (n (3 n-1)-27) \sqrt{n} v-\frac{2 n}{3}\right)+u_1^2 \left(2 u_2 v \left(\sqrt{n}+3 v\right)+\frac{2}{3} v \left(\sqrt{n} \left((3 n-1)
   v^2-3\right)-9 v\right)\right)+u_1 \left(\frac{1}{2} (1-3 n) v^4-3 u_2 v^2+3 v^2\right)$,\\ 
	$f_2(\mathbf{u},v)=u_1 \left(u_2 \left(4 n^{5/2} v^3-\frac{4}{3} n^{3/2} \left(v^2+3\right) v-6 n v^2\right)+n u_2^2 v
   \left(2 \sqrt{n}+3 v\right)\right)+\frac{1}{81} (1-3 n) n^3 u_1^4 u_2+u_1^3 \left(\frac{n^2 u_2^2}{3}+\frac{2}{27} n^2 u_2 \left(2 \sqrt{n} (3 n-1) v-9\right)\right)+u_1^2
   \left(u_2^2 \left(-2 n^{3/2} v-\frac{n^2}{3}\right)+u_2 \left(4 n^{3/2} v-2 n^3 v^2+\frac{2}{3} n^2 \left(v^2+1\right)\right)\right)-3 n u_2^2 v^2+n u_2 v^2 \left((1-3 n)
   v^2+6\right)$,\\
$g(\mathbf{u},v)=\frac{n^{3/2} (n (5 n (11 n-38)+39)+4320) u_1^5}{38880}+u_1^4 \left(\frac{1}{54} (11-3 n) n^{3/2} u_2+\frac{1}{54} (3 n-11) n^{3/2}-\frac{1}{648} (n (3 n-7) (5 n-1)+432) n
   v\right)+u_1^3 \left(\frac{1}{54} u_2 \left((3 n-5) n^{3/2}+9 (3 n-7) n v\right)+\frac{1}{432} \left(4 (7-3 n) n^{3/2}+\sqrt{n} (n (n (65 n-114)+17)+432) v^2+72 (7-3 n) n v\right)\right)+u_1^2
   \left(\frac{1}{2} \sqrt{n} u_2 \left(2-(n-1) v \left(\sqrt{n}+3 v\right)\right)-\frac{1}{72} (n-1) n (35 n-3) v^3+\frac{3}{2} (n-1) \sqrt{n} v^2+\frac{1}{3} (n-1) n v-\sqrt{n}\right)$\\$+u_1 \left(\frac{1}{2}
   u_2 \left((3 n+1) v^3+\sqrt{n} (3 n-1) v^2-2 \sqrt{n}-6 v\right)+\frac{1}{96} v \left(\sqrt{n} (n (75 n-38)-5) v^3-48 (3 n+1) v^2+24 (1-5 n) \sqrt{n} v+288\right)\right)+u_2 \left(3 v-\frac{1}{2} (3 n+1)
   v^3\right)+\frac{1}{10} v^3 \left(-5 n \left(n v^2-3\right)+v^2+5\right)$.
\end{widetext}
According to Theorem \ref{existenceCM}, there exists a 2-dimensional
invariant local center manifold $W^{c}\left( \mathbf{0}\right) $ of %
\eqref{HLcenterCase3_system517}, \newline
$W^{c}\left(\mathbf{0}\right) =\left\{\left(\mathbf{u},v\right) \in\mathbb{R}%
^{2}\times\mathbb{R}:v=h\left( \mathbf{u}\right)\right\}$,
satisfying\newline $\mathbf{h}\left(\mathbf{0}\right)= 0,\;D h\left(\mathbf{0}%
\right) =0,\;\left\vert \mathbf{u}\right\vert <\delta$ for $\delta$
sufficiently small. The restriction of (\ref{HLcenterCase3_system517}) to
the center manifold is $\frac{d\mathbf{u}}{dN}=\mathbf{f}\left(\mathbf{u}%
,h\left(\mathbf{u}\right)\right)$, where the function $h\left(\mathbf{u}%
\right)$ satisfies \eqref{MaineqcM}:%
\begin{equation}
Dh\left(\mathbf{u}\right) \cdot \mathbf{f}\left(\mathbf{u},h\left( \mathbf{u}%
\right) \right) +3 h\left( \mathbf{u}\right) -g\left(\mathbf{u}, h\left(%
\mathbf{u}\right) \right) =0.  \label{HLcenterCase3_system517h}
\end{equation}

According to Theorem \ref{approximationCM}, the system %
\eqref{HLcenterCase3_system517h} can be solved approximately by expanding $%
h\left(\mathbf{u}\right) $ in Taylor series at $\mathbf{u}=\mathbf{0}.$ We
propose the ansatz 
\begin{align}  \label{ansatz}
&h(u_1,u_2)=a_{1} u_1^2+a_{10} u_1 u_2^2+a_{11} u_1^2 u_2^2+a_{12} u_1^3
u_2^2 +a_{13} u_2^3  \notag \\
& +a_{14} u_1 u_2^3+a_{15} u_1^2 u_2^3+a_{16} u_2^4+a_{17} u_1 u_2^4 +a_{18}
u_2^5+a_{2} u_1^3  \notag \\
&+a_{3} u_1^4+a_{4} u_1^5+a_{5} u_1 u_2+a_{6} u_1^2 u_2 +a_{7} u_1^3
u_2+a_{8} u_1^4 u_2  \notag \\
& +a_{9} u_2^2 +\mathcal{O}(\left\vert \mathbf{u}\right\vert^6)
\end{align}

\begin{figure}[t!]
\includegraphics[width=0.5\textwidth]{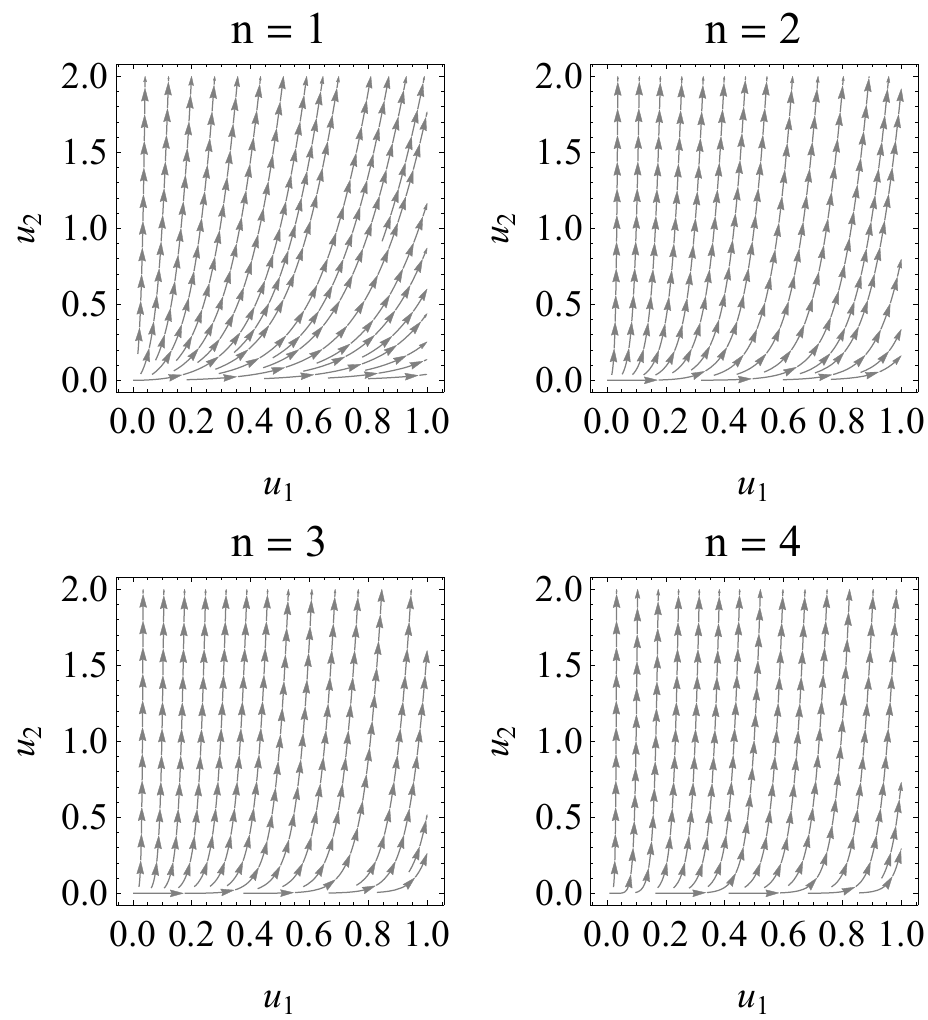}   
\caption{Numerical solutions of the system \eqref{Center2DCase3PLa}, \eqref{Center2DCase3PLb}.}
\label{fig:Center2DCase3PL}
\end{figure}

\begin{figure}[t!]
\centering
\includegraphics[width=0.5\textwidth]{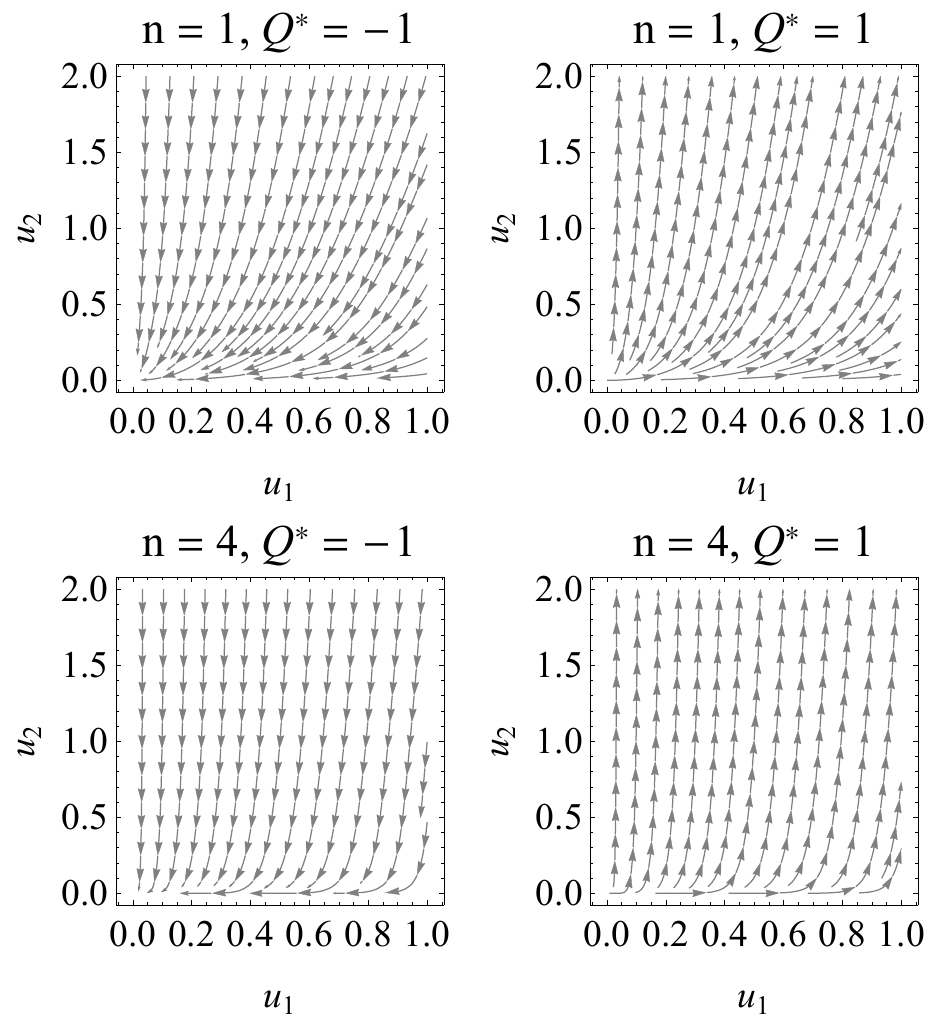}  
\caption{Numerical solutions of the system \eqref{Center2DCase3PLGenerala}, \eqref{Center2DCase3PLGeneralb}.}
\label{fig:Center2DCase3PLGeneral}
\end{figure}

By comparing the coefficients of the equal powers of $u_1, u_2$ we find the
non-null coefficients\newline $a_1= -\frac{\sqrt{n}}{3}$,
$a_2=-\frac{1}{324} \sqrt{n} (n (3 n-7)+108)$, \newline
$a_3= -\frac{1}{108} \sqrt{n} (n (3 n-7)+36)$, \newline
$a_4= -\frac{\sqrt{n} \left(n \left(n \left(65 n^2-570
n+7561\right)-15120\right)+38880\right)}{116640}$, \newline
$a_5=-\frac{\sqrt{n}}{3}, a_6= -\frac{\sqrt{n}}{3}, a_7= \frac{1}{108} \sqrt{%
n} (n (5 n+7)-36)$,\newline
$a_8= \frac{1}{36} (n-1) \sqrt{n} (5  n+12), a_{10}= -\frac{\sqrt{n}}{3},
a_{11}= -\frac{\sqrt{n}}{3}$,\newline
$a_{12}= \frac{1}{54} \sqrt{n} (n (15 n+7)-18), a_{14}= -\frac{\sqrt{n}}{3},
a_{15}=-\frac{\sqrt{n}}{3}, a_{17}= -\frac{\sqrt{n}}{3}$.\newline
Therefore,
$h(u_1,u_2)=-\frac{\sqrt{n} \left(n \left(n \left(65 n^2-570
n+7561\right)-15120\right)+38880\right) u_1^5}{116640}$
$+\frac{1}{36} (n-1) \sqrt{n} (5 n+12) u_1^4 u_2-\frac{1}{108} \sqrt{n} (n
(3 n-7)+36)  u_1^4+\frac{1}{54} \sqrt{n} (n (15 n+7)-18) u_1^3 u_2^2+\frac{1%
}{108} \sqrt{n} (n (5 n+7)-36) u_1^3 u_2-\frac{1}{324} \sqrt{n} (n (3
n-7)+108) u_1^3-\frac{1}{3} \sqrt{n} u_1^2  u_2^3-\frac{1}{3} \sqrt{n} u_1^2
u_2^2-\frac{1}{3} \sqrt{n} u_1^2 u_2-\frac{\sqrt{n} u_1^2}{3}-\frac{1}{3} 
\sqrt{n} u_1 u_2^4-\frac{1}{3} \sqrt{n} u_1  u_2^3-\frac{1}{3} \sqrt{n} u_1
u_2^2-\frac{1}{3} \sqrt{n} u_1 u_2$, \newline and the center manifold can be
expressed as \newline
$\theta = \frac{5}{18} n^{5/2} Q^2 T^3+\frac{7}{54} n^{3/2} Q^2 T^3-\frac{5}{%
36} n^{5/2} Q T^4-\frac{7}{36} n^{3/2} Q T^4-\frac{65}{108} n^{5/2} Q T^3-%
\frac{35}{108} n^{3/2} Q T^3-\frac{13 n^{9/2} T^5}{23328}+\frac{19 n^{7/2}
T^5}{3888}-\frac{7561 n^{5/2} T^5}{116640}+\frac{7}{54} n^{3/2} T^5+\frac{1}{%
9} n^{5/2} T^4+\frac{7}{27} n^{3/2} T^4+\frac{17}{54} n^{5/2} T^3+\frac{35}{%
162} n^{3/2} T^3-\frac{1}{3} \sqrt{n} Q^4 T+\frac{1}{3}  \sqrt{n} Q^3 T^2+%
\frac{5}{3} \sqrt{n} Q^3 T-\frac{1}{3} \sqrt{n} Q^2 T^3-\frac{4}{3} \sqrt{n}
Q^2 T^2-\frac{10}{3} \sqrt{n} Q^2 T+\frac{1}{3} \sqrt{n} Q T^4+\sqrt{n} Q
T^3+2 \sqrt{n} Q T^2+\frac{10}{3} \sqrt{n} Q  T-\frac{\sqrt{n} T^5}{3}-\frac{%
2 \sqrt{n} T^4}{3}-\sqrt{n} T^3-\frac{4 \sqrt{n} T^2}{3}-\frac{5 \sqrt{n} T}{%
3}$. 
The dynamics on the center manifold is given by  
\begin{align}
&\frac{d u_1}{dN}=\frac{1}{27} n u_1^3 \left(9 \left(u_2^2+u_2+1\right)-n
u_1^2\right), \label{Center2DCase3PLa}\\
&\frac{d u_2}{d N}=\frac{4}{3} n^2 u_1^2 u_2^3+n^2 u_1^2 (u_1+1) u_2^2 
\notag \\
&\;\;\,\;\;\; \;\;\,\;\;\; +\frac{2}{27} n^2 u_1^2 u_2 (u_1 (9-(n-9) u_1)+9). \label{Center2DCase3PLb}
\end{align}
For which the origin is unstable (see Figure \ref{fig:Center2DCase3PL}).
Using the Theorem \ref{stabilityCM}, we conclude that the center manifold of
origin for the system \eqref{Center2DCase3PLa}, \eqref{Center2DCase3PLb}, and the origin are unstable
(saddle).

\textbf{Proof Part (ii)}:
More general, by introducing the linear transformation 
\begin{align}
& (u_1, u_2, v)=\left(T,Q^{*}-Q,\frac{1}{3} \left(3 Q^{*} \theta +\sqrt{n}
T\right)\right),  \notag \\
& \;\;\;\;\;\;\; \;\;\;\;\;\;\; \;\;\;\;\;\;\; u_1\in[0,1], u_2\in[%
1-|Q^{*}|,1+|Q^{*}|]\subset [0,2],
\end{align}
and taking Taylor series near $(u_1,u_2,v)=(0,0,0)$ up to fifth order we
obtain the system \eqref{system517a}, \eqref{system517b}, \eqref{system517c} can be written into its Jordan canonical
form: %
\begin{equation}
\left(%
\begin{array}{c}
\frac{du_1}{dN} \\ 
\frac{du_2}{dN} \\ 
\frac{d v}{dN}%
\end{array}%
\right)=\left(%
\begin{array}{ccc}
0 & 0 & 0 \\ 
0 & 0 & 0 \\ 
0 & 0 & -3 Q^{*}%
\end{array}%
\right)\left(%
\begin{array}{c}
u_1 \\ 
u_2 \\ 
v%
\end{array}%
\right)+\left(%
\begin{array}{c}
f_1^{*} (\mathbf{u},v) \\ 
f_2^{*} (\mathbf{u},v) \\ 
g^{*} (\mathbf{u},v)%
\end{array}%
\right),  \label{HLcenterCase3_systemGeneral}
\end{equation}%
where 
\begin{widetext}
$f_1^{*} (\mathbf{u},v)=\frac{u_1 \left(\sqrt{n} u_1-3 v\right)^2 \left(-(3 n-1) \left(\sqrt{n} u_1-3 v\right)^2+54 {Q^{*}}^2 (u_1-1)^2+54 {Q^{*}} (2 u_1-1) u_2\right)}{162 {Q^{*}}^3}$,\\
$f_2^{*} (\mathbf{u},v)=-\frac{n
   \left(\sqrt{n} u_1-3 v\right)^2 \left({Q^{*}}^2 (u_1-1) \left((3 n-1) \left(\sqrt{n} u_1-3 v\right)^2-54 \left(u_2^2-1\right)\right)+2 (3 n-1) {Q^{*}} u_2 \left(\sqrt{n} u_1-3
   v\right)^2-(3 n-1) (u_1-1) \left(\sqrt{n} u_1-3 v\right)^2-54 {Q^{*}}^4 (u_1-1)+108 {Q^{*}}^3 (u_1-1) u_2\right)}{162 {Q^{*}}^4}$,\newline
	$g^{*} (\mathbf{u},v)=\frac{\left(1-5 n^2\right) v^5}{10 {Q^{*}}^3}+\frac{v \left(u_1 \left(n^2 ((38-15 n) n-7) u_1^3-108 n {Q^{*}}^2 u_1 (n (3 u_1-2)+u_1 (4 u_1-7)+2)+1944 {Q^{*}}^4\right)+108 {Q^{*}}
   u_2 \left(n u_1^2 (3 n (u_1-1)-7 u_1+3)-18 {Q^{*}}^2 (u_1-1)\right)\right)}{648 {Q^{*}}^3}+\frac{\sqrt{n} u_1 \left(u_1 \left(n^2 (5 n (11 n-38)+39) u_1^3+360 n
   {Q^{*}}^2 u_1 (n (6 u_1-3)+2 u_1 (6 u_1-11)+7)-38880 {Q^{*}}^4\right)+720 {Q^{*}} u_2 \left(n u_1^2 (-3 n (u_1-1)+11 u_1-5)+54 {Q^{*}}^2
   (u_1-1)\right)\right)}{38880 {Q^{*}}^3}$\\$+\frac{\sqrt{n} (n (75 n-38)-5) u_1 v^4}{96 {Q^{*}}^3}-\frac{v^3 \left(36 (3 n+1) {Q^{*}}^2 (u_1-1)-36 (3 n+1) {Q^{*}} (u_1-1) u_2+(n-1) n (35
   n-3) u_1^2\right)}{72 {Q^{*}}^3}$\\$+\frac{\sqrt{n} u_1 v^2 \left(108 {Q^{*}}^2 \left(6 (n-1) u_1-5 n+4 u_1^2+1\right)-216 {Q^{*}} u_2 (3 n (u_1-1)-3 u_1+1)+n (n (65 n-114)+17)
   u_1^2\right)}{432 {Q^{*}}^3}$
\end{widetext}

By Theorem \ref{existenceCM}, exists a 2-dimensional local center manifold
of \eqref{HLcenterCase3_system517}, $W^{c}\left(\mathbf{0}\right)
=\left\{\left(\mathbf{u},v\right) \in\mathbb{R}^{2}\times\mathbb{R}%
:v=h\left( \mathbf{u}\right)\right\}$,
satisfying $\mathbf{h}\left(\mathbf{0}\right)= 0,\;D h\left(\mathbf{0}%
\right) =0,\;\left\vert \mathbf{u}\right\vert <\delta$ for $\delta$
sufficiently small. The restriction of (\ref{HLcenterCase3_systemGeneral})
to the center manifold is \newline
$\frac{d\mathbf{u}}{dN}=\mathbf{f}\left(\mathbf{u},h\left(\mathbf{u}%
\right)\right)$, where the function $h\left(\mathbf{u}\right)$ that
satisfies \eqref{MaineqcM}:%
\begin{equation}
Dh\left(\mathbf{u}\right) \cdot \mathbf{f}\left(\mathbf{u},h\left( \mathbf{u}%
\right) \right) +3 Q^{*} h\left( \mathbf{u}\right) -g\left(\mathbf{u},
h\left(\mathbf{u}\right) \right) =0.  \label{HLcenterCase3_systemGeneralh}
\end{equation}
Replacing \eqref{ansatz} in \eqref{HLcenterCase3_systemGeneralh} we find the
non-null coefficients $a_1= -\frac{\sqrt{n}}{3}, a_2= \frac{\sqrt{n}
\left(12 n^2+(7-15 n) n {Q^{*}}^2-108 {Q^{*}}^4\right)}{324 {Q^{*}}^4}$,%
\newline
$a_3= \frac{\sqrt{n} \left(12 n^2+(7-15 n) n {Q^{*}}^2-36 {Q^{*}}^4\right)}{%
108 {Q^{*}}^4}$,\newline
\begin{small}
$a_4= \frac{\sqrt{n} \left(-3360 n^4+120 n^3 (55 n-27) {Q^{*}}^2+n^2
(5 (762-661 n) n+24839) {Q^{*}}^4-2160 n (15 n-7) {Q^{*}}^6-38880 {Q^{*}}%
^8\right)}{116640 {Q^{*}}^8}$,
\end{small} \newline 
$a_5= -\frac{\sqrt{n}}{3 {Q^{*}}}, a_6= -\frac{\sqrt{n}}{3 {Q^{*}}}, a_7= 
\frac{\sqrt{n} \left(20 n^2+(7-15 n) n {Q^{*}}^2-36 {Q^{*}}^4\right)}{108 {%
Q^{*}}^5}$,\newline
$a_8=  \frac{\sqrt{n} \left(20 n^2+(7-15 n) n {Q^{*}}^2-12 {Q^{*}}^4\right)}{%
36 {Q^{*}}^5}, a_{10}=-\frac{\sqrt{n}}{3 {Q^{*}}^2}, a_{11}= -\frac{\sqrt{n}%
}{3 {Q^{*}}^2}$,\newline
$a_{12}=  \frac{\sqrt{n} \left(30 n^2+(7-15 n) n {Q^{*}}^2-18 {Q^{*}}%
^4\right)}{54 {Q^{*}}^6}, a_{14}= -\frac{\sqrt{n}}{3 {Q^{*}}^3}$,\newline
$a_{15}= -\frac{\sqrt{n}}{3 {Q^{*}}^3}, a_{17}= -\frac{\sqrt{n}}{3 {Q^{*}}^4}
$. \newline 
Therefore,
\begin{widetext}
 	$h(u_1,u_2)=\frac{\sqrt{n} u_1^4 \left(12 n^2+(7-15 n) n {Q^{*}}^2-36 {Q^{*}}^4\right)}{108 {Q^{*}}^4}+\frac{\sqrt{n} u_1^3 \left(12 n^2+(7-15 n) n {Q^{*}}^2-108 {Q^{*}}^4\right)}{324 {Q^{*}}^4}+\frac{\sqrt{n}
   u_1^3 u_2^2 \left(30 n^2+(7-15 n) n {Q^{*}}^2-18 {Q^{*}}^4\right)}{54 {Q^{*}}^6}+\frac{\sqrt{n} u_1^4 u_2 \left(20 n^2+(7-15 n) n {Q^{*}}^2-12 {Q^{*}}^4\right)}{36
   {Q^{*}}^5}+\frac{\sqrt{n} u_1^3 u_2 \left(20 n^2+(7-15 n) n {Q^{*}}^2-36 {Q^{*}}^4\right)}{108 {Q^{*}}^5}$\\$+\frac{\sqrt{n} u_1^5 \left(-3360 n^4+120 n^3 (55 n-27) {Q^{*}}^2+n^2 (5 (762-661 n)
   n+24839) {Q^{*}}^4-2160 n (15 n-7) {Q^{*}}^6-38880 {Q^{*}}^8\right)}{116640 {Q^{*}}^8}-\frac{\sqrt{n} u_1 u_2^4}{3 {Q^{*}}^4}-\frac{\sqrt{n} u_1^2 u_2^3}{3 {Q^{*}}^3}-\frac{\sqrt{n}
   u_1 u_2^3}{3 {Q^{*}}^3}-\frac{\sqrt{n} u_1^2 u_2^2}{3 {Q^{*}}^2}-\frac{\sqrt{n} u_1 u_2^2}{3 {Q^{*}}^2}-\frac{\sqrt{n} u_1^2 u_2}{3 {Q^{*}}}-\frac{\sqrt{n}
   u_1 u_2}{3 {Q^{*}}}-\frac{\sqrt{n} u_1^2}{3}$. \newline
		Finally, the center manifold can be expressed as 	\newline
		$\theta = \frac{5 n^{5/2} Q^2 T^3}{9 {Q^{*}}^7}-\frac{5 n^{5/2} Q^2 T^3}{18 {Q^{*}}^5}+\frac{7 n^{3/2} Q^2 T^3}{54 {Q^{*}}^5}-\frac{5 n^{5/2} Q T^4}{9 {Q^{*}}^6}-\frac{35 n^{5/2} Q T^3}{27
   {Q^{*}}^6}+\frac{5 n^{5/2} Q T^4}{12 {Q^{*}}^4}-\frac{7 n^{3/2} Q T^4}{36 {Q^{*}}^4}+\frac{25 n^{5/2} Q T^3}{36 {Q^{*}}^4}-\frac{35 n^{3/2} Q T^3}{108 {Q^{*}}^4}-\frac{7 n^{9/2} T^5}{243 {Q^{*}}^9}+\frac{55
   n^{9/2} T^5}{972 {Q^{*}}^7}-\frac{n^{7/2} T^5}{36 {Q^{*}}^7}-\frac{661 n^{9/2} T^5}{23328 {Q^{*}}^5}+\frac{127 n^{7/2} T^5}{3888 {Q^{*}}^5}+\frac{24839 n^{5/2} T^5}{116640 {Q^{*}}^5}+\frac{2 n^{5/2} T^4}{3
   {Q^{*}}^5}+\frac{7 n^{5/2} T^3}{9 {Q^{*}}^5}-\frac{5 n^{5/2} T^5}{18 {Q^{*}}^3}+\frac{7 n^{3/2} T^5}{54 {Q^{*}}^3}-\frac{5 n^{5/2} T^4}{9 {Q^{*}}^3}+\frac{7 n^{3/2} T^4}{27 {Q^{*}}^3}-\frac{25 n^{5/2} T^3}{54
   {Q^{*}}^3}+\frac{35 n^{3/2} T^3}{162 {Q^{*}}^3}-\frac{\sqrt{n} Q^4 T}{3 {Q^{*}}^5}+\frac{\sqrt{n} Q^3 T^2}{3 {Q^{*}}^4}+\frac{5 \sqrt{n} Q^3 T}{3 {Q^{*}}^4}-\frac{\sqrt{n} Q^2 T^3}{3 {Q^{*}}^3}-\frac{4
   \sqrt{n} Q^2 T^2}{3 {Q^{*}}^3}-\frac{10 \sqrt{n} Q^2 T}{3 {Q^{*}}^3}+\frac{\sqrt{n} Q T^4}{3 {Q^{*}}^2}+\frac{\sqrt{n} Q T^3}{{Q^{*}}^2}+\frac{2 \sqrt{n} Q T^2}{{Q^{*}}^2}+\frac{10 \sqrt{n} Q T}{3
   {Q^{*}}^2}-\frac{\sqrt{n} T^5}{3 {Q^{*}}}-\frac{2 \sqrt{n} T^4}{3 {Q^{*}}}-\frac{\sqrt{n} T^3}{{Q^{*}}}-\frac{4 \sqrt{n} T^2}{3 {Q^{*}}}-\frac{5 \sqrt{n} T}{3 {Q^{*}}}$.\newline
The dynamics on the center manifold is given by 
\begin{align}
&\frac{d u_1}{dN}=\frac{n u_1^3 \left({Q^{*}}^2 \left(n (2 n-1) u_1^2+9 \left({Q^{*}}^2+{Q^{*}} u_2+u_2^2\right)\right)-2 n^2 u_1^2\right)}{27 {Q^{*}}^5}, \label{Center2DCase3PLGenerala}\\
&\frac{d u_2}{dN}=\frac{4 n^2 u_1^2 u_2^3}{3 {Q^{*}}^5}+\frac{n^2 u_1^2 (u_1+1) u_2^2}{{Q^{*}}^4}+\frac{2 n^2 u_1^2 u_2 \left({Q^{*}}^4 \left(u_1 \left(\left(-2 n^2+n+9\right)
   u_1+9\right)+9\right)-6 n^2 u_1^2+2 n (4 n-1) {Q^{*}}^2 u_1^2\right)}{27 {Q^{*}}^7}\nonumber \\
	& -\frac{n^2 \left({Q^{*}}^2-1\right) u_1^2 \left(-2 n^2 u_1^2 (3 u_1+1)+n (2 n-1) {Q^{*}}^2
   u_1^2 (3 u_1+1)+9 {Q^{*}}^4 (u_1+1) \left(u_1^2+1\right)\right)}{27 {Q^{*}}^6}. \label{Center2DCase3PLGeneralb}
\end{align}
\end{widetext}
In the Figure \ref{fig:Center2DCase3PLGeneral} are presented some numerical solutions of the system \eqref{Center2DCase3PLGenerala}, \eqref{Center2DCase3PLGeneralb} for $n=1,4$ and $Q^{*}=\pm 1$. The plot illustrates the generic feature that for $Q^{*}>0$ (respectively,  $Q^{*}<0$), the center manifold is unstable (respectively, stable), but in this case the third eigenvalue is $-3 Q^{*}<0$ (respectively, $-3 Q^{*}>0$). That is, the origin is a saddle.  $\blacksquare$

\subsection{E-models}

\label{Case3D}

\begin{figure*}[ht!]
\centering
\subfigure[]{\includegraphics[width=2.5in, height=2.5in,angle=-90]{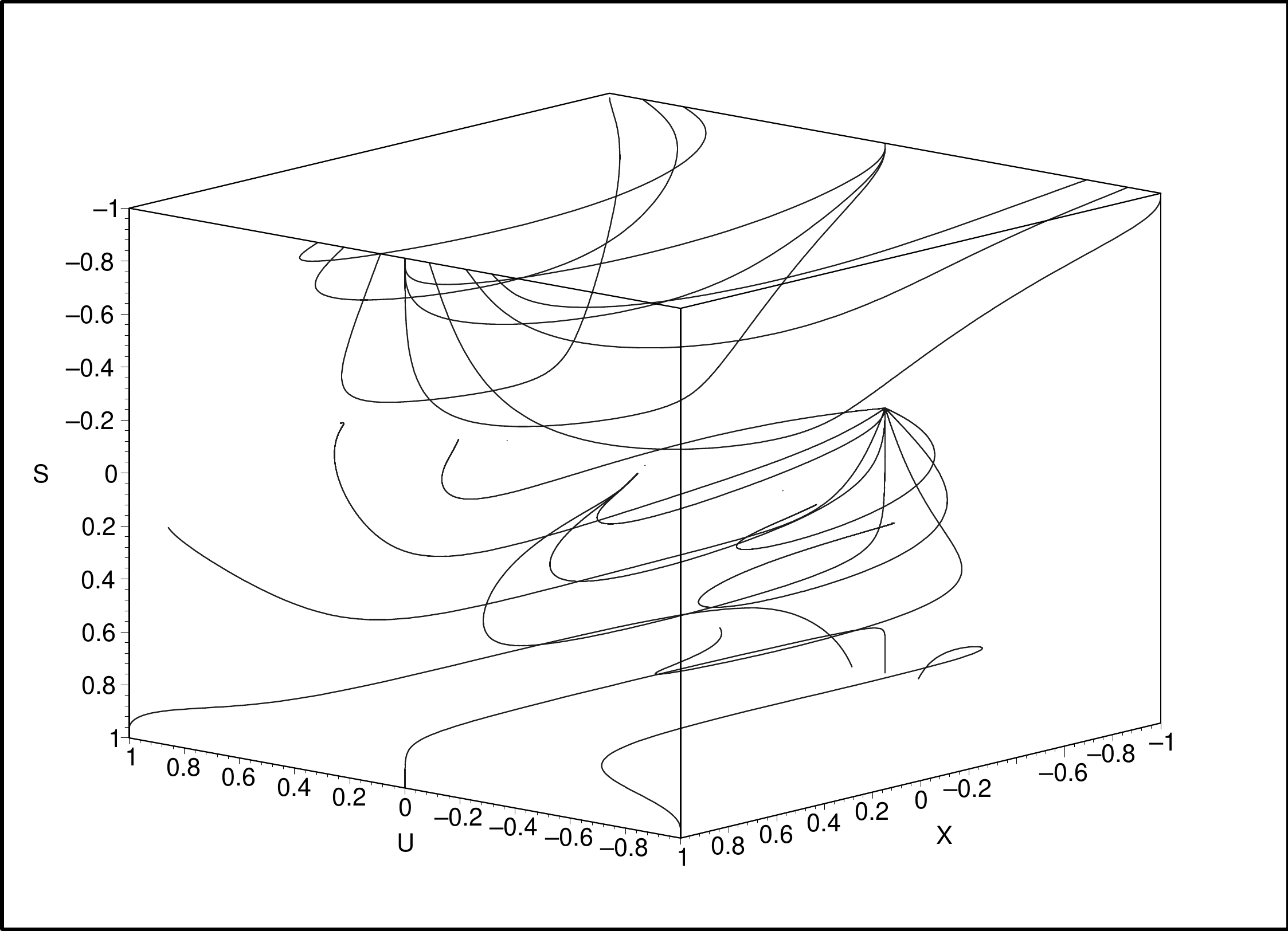}	} 
\hspace{2cm}  \subfigure[]{\includegraphics[width=2.5in,
height=2.5in,angle=-90]{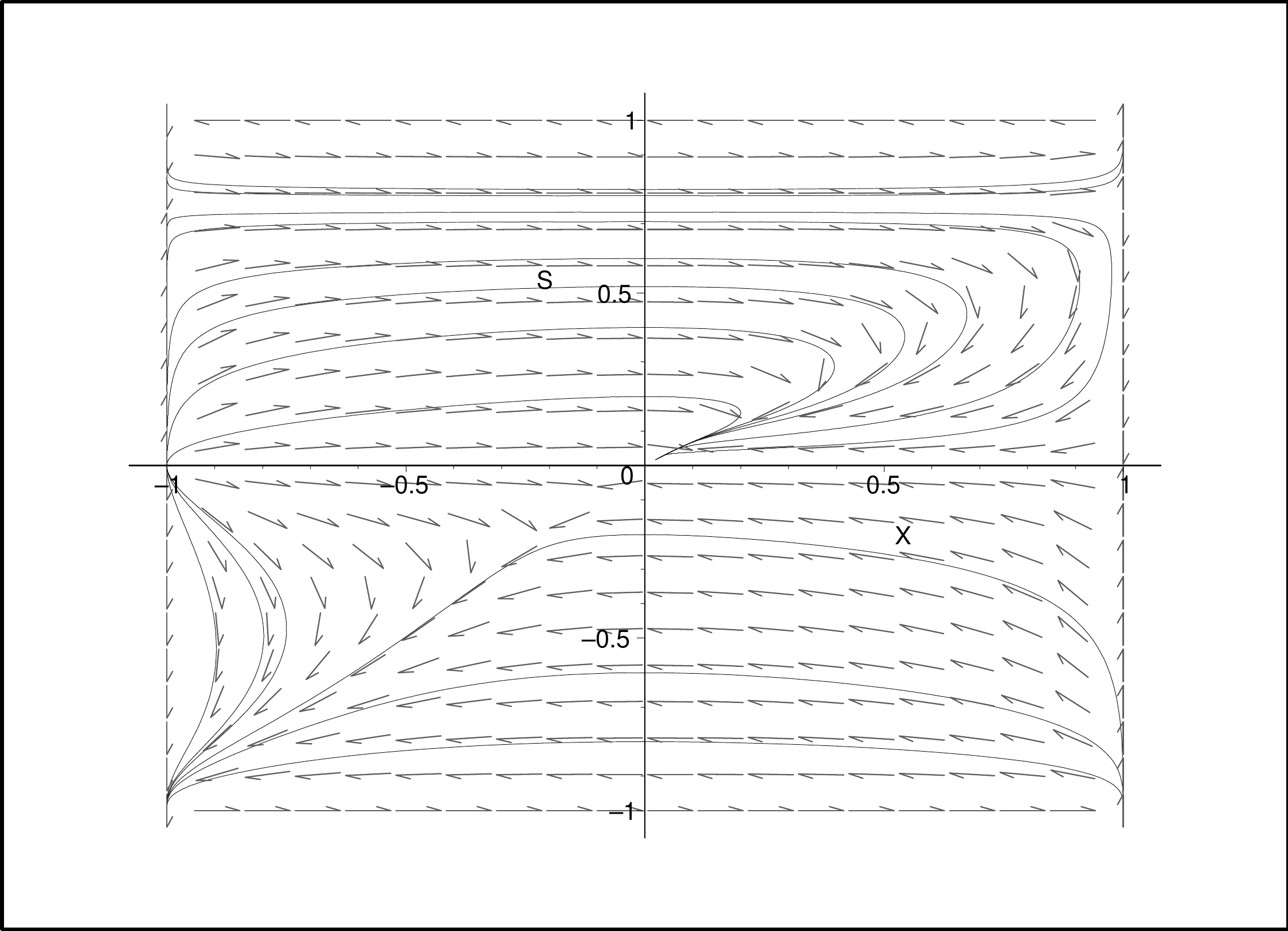}}  \subfigure[]{\includegraphics[width=2.5in,
height=2.5in,angle=-90]{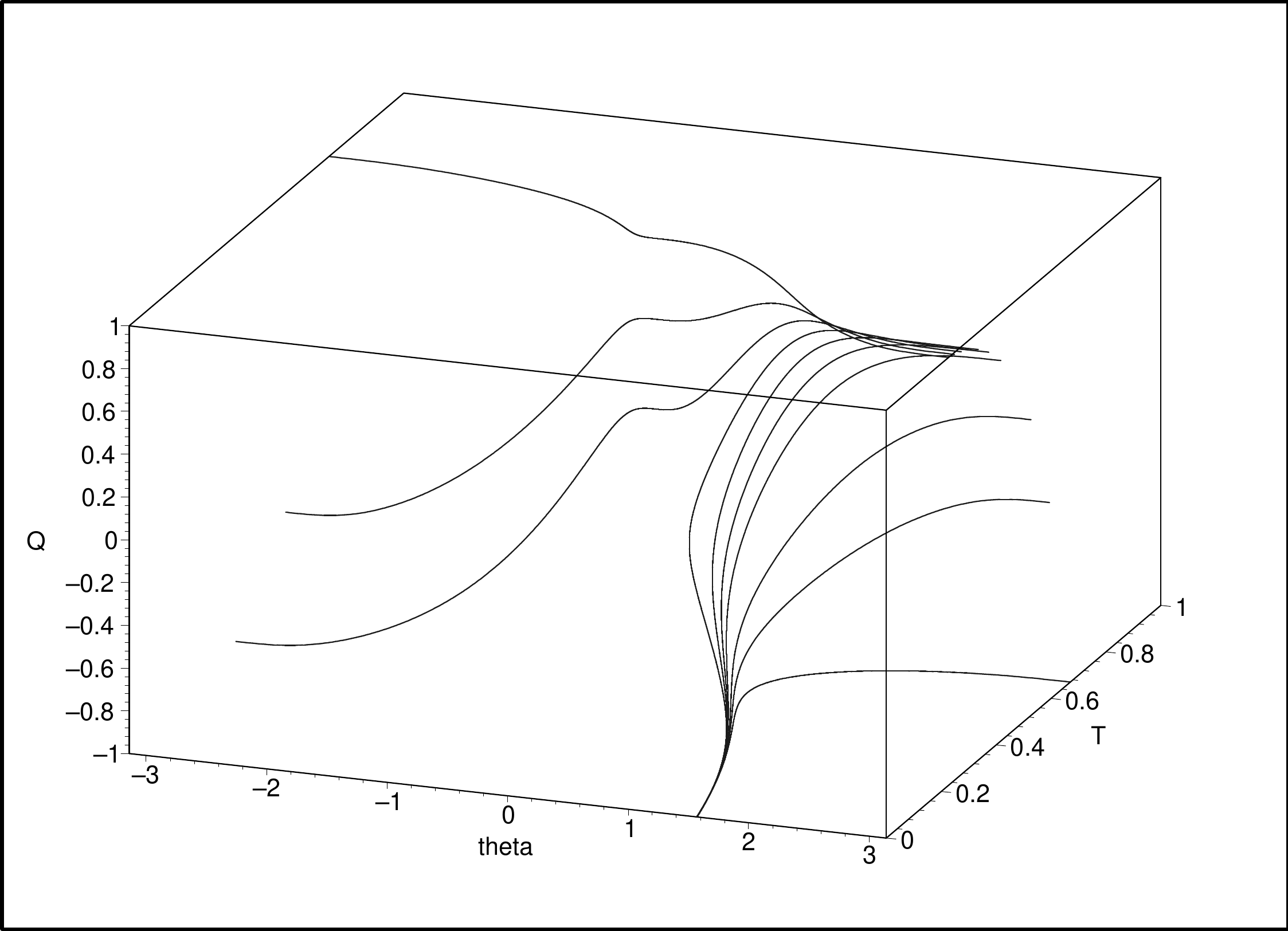}	} \hspace{2cm}  \subfigure[]{%
\includegraphics[width=2.5in, height=2.5in,angle=-90]{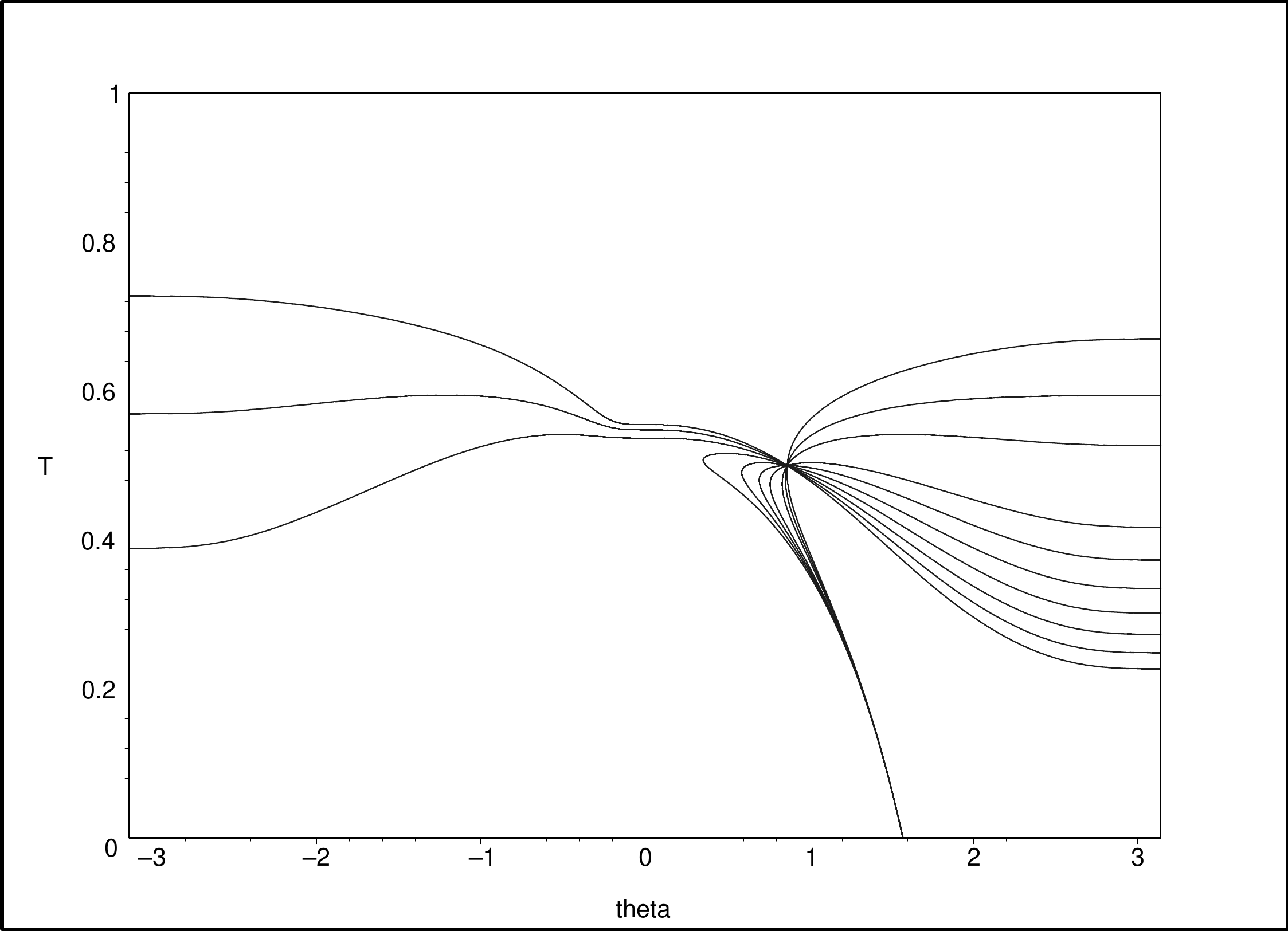}}  %
\subfigure[]{\includegraphics[width=2.5in, height=2.5in,angle=-90]{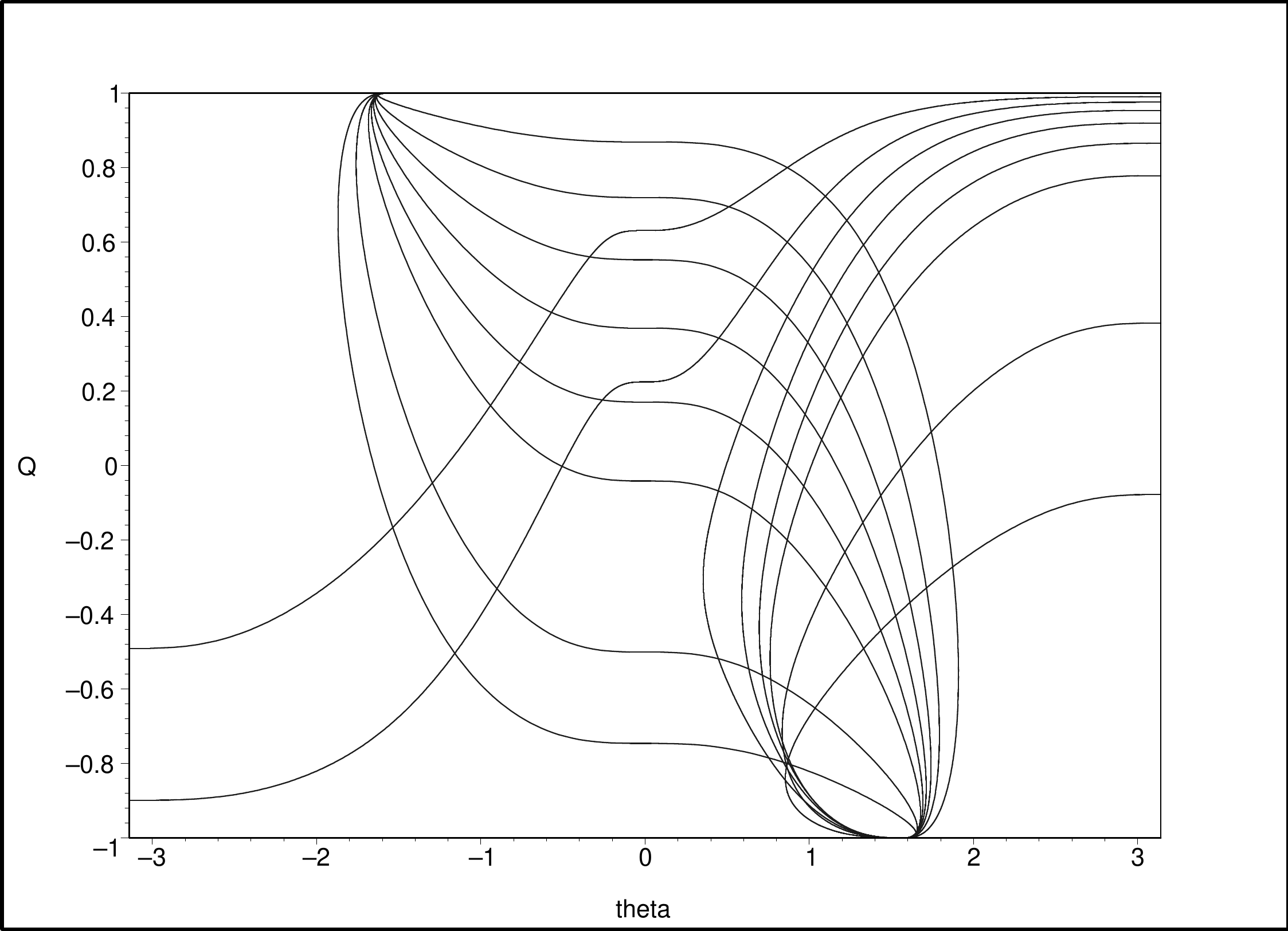}	} 
\hspace{2cm}  \subfigure[]{\includegraphics[width=2.5in,
height=2.5in,angle=-90]{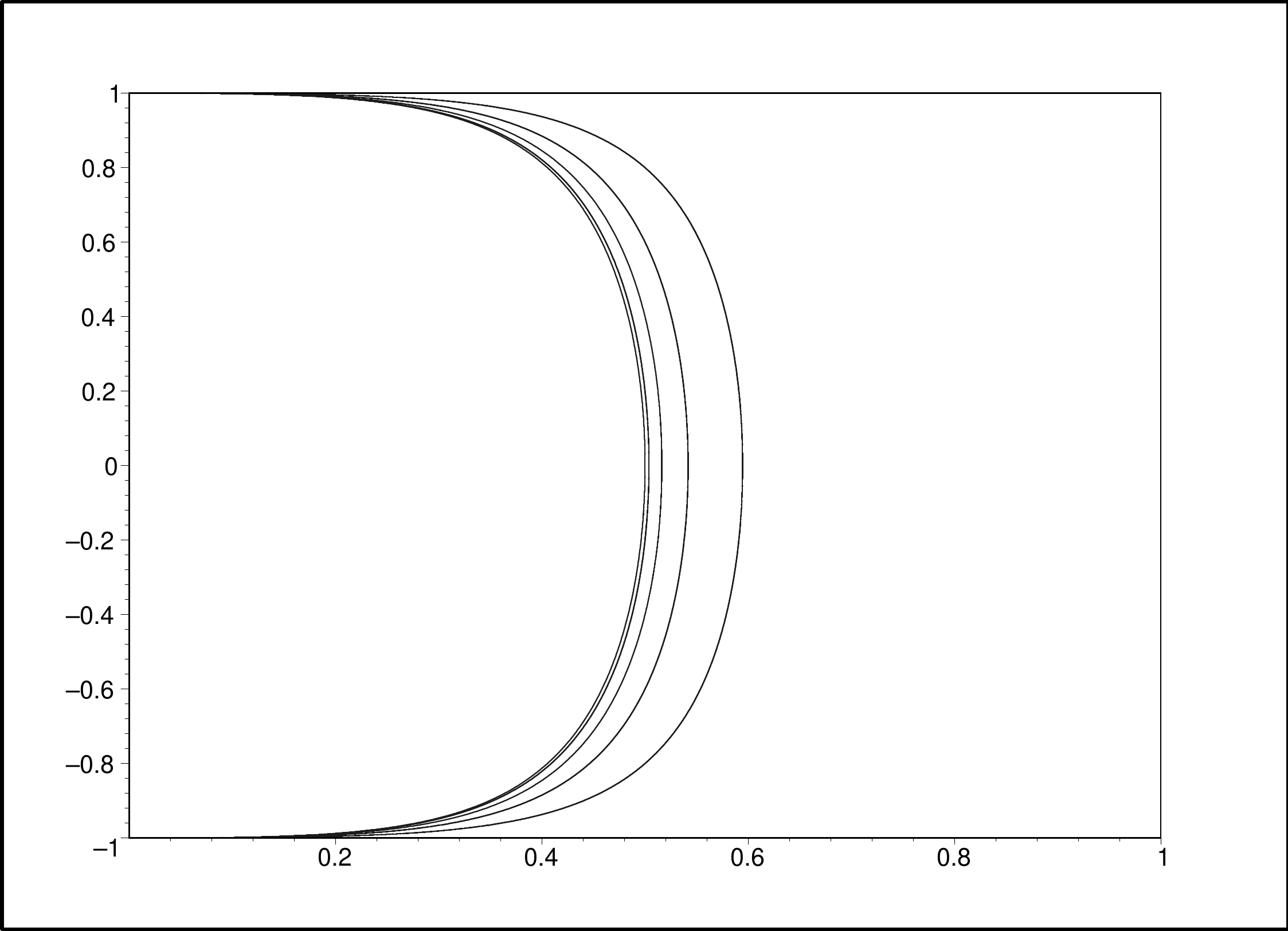}} 
\caption{\textit{{(a) Compact phase portrait of \eqref{SYSTEM94a}, \eqref{SYSTEM94b}, \eqref{SYSTEM94c} for the
choice $n=2, \protect\mu=1$. (b) Dynamics in the invariant set $Z=0$.
(c)-(f): Dynamics of \eqref{system527a}, \eqref{system527b}, \eqref{system527c} and some 2D projections of the
solutions for $n=4, \protect\mu=1$.}}}
\label{fig:Case3HLE}
\end{figure*}

In this example the dynamical system \eqref{Example3Sa}, \eqref{Example3Sb}, \eqref{Example3Sc} is reduced to 
\begin{align}
&\frac{dX}{d\tau}=\left(X^2-1\right) \left(3 \sqrt{1-U^2} X-\sqrt{6} \,\tan
\left(\frac{\pi S}{2}\right)\right), \label{SYSTEM94a}\\
&\frac{dU}{d\tau}=3 U \sqrt{1-U^2} X^2, \label{SYSTEM94b}\\
&\frac{dS}{d\tau}=\frac{X \left(-6 \mu \sin (\pi S)-\sqrt{6} \cos (\pi S)+%
\sqrt{6}\right)}{\pi n}, \label{SYSTEM94c}
\end{align}
defined on the compact phase space 
\begin{equation*}
\Big\{(X, U, S)\in\mathbb{R}^3: -1\leq X\leq 1, -1\leq U \leq 1, -1\leq S
\leq 1\Big\}.
\end{equation*}
The coordinates equilibrium points of the system \eqref{SYSTEM94a}, \eqref{SYSTEM94b}, \eqref{SYSTEM94c} and their stability
conditions are summarized as follows:

\begin{itemize}
\item $(x,U,S)=(0,U_c,0)$ with eigenvalues $\left\{0,0,-3 \sqrt{1-U_c^2}%
\right\}$.
Non-hyperbolic with a 1 dimensional stable manifold.

\item $(x,U,S)=(-1,0,0)$ with eigenvalues $\left\{3,6,\frac{6 \mu }{n}%
\right\}$. It is a source.

\item $(x,U,S)=\left(-1,0,\frac{2 \;\text{arctan}\;\left(\sqrt{6} \mu \right)%
}{\pi }\right)$ with eigenvalues\newline
$\left\{3,12 \mu +6,-\frac{6 \mu }{n}\right\}$. It is a saddle.

\item $(x,U,S)=(1,0,0)$ with eigenvalues $\left\{3,6,-\frac{6 \mu }{n}%
\right\}$. It is a saddle.

\item $(x,U,S)=\left(1,0,\frac{2 \;\text{arctan}\;\left(\sqrt{6} \mu \right)%
}{\pi }\right)$ with eigenvalues\newline
$\left\{3,6-12 \mu ,\frac{6 \mu }{n}\right\}$. It is a source for $\mu<\frac{%
1}{2}$.

\item $(x,U,S)=(-1, -1, 0)$ with eigenvalues $\left\{0,-\infty ,\frac{6 \mu 
}{n}\right\}$. Non-hyperbolic with a 1 dimensional stable manifold.

\item $(x,U,S)=\left(-1,-1,\frac{2 \;\text{arctan}\;\left(\sqrt{6} \mu
\right)}{\pi }\right)$ with eigenvalues \newline
$\left\{-\infty ,12 \mu ,-\frac{6 \mu }{n}\right\}$. It is a saddle.

\item $(x,U,S)=(-1, 1, 0)$ with eigenvalues $\left\{0,-\infty ,\frac{6 \mu }{%
n}\right\}$. Non-hyperbolic. Behaves as saddle.

\item $(x,U,S)=\left(-1,1,\frac{2 \;\text{arctan}\;\left(\sqrt{6} \mu \right)%
}{\pi }\right)$ with eigenvalues \newline
$\left\{-\infty ,12 \mu ,-\frac{6 \mu }{n}\right\}$. It is a saddle.

\item $(x,U,S)=(1, -1, 0)$ with eigenvalues $\left\{0,-\infty ,-\frac{6 \mu 
}{n}\right\}$. Non-hyperbolic with a 2 dimensional stable manifold.

\item $(x,U,S)=\left(1,-1,\frac{2 \;\text{arctan}\;\left(\sqrt{6} \mu \right)%
}{\pi }\right)$ with eigenvalues \newline
$\left\{-\infty ,-12 \mu ,\frac{6 \mu }{n}\right\}$. It is a saddle.

\item $(X,Z,S)=(1, 1, 0)$. The eigenvalues are $\left\{0,-\infty ,-\frac{6
\mu }{n}\right\}$. Nonhyperbolic with a 2 dimensional stable manifold.

\item $(x,U,S)=\left(1,1,\frac{2 \;\text{arctan}\;\left(\sqrt{6} \mu \right)%
}{\pi }\right)$ with eigenvalues \newline
$\left\{-\infty ,-12 \mu ,\frac{6 \mu }{n}\right\}$. It is a saddle.
\end{itemize}

\subsubsection{Alternative compactification}

Introducing the new variables 
\begin{align}
&\hat{H}=\sqrt{H^2+\frac{\mu^2 \Lambda^2}{16(3\lambda-1)^2}}, \\
&Q=\frac{H}{\bar{H}}, \\
& \Sigma =\frac{\dot\phi}{2 \sqrt{6} \bar{H}}, \\
&Y=\left(\frac{V(\phi)}{6(3\lambda-1)\bar{H}^2}\right)^{\frac{1}{2 n}}=\hat{%
T}\left(1-e^{-\sqrt{\frac{2}{3 \alpha}} \phi}\right) , \\
&\hat{T}=\left[\frac{V_0}{6(3\lambda-1)\hat{H}^2}\right]^{\frac{1}{2n}},
\end{align}
such that 
\begin{align}
&\dot\phi = \frac{2 \Sigma \sqrt{V_0} \hat{T}^{-n}}{\sqrt{3 \lambda -1}}, \\
&\phi=-\frac{\sqrt{\frac{2}{3}} n \ln \left(1-\frac{Y}{\hat{T}}\right)}{\mu }%
, \\
&\Lambda=\frac{2 \sqrt{\frac{2}{3}} \sqrt{3 \lambda -1} \sqrt{1-Q^2} \sqrt{%
V_0} \hat{T}^{-n}}{\mu }, \\
&H= \frac{Q\sqrt{V_0} \hat{T}^{-n}}{\sqrt{18 \lambda -6}}, \\
&\Sigma + Y^{2 n}=1, 
\end{align}
and the new time variable $\frac{dM}{d t}=H/Q$, we obtain the dynamical
system becomes 

\begin{align}
&\frac{d \Sigma }{dM}=6 \mu Y^{2 n-1} (Y-\hat{T})+Q \Sigma\left((q+1)
Q^2-3\right), \\
&\frac{d Y}{dM}=\frac{(q+1) Q^3 Y+6 \mu \Sigma (\hat{T}-Y)}{n}, \\
&\frac{d \hat{T}}{dM}=\frac{(q+1) Q^3 \hat{T}}{n}, \\
&\frac{d Q}{dM}=(q+1) Q^2 \left(Q^2-1\right).
\end{align}
where 
\begin{equation}
Q^2 q= \frac{1}{2} \left(-3 Y^{2 n}-2 Q^2+3 \Sigma ^2+3\right)=3 \Sigma
^2-Q^2.
\end{equation}
Introducing the complementary global transformation 
\begin{align}
&\Sigma =F(\theta ) \sin (\theta ), \quad Y=\cos (\theta ) ,  \notag \\
&F(\theta )=\sqrt{\frac{1-\cos ^{2 n}(\theta )}{1-\cos ^2(\theta )}},
\end{align}
we obtain the regular unconstrained 3D dynamical system 
\begin{align}
&\frac{d\theta}{dM}=-\frac{3 Q F(\theta )^2 \sin (\theta ) \cos (\theta )}{n}
\notag\\
& -\frac{6 \mu F(\theta ) (\hat{T}-\cos (\theta ))}{n}, \end{align}
\begin{align}
&\frac{d \hat{T}}{dM}= \frac{3 Q\hat{T} \left(1-\cos ^{2 n}(\theta )\right)}{%
n}, \\
&\frac{d Q}{dM}=3 \left(Q^2-1\right) \left(1-\cos^{2n}(\theta)\right).
\end{align}
\begin{figure}[t!]
\includegraphics[width=0.5\textwidth]{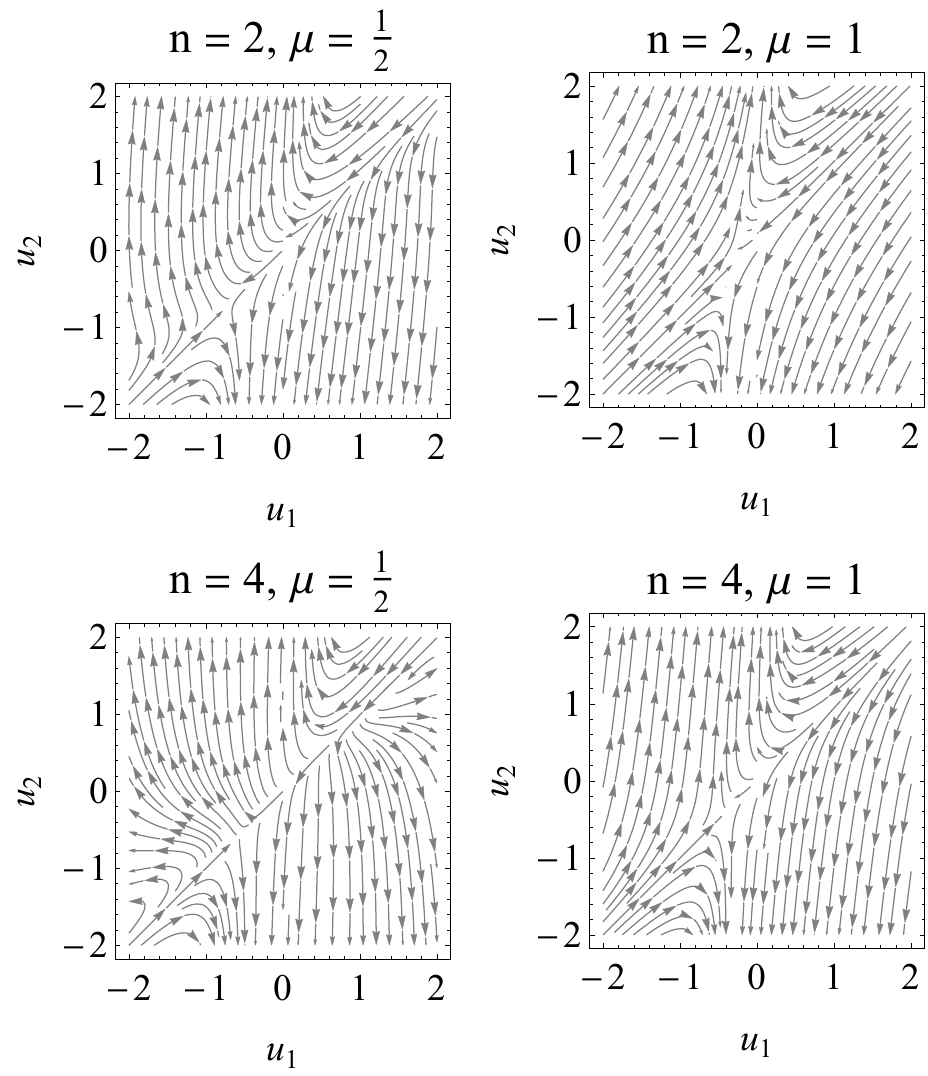}  
\caption{Numerical solutions of the system \eqref{Center2DCase3EM} shows the
instability of the origin $(u_1,u_2)=(0,0)$.}
\label{fig:Center2DCase3EM}
\end{figure}
where the deceleration parameter is expressed as
\begin{equation}
Q^2 q=-3 \cos ^{2 n}(\theta )-Q^2+3.
\end{equation}
Introducing the new compact variable $T=\frac{\hat{T}}{1+\hat{T}}$, and the
new time derivative $\frac{d{\bar{\tau}}}{d M}=1+\hat{T}=(1-T)^{-1}$, we
obtain: 
\begin{align}
&\frac{d\theta}{d\bar{\tau}}=-\frac{6 \mu F(\theta ) ((T-1) \cos (\theta )+T)%
}{n}  \notag \\
&+\frac{3 Q (T-1) F(\theta )^2 \sin (2 \theta )}{2 n}, \label{system527a}\\
&\frac{d T}{d\bar{\tau}}=\frac{3 Q (T-1)^2 T \left(1-\cos ^{2 n}(\theta
)\right)}{n}, \label{system527b}\\
&\frac{d Q}{d\bar{\tau}}=-3 \left(Q^2-1\right) (T-1) \left(1-\cos ^{2
n}(\theta )\right). \label{system527c}
\end{align}

The coordinates equilibrium points of the system \eqref{system527a}, \eqref{system527b}, \eqref{system527c} and their stability
conditions are summarized as follows:

\begin{itemize}
\item ${}_\pm M_{\pm}$: $\hat{T}=T=0; \Sigma=\pm 1, Q=\pm 1, Y=0; \theta=\pm 
\frac{\pi}{2}+2 k \pi, k\in\mathbb{Z}$, were we have used the Kernel: ${}_{%
\text{sign}(Q)} M_{\text{sign}(\Sigma)}$. The eigenvalues of ${}_{-}M_{+}$
are $\left\{-\frac{3}{n},-6,-\frac{6 \mu +3}{n}\right\}$. It is a sink. The
eigenvalues of ${}_{-}M_{-}$ are $\left\{-\frac{3}{n},-6,\frac{6 \mu -3}{n}%
\right\}$. It is a sink for $\mu<\frac{1}{2}$ or a saddle otherwise. The
eigenvalues of ${}_{+}M_{+}$ are $\left\{\frac{3}{n},6,\frac{3-6 \mu }{n}%
\right\}$. It is a source for $\mu<\frac{1}{2}$ or a saddle otherwise. The
eigenvalues of ${}_{+}M_{-}$ are $\left\{\frac{3}{n},6,\frac{6 \mu +3}{n}%
\right\}$. It is a source.

\item $dS^{\pm}$: $\hat{T}=1, T=\frac{1}{2}; \Sigma=0, Q=\pm 1 ,
Y=1;\theta=2 k \pi, k\in\mathbb{Z}$. The eigenvalues of $dS^{+}$ are $%
\left\{0,0,-\frac{3}{2}\right\}$. Nonhyperbolic with 1D stable manifold. The
eigenvalues of $dS^{-}$ are $\left\{0,0, \frac{3}{2}\right\}$. Nonhyperbolic
with 1D unstable manifold. More generally, we have $dS(Q^{*})$: $\hat{T}=1,
T=\frac{1}{2}; \Sigma=0, Q=Q^{*}, Y=1;\theta=2 k \pi, k\in\mathbb{Z}$, with
eigenvalues\newline
$\left\{0,0,-\frac{3}{2} Q^{*}\right\}$.

\item $PL^{\pm}$: $\hat{T}=T=0; \Sigma=\pm 2\mu, Q=\pm 1; Y=-(1-4\mu^2)^{%
\frac{1}{2n}}$;
$\theta=\pm\arccos Y$. Exists for $\mu<1/2$. The eigenvalues are
$\left\{\frac{12 \mu ^2}{n},24 \mu ^2,12 \mu ^2-3\right\}$. It is a saddle. 
\end{itemize}

In the figure \ref{fig:Case3HLE} it is shown (a) a compact phase portrait of the system %
\eqref{SYSTEM94a}, \eqref{SYSTEM94b}, \eqref{SYSTEM94c} for the choice $n=2, \mu=1$. (b) Dynamics in the invariant
set $Z=0$. In Fig. \ref{fig:Case3HLE} (c)-(f) it is presented the dynamics
of the system \eqref{system527a}, \eqref{system527b}, \eqref{system527c} and some 2D projections of the solutions for $n=4, \mu=1
$. 

\subsubsection{Center manifold of the de Sitter solution: for (i) $dS^{+}$:$(%
\protect\theta, T, Q)=\left(0, \frac{1}{2}, 1\right)$, (ii) For $%
dS^{+}(Q^{*}): (\protect\theta, T, Q)=\left(0, \frac{1}{2}, Q^{*}\right) ,
Q^{*}\neq \pm 1,0$, for Ho\v{r}ava-Lifshitz model with flat universe with $%
\Lambda \neq 0$ and E-model under the detailed-balance.}

\label{PROPOSITION10}

\begin{prop}
\label{centerP3EMCase3DS} (i) The point $dS^+$: $(\theta, T, Q)=(0, \frac{1}{%
2}, 1)$ of the system \eqref{system527a}, \eqref{system527b}, \eqref{system527c} is unstable (saddle). (ii) The line
of fixed points $dS_{+}(Q^{*})$ :$(\theta, T, Q)=\left(0, \frac{1}{2},
Q^{*}\right) , Q^{*}\neq \pm 1, 0$ of the system \eqref{system527a}, \eqref{system527b}, \eqref{system527c} is
unstable (saddle).
\end{prop}

\textbf{Proof Part (i)}:

Taking the linear transformation 
\begin{small} 
\begin{align}
& (u_1, u_2, v)=  
 \left(\frac{\mu (4-8 T)}{\sqrt{n}},\frac{\mu (4-8 T)}{\sqrt{n}}+Q-1,\theta
+\frac{4 \mu (2 T-1)}{\sqrt{n}}\right),
\end{align}%
\end{small}
 and taking Taylor series near $(u_1,u_2,v)=(0,0,0)$ up to fifth order we
obtain the system \eqref{system527a}, \eqref{system527b}, \eqref{system527c} can be written into its Jordan canonical
form: %
\begin{equation}
\left(%
\begin{array}{c}
\frac{du_1}{dN} \\ 
\frac{du_2}{dN} \\ 
\frac{d v}{dN}%
\end{array}%
\right)=\left(%
\begin{array}{ccc}
0 & 0 & 0 \\ 
0 & 0 & 0 \\ 
0 & 0 & -\frac{3}{2}%
\end{array}%
\right)\left(%
\begin{array}{c}
u_1 \\ 
u_2 \\ 
v%
\end{array}%
\right)+\left(%
\begin{array}{c}
f_1(\mathbf{u},v) \\ 
f_2(\mathbf{u},v) \\ 
g(\mathbf{u},v)%
\end{array}%
\right),  \label{HLcenterCase3_system527}
\end{equation}%
where
\begin{widetext}
\begin{small}
$f_1(\mathbf{u},v)=\frac{u_1^5 v \left(-15 n^{3/2}-32 \mu ^3 (15 (n-1) n+4)+80 \mu ^2 (1-3 n) \sqrt{n}+20 \mu  (1-3 n) n\right)}{160 \mu ^2 \sqrt{n}}+\frac{u_1^4 v^2 \left(-3 n^{3/2}-32 \mu ^3 (15 (n-1) n+4)+48 \mu ^2 (1-3 n)
   \sqrt{n}+12 \mu  (1-3 n) n\right)}{64 \mu ^2 \sqrt{n}}+\frac{u_1^2 \left(3 n^{3/2} u_1^3-32 \mu ^3 \left((3 n-1) u_1^2-6\right) (u_1-u_2-1)+12 \mu  n u_1^2 (-u_1+u_2+1)+8
   \mu ^2 \sqrt{n} u_1 (u_1 ((3 n-1) u_1+6)-6 (u_2+1))\right)}{64 \mu ^2 \sqrt{n}}$\\$+\frac{u_1 v^3 \left(-16 \mu ^2 \left(15 n^2 u_1^2-3 n \left(5 u_1^2-3 u_1+3
   u_2+3\right)+4 u_1^2-3 u_1+3 u_2+3\right)+3 (1-3 n) n u_1^2-12 \mu  \sqrt{n} (3 n-1) u_1 (u_1-u_2-1)\right)}{24 \mu  \sqrt{n}}+\frac{u_1^2 v^4 \left(-3 n^2-16 \mu ^2
   (15 (n-1) n+4)+4 \mu  (1-3 n) \sqrt{n}+n\right)}{32 \mu  \sqrt{n}}+\frac{u_1^4 v \left(16 \mu ^2 (3 n-1)+3 n\right) \left(\sqrt{n} (u_2+1)-4 \mu \right)}{32 \mu ^2 \sqrt{n}}+\frac{3 u_1^3 v^2 \left(16
   \mu ^2 (3 n-1)+n\right) \left(\sqrt{n} (u_2+1)-4 \mu \right)}{64 \mu ^2 \sqrt{n}}+\frac{1}{8} u_1^3 v \left(\frac{(u_2+1) \left(16 \mu ^2 (3 n-1)+3 n\right)}{\mu  \sqrt{n}}+12\right)+\frac{3 u_1^2
   v^2 \left(4 \mu  \sqrt{n}+16 \mu ^2 (3 n-1) (u_2+1)+n (u_2+1)\right)}{16 \mu  \sqrt{n}}+u_1^2 \left(\frac{6 \mu  v}{\sqrt{n}}-\frac{3}{2} (u_2+1) v\right)+\frac{(3 n-1) u_1 v^4
   \left(\sqrt{n} (u_2+1)-4 \mu \right)}{8 \sqrt{n}}+\frac{3}{4} u_1 v^2 \left(\frac{4 \mu }{\sqrt{n}}-u_2-1\right)-\frac{6 \mu  u_1 (u_2+1) v}{\sqrt{n}}-\frac{\mu  (15 (n-1) n+4) u_1
   v^5}{5 \sqrt{n}}+\frac{\mu  (3 n-1) (u_2+1) v^4}{2 \sqrt{n}}-\frac{3 \mu  (u_2+1) v^2}{\sqrt{n}},
	$\\
$f_2(\mathbf{u},v)=\frac{1}{64} \left(8 (12 n-1) n+\frac{3 n}{\mu ^2}+\frac{12 (2 n-1) \sqrt{n}}{\mu }-8+\frac{32 (1-3 n) \mu }{\sqrt{n}}\right) u_1^5$\\$+\frac{1}{160} v \left(-80 (6 n+1) n-\frac{15 n}{\mu ^2}+\frac{20 (n (12 n-7)+1)
   \sqrt{n}}{\mu }+80-\frac{32 (15 (n-1) n+4) \mu }{\sqrt{n}}\right) u_1^5$\\$+\frac{\left(8 (3 n-1) (u_2+1) \mu ^2+4 \sqrt{n} (2 n (-3 n u_2+u_2+3)+3) \mu -3 n (4 n-1) (u_2+1)\right)
   u_1^4}{16 \sqrt{n} \mu }$\\$+\frac{v \left(64 (1-3 n) \mu ^3+16 \sqrt{n} (3 n-1) (4 n+1) (u_2+1) \mu ^2+4 n (2 n ((2-6 n) u_2+3)-3) \mu +3 n^{3/2} (u_2+1)\right) u_1^4}{32 \sqrt{n} \mu
   ^2}$\\$+\frac{v^2 \left(-32 (15 (n-1) n+4) \mu ^3-48 \sqrt{n} \left(6 n^2+n-1\right) \mu ^2+12 n (n (12 n-7)+1) \mu -3 n^{3/2}\right) u_1^4}{64 \sqrt{n} \mu ^2}+\frac{3}{8} \left(\frac{u_2 (u_2+2)
   n^{3/2}}{\mu }-2 (4 n+1) (u_2+1)+\frac{8 \mu }{\sqrt{n}}\right) u_1^3+\frac{v^3 \left(-16 (15 (n-1) n+4) \mu ^2-12 \sqrt{n} \left(6 n^2+n-1\right) \mu +3 n (n (12 n-7)+1)\right) u_1^3}{24 \sqrt{n} \mu
   }+\frac{v \left(16 (3 n-1) (u_2+1) \mu ^2+4 \sqrt{n} (2 n (3-(3 n-1) u_2 (u_2+2))+3) \mu -3 n (4 n-1) (u_2+1)\right) u_1^3}{8 \sqrt{n} \mu }+\frac{3 v^2 \left(64 (1-3 n) \mu ^3+16 \sqrt{n}
   (3 n-1) (4 n+1) (u_2+1) \mu ^2+4 n (4 (1-3 n) u_2 n+2 n-1) \mu +n^{3/2} (u_2+1)\right) u_1^3}{64 \sqrt{n} \mu ^2}+\frac{3}{4} v \left(\frac{u_2 (u_2+2) n^{3/2}}{\mu }-2 (4 n+1)
   (u_2+1)+\frac{8 \mu }{\sqrt{n}}\right) u_1^2+\left(\frac{3}{2} n u_2 (u_2+2)-\frac{3 (u_2+1) \mu }{\sqrt{n}}\right) u_1^2-\frac{(3 n-1) v^3 \left(u_2 n^2-(4 n+1) (u_2+1)
   \mu  \sqrt{n}+4 \mu ^2\right) u_1^2}{2 \sqrt{n} \mu }+\frac{v^4 \left(-16 (15 (n-1) n+4) \mu ^2-4 \sqrt{n} \left(6 n^2+n-1\right) \mu +n (n (12 n-7)+1)\right) u_1^2}{32 \sqrt{n} \mu }+\frac{3 v^2 \left(16 (3
   n-1) (u_2+1) \mu ^2+4 \sqrt{n} \left(-6 u_2 (u_2+2) n^2+2 (u_2+1)^2 n+1\right) \mu -n (4 n-1) (u_2+1)\right) u_1^2}{16 \sqrt{n} \mu }-\frac{(15 (n-1) n+4) v^5 \mu  u_1}{5
   \sqrt{n}}+\frac{(3 n-1) v^3 \left(2 (u_2+1) \mu -n^{3/2} u_2 (u_2+2)\right) u_1}{\sqrt{n}}+\left(3 n u_2 (u_2+2) v-\frac{6 (u_2+1) v \mu }{\sqrt{n}}\right) u_1-\frac{(3
   n-1) v^4 \left(u_2 n^2-(4 n+1) (u_2+1) \mu  \sqrt{n}+4 \mu ^2\right) u_1}{8 \sqrt{n} \mu }+\frac{3 v^2 \left(u_2 (u_2+2) n^2-2 (4 n+1) (u_2+1) \mu  \sqrt{n}+8 \mu ^2\right)
   u_1}{8 \sqrt{n} \mu }+\frac{3}{2} n u_2 (u_2+2) v^2-\frac{3 (u_2+1) v^2 \mu }{\sqrt{n}}-\frac{(3 n-1) v^4 \left(n^{3/2} u_2 (u_2+2)-2 (u_2+1) \mu \right)}{4 \sqrt{n}}$,\\
$g(\mathbf{u},v)=\frac{u_1^4 v \left(-3 n^{3/2}+64 \mu ^3 (3 n-1)-2 \mu ^2 \sqrt{n} (n (15 n+28)-13)+6 \mu  (1-3 n) n\right)}{32 \mu ^2 \sqrt{n}}+\frac{3 u_1^2 v^2 \left(3 n^2 (u_2+1)-4 \mu  (3 n+2) \sqrt{n}+4 \mu ^2 (-3 n
   (4 u_2+3)+4 u_2+2)\right)}{16 \mu  \sqrt{n}}+\frac{1}{20} \left(1-5 n^2\right) v^5-\frac{u_1^3 v^2 \left(3 n^{3/2}+36 \mu  n^2+192 \mu ^3 (1-3 n)+2 \mu ^2 \sqrt{n} (n (65 n+84)-43)\right)}{64 \mu ^2
   \sqrt{n}}$\\$+\frac{1}{320} u_1^5 \left(-\frac{15 n}{\mu ^2}+\frac{20 \sqrt{n} (2-3 n)}{\mu }+\frac{160 \mu  (3 n-1)}{\sqrt{n}}-5 n (11 n+28)+61\right)+\frac{u_1^4 \left(-4 \mu  \sqrt{n} (3 n+4)+2 \mu ^2 (-3 n (4
   u_2+3)+4 u_2+2)+n (3 n-2) (u_2+1)\right)}{16 \mu  \sqrt{n}}+u_1^3 \left(-\frac{3 \mu }{\sqrt{n}}+\frac{3 \sqrt{n}}{8 \mu }+\frac{3}{8} n (2 u_2+1)+u_2+1\right)+\frac{u_1^3 v
   \left(-36 \mu  \sqrt{n} (n+1)+8 \mu ^2 (-3 n (4 u_2+3)+4 u_2+2)+3 n (3 n-1) (u_2+1)\right)}{16 \mu  \sqrt{n}}+\frac{3}{8} u_1^2 \left(-\frac{\sqrt{n} (u_2+1)}{\mu }+\frac{4 (\mu +2 \mu 
   u_2)}{\sqrt{n}}+4\right)+\frac{3}{8} u_1^2 v \left(\frac{\sqrt{n}}{\mu }-\frac{16 \mu }{\sqrt{n}}+6 n u_2+4 n+6 u_2+6\right)-\frac{u_1^2 v^3 \left(32 \mu ^2 (1-3 n)+\mu  \sqrt{n} (7 n (5
   n+4)-17)+n (3 n+1)\right)}{16 \mu  \sqrt{n}}+v^4 \left(\frac{\mu  (3 n (4 u_1-4 u_2-3)-4 u_1+4 u_2+2)}{8 \sqrt{n}}-\frac{1}{64} (n (75 n+28)-25) u_1\right)+\frac{u_1 v^3 \left(-4 \mu 
   \sqrt{n} (3 n+1)+8 \mu ^2 (-3 n (4 u_2+3)+4 u_2+2)+n (3 n+1) (u_2+1)\right)}{16 \mu  \sqrt{n}}+\frac{3}{8} u_1 v^2 \left(-\frac{8 \mu }{\sqrt{n}}+6 n u_2+5 n+4 u_2+4\right)+\frac{3}{8}
   u_1 v \left(-\frac{\sqrt{n} (u_2+1)}{\mu }+\frac{8 (\mu +2 \mu  u_2)}{\sqrt{n}}+4\right)+\frac{1}{4} (3 n+1) (u_2+1) v^3+\frac{3 \mu  (2 u_2+1) v^2}{2 \sqrt{n}}-\frac{3 u_1
   u_2}{2}-\frac{3 u_2 v}{2}$.
	\end{small}
\end{widetext}

According to Theorem \ref{existenceCM}, there exists a 2-dimensional
invariant local center manifold $W^{c}\left( \mathbf{0}\right) $ of %
\eqref{HLcenterCase3_system527}, 
$W^{c}\left(\mathbf{0}\right) =\left\{\left(\mathbf{u},v\right) \in\mathbb{R}%
^{2}\times\mathbb{R}:v=h\left( \mathbf{u}\right)\right\}$,
satisfying $\mathbf{h}\left(\mathbf{0}\right)= 0,\;D h\left(\mathbf{0}%
\right) =0,\;\left\vert \mathbf{u}\right\vert <\delta$ for $\delta$
sufficiently small. The restriction of (\ref{HLcenterCase3_system527}) to
the center manifold is 
\begin{equation}
\frac{d\mathbf{u}}{dN}=\mathbf{f}\left(\mathbf{u},h\left(\mathbf{u}%
\right)\right),  \label{HLcenterCase3_system527rest}
\end{equation}
where the function $h\left(\mathbf{u}\right)$ that defines the local center
manifold satisfies \eqref{MaineqcM}:%
\begin{equation}
Dh\left(\mathbf{u}\right) \cdot \mathbf{f}\left(\mathbf{u},h\left( \mathbf{u}%
\right) \right) +\frac{3}{2} h\left( \mathbf{u}\right) -g\left(\mathbf{u},
h\left(\mathbf{u}\right) \right) =0.  \label{HLcenterCase3_system527h}
\end{equation}

According to Theorem \ref{approximationCM}, the system %
\eqref{HLcenterCase3_system527h} can be solved approximately by expanding $%
h\left(\mathbf{u}\right) $ in Taylor series at $\mathbf{u}=\mathbf{0},$ We
propose the ansatz 
\begin{align}  \label{ansatz2}
&h(u_1,u_2)=a_{1} u_1^2+a_{10} u_1 u_2^2+a_{11} u_1^2 u_2^2+a_{12} u_1^3
u_2^2 +a_{13} u_2^3  \notag \\
& +a_{14} u_1 u_2^3+a_{15} u_1^2 u_2^3+a_{16} u_2^4+a_{17} u_1 u_2^4 +a_{18}
u_2^5+a_{2} u_1^3  \notag \\
&+a_{3} u_1^4+a_{4} u_1^5+a_{5} u_1 u_2+a_{6} u_1^2 u_2 +a_{7} u_1^3
u_2+a_{8} u_1^4 u_2  \notag \\
& +a_{9} u_2^2 +\mathcal{O}(\left\vert \mathbf{u}\right\vert^6)
\end{align}
By comparing the coefficients of the equal powers of $u_1, u_2$ we find the
non-null coefficients
\begin{widetext}
$a_1= \frac{\mu }{\sqrt{n}}-\frac{\sqrt{n}}{4 \mu }+1, a_2= \frac{6 \mu ^2}{n}+\frac{1}{16} \left(\frac{1}{\mu ^2}+4\right) n+\frac{3 \mu }{\sqrt{n}}-\frac{\sqrt{n}}{4 \mu }-\frac{1}{12}$,\\
$a_3=
   \frac{3648 \mu ^6-16 \mu ^3 n^{3/2} (5 n+9)+4 \mu  n^{5/2}-n^3-12 \mu ^2 (n-2) n^2+1920 \mu ^5 \sqrt{n}+48 \mu ^4 n (3 n-8)}{64 \mu ^3 n^{3/2}}$,\\
	$a_4= \frac{542208 \mu ^8-192 \mu ^5 n^{3/2} (59 n+200)+48 \mu ^3
   n^{5/2} (15 n+26)-12 \mu  n^{7/2}+3 n^4+8 \mu ^4 n^3 (13 n-1050)+24 \mu ^2 n^3 (3 n-7)+306432 \mu ^7 \sqrt{n}+192 \mu ^6 n (121 n-327)}{768 \mu ^4 n^2}+\frac{3121}{480}, a_5= -1, a_6= -\frac{3 \mu
   }{\sqrt{n}}+\frac{\sqrt{n}}{4 \mu }-2, a_7=\frac{1}{16} \left(-\frac{480 \mu ^2}{n}+\left(20-\frac{1}{\mu ^2}\right) n-\frac{192 \mu }{\sqrt{n}}+\frac{8 \sqrt{n}}{\mu }+4\right), a_8= -\frac{399 \mu
   ^3}{n^{3/2}}+\frac{n^{3/2}}{64 \mu ^3}-\frac{180 \mu ^2}{n}-\frac{n}{8 \mu ^2}+\frac{\mu  (59 n+120)}{4 \sqrt{n}}-\frac{3 \sqrt{n} (5 n+6)}{16 \mu }+15 n+9, a_{10}=1, a_{11}= \frac{6 \mu
   }{\sqrt{n}}-\frac{\sqrt{n}}{4 \mu }+3, a_{12}= \frac{1}{16} \left(\frac{1440 \mu ^2}{n}+\left(\frac{1}{\mu ^2}-120\right) n+\frac{480 \mu }{\sqrt{n}}-\frac{12 \sqrt{n}}{\mu }-8\right), a_{14}=
   -1, a_{15}= \frac{n-40 \mu ^2}{4 \mu  \sqrt{n}}-4, a_{17}=1$.\\
				Therefore, \\
	$h(u_1,u_2)=u_1^4 u_2 \left(-\frac{399 \mu ^3}{n^{3/2}}+\frac{n^{3/2}}{64 \mu ^3}-\frac{180 \mu ^2}{n}-\frac{n}{8 \mu ^2}+\frac{\mu  (59 n+120)}{4 \sqrt{n}}-\frac{3 \sqrt{n} (5 n+6)}{16 \mu }+15 n+9\right)$\\$+\frac{u_1^4
   \left(3648 \mu ^6-16 \mu ^3 n^{3/2} (5 n+9)+4 \mu  n^{5/2}-n^3-12 \mu ^2 (n-2) n^2+1920 \mu ^5 \sqrt{n}+48 \mu ^4 n (3 n-8)\right)}{64 \mu ^3 n^{3/2}}$\\$+u_1^5 \left(\frac{542208 \mu ^8-192 \mu ^5 n^{3/2} (59
   n+200)+48 \mu ^3 n^{5/2} (15 n+26)-12 \mu  n^{7/2}+3 n^4+8 \mu ^4 n^3 (13 n-1050)+24 \mu ^2 n^3 (3 n-7)+306432 \mu ^7 \sqrt{n}+192 \mu ^6 n (121 n-327)}{768 \mu ^4 n^2}+\frac{3121}{480}\right)+u_1^3 \left(\frac{6
   \mu ^2}{n}+\frac{1}{16} \left(\frac{1}{\mu ^2}+4\right) n+\frac{3 \mu }{\sqrt{n}}-\frac{\sqrt{n}}{4 \mu }-\frac{1}{12}\right)+\frac{1}{16} u_1^3 u_2^2 \left(\frac{1440 \mu ^2}{n}+\left(\frac{1}{\mu
   ^2}-120\right) n+\frac{480 \mu }{\sqrt{n}}-\frac{12 \sqrt{n}}{\mu }-8\right)+\frac{1}{16} u_1^3 u_2 \left(-\frac{480 \mu ^2}{n}+\left(20-\frac{1}{\mu ^2}\right) n-\frac{192 \mu }{\sqrt{n}}+\frac{8
   \sqrt{n}}{\mu }+4\right)+u_1^2 \left(\frac{\mu }{\sqrt{n}}-\frac{\sqrt{n}}{4 \mu }+1\right)+u_1^2 u_2^3 \left(\frac{n-40 \mu ^2}{4 \mu  \sqrt{n}}-4\right)+u_1^2 u_2^2 \left(\frac{6 \mu
   }{\sqrt{n}}-\frac{\sqrt{n}}{4 \mu }+3\right)+u_1^2 u_2 \left(-\frac{3 \mu }{\sqrt{n}}+\frac{\sqrt{n}}{4 \mu }-2\right)+u_1 u_2^4-u_1 u_2^3+u_1 u_2^2-u_1 u_2$. \\
	Finally, the center manifold can be expressed as 	\\
	$\theta =-\frac{23134208 T^5 \mu ^9}{n^{9/2}}+\frac{57835520 T^4 \mu ^9}{n^{9/2}}-\frac{57835520 T^3 \mu ^9}{n^{9/2}}+\frac{28917760 T^2 \mu ^9}{n^{9/2}}-\frac{7229440 T \mu ^9}{n^{9/2}}+\frac{722944 \mu ^9}{n^{9/2}}-\frac{991232
   T^5 \mu ^7}{n^{5/2}}+\frac{5627904 T^5 \mu ^7}{n^{7/2}}-\frac{1634304 Q T^4 \mu ^7}{n^{7/2}}+\frac{2478080 T^4 \mu ^7}{n^{5/2}}-\frac{12201984 T^4 \mu ^7}{n^{7/2}}+\frac{3268608 Q T^3 \mu ^7}{n^{7/2}}-\frac{2478080 T^3
   \mu ^7}{n^{5/2}}+\frac{10334208 T^3 \mu ^7}{n^{7/2}}-\frac{2451456 Q T^2 \mu ^7}{n^{7/2}}+\frac{1239040 T^2 \mu ^7}{n^{5/2}}-\frac{4233216 T^2 \mu ^7}{n^{7/2}}-\frac{102144 Q \mu ^7}{n^{7/2}}+\frac{817152 Q T \mu
   ^7}{n^{7/2}}-\frac{309760 T \mu ^7}{n^{5/2}}+\frac{824832 T \mu ^7}{n^{7/2}}+\frac{30976 \mu ^7}{n^{5/2}}-\frac{59136 \mu ^7}{n^{7/2}}-\frac{13312 T^5 \mu ^5}{3 \sqrt{n}}+\frac{112640 T^5 \mu ^5}{n^{3/2}}-\frac{5899264
   T^5 \mu ^5}{15 n^{5/2}}+\frac{60416 Q T^4 \mu ^5}{n^{3/2}}+\frac{245760 Q T^4 \mu ^5}{n^{5/2}}+\frac{33280 T^4 \mu ^5}{3 \sqrt{n}}-\frac{332800 T^4 \mu ^5}{n^{3/2}}+\frac{2064896 T^4 \mu ^5}{3 n^{5/2}}-\frac{46080 Q^2
   T^3 \mu ^5}{n^{5/2}}-\frac{120832 Q T^3 \mu ^5}{n^{3/2}}-\frac{384000 Q T^3 \mu ^5}{n^{5/2}}-\frac{33280 T^3 \mu ^5}{3 \sqrt{n}}+\frac{384000 T^3 \mu ^5}{n^{3/2}}-\frac{1373696 T^3 \mu ^5}{3 n^{5/2}}+\frac{5760 Q^2 \mu
   ^5}{n^{5/2}}+\frac{69120 Q^2 T^2 \mu ^5}{n^{5/2}}+\frac{90624 Q T^2 \mu ^5}{n^{3/2}}+\frac{207360 Q T^2 \mu ^5}{n^{5/2}}+\frac{16640 T^2 \mu ^5}{3 \sqrt{n}}-\frac{217600 T^2 \mu ^5}{n^{3/2}}+\frac{438016 T^2 \mu ^5}{3
   n^{5/2}}+\frac{3776 Q \mu ^5}{n^{3/2}}+\frac{1920 Q \mu ^5}{n^{5/2}}-\frac{34560 Q^2 T \mu ^5}{n^{5/2}}-\frac{30208 Q T \mu ^5}{n^{3/2}}-\frac{42240 Q T \mu ^5}{n^{5/2}}-\frac{4160 T \mu ^5}{3 \sqrt{n}}+\frac{60800 T \mu
   ^5}{n^{3/2}}-\frac{71488 T \mu ^5}{3 n^{5/2}}+\frac{416 \mu ^5}{3 \sqrt{n}}-\frac{6720 \mu ^5}{n^{3/2}}+\frac{28832 \mu ^5}{15 n^{5/2}}-\frac{3072 T^5 \mu ^3}{\sqrt{n}}+\frac{9216 T^5 \mu ^3}{n^{3/2}}-\frac{3840 Q T^4
   \mu ^3}{\sqrt{n}}-\frac{7680 Q T^4 \mu ^3}{n^{3/2}}+\frac{10752 T^4 \mu ^3}{\sqrt{n}}-\frac{12800 T^4 \mu ^3}{n^{3/2}}-\frac{160 Q^3 \mu ^3}{n^{3/2}}+\frac{3840 Q^2 T^3 \mu ^3}{\sqrt{n}}+\frac{3328 Q^2 T^3 \mu
   ^3}{n^{3/2}}-\frac{640 Q T^3 \mu ^3}{\sqrt{n}}+\frac{7040 Q T^3 \mu ^3}{n^{3/2}}-\frac{9472 T^3 \mu ^3}{\sqrt{n}}+\frac{24320 T^3 \mu ^3}{3 n^{3/2}}-\frac{480 Q^2 \mu ^3}{\sqrt{n}}+\frac{160 Q^2 \mu
   ^3}{n^{3/2}}-\frac{640 Q^3 T^2 \mu ^3}{n^{3/2}}-\frac{5760 Q^2 T^2 \mu ^3}{\sqrt{n}}-\frac{2688 Q^2 T^2 \mu ^3}{n^{3/2}}+\frac{6720 Q T^2 \mu ^3}{\sqrt{n}}-\frac{1920 Q T^2 \mu ^3}{n^{3/2}}+\frac{1920 T^2 \mu
   ^3}{\sqrt{n}}-\frac{3200 T^2 \mu ^3}{n^{3/2}}+\frac{800 Q \mu ^3}{\sqrt{n}}-\frac{160 Q \mu ^3}{n^{3/2}}+\frac{640 Q^3 T \mu ^3}{n^{3/2}}+\frac{2880 Q^2 T \mu ^3}{\sqrt{n}}+\frac{192 Q^2 T \mu ^3}{n^{3/2}}-\frac{4320 Q T
   \mu ^3}{\sqrt{n}}+\frac{480 Q T \mu ^3}{n^{3/2}}+\frac{768 T \mu ^3}{\sqrt{n}}+\frac{640 T \mu ^3}{n^{3/2}}-\frac{256 \mu ^3}{\sqrt{n}}-\frac{64 \mu ^3}{3 n^{3/2}}-\frac{128 T^5 \mu }{\sqrt{n}}+\frac{4 Q^4 \mu
   }{\sqrt{n}}+\frac{64 Q T^4 \mu }{\sqrt{n}}+\frac{192 T^4 \mu }{\sqrt{n}}-\frac{16 Q^3 \mu }{\sqrt{n}}-\frac{32 Q^2 T^3 \mu }{\sqrt{n}}-\frac{32 Q T^3 \mu }{\sqrt{n}}-\frac{160 T^3 \mu }{\sqrt{n}}+\frac{28 Q^2 \mu
   }{\sqrt{n}}+\frac{16 Q^3 T^2 \mu }{\sqrt{n}}-\frac{16 Q^2 T^2 \mu }{\sqrt{n}}+\frac{48 Q T^2 \mu }{\sqrt{n}}+\frac{48 T^2 \mu }{\sqrt{n}}-\frac{24 Q \mu }{\sqrt{n}}-\frac{8 Q^4 T \mu }{\sqrt{n}}+\frac{24 Q^3 T \mu
   }{\sqrt{n}}-\frac{40 Q^2 T \mu }{\sqrt{n}}+\frac{24 Q T \mu }{\sqrt{n}}-\frac{24 T \mu }{\sqrt{n}}+\frac{12 \mu }{\sqrt{n}}$.

	The dynamics on the center manifold  is given by
\begin{align}	\label{Center2DCase3EM}
&\frac{d u_1}{dN}=u_1^4 \left(\frac{3 \mu  \left(-13 \mu ^2-6 \mu  \sqrt{n}+n\right)}{n^{3/2}}+u_2 \left(\frac{3 \mu  \left(65 \mu ^2+24 \mu  \sqrt{n}-n (4
   n+3)\right)}{n^{3/2}}-\frac{3}{2}\right)+\frac{3}{4}\right)\nonumber \\
	& +\frac{3 u_1^5 \left(-504 \mu ^4+4 \mu  n^{3/2} (4 n+5)-3 n^2-260 \mu ^3 \sqrt{n}+\mu ^2 (53-12 n) n\right)}{4 n^2}\nonumber \\
	& +u_1^3 \left(-\frac{3 \mu  \left(2
   \mu +\sqrt{n}\right)}{n}+u_2^2 \left(\frac{3}{4}-\frac{9 \mu  \left(4 \mu +\sqrt{n}\right)}{n}\right)+u_2 \left(\frac{6 \mu  \left(3 \mu +\sqrt{n}\right)}{n}-\frac{3}{4}\right)+\frac{3}{4}\right)\nonumber \\
	& +u_1^2
   \left(-\frac{3 \mu }{\sqrt{n}}+\frac{3 \mu  u_2^3}{\sqrt{n}}-\frac{3 \mu  u_2^2}{\sqrt{n}}+\frac{3 \mu  u_2}{\sqrt{n}}\right),\\
&\frac{d u_2}{d N}=u_1^4 \Big(\frac{3}{4} \left(-\frac{52 \mu
   ^3}{n^{3/2}}+\frac{n^{3/2}}{\mu }-\frac{24 \mu ^2}{n}+\frac{\mu  (4-8 n)}{\sqrt{n}}-6 n+1\right) \nonumber \\
		& +u_2 \left(\frac{195 \mu ^3}{n^{3/2}}-\frac{9 n^{3/2}}{4 \mu }+\frac{3 n^2}{16 \mu ^2}+\frac{3 \mu ^2 (13
   n+24)}{n}+\frac{3 \mu  (10 n-3)}{\sqrt{n}}+12 n-\frac{3}{2}\right)\Big)\nonumber \\
	& +u_1^3 \left(u_2^2 \left(\frac{9 n^{3/2}}{8 \mu }-\frac{36 \mu ^2}{n}-\frac{3 \mu  (7 n+3)}{\sqrt{n}}-18
   n+\frac{3}{4}\right)+\frac{3}{4} u_2 \left(-\frac{n^{3/2}}{\mu }+\frac{24 \mu ^2}{n}+\frac{8 \mu  (n+1)}{\sqrt{n}}+12 n-1\right)-\frac{3 \left(2 \mu ^2+n^2+\mu  \sqrt{n}\right)}{n}+\frac{3}{4}\right)\nonumber \\& +u_1^5 \left(\frac{-6048 \mu ^6+48 \mu ^3 (5-3 n) n^{3/2}+18 \mu  n^{7/2}-3 n^4-3120 \mu ^5 \sqrt{n}+12 \mu ^4 (53-64 n) n}{16 \mu ^2 n^2}-\frac{9}{4}\right)\nonumber \\
	& +u_1^2 \left(-\frac{3 \mu }{\sqrt{n}}+u_2^3 \left(\frac{3
   \mu }{\sqrt{n}}+6 n\right)+u_2^2 \left(-\frac{3 \mu }{\sqrt{n}}-\frac{9 n}{2}\right)+u_2 \left(\frac{3 \mu }{\sqrt{n}}+3 n\right)\right).
	\end{align}
		\end{widetext}

For which the origin is unstable (see Figure \ref{fig:Center2DCase3EM}).
Using the Theorem \ref{stabilityCM}, we conclude that the center manifold of
origin for the system \eqref{Center2DCase3PLa}, \eqref{Center2DCase3PLb}, and the origin are unstable
(saddle). $\blacksquare$ 

\textbf{Proof Part (ii)}:

More general, by introducing the linear transformation 
\begin{align}
& (u_1, u_2, v)=  \notag \\
& \left(\frac{\mu (4-8 T)}{\sqrt{n}},\frac{\mu (4-8 T)}{\sqrt{n}}%
+Q-Q^{*},\theta Q^{*} +\frac{4 \mu (2 T-1)}{\sqrt{n}}\right),
\end{align}%
and taking Taylor series near $(u_1,u_2,v)=(0,0,0)$, and taking Taylor
series near $(u_1,u_2,v)=(0,0,0)$ up to fifth order we obtain the system %
\eqref{system517a}, \eqref{system517b}, \eqref{system517c} can be written into its Jordan canonical form: %
\begin{equation}
\left(%
\begin{array}{c}
\frac{du_1}{dN} \\ 
\frac{du_2}{dN} \\ 
\frac{d v}{dN}%
\end{array}%
\right)=\left(%
\begin{array}{ccc}
0 & 0 & 0 \\ 
0 & 0 & 0 \\ 
0 & 0 & -\frac{3}{2} Q^{*}%
\end{array}%
\right)\left(%
\begin{array}{c}
u_1 \\ 
u_2 \\ 
v%
\end{array}%
\right)+\left(%
\begin{array}{c}
f_1^{*} (\mathbf{u},v) \\ 
f_2^{*} (\mathbf{u},v) \\ 
g^{*} (\mathbf{u},v)%
\end{array}%
\right),  \label{HLcenterCase3_systemGeneralEM}
\end{equation}%
where
\begin{widetext}
$f_1^{*} (\mathbf{u},v)=\frac{(u_1+v)^2 \left(3 n^{3/2} {Q^{*}}^3 u_1^3+12 \mu  n {Q^{*}}^2 u_1^2 ({Q^{*}}-u_1+u_2)-32 \mu ^3 ({Q^{*}}-u_1+u_2) \left(6 {Q^{*}}^2-(3 n-1) (u_1+v)^2\right)+8
   \mu ^2 \sqrt{n} {Q^{*}} u_1 \left((3 n-1) (u_1+v)^2-6 {Q^{*}}^2+6 {Q^{*}} (u_1-u_2)\right)\right)}{64 \mu ^2 \sqrt{n} {Q^{*}}^4}$,\\
$f_2^{*} (\mathbf{u},v)=-\frac{3 \sqrt{n} \mu  u_1^5}{2 {Q^{*}}^4}+\frac{\mu  u_1^5}{2 \sqrt{n} {Q^{*}}^4}-\frac{3 n^{5/2} u_1^5}{16 {Q^{*}}^2 \mu }+\frac{7 n^{3/2} u_1^5}{16 {Q^{*}}^2 \mu }-\frac{3 \sqrt{n}
   u_1^5}{16 {Q^{*}}^2 \mu }+\frac{3 n^{5/2} u_1^5}{16 {Q^{*}}^4 \mu }-\frac{n^{3/2} u_1^5}{16 {Q^{*}}^4 \mu }+\frac{3 n u_1^5}{64 {Q^{*}} \mu ^2}+\frac{3 n^2 u_1^5}{2
   {Q^{*}}^3}-\frac{n u_1^5}{8 {Q^{*}}^3}-\frac{u_1^5}{8 {Q^{*}}^3}-\frac{3 n^2 u_2 u_1^4}{2 {Q^{*}}^3}+\frac{n u_2 u_1^4}{2 {Q^{*}}^3}+\frac{6 n^2 v
   u_1^4}{{Q^{*}}^3}-\frac{n v u_1^4}{2 {Q^{*}}^3}-\frac{v u_1^4}{2 {Q^{*}}^3}+\frac{3 \sqrt{n} u_2 \mu  u_1^4}{2 {Q^{*}}^4}-\frac{u_2 \mu  u_1^4}{2 \sqrt{n}
   {Q^{*}}^4}-\frac{6 \sqrt{n} v \mu  u_1^4}{{Q^{*}}^4}+\frac{2 v \mu  u_1^4}{\sqrt{n} {Q^{*}}^4}+\frac{3 \sqrt{n} \mu  u_1^4}{2 {Q^{*}}^3}-\frac{\mu  u_1^4}{2 \sqrt{n} {Q^{*}}^3}-\frac{3
   n^{3/2} u_2 u_1^4}{4 {Q^{*}}^2 \mu }+\frac{3 \sqrt{n} u_2 u_1^4}{16 {Q^{*}}^2 \mu }-\frac{3 n^{5/2} v u_1^4}{4 {Q^{*}}^2 \mu }+\frac{n^{3/2} v u_1^4}{{Q^{*}}^2 \mu }-\frac{3
   \sqrt{n} v u_1^4}{8 {Q^{*}}^2 \mu }+\frac{3 n^{5/2} v u_1^4}{4 {Q^{*}}^4 \mu }-\frac{n^{3/2} v u_1^4}{4 {Q^{*}}^4 \mu }-\frac{3 n^{3/2} u_1^4}{4 {Q^{*}} \mu }+\frac{3 \sqrt{n}
   u_1^4}{16 {Q^{*}} \mu }-\frac{3 n^2 u_1^4}{4 {Q^{*}}^2}+\frac{7 n u_1^4}{4 {Q^{*}}^2}+\frac{3 u_1^4}{4 {Q^{*}}^2}+\frac{3 n v u_1^4}{32 {Q^{*}} \mu ^2}+\frac{3 n^2 u_1^4}{4
   {Q^{*}}^4}-\frac{n u_1^4}{4 {Q^{*}}^4}+\frac{9 n^2 v^2 u_1^3}{{Q^{*}}^3}-\frac{3 n v^2 u_1^3}{4 {Q^{*}}^3}-\frac{3 v^2 u_1^3}{4 {Q^{*}}^3}-\frac{3 n u_2
   u_1^3}{{Q^{*}}^2}-\frac{3 u_2 u_1^3}{4 {Q^{*}}^2}-\frac{6 n^2 u_2 v u_1^3}{{Q^{*}}^3}+\frac{2 n u_2 v u_1^3}{{Q^{*}}^3}-\frac{3 n^2 v u_1^3}{{Q^{*}}^2}+\frac{4 n
   v u_1^3}{{Q^{*}}^2}+\frac{3 v u_1^3}{2 {Q^{*}}^2}+\frac{3 n^2 v u_1^3}{{Q^{*}}^4}-\frac{n v u_1^3}{{Q^{*}}^4}-\frac{9 \sqrt{n} v^2 \mu  u_1^3}{{Q^{*}}^4}+\frac{3 v^2 \mu 
   u_1^3}{\sqrt{n} {Q^{*}}^4}+\frac{6 \sqrt{n} u_2 v \mu  u_1^3}{{Q^{*}}^4}-\frac{2 u_2 v \mu  u_1^3}{\sqrt{n} {Q^{*}}^4}+\frac{6 \sqrt{n} v \mu  u_1^3}{{Q^{*}}^3}-\frac{2 v \mu 
   u_1^3}{\sqrt{n} {Q^{*}}^3}+\frac{3 \mu  u_1^3}{\sqrt{n} {Q^{*}}^2}-\frac{3 n u_1^3}{{Q^{*}}}-\frac{3 u_1^3}{4 {Q^{*}}}+\frac{3 n^{3/2} u_2^2 u_1^3}{8 {Q^{*}}^2 \mu }-\frac{9
   n^{5/2} v^2 u_1^3}{8 {Q^{*}}^2 \mu }+\frac{3 n^{3/2} v^2 u_1^3}{4 {Q^{*}}^2 \mu }-\frac{3 \sqrt{n} v^2 u_1^3}{16 {Q^{*}}^2 \mu }+\frac{9 n^{5/2} v^2 u_1^3}{8 {Q^{*}}^4 \mu }-\frac{3
   n^{3/2} v^2 u_1^3}{8 {Q^{*}}^4 \mu }+\frac{3 n^{3/2} u_1^3}{8 \mu }+\frac{3 n^{3/2} u_2 u_1^3}{4 {Q^{*}} \mu }-\frac{3 n^{3/2} u_2 v u_1^3}{2 {Q^{*}}^2 \mu }+\frac{3 \sqrt{n}
   u_2 v u_1^3}{8 {Q^{*}}^2 \mu }-\frac{3 n^{3/2} v u_1^3}{2 {Q^{*}} \mu }+\frac{3 \sqrt{n} v u_1^3}{8 {Q^{*}} \mu }-\frac{3 n^{3/2} u_1^3}{8 {Q^{*}}^2 \mu }+\frac{3 n v^2
   u_1^3}{64 {Q^{*}} \mu ^2}+\frac{6 n^2 v^3 u_1^2}{{Q^{*}}^3}-\frac{n v^3 u_1^2}{2 {Q^{*}}^3}-\frac{v^3 u_1^2}{2 {Q^{*}}^3}+\frac{3 n u_2^2 u_1^2}{2 {Q^{*}}^2}-\frac{9 n^2
   u_2 v^2 u_1^2}{{Q^{*}}^3}+\frac{3 n u_2 v^2 u_1^2}{{Q^{*}}^3}-\frac{9 n^2 v^2 u_1^2}{2 {Q^{*}}^2}+\frac{3 n v^2 u_1^2}{{Q^{*}}^2}+\frac{3 v^2 u_1^2}{4
   {Q^{*}}^2}+\frac{9 n^2 v^2 u_1^2}{2 {Q^{*}}^4}-\frac{3 n v^2 u_1^2}{2 {Q^{*}}^4}+\frac{3 n u_1^2}{2}+\frac{3 n u_2 u_1^2}{{Q^{*}}}-\frac{6 n u_2 v
   u_1^2}{{Q^{*}}^2}-\frac{3 u_2 v u_1^2}{2 {Q^{*}}^2}-\frac{6 n v u_1^2}{{Q^{*}}}-\frac{3 v u_1^2}{2 {Q^{*}}}-\frac{6 \sqrt{n} v^3 \mu  u_1^2}{{Q^{*}}^4}+\frac{2 v^3 \mu 
   u_1^2}{\sqrt{n} {Q^{*}}^4}+\frac{9 \sqrt{n} u_2 v^2 \mu  u_1^2}{{Q^{*}}^4}-\frac{3 u_2 v^2 \mu  u_1^2}{\sqrt{n} {Q^{*}}^4}+\frac{9 \sqrt{n} v^2 \mu  u_1^2}{{Q^{*}}^3}-\frac{3
   v^2 \mu  u_1^2}{\sqrt{n} {Q^{*}}^3}-\frac{3 u_2 \mu  u_1^2}{\sqrt{n} {Q^{*}}^2}+\frac{6 v \mu  u_1^2}{\sqrt{n} {Q^{*}}^2}-\frac{3 \mu  u_1^2}{\sqrt{n} {Q^{*}}}-\frac{3 n^{5/2} v^3
   u_1^2}{4 {Q^{*}}^2 \mu }+\frac{n^{3/2} v^3 u_1^2}{4 {Q^{*}}^2 \mu }+\frac{3 n^{5/2} v^3 u_1^2}{4 {Q^{*}}^4 \mu }-\frac{n^{3/2} v^3 u_1^2}{4 {Q^{*}}^4 \mu }-\frac{3 n^{3/2} u_2 v^2
   u_1^2}{4 {Q^{*}}^2 \mu }+\frac{3 \sqrt{n} u_2 v^2 u_1^2}{16 {Q^{*}}^2 \mu }-\frac{3 n^{3/2} v^2 u_1^2}{4 {Q^{*}} \mu }+\frac{3 \sqrt{n} v^2 u_1^2}{16 {Q^{*}} \mu }+\frac{3 n^{3/2}
   u_2^2 v u_1^2}{4 {Q^{*}}^2 \mu }+\frac{3 n^{3/2} v u_1^2}{4 \mu }+\frac{3 n^{3/2} u_2 v u_1^2}{2 {Q^{*}} \mu }-\frac{3 n^{3/2} v u_1^2}{4 {Q^{*}}^2 \mu }-\frac{3 n u_1^2}{2
   {Q^{*}}^2}+\frac{3 n^2 v^4 u_1}{2 {Q^{*}}^3}-\frac{n v^4 u_1}{8 {Q^{*}}^3}-\frac{v^4 u_1}{8 {Q^{*}}^3}-\frac{6 n^2 u_2 v^3 u_1}{{Q^{*}}^3}+\frac{2 n u_2 v^3
   u_1}{{Q^{*}}^3}-\frac{3 n^2 v^3 u_1}{{Q^{*}}^2}+\frac{n v^3 u_1}{{Q^{*}}^2}+\frac{3 n^2 v^3 u_1}{{Q^{*}}^4}-\frac{n v^3 u_1}{{Q^{*}}^4}-\frac{3 n u_2 v^2
   u_1}{{Q^{*}}^2}-\frac{3 u_2 v^2 u_1}{4 {Q^{*}}^2}-\frac{3 n v^2 u_1}{{Q^{*}}}-\frac{3 v^2 u_1}{4 {Q^{*}}}+\frac{3 n u_2^2 v u_1}{{Q^{*}}^2}+3 n v u_1+\frac{6 n
   u_2 v u_1}{{Q^{*}}}-\frac{3 n v u_1}{{Q^{*}}^2}-\frac{3 \sqrt{n} v^4 \mu  u_1}{2 {Q^{*}}^4}+\frac{v^4 \mu  u_1}{2 \sqrt{n} {Q^{*}}^4}+\frac{6 \sqrt{n} u_2 v^3 \mu 
   u_1}{{Q^{*}}^4}-\frac{2 u_2 v^3 \mu  u_1}{\sqrt{n} {Q^{*}}^4}+\frac{6 \sqrt{n} v^3 \mu  u_1}{{Q^{*}}^3}-\frac{2 v^3 \mu  u_1}{\sqrt{n} {Q^{*}}^3}+\frac{3 v^2 \mu  u_1}{\sqrt{n}
   {Q^{*}}^2}-\frac{6 u_2 v \mu  u_1}{\sqrt{n} {Q^{*}}^2}-\frac{6 v \mu  u_1}{\sqrt{n} {Q^{*}}}-\frac{3 n^{5/2} v^4 u_1}{16 {Q^{*}}^2 \mu }+\frac{n^{3/2} v^4 u_1}{16 {Q^{*}}^2 \mu
   }+\frac{3 n^{5/2} v^4 u_1}{16 {Q^{*}}^4 \mu }-\frac{n^{3/2} v^4 u_1}{16 {Q^{*}}^4 \mu }+\frac{3 n^{3/2} u_2^2 v^2 u_1}{8 {Q^{*}}^2 \mu }+\frac{3 n^{3/2} v^2 u_1}{8 \mu }+\frac{3
   n^{3/2} u_2 v^2 u_1}{4 {Q^{*}} \mu }-\frac{3 n^{3/2} v^2 u_1}{8 {Q^{*}}^2 \mu }-\frac{3 n^2 u_2 v^4}{2 {Q^{*}}^3}+\frac{n u_2 v^4}{2 {Q^{*}}^3}-\frac{3 n^2 v^4}{4
   {Q^{*}}^2}+\frac{n v^4}{4 {Q^{*}}^2}+\frac{3 n^2 v^4}{4 {Q^{*}}^4}-\frac{n v^4}{4 {Q^{*}}^4}+\frac{3 n u_2^2 v^2}{2 {Q^{*}}^2}+\frac{3 n v^2}{2}+\frac{3 n u_2 v^2}{{Q^{*}}}-\frac{3 n v^2}{2
   {Q^{*}}^2}+\frac{3 \sqrt{n} u_2 v^4 \mu }{2 {Q^{*}}^4}-\frac{u_2 v^4 \mu }{2 \sqrt{n} {Q^{*}}^4}+\frac{3 \sqrt{n} v^4 \mu }{2 {Q^{*}}^3}-\frac{v^4 \mu }{2 \sqrt{n} {Q^{*}}^3}-\frac{3 u_2 v^2
   \mu }{\sqrt{n} {Q^{*}}^2}-\frac{3 v^2 \mu }{\sqrt{n} {Q^{*}}}$,\\
		$g^{*} (\mathbf{u},v)=\frac{3 \sqrt{n} \mu  u_1^5}{2 {Q^{*}}^4}-\frac{\mu  u_1^5}{2 \sqrt{n} {Q^{*}}^4}-\frac{3 n^{3/2} u_1^5}{16 {Q^{*}}^2 \mu }+\frac{\sqrt{n} u_1^5}{8 {Q^{*}}^2 \mu }-\frac{3 n u_1^5}{64
   {Q^{*}} \mu ^2}-\frac{11 n^2 u_1^5}{64 {Q^{*}}^3}-\frac{7 n u_1^5}{16 {Q^{*}}^3}+\frac{61 u_1^5}{320 {Q^{*}}^3}-\frac{15 n^2 v u_1^4}{16 {Q^{*}}^3}-\frac{7 n v u_1^4}{4
   {Q^{*}}^3}+\frac{13 v u_1^4}{16 {Q^{*}}^3}-\frac{3 \sqrt{n} u_2 \mu  u_1^4}{2 {Q^{*}}^4}+\frac{u_2 \mu  u_1^4}{2 \sqrt{n} {Q^{*}}^4}+\frac{6 \sqrt{n} v \mu 
   u_1^4}{{Q^{*}}^4}-\frac{2 v \mu  u_1^4}{\sqrt{n} {Q^{*}}^4}-\frac{9 \sqrt{n} \mu  u_1^4}{8 {Q^{*}}^3}+\frac{\mu  u_1^4}{4 \sqrt{n} {Q^{*}}^3}+\frac{3 n^{3/2} u_2 u_1^4}{16
   {Q^{*}}^2 \mu }-\frac{\sqrt{n} u_2 u_1^4}{8 {Q^{*}}^2 \mu }-\frac{9 n^{3/2} v u_1^4}{16 {Q^{*}}^2 \mu }+\frac{3 \sqrt{n} v u_1^4}{16 {Q^{*}}^2 \mu }+\frac{3 n^{3/2} u_1^4}{16
   {Q^{*}} \mu }-\frac{\sqrt{n} u_1^4}{8 {Q^{*}} \mu }-\frac{3 n u_1^4}{4 {Q^{*}}^2}-\frac{u_1^4}{{Q^{*}}^2}-\frac{3 n v u_1^4}{32 {Q^{*}} \mu ^2}-\frac{65 n^2 v^2 u_1^3}{32
   {Q^{*}}^3}-\frac{21 n v^2 u_1^3}{8 {Q^{*}}^3}+\frac{43 v^2 u_1^3}{32 {Q^{*}}^3}+\frac{3 n u_2 u_1^3}{4 {Q^{*}}^2}+\frac{u_2 u_1^3}{{Q^{*}}^2}-\frac{9 n v u_1^3}{4
   {Q^{*}}^2}-\frac{9 v u_1^3}{4 {Q^{*}}^2}+\frac{9 \sqrt{n} v^2 \mu  u_1^3}{{Q^{*}}^4}-\frac{3 v^2 \mu  u_1^3}{\sqrt{n} {Q^{*}}^4}-\frac{6 \sqrt{n} u_2 v \mu 
   u_1^3}{{Q^{*}}^4}+\frac{2 u_2 v \mu  u_1^3}{\sqrt{n} {Q^{*}}^4}-\frac{9 \sqrt{n} v \mu  u_1^3}{2 {Q^{*}}^3}+\frac{v \mu  u_1^3}{\sqrt{n} {Q^{*}}^3}-\frac{3 \mu  u_1^3}{\sqrt{n}
   {Q^{*}}^2}+\frac{3 n u_1^3}{8 {Q^{*}}}+\frac{u_1^3}{{Q^{*}}}-\frac{9 n^{3/2} v^2 u_1^3}{16 {Q^{*}}^2 \mu }+\frac{9 n^{3/2} u_2 v u_1^3}{16 {Q^{*}}^2 \mu }-\frac{3 \sqrt{n}
   u_2 v u_1^3}{16 {Q^{*}}^2 \mu }+\frac{9 n^{3/2} v u_1^3}{16 {Q^{*}} \mu }-\frac{3 \sqrt{n} v u_1^3}{16 {Q^{*}} \mu }+\frac{3 \sqrt{n} u_1^3}{8 \mu }-\frac{3 n v^2 u_1^3}{64
   {Q^{*}} \mu ^2}-\frac{35 n^2 v^3 u_1^2}{16 {Q^{*}}^3}-\frac{7 n v^3 u_1^2}{4 {Q^{*}}^3}+\frac{17 v^3 u_1^2}{16 {Q^{*}}^3}-\frac{9 n v^2 u_1^2}{4 {Q^{*}}^2}-\frac{3 v^2 u_1^2}{2
   {Q^{*}}^2}+\frac{9 n u_2 v u_1^2}{4 {Q^{*}}^2}+\frac{9 u_2 v u_1^2}{4 {Q^{*}}^2}+\frac{3 n v u_1^2}{2 {Q^{*}}}+\frac{9 v u_1^2}{4 {Q^{*}}}+\frac{6 \sqrt{n} v^3 \mu 
   u_1^2}{{Q^{*}}^4}-\frac{2 v^3 \mu  u_1^2}{\sqrt{n} {Q^{*}}^4}-\frac{9 \sqrt{n} u_2 v^2 \mu  u_1^2}{{Q^{*}}^4}+\frac{3 u_2 v^2 \mu  u_1^2}{\sqrt{n} {Q^{*}}^4}-\frac{27 \sqrt{n}
   v^2 \mu  u_1^2}{4 {Q^{*}}^3}+\frac{3 v^2 \mu  u_1^2}{2 \sqrt{n} {Q^{*}}^3}+\frac{3 u_2 \mu  u_1^2}{\sqrt{n} {Q^{*}}^2}-\frac{6 v \mu  u_1^2}{\sqrt{n} {Q^{*}}^2}+\frac{3 \mu 
   u_1^2}{2 \sqrt{n} {Q^{*}}}-\frac{3 n^{3/2} v^3 u_1^2}{16 {Q^{*}}^2 \mu }-\frac{\sqrt{n} v^3 u_1^2}{16 {Q^{*}}^2 \mu }+\frac{9 n^{3/2} u_2 v^2 u_1^2}{16 {Q^{*}}^2 \mu }+\frac{9
   n^{3/2} v^2 u_1^2}{16 {Q^{*}} \mu }-\frac{3 \sqrt{n} {Q^{*}} u_1^2}{8 \mu }-\frac{3 \sqrt{n} u_2 u_1^2}{8 \mu }+\frac{3 \sqrt{n} v u_1^2}{8 \mu }+\frac{3 u_1^2}{2}-\frac{75 n^2 v^4
   u_1}{64 {Q^{*}}^3}-\frac{7 n v^4 u_1}{16 {Q^{*}}^3}+\frac{25 v^4 u_1}{64 {Q^{*}}^3}-\frac{3 n v^3 u_1}{4 {Q^{*}}^2}-\frac{v^3 u_1}{4 {Q^{*}}^2}+\frac{9 n u_2 v^2 u_1}{4
   {Q^{*}}^2}+\frac{3 u_2 v^2 u_1}{2 {Q^{*}}^2}+\frac{15 n v^2 u_1}{8 {Q^{*}}}+\frac{3 v^2 u_1}{2 {Q^{*}}}-\frac{3 u_2 u_1}{2}+\frac{3 v u_1}{2}+\frac{3 \sqrt{n} v^4 \mu 
   u_1}{2 {Q^{*}}^4}-\frac{v^4 \mu  u_1}{2 \sqrt{n} {Q^{*}}^4}-\frac{6 \sqrt{n} u_2 v^3 \mu  u_1}{{Q^{*}}^4}+\frac{2 u_2 v^3 \mu  u_1}{\sqrt{n} {Q^{*}}^4}-\frac{9 \sqrt{n} v^3 \mu
    u_1}{2 {Q^{*}}^3}+\frac{v^3 \mu  u_1}{\sqrt{n} {Q^{*}}^3}-\frac{3 v^2 \mu  u_1}{\sqrt{n} {Q^{*}}^2}+\frac{6 u_2 v \mu  u_1}{\sqrt{n} {Q^{*}}^2}+\frac{3 v \mu  u_1}{\sqrt{n}
   {Q^{*}}}+\frac{3 n^{3/2} u_2 v^3 u_1}{16 {Q^{*}}^2 \mu }+\frac{\sqrt{n} u_2 v^3 u_1}{16 {Q^{*}}^2 \mu }+\frac{3 n^{3/2} v^3 u_1}{16 {Q^{*}} \mu }+\frac{\sqrt{n} v^3 u_1}{16
   {Q^{*}} \mu }-\frac{3 \sqrt{n} {Q^{*}} v u_1}{8 \mu }-\frac{3 \sqrt{n} u_2 v u_1}{8 \mu }-\frac{n^2 v^5}{4 {Q^{*}}^3}+\frac{v^5}{20 {Q^{*}}^3}+\frac{3 n u_2 v^3}{4
   {Q^{*}}^2}+\frac{u_2 v^3}{4 {Q^{*}}^2}+\frac{3 n v^3}{4 {Q^{*}}}+\frac{v^3}{4 {Q^{*}}}-\frac{3 u_2 v}{2}-\frac{3 \sqrt{n} u_2 v^4 \mu }{2 {Q^{*}}^4}+\frac{u_2 v^4 \mu }{2 \sqrt{n}
   {Q^{*}}^4}-\frac{9 \sqrt{n} v^4 \mu }{8 {Q^{*}}^3}+\frac{v^4 \mu }{4 \sqrt{n} {Q^{*}}^3}+\frac{3 u_2 v^2 \mu }{\sqrt{n} {Q^{*}}^2}+\frac{3 v^2 \mu }{2 \sqrt{n} {Q^{*}}}$.
\end{widetext}

 By Theorem \ref{existenceCM}, exists a 2-dimensional local center
manifold of \eqref{HLcenterCase3_system527}, $W^{c}\left(\mathbf{0}\right)
=\left\{\left(\mathbf{u},v\right) \in\mathbb{R}^{2}\times\mathbb{R}%
:v=h\left( \mathbf{u}\right)\right\}$,
satisfying $\mathbf{h}\left(\mathbf{0}\right)= 0,\;D h\left(\mathbf{0}%
\right) =0,\;\left\vert \mathbf{u}\right\vert <\delta$ for $\delta$
sufficiently small. The restriction of (\ref{HLcenterCase3_systemGeneralEM})
to the center manifold is \newline
$\frac{d\mathbf{u}}{dN}=\mathbf{f}\left(\mathbf{u},h\left(\mathbf{u}%
\right)\right)$, where the function $h\left(\mathbf{u}\right)$ that
satisfies \eqref{MaineqcM}:%
\begin{equation}
Dh\left(\mathbf{u}\right) \cdot \mathbf{f}\left(\mathbf{u},h\left( \mathbf{u}%
\right) \right) +\frac{3}{2} Q^{*} h\left( \mathbf{u}\right) -g\left(\mathbf{%
u}, h\left(\mathbf{u}\right) \right) =0.
\label{HLcenterCase3_systemGeneralEMh}
\end{equation}
Replacing \eqref{ansatz} in \eqref{HLcenterCase3_systemGeneralEMh} we find
the non-null coefficients
\begin{widetext}  
	$a_1= \frac{\frac{\mu }{\sqrt{n}}+{Q^{*}}}{{Q^{*}}^2}-\frac{\sqrt{n}}{4 \mu }, a_2= \frac{6 \mu ^2}{n {Q^{*}}^4}+\frac{3 \mu }{\sqrt{n} {Q^{*}}^3}+\frac{1}{16} n \left(\frac{1}{\mu ^2}+\frac{4
   \left(5 {Q^{*}}^2-4\right)}{{Q^{*}}^4}\right)-\frac{\sqrt{n}}{4 \mu  {Q^{*}}}-\frac{1}{12 {Q^{*}}^2}$,\\
	$a_3= \frac{3648 \mu ^6+4 \mu  n^{5/2} {Q^{*}}^5+16 \mu ^3 n^{3/2} {Q^{*}} \left(3 (5 n-3)
   {Q^{*}}^2-20 n\right)-n^3 {Q^{*}}^6+12 \mu ^2 n^2 {Q^{*}}^2 \left(n \left(4-5 {Q^{*}}^2\right)+2 {Q^{*}}^2\right)+16 \mu ^4 n \left((61 n-24) {Q^{*}}^2-52 n\right)+1920 \mu ^5 \sqrt{n} {Q^{*}}}{64 \mu ^3
   n^{3/2} {Q^{*}}^6}$,\\
	$a_4=-\frac{n^{3/2}}{64 \mu ^3 {Q^{*}}}+\frac{n^2}{256 \mu ^4}+\frac{-201 \mu ^2+\frac{706 \mu ^4}{n^2}+7 n^2}{{Q^{*}}^8}+\frac{(19-55 n) n^2+\mu ^2 (925 n-327)}{4 n {Q^{*}}^6}+\frac{5 \left(3 n^2+\mu ^2 (61
   n-40)\right)}{4 \mu  \sqrt{n} {Q^{*}}^5}+\frac{399 \mu ^3-91 \mu  n^2}{n^{3/2} {Q^{*}}^7}+\frac{5 n \left(\left(661-\frac{36}{\mu ^2}\right) n-1506\right)+3121}{480 {Q^{*}}^4}+\frac{(26-45 n) \sqrt{n}}{16 \mu 
   {Q^{*}}^3}+\frac{(15 n-7) n}{32 \mu ^2 {Q^{*}}^2}$, $a_5= -\frac{1}{{Q^{*}}}, a_6= \frac{-\frac{12 \mu }{\sqrt{n}}+\frac{\sqrt{n} {Q^{*}}^2}{\mu }-8
   {Q^{*}}}{4 {Q^{*}}^3}, a_7= -\frac{\frac{480 \mu ^2}{n}-\frac{8 \sqrt{n} {Q^{*}}^3}{\mu }+n \left(\frac{{Q^{*}}^4}{\mu ^2}+60 {Q^{*}}^2-80\right)+\frac{192 \mu  {Q^{*}}}{\sqrt{n}}-4 {Q^{*}}^2}{16
   {Q^{*}}^5}$,\\$a_8= \frac{-25536 \mu ^6-8 \mu  n^{5/2} {Q^{*}}^5+192 \mu ^3 n^{3/2} {Q^{*}} \left(3 {Q^{*}}^2-5 n \left({Q^{*}}^2-2\right)\right)+n^3 {Q^{*}}^6+12 \mu ^2 n^2 {Q^{*}}^2 \left(3 (5 n-2)
   {Q^{*}}^2-20 n\right)+16 \mu ^4 n \left(5 (24-61 n) {Q^{*}}^2+364 n\right)-11520 \mu ^5 \sqrt{n} {Q^{*}}}{64 \mu ^3 n^{3/2} {Q^{*}}^7}$, 
	$a_{10}=\frac{1}{{Q^{*}}^2}, a_{11}= \frac{6 \mu
   }{\sqrt{n} {Q^{*}}^4}-\frac{\sqrt{n}}{4 \mu  {Q^{*}}^2}+\frac{3}{{Q^{*}}^3}, a_{12}= \frac{\frac{1440 \mu ^2}{n}+\frac{n {Q^{*}}^4}{\mu ^2}-\frac{12 \sqrt{n} {Q^{*}}^3}{\mu }+120 n
   \left({Q^{*}}^2-2\right)+\frac{480 \mu  {Q^{*}}}{\sqrt{n}}-8 {Q^{*}}^2}{16 {Q^{*}}^6}$, $a_{14}= -\frac{1}{{Q^{*}}^3}, a_{15}= \frac{-\frac{40 \mu }{\sqrt{n}}+\frac{\sqrt{n}
   {Q^{*}}^2}{\mu }-16 {Q^{*}}}{4 {Q^{*}}^5}, a_{17}= \frac{1}{{Q^{*}}^4}$. \\	
	
	Therefore,\\
	$h(u_1,u_2)=\frac{706 \mu ^4 u_1^5}{n^2 {Q^{*}}^8}+\frac{399 \mu ^3 u_1^5}{n^{3/2} {Q^{*}}^7}-\frac{327 \mu ^2 u_1^5}{4 n {Q^{*}}^6}+\frac{925 \mu ^2 u_1^5}{4 {Q^{*}}^6}-\frac{201 \mu ^2
   u_1^5}{{Q^{*}}^8}+\frac{305 \sqrt{n} \mu  u_1^5}{4 {Q^{*}}^5}-\frac{50 \mu  u_1^5}{\sqrt{n} {Q^{*}}^5}-\frac{91 \sqrt{n} \mu  u_1^5}{{Q^{*}}^7}-\frac{45 n^{3/2} u_1^5}{16 {Q^{*}}^3
   \mu }+\frac{13 \sqrt{n} u_1^5}{8 {Q^{*}}^3 \mu }+\frac{15 n^{3/2} u_1^5}{4 {Q^{*}}^5 \mu }+\frac{15 n^2 u_1^5}{32 {Q^{*}}^2 \mu ^2}-\frac{7 n u_1^5}{32 {Q^{*}}^2 \mu ^2}-\frac{3 n^2
   u_1^5}{8 {Q^{*}}^4 \mu ^2}-\frac{n^{3/2} u_1^5}{64 {Q^{*}} \mu ^3}+\frac{661 n^2 u_1^5}{96 {Q^{*}}^4}-\frac{251 n u_1^5}{16 {Q^{*}}^4}+\frac{3121 u_1^5}{480 {Q^{*}}^4}+\frac{n^2
   u_1^5}{256 \mu ^4}-\frac{55 n^2 u_1^5}{4 {Q^{*}}^6}+\frac{19 n u_1^5}{4 {Q^{*}}^6}+\frac{7 n^2 u_1^5}{{Q^{*}}^8}-\frac{399 u_2 \mu ^3 u_1^4}{n^{3/2} {Q^{*}}^7}+\frac{57 \mu ^3
   u_1^4}{n^{3/2} {Q^{*}}^6}-\frac{180 u_2 \mu ^2 u_1^4}{n {Q^{*}}^6}+\frac{30 \mu ^2 u_1^4}{n {Q^{*}}^5}-\frac{15 n u_2 u_1^4}{{Q^{*}}^4}+\frac{9 u_2
   u_1^4}{{Q^{*}}^4}+\frac{30 n u_2 u_1^4}{{Q^{*}}^6}-\frac{305 \sqrt{n} u_2 \mu  u_1^4}{4 {Q^{*}}^5}+\frac{30 u_2 \mu  u_1^4}{\sqrt{n} {Q^{*}}^5}+\frac{91 \sqrt{n}
   u_2 \mu  u_1^4}{{Q^{*}}^7}+\frac{61 \sqrt{n} \mu  u_1^4}{4 {Q^{*}}^4}-\frac{6 \mu  u_1^4}{\sqrt{n} {Q^{*}}^4}-\frac{13 \sqrt{n} \mu  u_1^4}{{Q^{*}}^6}+\frac{45 n^{3/2} u_2
   u_1^4}{16 {Q^{*}}^3 \mu }-\frac{9 \sqrt{n} u_2 u_1^4}{8 {Q^{*}}^3 \mu }-\frac{15 n^{3/2} u_2 u_1^4}{4 {Q^{*}}^5 \mu }-\frac{15 n^{3/2} u_1^4}{16 {Q^{*}}^2 \mu }+\frac{3
   \sqrt{n} u_1^4}{8 {Q^{*}}^2 \mu }+\frac{3 n^{3/2} u_1^4}{4 {Q^{*}}^4 \mu }-\frac{n u_2 u_1^4}{8 {Q^{*}}^2 \mu ^2}+\frac{n u_1^4}{16 {Q^{*}} \mu ^2}+\frac{15 n u_1^4}{4
   {Q^{*}}^3}-\frac{9 u_1^4}{4 {Q^{*}}^3}-\frac{n^{3/2} u_1^4}{64 \mu ^3}+\frac{n^{3/2} u_2 u_1^4}{64 {Q^{*}} \mu ^3}-\frac{5 n u_1^4}{{Q^{*}}^5}+\frac{15 n u_2^2 u_1^3}{2
   {Q^{*}}^4}-\frac{u_2^2 u_1^3}{2 {Q^{*}}^4}-\frac{15 n u_2^2 u_1^3}{{Q^{*}}^6}+\frac{90 u_2^2 \mu ^2 u_1^3}{n {Q^{*}}^6}-\frac{30 u_2 \mu ^2 u_1^3}{n
   {Q^{*}}^5}+\frac{6 \mu ^2 u_1^3}{n {Q^{*}}^4}-\frac{15 n u_2 u_1^3}{4 {Q^{*}}^3}+\frac{u_2 u_1^3}{4 {Q^{*}}^3}+\frac{5 n u_2 u_1^3}{{Q^{*}}^5}+\frac{30 u_2^2 \mu 
   u_1^3}{\sqrt{n} {Q^{*}}^5}-\frac{12 u_2 \mu  u_1^3}{\sqrt{n} {Q^{*}}^4}+\frac{3 \mu  u_1^3}{\sqrt{n} {Q^{*}}^3}-\frac{3 \sqrt{n} u_2^2 u_1^3}{4 {Q^{*}}^3 \mu }+\frac{\sqrt{n}
   u_2 u_1^3}{2 {Q^{*}}^2 \mu }-\frac{\sqrt{n} u_1^3}{4 {Q^{*}} \mu }+\frac{5 n u_1^3}{4 {Q^{*}}^2}-\frac{u_1^3}{12 {Q^{*}}^2}+\frac{n u_2^2 u_1^3}{16 {Q^{*}}^2 \mu
   ^2}+\frac{n u_1^3}{16 \mu ^2}-\frac{n u_2 u_1^3}{16 {Q^{*}} \mu ^2}-\frac{n u_1^3}{{Q^{*}}^4}-\frac{4 u_2^3 u_1^2}{{Q^{*}}^4}+\frac{3 u_2^2
   u_1^2}{{Q^{*}}^3}-\frac{2 u_2 u_1^2}{{Q^{*}}^2}-\frac{10 u_2^3 \mu  u_1^2}{\sqrt{n} {Q^{*}}^5}+\frac{6 u_2^2 \mu  u_1^2}{\sqrt{n} {Q^{*}}^4}-\frac{3 u_2 \mu 
   u_1^2}{\sqrt{n} {Q^{*}}^3}+\frac{\mu  u_1^2}{\sqrt{n} {Q^{*}}^2}+\frac{u_1^2}{{Q^{*}}}+\frac{\sqrt{n} u_2^3 u_1^2}{4 {Q^{*}}^3 \mu }-\frac{\sqrt{n} u_2^2 u_1^2}{4
   {Q^{*}}^2 \mu }+\frac{\sqrt{n} u_2 u_1^2}{4 {Q^{*}} \mu }-\frac{\sqrt{n} u_1^2}{4 \mu }+\frac{u_2^4 u_1}{{Q^{*}}^4}-\frac{u_2^3 u_1}{{Q^{*}}^3}+\frac{u_2^2
   u_1}{{Q^{*}}^2}-\frac{u_2 u_1}{{Q^{*}}}$. 
\\
	Finally, the center manifold can be expressed as \\	
		$\theta =-\frac{23134208 T^5 \mu ^9}{n^{9/2} {Q^{*}}^9}+\frac{57835520 T^4 \mu ^9}{n^{9/2} {Q^{*}}^9}-\frac{57835520 T^3 \mu ^9}{n^{9/2} {Q^{*}}^9}+\frac{28917760 T^2 \mu ^9}{n^{9/2}
   {Q^{*}}^9}-\frac{7229440 T \mu ^9}{n^{9/2} {Q^{*}}^9}+\frac{722944 \mu ^9}{n^{9/2} {Q^{*}}^9}-\frac{7577600 T^5 \mu ^7}{n^{5/2} {Q^{*}}^7}+\frac{5627904 T^5 \mu ^7}{n^{7/2} {Q^{*}}^7}+\frac{6586368 T^5 \mu
   ^7}{n^{5/2} {Q^{*}}^9}+\frac{18944000 T^4 \mu ^7}{n^{5/2} {Q^{*}}^7}-\frac{12201984 T^4 \mu ^7}{n^{7/2} {Q^{*}}^7}-\frac{1634304 Q T^4 \mu ^7}{n^{7/2} {Q^{*}}^8}-\frac{16465920 T^4 \mu ^7}{n^{5/2}
   {Q^{*}}^9}-\frac{18944000 T^3 \mu ^7}{n^{5/2} {Q^{*}}^7}+\frac{10334208 T^3 \mu ^7}{n^{7/2} {Q^{*}}^7}+\frac{3268608 Q T^3 \mu ^7}{n^{7/2} {Q^{*}}^8}+\frac{16465920 T^3 \mu ^7}{n^{5/2} {Q^{*}}^9}+\frac{9472000
   T^2 \mu ^7}{n^{5/2} {Q^{*}}^7}-\frac{4233216 T^2 \mu ^7}{n^{7/2} {Q^{*}}^7}-\frac{2451456 Q T^2 \mu ^7}{n^{7/2} {Q^{*}}^8}-\frac{8232960 T^2 \mu ^7}{n^{5/2} {Q^{*}}^9}-\frac{2368000 T \mu ^7}{n^{5/2}
   {Q^{*}}^7}+\frac{824832 T \mu ^7}{n^{7/2} {Q^{*}}^7}+\frac{817152 Q T \mu ^7}{n^{7/2} {Q^{*}}^8}+\frac{2058240 T \mu ^7}{n^{5/2} {Q^{*}}^9}+\frac{236800 \mu ^7}{n^{5/2} {Q^{*}}^7}-\frac{59136 \mu ^7}{n^{7/2}
   {Q^{*}}^7}-\frac{102144 Q \mu ^7}{n^{7/2} {Q^{*}}^8}-\frac{205824 \mu ^7}{n^{5/2} {Q^{*}}^9}-\frac{676864 T^5 \mu ^5}{3 \sqrt{n} {Q^{*}}^5}+\frac{759808 T^5 \mu ^5}{n^{3/2} {Q^{*}}^5}-\frac{5899264 T^5 \mu
   ^5}{15 n^{5/2} {Q^{*}}^5}+\frac{450560 T^5 \mu ^5}{\sqrt{n} {Q^{*}}^7}-\frac{647168 T^5 \mu ^5}{n^{3/2} {Q^{*}}^7}-\frac{229376 T^5 \mu ^5}{\sqrt{n} {Q^{*}}^9}+\frac{1692160 T^4 \mu ^5}{3 \sqrt{n}
   {Q^{*}}^5}-\frac{1524736 T^4 \mu ^5}{n^{3/2} {Q^{*}}^5}+\frac{2064896 T^4 \mu ^5}{3 n^{5/2} {Q^{*}}^5}-\frac{312320 Q T^4 \mu ^5}{n^{3/2} {Q^{*}}^6}+\frac{245760 Q T^4 \mu ^5}{n^{5/2} {Q^{*}}^6}-\frac{1126400
   T^4 \mu ^5}{\sqrt{n} {Q^{*}}^7}+\frac{1191936 T^4 \mu ^5}{n^{3/2} {Q^{*}}^7}+\frac{372736 Q T^4 \mu ^5}{n^{3/2} {Q^{*}}^8}+\frac{573440 T^4 \mu ^5}{\sqrt{n} {Q^{*}}^9}-\frac{1692160 T^3 \mu ^5}{3 \sqrt{n}
   {Q^{*}}^5}+\frac{1149952 T^3 \mu ^5}{n^{3/2} {Q^{*}}^5}-\frac{1373696 T^3 \mu ^5}{3 n^{5/2} {Q^{*}}^5}+\frac{624640 Q T^3 \mu ^5}{n^{3/2} {Q^{*}}^6}-\frac{384000 Q T^3 \mu ^5}{n^{5/2} {Q^{*}}^6}-\frac{46080 Q^2
   T^3 \mu ^5}{n^{5/2} {Q^{*}}^7}+\frac{1126400 T^3 \mu ^5}{\sqrt{n} {Q^{*}}^7}-\frac{765952 T^3 \mu ^5}{n^{3/2} {Q^{*}}^7}-\frac{745472 Q T^3 \mu ^5}{n^{3/2} {Q^{*}}^8}-\frac{573440 T^3 \mu ^5}{\sqrt{n}
   {Q^{*}}^9}+\frac{846080 T^2 \mu ^5}{3 \sqrt{n} {Q^{*}}^5}-\frac{387584 T^2 \mu ^5}{n^{3/2} {Q^{*}}^5}+\frac{438016 T^2 \mu ^5}{3 n^{5/2} {Q^{*}}^5}-\frac{468480 Q T^2 \mu ^5}{n^{3/2} {Q^{*}}^6}+\frac{207360 Q
   T^2 \mu ^5}{n^{5/2} {Q^{*}}^6}+\frac{69120 Q^2 T^2 \mu ^5}{n^{5/2} {Q^{*}}^7}-\frac{563200 T^2 \mu ^5}{\sqrt{n} {Q^{*}}^7}+\frac{169984 T^2 \mu ^5}{n^{3/2} {Q^{*}}^7}+\frac{559104 Q T^2 \mu ^5}{n^{3/2}
   {Q^{*}}^8}+\frac{286720 T^2 \mu ^5}{\sqrt{n} {Q^{*}}^9}-\frac{211520 T \mu ^5}{3 \sqrt{n} {Q^{*}}^5}+\frac{50048 T \mu ^5}{n^{3/2} {Q^{*}}^5}-\frac{71488 T \mu ^5}{3 n^{5/2} {Q^{*}}^5}+\frac{156160 Q T \mu
   ^5}{n^{3/2} {Q^{*}}^6}-\frac{42240 Q T \mu ^5}{n^{5/2} {Q^{*}}^6}-\frac{34560 Q^2 T \mu ^5}{n^{5/2} {Q^{*}}^7}+\frac{140800 T \mu ^5}{\sqrt{n} {Q^{*}}^7}+\frac{10752 T \mu ^5}{n^{3/2} {Q^{*}}^7}-\frac{186368 Q
   T \mu ^5}{n^{3/2} {Q^{*}}^8}-\frac{71680 T \mu ^5}{\sqrt{n} {Q^{*}}^9}+\frac{21152 \mu ^5}{3 \sqrt{n} {Q^{*}}^5}-\frac{320 \mu ^5}{n^{3/2} {Q^{*}}^5}+\frac{28832 \mu ^5}{15 n^{5/2} {Q^{*}}^5}-\frac{19520 Q \mu
   ^5}{n^{3/2} {Q^{*}}^6}+\frac{1920 Q \mu ^5}{n^{5/2} {Q^{*}}^6}+\frac{5760 Q^2 \mu ^5}{n^{5/2} {Q^{*}}^7}-\frac{14080 \mu ^5}{\sqrt{n} {Q^{*}}^7}-\frac{6400 \mu ^5}{n^{3/2} {Q^{*}}^7}+\frac{23296 Q \mu
   ^5}{n^{3/2} {Q^{*}}^8}+\frac{7168 \mu ^5}{\sqrt{n} {Q^{*}}^9}-\frac{15360 T^5 \mu ^3}{\sqrt{n} {Q^{*}}^3}+\frac{9216 T^5 \mu ^3}{n^{3/2} {Q^{*}}^3}+\frac{12288 T^5 \mu ^3}{\sqrt{n} {Q^{*}}^5}+\frac{23040 T^4
   \mu ^3}{\sqrt{n} {Q^{*}}^3}-\frac{12800 T^4 \mu ^3}{n^{3/2} {Q^{*}}^3}+\frac{11520 Q T^4 \mu ^3}{\sqrt{n} {Q^{*}}^4}-\frac{7680 Q T^4 \mu ^3}{n^{3/2} {Q^{*}}^4}-\frac{12288 T^4 \mu ^3}{\sqrt{n}
   {Q^{*}}^5}-\frac{15360 Q T^4 \mu ^3}{\sqrt{n} {Q^{*}}^6}-\frac{14080 T^3 \mu ^3}{\sqrt{n} {Q^{*}}^3}+\frac{24320 T^3 \mu ^3}{3 n^{3/2} {Q^{*}}^3}-\frac{13440 Q T^3 \mu ^3}{\sqrt{n} {Q^{*}}^4}+\frac{7040 Q T^3
   \mu ^3}{n^{3/2} {Q^{*}}^4}-\frac{3840 Q^2 T^3 \mu ^3}{\sqrt{n} {Q^{*}}^5}+\frac{3328 Q^2 T^3 \mu ^3}{n^{3/2} {Q^{*}}^5}+\frac{4608 T^3 \mu ^3}{\sqrt{n} {Q^{*}}^5}+\frac{12800 Q T^3 \mu ^3}{\sqrt{n}
   {Q^{*}}^6}+\frac{7680 Q^2 T^3 \mu ^3}{\sqrt{n} {Q^{*}}^7}+\frac{5760 T^2 \mu ^3}{\sqrt{n} {Q^{*}}^3}-\frac{3200 T^2 \mu ^3}{n^{3/2} {Q^{*}}^3}+\frac{2880 Q T^2 \mu ^3}{\sqrt{n} {Q^{*}}^4}-\frac{1920 Q T^2 \mu
   ^3}{n^{3/2} {Q^{*}}^4}+\frac{5760 Q^2 T^2 \mu ^3}{\sqrt{n} {Q^{*}}^5}-\frac{2688 Q^2 T^2 \mu ^3}{n^{3/2} {Q^{*}}^5}-\frac{3840 T^2 \mu ^3}{\sqrt{n} {Q^{*}}^5}-\frac{640 Q^3 T^2 \mu ^3}{n^{3/2}
   {Q^{*}}^6}+\frac{3840 Q T^2 \mu ^3}{\sqrt{n} {Q^{*}}^6}-\frac{11520 Q^2 T^2 \mu ^3}{\sqrt{n} {Q^{*}}^7}-\frac{1920 T \mu ^3}{\sqrt{n} {Q^{*}}^3}+\frac{640 T \mu ^3}{n^{3/2} {Q^{*}}^3}+\frac{1440 Q T \mu
   ^3}{\sqrt{n} {Q^{*}}^4}+\frac{480 Q T \mu ^3}{n^{3/2} {Q^{*}}^4}-\frac{2880 Q^2 T \mu ^3}{\sqrt{n} {Q^{*}}^5}+\frac{192 Q^2 T \mu ^3}{n^{3/2} {Q^{*}}^5}+\frac{2688 T \mu ^3}{\sqrt{n} {Q^{*}}^5}+\frac{640 Q^3 T
   \mu ^3}{n^{3/2} {Q^{*}}^6}-\frac{5760 Q T \mu ^3}{\sqrt{n} {Q^{*}}^6}+\frac{5760 Q^2 T \mu ^3}{\sqrt{n} {Q^{*}}^7}+\frac{320 \mu ^3}{\sqrt{n} {Q^{*}}^3}-\frac{64 \mu ^3}{3 n^{3/2} {Q^{*}}^3}-\frac{480 Q \mu
   ^3}{\sqrt{n} {Q^{*}}^4}-\frac{160 Q \mu ^3}{n^{3/2} {Q^{*}}^4}+\frac{480 Q^2 \mu ^3}{\sqrt{n} {Q^{*}}^5}+\frac{160 Q^2 \mu ^3}{n^{3/2} {Q^{*}}^5}-\frac{576 \mu ^3}{\sqrt{n} {Q^{*}}^5}-\frac{160 Q^3 \mu
   ^3}{n^{3/2} {Q^{*}}^6}+\frac{1280 Q \mu ^3}{\sqrt{n} {Q^{*}}^6}-\frac{960 Q^2 \mu ^3}{\sqrt{n} {Q^{*}}^7}-\frac{128 T^5 \mu }{\sqrt{n} {Q^{*}}}+\frac{192 T^4 \mu }{\sqrt{n} {Q^{*}}}+\frac{64 Q T^4 \mu
   }{\sqrt{n} {Q^{*}}^2}-\frac{160 T^3 \mu }{\sqrt{n} {Q^{*}}}-\frac{32 Q T^3 \mu }{\sqrt{n} {Q^{*}}^2}-\frac{32 Q^2 T^3 \mu }{\sqrt{n} {Q^{*}}^3}+\frac{48 T^2 \mu }{\sqrt{n} {Q^{*}}}+\frac{48 Q T^2 \mu }{\sqrt{n}
   {Q^{*}}^2}-\frac{16 Q^2 T^2 \mu }{\sqrt{n} {Q^{*}}^3}+\frac{16 Q^3 T^2 \mu }{\sqrt{n} {Q^{*}}^4}-\frac{24 T \mu }{\sqrt{n} {Q^{*}}}+\frac{24 Q T \mu }{\sqrt{n} {Q^{*}}^2}-\frac{40 Q^2 T \mu }{\sqrt{n}
   {Q^{*}}^3}+\frac{24 Q^3 T \mu }{\sqrt{n} {Q^{*}}^4}-\frac{8 Q^4 T \mu }{\sqrt{n} {Q^{*}}^5}+\frac{12 \mu }{\sqrt{n} {Q^{*}}}-\frac{24 Q \mu }{\sqrt{n} {Q^{*}}^2}+\frac{28 Q^2 \mu }{\sqrt{n}
   {Q^{*}}^3}-\frac{16 Q^3 \mu }{\sqrt{n} {Q^{*}}^4}+\frac{4 Q^4 \mu }{\sqrt{n} {Q^{*}}^5}$. 
\\
The dynamics on the center manifold is given by\\
\begin{small}
\begin{align}
&\frac{d u_1}{dN}=-\frac{39 \mu ^3 u_1^4 ({Q^{*}}+5 u_1-5 u_2)}{n^{3/2} {Q^{*}}^6}-\frac{378 \mu ^4 u_1^5}{n^2 {Q^{*}}^7}-\frac{3 \mu ^2 u_1^3 \left({Q^{*}}^2 \left((124 n-53) u_1^2-96 u_1
   u_2+48 u_2^2\right)-112 n u_1^2+8 {Q^{*}}^4+24 {Q^{*}}^3 (u_1-u_2)\right)}{4 n {Q^{*}}^7}\nonumber \\
	& +\frac{3 u_1^3 \left({Q^{*}}^2 \left((6 n-3) u_1^2-2 u_1
   u_2+u_2^2\right)-6 n u_1^2+{Q^{*}}^4+{Q^{*}}^3 (u_1-u_2)\right)}{4 {Q^{*}}^5}\nonumber\\
	& -\frac{3 \mu  u_1^2 \left({Q^{*}}^3 \left((2 n-1) u_1^2-2 u_1
   u_2+u_2^2\right)+{Q^{*}}^2 (u_1-u_2) \left((6 n-5) u_1^2-2 u_1 u_2+u_2^2\right)-2 n {Q^{*}} u_1^2+10 n u_1^2 (u_2-u_1)+{Q^{*}}^5+{Q^{*}}^4
   (u_1-u_2)\right)}{\sqrt{n} {Q^{*}}^6}, \label{Case3HLGeneralEMa}\\
&\frac{d u_2}{dN}=-\frac{3 n^{5/2} \left({Q^{*}}^2-1\right) u_1^5}{128 \mu ^3 {Q^{*}}^2}+\mu ^3 \left(\frac{3 u_1^5 \left(63 n \left({Q^{*}}^2-1\right)-65 {Q^{*}}^2\right)}{n^{3/2} {Q^{*}}^8}-\frac{39 u_1^4
   ({Q^{*}}-5 u_2)}{n^{3/2} {Q^{*}}^6}\right) \nonumber \\
	& +\frac{\frac{3 n^{3/2} u_1^4 ({Q^{*}}-3 u_2)}{4 {Q^{*}}^4}+\frac{3 n^{3/2} u_1^3 \left(-{Q^{*}}^4+{Q^{*}}^2-2 {Q^{*}} u_2+3
   u_2^2\right)}{8 {Q^{*}}^4}-\frac{9 n^{3/2} u_1^5 \left(4 n \left({Q^{*}}^2-1\right)^2-3 {Q^{*}}^4+{Q^{*}}^2\right)}{16 {Q^{*}}^6}}{\mu }-\frac{378 \mu ^4 u_1^5}{n^2 {Q^{*}}^7}+\frac{\frac{3
   n^2 u_1^4 \left({Q^{*}}^3-{Q^{*}}+2 u_2\right)}{32 {Q^{*}}^3}-\frac{3 n^2 u_1^5}{16 {Q^{*}}^3}}{\mu ^2}\nonumber \\
	& +\frac{u_1^5 \left(72 n^2+3 (2 n+1) (4 n-3) {Q^{*}}^4+6 (1-16 n) n
   {Q^{*}}^2\right)}{4 {Q^{*}}^7} \nonumber \\
	& +\frac{3 u_1^4 \left(4 n^2 \left({Q^{*}}^2-1\right) \left({Q^{*}} \left({Q^{*}}^2-2 {Q^{*}} u_2-1\right)+6 u_2\right)-2 n {Q^{*}}^2 \left(2 {Q^{*}}^3-4
   {Q^{*}}^2 u_2+{Q^{*}}-4 u_2\right)+{Q^{*}}^4 ({Q^{*}}-2 u_2)\right)}{4 {Q^{*}}^7} \nonumber \\
	& +\mu  \Big(\frac{u_1^5 \left(336 n^2+3 (n (124 n-125)+40) {Q^{*}}^4+3 (101-236 n) n
   {Q^{*}}^2\right)}{8 \sqrt{n} {Q^{*}}^8}+\frac{u_1^4 \left(3 {Q^{*}}^2 ({Q^{*}}-3 u_2)-6 n ({Q^{*}}-5 u_2)\right)}{\sqrt{n} {Q^{*}}^6}\nonumber \\
	& -\frac{3 u_1^2 ({Q^{*}}-u_2)
   \left({Q^{*}}^2+u_2^2\right)}{\sqrt{n} {Q^{*}}^4}+\frac{3 u_1^3 \left(n \left({Q^{*}}^4-2 {Q^{*}}^3 u_2+{Q^{*}}^2 \left(3 u_2^2-1\right)+4 {Q^{*}} u_2-10
   u_2^2\right)-{Q^{*}}^2 \left({Q^{*}}^2-2 {Q^{*}} u_2+3 u_2^2\right)\right)}{\sqrt{n} {Q^{*}}^6}\Big) \nonumber \\
	& +\frac{3 u_1^3 \left({Q^{*}}^2 \left({Q^{*}}^2-{Q^{*}}
   u_2+u_2^2\right)-4 n \left({Q^{*}}^2-3 {Q^{*}} u_2+6 u_2^2\right)\right)}{4 {Q^{*}}^5}\nonumber \\
	& +\mu ^2 \left(\frac{3 u_1^5 \left(\left(\frac{53}{n}-20\right) {Q^{*}}^2-44\right)}{4
   {Q^{*}}^7}+\frac{3 u_1^4 \left(13 n \left({Q^{*}} \left({Q^{*}}^2-4 {Q^{*}} u_2-1\right)+6 u_2\right)-12 {Q^{*}}^2 ({Q^{*}}-4 u_2)\right)}{2 n {Q^{*}}^7}-\frac{6 u_1^3
   \left({Q^{*}}^2-3 {Q^{*}} u_2+6 u_2^2\right)}{n {Q^{*}}^5}\right)\nonumber\\
	& +\frac{3 n u_1^2 \left({Q^{*}}^5-{Q^{*}}^3+2 {Q^{*}}^2 u_2-3 {Q^{*}} u_2^2+4 u_2^3\right)}{2 {Q^{*}}^5}. \label{Case3HLGeneralEMb}
\end{align}

\end{small}
\end{widetext}

\begin{figure*}[t]
\centering
\includegraphics[width=0.5\textwidth]{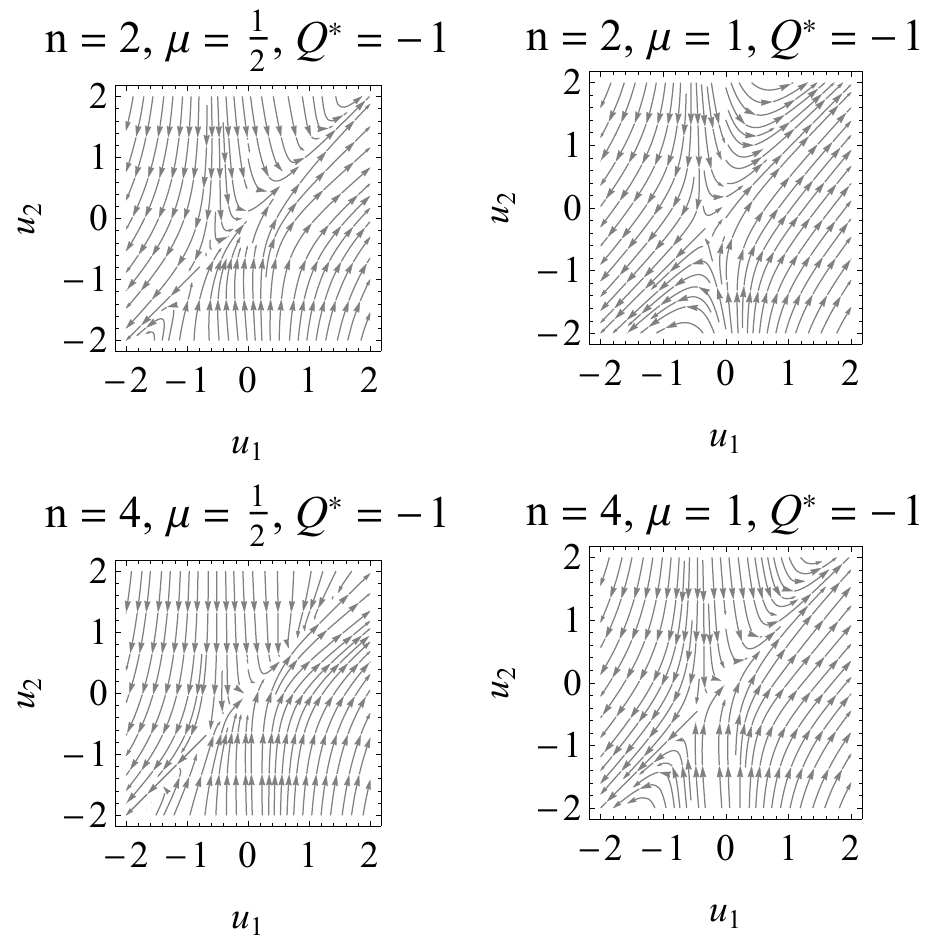}  %
\includegraphics[width=0.47\textwidth]{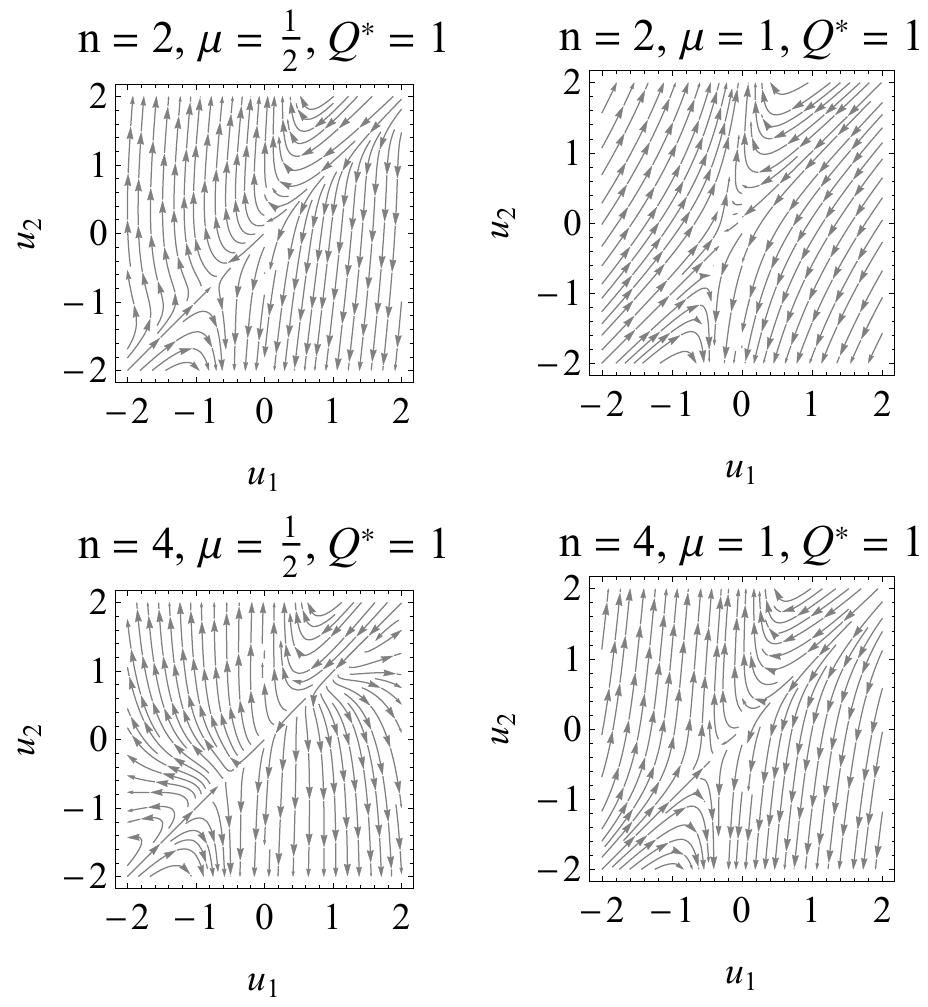} 
\caption{Numerical solutions of the system \eqref{Case3HLGeneralEMa},\eqref{Case3HLGeneralEMb}.}
\label{fig:Center2DCase3PLGeneralEMA}
\end{figure*}

In the Figure \ref{fig:Center2DCase3PLGeneralEMA} are presented some
numerical solutions of the system \eqref{Case3HLGeneralEMa}, \eqref{Case3HLGeneralEMb} for $n=2,4, \mu=%
\frac{1}{2}, 1$ and $Q^{*}=\pm 1$. The plot illustrates the generic feature
that for $Q^{*}>0$ (respectively, $Q^{*}<0$), the center manifold is
unstable (respectively, stable), but in this case the third eigenvalue is $-%
\frac{3}{2} Q^{*}<0$ (respectively, $-\frac{3}{2} Q^{*}>0$). This means the
the origin is a saddle. Using the Theorem \ref{stabilityCM}, we conclude
that the center manifold of origin for the system \eqref{Center2DCase3PLa}, \eqref{Center2DCase3PLb},
and the origin are unstable (saddle). $\blacksquare$ 

\section{\protect Case 4: $k\neq 0, \Lambda \neq 0$ under the
detailed-balance condition}

\label{Case4} In this case the autonomous system writes: 

\begin{align}
\frac{dx}{dN}& =\sqrt{6} s \left[-x^2+(u-z)^2+1\right] +x \left[3 x^2+2
(u-z) z-3\right],  \label{GenHLeqxcase4} \\
\frac{dz}{dN}&=z \left[3 x^2+2 (u-z) z-2\right],  \label{GenHLeqzcase4} \\
\frac{du}{dN}&= u \left[3 x^2+2 (u-z) z\right],  \label{GenHLequcase4} \\
\frac{ds}{dN}&= -2 \sqrt{6} x f(s).  \label{GenHLeqscase4}
\end{align}
where the prime means derivative with respect to $\ln a$, defined on the
phase space
$\{(x,z,u,s)\in\mathbb{R}^3: x^2-(u-k z)^2 \leq 1\}.$ 
 The equilibrium points/curves at the finite region of the phase
space are presented in Table \ref{HLcrit4}, where is shown the existence and
stability conditions. We proceed to the discussion of the more relevant
features of them. 

\begin{itemize}
\item  $P_{14}(\hat{s}): (x,z,u,s)=\left(1, 0, 0, \hat{s}\right)$,
where $f(\hat{s})=0$. It is a source for $f^{\prime }\left(\hat{s}\right)<0, 
\hat{s}<\sqrt{\frac{3}{2}}$. 

\item  $P_{15}(\hat{s}): (x,z,s)=\left(-1, 0, \hat{s}\right)$, where $%
f(\hat{s})=0$. It is a source for $f^{\prime }\left(\hat{s}\right)>0, \hat{s}%
>-\sqrt{\frac{3}{2}}$. 

\item  $P_{16}(\hat{s}): (x,z,u,s)=\left(\sqrt{\frac{2}{3}} \hat{s},
0, 0, \hat{s}\right)$, where \newline
$f(\hat{s})=0$, $-\sqrt{\frac{3}{2}}\leq \hat{s}\leq \sqrt{\frac{3}{2}}$.
This point is reduced to $P_{16}$ studied in \cite{Leon:2009rc}. It is
generically a saddle. 

\item $P_{16}^{0}(u_c): (x,z,u,s)=(0,0,u_c,0)$. This point is new,
and it was not found in \cite{Leon:2009rc}. It is nonhyperbolic. 

\item $P_{17,18}(\hat{s}): \left(\frac{\sqrt{\frac{2}{3}}}{\hat{s}},
\pm \frac{\sqrt{1-\hat{s}^2}}{\hat{s}},0, \hat{s}\right)$, where $f(\hat{s}%
)=0$, $-1\leq \hat{s}\leq 1, \hat{s}\neq 0$. It is generically a saddle. 

\item  There are four lines of equilibrium points $P_{19,20}(s_c):
(x,z,u,s)=(0, 0, \pm i, s_c), s_c\in \mathbb{R}$ and $P_{21,22}(s_c):
(x,z,u,s)=(0, \pm i, 0, s_c), s_c\in \mathbb{R}$ which are not considered
since they are complex valued. 
\end{itemize}

\begin{table*}[ht]
\caption{Case 4: Equilibrium points at the finite region of the system 
\eqref{GenHLeqxcase4}, \eqref{GenHLeqzcase4}, \eqref{GenHLequcase4}, 
\eqref{GenHLeqscase4}.}
\label{HLcrit4}{\centering
\begin{tabular*}{\textwidth}{@{\extracolsep{\fill}}lrrrl}
\hline
Equil. & \multicolumn{1}{c}{$(x,z,u,s)$} & \multicolumn{1}{c}{Existence} & 
\multicolumn{1}{c}{Eigenvalues} & \multicolumn{1}{c}{Stability} \\ 
Points &  &  &  &  \\ \hline
$P_{14}(\hat{s})$ & $\left(1, 0, 0, \hat{s}\right)$ & $f(\hat{s})=0$ & $6-2 
\sqrt{6} \hat{s}, 1, 3, -2 \sqrt{6} f^{\prime }\left(\hat{s}\right)$ & 
nonhyperbolic for \\ 
&  &  &  & $f^{\prime }\left(\hat{s}\right)=0$, or $\hat{s}=\sqrt{\frac{3}{2}%
}$. \\ 
&  &  &  & source for $f^{\prime }\left(\hat{s}\right)<0, \hat{s}<\sqrt{%
\frac{3}{2}}$. \\ 
&  &  &  & saddle otherwise. \\ \hline
$P_{15}(\hat{s})$ & $\left(-1, 0, \hat{s}\right)$ & $f(\hat{s})=0$ & $6+2 
\sqrt{6} \hat{s}, 1, 3, 2 \sqrt{6} f^{\prime }\left(\hat{s}\right)$ & 
nonhyperbolic for \\ 
&  &  &  & $f^{\prime }\left(\hat{s}\right)=0$, or $\hat{s}=-\sqrt{\frac{3}{2%
}}$. \\ 
&  &  &  & source for $f^{\prime }\left(\hat{s}\right)>0, \hat{s}>-\sqrt{%
\frac{3}{2}}$. \\ 
&  &  &  & saddle otherwise. \\ \hline
$P_{16}(\hat{s})$ & $\left(\sqrt{\frac{2}{3}} \hat{s}, 0, 0, \hat{s}\right)$
& $-\sqrt{\frac{3}{2}}\leq \hat{s}\leq \sqrt{\frac{3}{2}}$ & $2 \hat{s}^2-3,
2 \left(\hat{s}^2-1\right), 2 \hat{s}^2, - 4 \hat{s} f^{\prime }\left(\hat{s}%
\right)$ & nonhyperbolic for \\ 
&  &  &  & $f^{\prime }\left(\hat{s}\right)=0$, or \\ 
&  &  &  & $\hat{s}\in \left\{-\sqrt{\frac{3}{2}},-1, 0, 1, \sqrt{\frac{3}{2}%
}\right\}$ \\ 
&  &  &  & saddle otherwise. \\ \hline
$P_{16}^{0}(u_c)$ & $(0,0,u_c,0)$ & always & $-2,0,-\frac{3}{2}\pm \frac{1}{2%
} \sqrt{9-48 f(0) \left(u_c^2+1\right)}$ & nonhyperbolic. \\ \hline
$P_{17,18}(\hat{s})$ & $\left(\frac{\sqrt{\frac{2}{3}}}{\hat{s}}, \pm \frac{%
\sqrt{1-\hat{s}^2}}{\hat{s}},0, \hat{s}\right)$ & $-1\leq \hat{s}\leq 1, 
\hat{s}\neq 0$ & $2, -\frac{1}{2}\pm \frac{\sqrt{16 \hat{s}^2-15 \hat{s}^4}}{%
2 \hat{s}^2}, -\frac{4 f^{\prime }\left(\hat{s}\right)}{\hat{s}}$ & 
nonhyperbolic for \\ 
&  &  &  & $f^{\prime }\left(\hat{s}\right)=0$, or $\hat{s}\in \left\{-1,
1\right\}$. \\ 
&  &  &  & saddle otherwise. \\ \hline
\end{tabular*}
}
\end{table*}
\begin{table*}[t]
\caption{Equilibrium points at the finite region of the system 
\eqref{GenHLext2}, \eqref{GenHLext3}, \eqref{GenHLext4}, \eqref{GenHLextx}, 
\eqref{GenHLexty}, \eqref{GenHLexts} for arbitrary potentials.}
\label{HLcrit5}{\centering
\begin{tabular*}{\textwidth}{@{\extracolsep{\fill}}lrrrl}
\hline
Equil. & \multicolumn{1}{c}{$(x_2,x_3,x_4,x,y,s)$} & \multicolumn{1}{c}{
Existence} & \multicolumn{1}{c}{Eigenvalues} & \multicolumn{1}{c}{Stability}
\\ 
Points &  &  &  &  \\ \hline
$P_{23}(\hat{s})$ & $\left(0,0,1-x_c^2,x_c,0,\hat{s}\right)$ & $f(\hat{s})=0$
& $6, 0, 2, 4, 3-\sqrt{6} \hat{s} x_c, -2 \sqrt{6} x_c f^{\prime }\left(\hat{%
s}\right)$ & 5D unstable manifold for \\ 
&  &  &  & $x_c f^{\prime }\left(\hat{s}\right)<0, \hat{s}x_c<\sqrt{\frac{3}{%
2}}$. \\ \hline
$P_{24,25}(\hat{s})$ & $\left(0,0,0,\pm 1,0,\hat{s}\right)$ & $f(\hat{s})=0$
& $6, 0, 2, 4, 3\mp \sqrt{6} \hat{s}, \mp 2 \sqrt{6} f^{\prime }\left(\hat{s}%
\right)$ & 5D unstable manifold for \\ 
&  &  &  & $\pm f^{\prime }\left(\hat{s}\right)<0, \pm\hat{s}<\sqrt{\frac{3}{%
2}}$. \\ \hline
$P_{26}(y_c)$ & $\left(0,0,0,0,y_c,0\right)$ & always & $-6,-4,-2,0$, & 5D
stable manifold \\ 
&  &  & $-\frac{1}{2} \left(3+\sqrt{9-48 f(0) y_c^2}\right),-\frac{1}{2}
\left(3-\sqrt{9-48 f(0) y_c^2}\right)$ & for $f(0)>0$. \\ \hline
$P_{26}^{0}$ & $\left(0,0,0,0,0,0\right)$ & always & $-6, -4, -3, -2, 0, 0$
& 4D stable manifold. \\ \hline
$P_{27,28}(\hat{s})$ & $\left(0,0,0,\sqrt{\frac{2}{3}} \hat{s},\pm\sqrt{ 1-%
\frac{2 \hat{s}^2}{3}},\hat{s}\right)$ & $f(\hat{s})=0$ & $4 \hat{s}^2, 2 
\hat{s}^2-3, 2 \left(2 \hat{s}^2-3\right)$, & saddle. \\ 
&  &  & $-4 \left(1-\hat{s}^2\right), 2 \left(2 \hat{s}^2-1\right), -4 \hat{s%
} f^{\prime }\left(\hat{s}\right)$ &  \\ \hline
$P_{29}(s_c)$ & $(1,0,0,0,0,s_c)$ & always & $- 4,-2,-2,2,1,0$ & saddle. \\ 
\hline
$P_{30,31}(\hat{s})$ & $\left(1-\frac{1}{2 \hat{s}^2}, 0, 0, \frac{1}{\sqrt{
6} \hat{s}}, \pm\frac{1}{\sqrt{3} \hat{s}}, \hat{s}\right)$ & $\hat{s}\neq 0$%
, $f(\hat{s})=0$ & $-4, -2, 2, -\frac{\hat{s}+\sqrt{2-3 \hat{s}^2}}{\hat{s}}%
, \frac{\sqrt{2-3 \hat{s}^2}-\hat{s}}{\hat{s}}, -\frac{2 f^{\prime }\left(%
\hat{s}\right)}{\hat{s}}$ & saddle. \\ \hline
$P_{32}(s_c)$ & $\left(0,1,0,0,0,s_c\right)$ & all $s_c$ & $4, -2, 2, 2, -1,
0$ & saddle. \\ \hline
$P_{33,34}(\hat{s})$ & $\left(0, 1-\frac{1}{\hat{s}^2}, 0, \frac{\sqrt{\frac{%
2}{3}}}{\hat{s}}, \pm\frac{1}{\sqrt{3} \hat{s}}, \hat{s}\right)$ & $\hat{s}%
\neq 0$, $f(\hat{s})=0$ & $4, -2, 2, -\frac{\hat{s}+\sqrt{16-15 \hat{s}^2}}{%
2 \hat{s}}, \frac{\sqrt{16-15 \hat{s}^2}-\hat{s}}{2 \hat{s}}, -\frac{4
f^{\prime }\left(\hat{s}\right)}{\hat{s}}$ & saddle. \\ \hline
$P_{35}(\hat{s})$ & $\left(0, 0, 1-\frac{3}{2 \hat{s}^2}, \frac{\sqrt{\frac{3%
}{2}}}{\hat{s}}, 0, \hat{s}\right)$ & $f(\hat{s})=0$ & $0, 0, 2, 4, 6, -%
\frac{6 f^{\prime }\left(\hat{s}\right)}{\hat{s}}$ & 4D unstable manifold \\ 
&  &  &  & for $\hat{s}f^{\prime }\left(\hat{s}\right)<0$. \\ \hline
\end{tabular*}%
}
\end{table*}

\section{\protect Beyond the detailed-balance condition}
\label{Case5} 

Using the variables \eqref {HLauxilliaryc} the corresponding
autonomous system is found to be:  
\begin{align}
&\frac{dx_2}{dN}=2 x_2 \left(3 x^2+x_2+2 x_3+3 x_4-1\right),
\label{GenHLext2} \\
&\frac{dx_3}{dN}=2x_3 \left(3 x^2+x_2+2 x_3+3 x_4-2\right),
\label{GenHLext3} \\
&\frac{dx_4}{dN}=2 x_4 \left(3x^2+x_2+2 x_3+3 x_4-3\right),
\label{GenHLext4} \\
&\frac{dx}{dN}= 3 x^3+(x_2+2 x_3+3 x_4-3) x+\sqrt{6} s y^2,
\label{GenHLextx} \\
&\frac{dy}{dN}=\left(3 x^2-\sqrt{6} s x+x_2+2 x_3+3 x_4\right) y,
\label{GenHLexty} \\
&\frac{ds}{dN}=-2 \sqrt{6} x f(s) ,  \label{GenHLexts}
\end{align}
defining a dynamical system in $\mathbb{R}^6$. 

\subsection{\protect Arbitrary potentials}
 \label{Case5A} 
The stability of the hyperbolic equilibrium points is
given by analyzing the signs of the reals parts of the eigenvalues of the
matrix of linear perturbations $\mathbf{Q}$ evaluated at each equilibrium
point. The results are shown in Table \ref{HLcrit5}. The equilibrium
points/curves at the finite region of the phase space are the following: 

\begin{itemize}
\item $P_{23}(\hat{s}): (x_2, x_3, x_4, x, y,
s)=\left(0,0,1-x_c^2,x_c,0,\hat{s}\right)$. Exists for $\hat{s}$ such that $%
f(\hat{s})=0$. The eigenvalues are
$6, 0, 2, 4, 3-\sqrt{6} \hat{s} x_c,  -2 \sqrt{6} x_c f^{\prime }\left(\hat{s%
}\right)$.

\item $P_{24,25}(\hat{s}): (x_2, x_3, x_4, x, y, s)=\left(0,0,0,\pm
1,0,\hat{s}\right)$. Exists for $\hat{s}$ such that $f(\hat{s})=0$. The
eigenvalues are
$6, 0, 2, 4, 3\mp \sqrt{6} \hat{s}, \mp 2 \sqrt{6} f^{\prime }\left(\hat{s}%
\right)$. 

\item $P_{26}(y_c): (x_2, x_3, x_4, x, y,
s)=\left(0,0,0,0,y_c,0\right)$. Always exists. The eigenvalues are  $%
-6,-4,-2,0$,\newline
$-\frac{1}{2} \left(3+ \sqrt{9-48 f(0) y_c^2}\right),-\frac{1}{2} \left(3- \sqrt{9-48 f(0) y_c^2}\right)$. 

\item  $P_{26}^{0}: (x_2, x_3, x_4, x, y, s)=\left(0,0,0,0,0,0\right)$%
. Always exists. 
The eigenvalues are $-6, -4, -3, -2, 0, 0$.

\item  $P_{27,28}(\hat{s}): (x_2, x_3, x_4, x, y, s)$
$=\left(0,0,0,\sqrt{\frac{2}{3}} \hat{s},\pm\sqrt{ 1-\frac{2 \hat{s}^2}{3}},%
\hat{s}\right)$. Exists for $\hat{s}$ such that $f(\hat{s})=0$. The
eigenvalues are \newline
$4 \hat{s}^2, 2 \hat{s}^2-3, 2 \left(2 \hat{s}^2-3\right), -4 \left(1-\hat{s}%
^2\right), 2 \left(2 \hat{s}^2-1\right), -4 \hat{s} f^{\prime }\left(\hat{s}%
\right)$.

\item $P_{29}(s_c): (x_2, x_3,x_4,x,y,s)=(1,0,0,0,0,s_c)$. It always
exists. The eigenvalues are $- 4,-2,-2,2,1,0$.

\item $P_{30,31}(\hat{s}): (x_2, x_3,x_4,x,y,s)=\left(1-\frac{1}{2 
\hat{s}^2}, 0, 0, \frac{1}{\sqrt{ 6} \hat{s}}, \pm\frac{1}{\sqrt{3} \hat{s}}%
, \hat{s}\right)$. Exists for $\hat{s}\neq 0$ such that $f(\hat{s})=0$. The
eigenvalues are
$-4, -2, 2, -\frac{\hat{s}+\sqrt{2-3 \hat{s}^2}}{\hat{s}}, \frac{\sqrt{2-3 
\hat{s}^2}-\hat{s}}{\hat{s}}, -\frac{2 f^{\prime }\left(\hat{s}\right)}{\hat{%
s}}$. 

\item  $P_{32}(s_c): (x_2, x_3,x_4,x,y,s)=\left(0,1,0,0,0,s_c\right)$%
. Exists for all $s_c$. The eigenvalues are $4, -2, 2, 2, -1, 0$. 

\item  $P_{33,34}(\hat{s}): (x_2, x_3,x_4,x,y,s)=\left(0, 1-\frac{1}{%
\hat{s}^2}, 0, \frac{\sqrt{\frac{2}{3}}}{\hat{s}}, \pm\frac{1}{\sqrt{3} \hat{%
s}}, \hat{s}\right)$. Exists for $\hat{s}\neq 0$ such that $f(\hat{s})=0$.
The eigenvalues are
$4, -2, 2, -\frac{\hat{s}+\sqrt{16-15 \hat{s}^2}}{2 \hat{s}}, \frac{\sqrt{%
16-15 \hat{s}^2}-\hat{s}}{2 \hat{s}}, -\frac{4 f^{\prime }\left(\hat{s}%
\right)}{\hat{s}} $. 

\item  $P_{35}(\hat{s}): (x_2, x_3,x_4,x,y,s)=\left(0, 0, 1-\frac{3}{%
2 \hat{s}^2}, \frac{\sqrt{\frac{3}{2}}}{\hat{s}}, 0, \hat{s}\right)$. Exists
for $\hat{s}$ such that $f(\hat{s})=0$. The eigenvalues are \newline
$0, 0, 2, 4, 6, -\frac{6 f^{\prime }\left(\hat{s}\right)}{\hat{s}}$.
\end{itemize}

 Owing to the fact that the dynamical system \eqref{GenHLext2}, %
\eqref{GenHLext3}, \eqref{GenHLext4}, \eqref{GenHLextx}, \eqref{GenHLexty}, %
\eqref{GenHLexts}, is unbounded, we use the compact variables 
\begin{align}
&X_2=\frac{x_2}{\sqrt{1+r^2}}, \quad X_3=\frac{x_3}{\sqrt{1+r^2}}, \quad X_4=%
\frac{x_4}{\sqrt{ 1+r^2}},  \notag \\
&X=\frac{x}{\sqrt{1+r^2}}, \quad Y=\frac{y}{\sqrt{1+r^2}}, \quad S=\frac{2}{%
\pi}\arctan(s),  \notag \\
& r=\sqrt{x_2^2+x_3^2+x_4^2+x^2+y^2},
\end{align}
and the time rescaling 
\begin{equation}
f^{\prime }=\left(1-X_2^2-X_3^2-X_4^2-X^2-Y^2\right) df/dN,
\end{equation}
such that we obtain the system 
\begin{widetext}
\begin{align}
&\frac{dX_2}{d\tau}=3 X^2
   X_2 \left(X^2+Y^2\right)  +P X_2 \left(X^2+Y^2\right) (X_2+2 X_3+3 X_4)  +P^2 X_2 \left(7 X^2+2 X_3^2+4 X_4^2-2 Y^2\right)
	 \nonumber \\ 	
	& +2 P^3 X_2 (X_2+2 X_3+3 X_4-P),\\
&\frac{dX_3}{d\tau}=  3 X^2 X_3 \left(X^2+Y^2\right) +P X_3 \left(X^2+Y^2\right) (X_2+2
   X_3+3 X_4) +P^2 X_3 \left(5 X^2-2 X_2^2+2 X_4^2-4 Y^2\right)
	 \nonumber \\ 	
	& +2 P^3 X_3 (X_2+2 X_3+3 X_4-2 P),\\
&\frac{dX_4}{d\tau}=3 X^2 X_4 \left(X^2+Y^2\right)   +P X_4 \left(X^2+Y^2\right) (X_2+2 X_3+3 X_4) +P^2 X_4 \left(3 X^2-2 \left(2 X_2^2+X_3^2+3 Y^2\right)\right)
    \nonumber \\ 	
	& +2 P^3 X_4 (X_2+2 X_3+3 X_4-3 P),
\\
&\frac{dX}{d\tau}=-3 X^3 \left(X_2^2+X_3^2+X_4^2\right) +P \Big[-X (X_2+2 X_3+3 X_4) \left(X_2^2+X_3^2+X_4^2\right)  +\sqrt{6} Y^2 \tan\left(\frac{\pi S}{2}\right)\Big]\nonumber \\
&+P^2 X \left(3 X^2-X_2^2+X_3^2+3 X_4^2-3 Y^2\right)  +P^3 X (X_2+2 X_3+3 X_4-3 P),\\
&\frac{dY}{d\tau}=-3 X^2
   Y \left(X_2^2+X_3^2+X_4^2\right)+P Y \Big[(-X_2-2 X_3-3 X_4) \left(X_2^2+X_3^2+X_4^2\right)  -\sqrt{6} X \tan\left(\frac{\pi S}{2}\right)\Big]  \nonumber \\ 	
	& +2 P^2 Y \left(3 X^2+X_2^2+2 X_3^2+3 X_4^2\right) + P^3 Y (X_2+2 X_3+3   X_4),\\
&\frac{dS}{d\tau}= -\frac{2 \sqrt{6} P X (\cos (\pi  S)+1) f\left(\tan \left(\frac{\pi S}{2}\right)\right)}{\pi },
\end{align}
\end{widetext}
where $P=\sqrt{1-X_2^2-X_3^2-X_4^2-X^2-Y^2}$. \newline
The invariant sets at infinity are the cylinders $X=0,
X_2^2+X_3^2+X_4^2+Y^2=1, S\in [-1, 1]$ and $X_2=X_3=X_4=0, X^2+Y^2=1, S\in
[-1, 1]$. \newline
After taking a time rescaling, the dynamics on the invariant set $%
X_2=X_3=X_4=0$ is given by 
\begin{align}
&\frac{dX}{d\tau}=\sqrt{6} Y^2 \,\tan\left(\frac{\pi S}{2}\right) +3 X
\left(2 X^2-1\right) \sqrt{1-X^2-Y^2}, \label{Example63Sa}\\
&\frac{dY}{d\tau}=-X Y \Big(\sqrt{6} \,\tan\left(\frac{\pi S}{2}\right) - 6
X \sqrt{1-X^2-Y^2}\Big), \label{Example63Sb} \\
&\frac{dS}{d\tau}= -\frac{2 \sqrt{6} X (\cos (\pi S)+1) f\left(\tan \left(%
\frac{\pi S}{2}\right)\right)}{\pi }.  \label{Example63Sc}
\end{align}

 This invariant set is relevant concerning the future asymptotic
dynamics. It contains the de Sitter solutions $P_{26}(y_c)$, and some
relevant invariant sets at the infinity region. 

\subsection{\protect Exponential Potential}
\label{Case5B} 

 In this case we obtain the autonomous system \eqref{GenHLext2}, \eqref{GenHLext3}, \eqref{GenHLext4}, \eqref{GenHLextx}, 
\eqref{GenHLexty}, \eqref{GenHLexts} transforms to: 
\begin{align}
&\frac{dx_2}{dN}=2 x_2 \left(3 x^2+x_2+2 x_3+3 x_4-1\right),  \label{HLext2}
\\
&\frac{dx_3}{dN}=2x_3 \left(3 x^2+x_2+2 x_3+3 x_4-2\right),  \label{HLext3}
\\
&\frac{dx_4}{dN}=2 x_4 \left(3x^2+x_2+2 x_3+3 x_4-3\right),  \label{HLext4}
\\
&\frac{dx}{dN}= 3 x^3+(x_2+2 x_3+3 x_4-3) x+\sqrt{6} s y^2,  \label{HLextx}
\\
&\frac{dy}{dN}=\left(3 x^2-\sqrt{6} s x+x_2+2 x_3+3 x_4\right) y, \label{HLexty}\\
& 1=x_1+x_2+x_3+x_4+x^2+y^2
\end{align}
The equilibrium points of the system \eqref{HLext2}, \eqref{HLext3}, \eqref{HLext4}, \eqref{HLextx}, \eqref{HLexty} and their behavior \cite%
{Leon:2009rc} is summarized as follows. The point $P_{35}$ studied in \cite%
{Leon:2009rc} is omitted. It is a special case of the line $P_{23}$. The
stability of the hyperbolic equilibrium points of this system
have been extensively  studied in \cite{Leon:2009rc}. Now we summarize the
previous findings. The  unstable equilibrium points $P_{27,28}, P_{29}$
correspond to dark matter domination, and the unstable point $P_{32}$
corresponds to an unphysical dark-energy dominated universe, and the
unstable $P_{30,31}, P_{33,34}$ which have physical $w_M$, $w_{DE}$ but
dependent on the specific dark-matter form. The system admits also the line
of equilibrium points $P_{23}$, which is nonhyperbolic with positive
non-null eigenvalues, thus unstable, with furthermore unphysical
cosmological quantities. Additionally, points $P_{24,25}$ are also
dark-matter dominated, unstable nonhyperbolic ones. Due to $\sigma_3$ has an
arbitrary sign, $P_{33,34}$ could also correspond to an oscillatory
universe, for a wide region of the parameters $\sigma_3$ and $s$. However,
this oscillatory behavior has a small probability to be the late-time state
of the universe because it is not stable (with at least two positive
eigenvalues). The scenario at hand admits a final equilibrium point, namely $%
P_{26}$, representing a de Sitter solution. As we show in detail in the next
section \ref{stabilityP26} by using Center Manifold Theory, it is indeed a
locally asymptotically stable and thus it can be a late-time attractor of HL
universe beyond the detailed-balance. 

\subsubsection{\protect Stability Analysis of the \emph{de Sitter}
Solution in Ho\v{r}ava-Lifshitz cosmology without detailed-balance for the
exponential potential.}
\label{stabilityP26} 

In order to analyze the stability of \emph{de Sitter} solution we
can use center manifold theorem. 

\begin{prop}
\label{centerP26} The origin for the system \eqref{HLexty} is
locally asymptotically stable. 
\end{prop}

 \textbf{Proof}. Defining $(u,v_1,v_2,v_3,v_4):=(y, x_4, x_3, x, x_2)$%
, the linear part of the vector field is transformed into its Jordan
canonical form: 
\begin{equation}
\left(%
\begin{array}{c}
\frac{du}{d\tau} \\ 
\frac{dv_1}{d\tau} \\ 
\frac{dv_2}{d\tau} \\ 
\frac{dv_3}{d\tau} \\ 
\frac{dv_4}{d\tau}%
\end{array}%
\right)=\left(%
\begin{array}{ccccc}
0 & 0 & 0 & 0 & 0 \\ 
0 & -6 & 0 & 0 & 0 \\ 
0 & 0 & -4 & 0 & 0 \\ 
0 & 0 & 0 & -3 & 0 \\ 
0 & 0 & 0 & 0 & -2%
\end{array}%
\right)\left(%
\begin{array}{c}
u \\ 
v_1 \\ 
v_2 \\ 
v_3 \\ 
v_4%
\end{array}%
\right)+\left(%
\begin{array}{c}
f(u,\mathbf{v}) \\ 
g_1(u,\mathbf{v}) \\ 
g_2(u,\mathbf{v}) \\ 
g_3(u,\mathbf{v}) \\ 
g_4(u,\mathbf{v})%
\end{array}%
\right),  \label{HLcenter2}
\end{equation}%
where $f(u,\mathbf{v})=u \left(3 v_1+2 v_2-\sqrt{6} s v_3+4  v_4\right),$ 
\newline
$g_1(u,\mathbf{v})=2 v_1 \left(3 v_3^2+3 v_1+2  v_2+v_4\right),$ \newline
$g_2(u,\mathbf{v})=2 v_2 \left(3 v_3^2+3 v_1+2  v_2+v_4\right),$ \newline
$g_3(u,\mathbf{v})=\sqrt{6} s u^2+v_3 \left(3 v_3^2+3  v_1+2 v_2+v_4\right),$\newline
and
$g_4(u,\mathbf{v})=2 v_4 \left(3 v_3^2+3  v_1+2 v_2+v_4\right).$ 
\begin{figure*}[t!]
\centering
\includegraphics[scale=1.5]{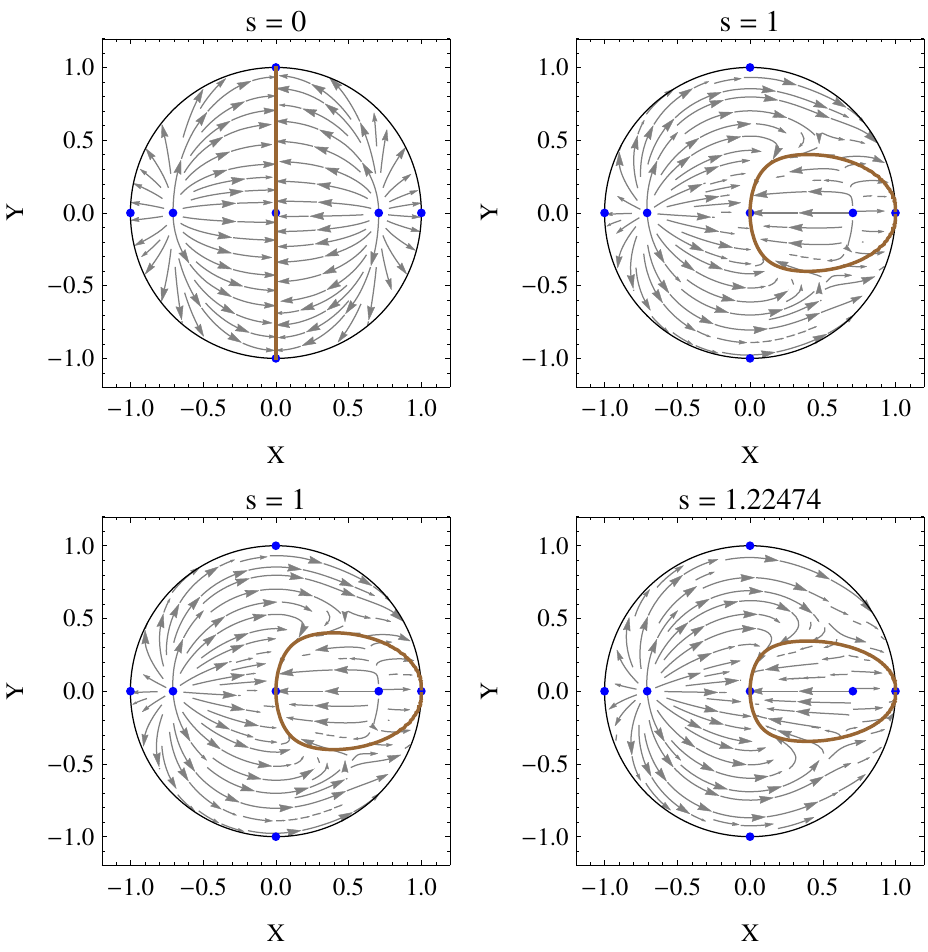} 
\caption{\textit{{(Color online) Compact phase portrait of the system \eqref{system71a}, \eqref{system71b},
for different choices of the parameter $s$.}}}
\label{fig:Case5HL}
\end{figure*}
 The system \eqref{HLcenter2} is written in diagonal form 
\begin{align}
\frac{du}{d\tau} & =Cu+f\left( u,\mathbf{v}\right)  \notag \\
\frac{d\mathbf{v}}{d\tau} & =P\mathbf{v}+\mathbf{g}\left( u,\mathbf{v}%
\right) ,  \label{HLcenter3}
\end{align}
where $\left( u,\mathbf{v}\right) \in\mathbb{R}\times\mathbb{R}^{4},$ $C$ is
the zero $1\times1$ matrix, $P$ is a $4\times 4$ matrix with negative
eigenvalues and $f,\mathbf{g}$ vanish at $\mathbf{0}$ and have vanishing
derivatives at $\mathbf{0.}$ 
According to Theorem \ref{existenceCM}, there exists a 1-dimensional
invariant local center manifold $W^{c}\left( \mathbf{0}\right) $ of %
\eqref{HLcenter3} tangent to the center subspace (the $\mathbf{v}=\mathbf{0}$
space) at $\mathbf{0}$. Moreover,\newline
$W^{c}\left( \mathbf{0}\right) =\left\{ \left( u,\mathbf{v}\right) \in%
\mathbb{R}\times\mathbb{R}^{4}:\mathbf{v}=\mathbf{h}\left( u\right), \mathbf{%
h}\left( 0\right) =\mathbf{0},\;D\mathbf{h}\left( 0\right) =\mathbf{0}
\right\}$, $\left\vert u\right\vert <\delta$ for $\delta$ sufficiently
small. The restriction of (\ref{HLcenter3}) to the center manifold is $\frac{%
du}{d\tau}=f\left( u,\mathbf{h}\left( u\right) \right)$, where the function $%
\mathbf{h}\left( u\right) $ that defines the local center manifold satisfies%
\begin{equation}
D\mathbf{h}\left( u\right) \left[ f\left( u,\mathbf{h}\left( u\right)
\right) \right] -P\mathbf{h}\left( u\right) -\mathbf{g}\left( u,\mathbf{h}%
\left( u\right) \right) =0.  \label{HLh}
\end{equation}
The vectorial equation \eqref{HLh} can be written as the system of ODE 
\label{HLh2}
\begin{align}
&2 h_1 \left(3 h_1+2 h_2+3 h_3^2+h_4-3\right)  \notag \\
& -u h_1^{\prime }(u) \left(-\sqrt{6} s h_3+3 h_1+2 h_2+3 h_3^2+h_4\right)=0,
\\
& 2h_2 \left(3 h_1+2 h_2+3 h_3^2+h_4-2\right)  \notag \\
&-u h_2^{\prime }(u) \left(-\sqrt{6} s h_3+3 h_1+2 h_2+3 h_3^2+h_4\right)=0,
\\
& \sqrt{6} s u^2+h_3 \left(3 h_1+2 h_2+3 h_3^2+h_4-3\right)  \notag \\
& -u h_3^{\prime }(u) \left(-\sqrt{6} s h_3+3 h_1+2 h_2+3 h_3^2+h_4\right)
=0, \\
& 2 h_4 \left(3 h_1+2 h_2+3 h_3^2+h_4-1\right)  \notag \\
& -u h_4^{\prime }(u) \left(-\sqrt{6} s h_3+3 h_1+2 h_2+3 h_3^2+h_4\right)=0.
\end{align}
 The equation \eqref{HLh2} can be solved approximately by expanding $%
\mathbf{h}\left( u\right) $ in Taylor series at $u=0.$ Due to $\mathbf{h}%
\left( 0\right) =\mathbf{0\ } $and $D\mathbf{h}\left( 0\right) =\mathbf{0},$
we substitute 
\begin{equation}
\mathbf{h}\left( u\right) :=\left(%
\begin{array}{c}
h_{1}\left(u\right) \\ 
h_{2}\left(u\right) \\ 
h_{3}\left(u\right) \\ 
h_{4}\left(u\right)%
\end{array}%
\right)=\left(%
\begin{array}{c}
\sum_{j=1}^{10} a_j u^{j+1} +O\left( u^{12}\right) \\ 
\sum_{j=1}^{10} b_j u^{j+1} +O\left( u^{12}\right) \\ 
\sum_{j=1}^{10} c_j u^{j+1} +O\left( u^{12}\right) \\ 
\sum_{j=1}^{10} d_j u^{j+1} +O\left( u^{12}\right)%
\end{array}%
\right),
\end{equation}%
into (\ref{HLh}) and set the coefficients of like powers of $u$ equal to
zero to find the unknowns $a_{1},b_{1},c_{1},d_{1},...$. The non-zero
coefficients are
$c_1= \sqrt{\frac{2}{3}} s, c_3=  \frac{4}{3} \sqrt{\frac{2}{3}} s^3, c_5=  
\frac{2}{3} \sqrt{\frac{2}{3}} s^3 \left(8 s^2-1\right)$, \newline
$c_7=\frac{8}{27} \sqrt{\frac{2}{3}} s^5 \left(112  s^2-27\right)$,\newline $c_9= 
\frac{4}{81}  \sqrt{\frac{2}{3}} s^5 \left(5440 s^4-1872 s^2+63\right)$. 

Therefore, the dynamics on the center manifold is governed by the
gradient-like equation 
\begin{align}
&\frac{dU}{d\tau}=-\nabla U(u), \quad U(u)=\frac{s^2 u^4}{2}+\frac{1}{9} s^2
\left(4 s^2-3\right) u^6  \notag \\
&+\frac{4}{135} s^4 \left(224 s^4-138 s^2+9\right) u^{10}+\frac{1}{6} s^4
\left(8 s^2-5\right) u^8  \notag \\
&+\frac{2}{243} s^6 \left(5440 s^4-3504 s^2+423\right) u^{12},
\label{HLrest1}
\end{align}
for which the origin is a degenerate minimum. Using the Theorem \ref%
{stabilityCM}, we conclude that the origin $u=0$ of \eqref{HLrest1} is
locally asymptotically stable. Hence, $(0,\mathbf{v})=(0,\mathbf{0})$ is
locally asymptotically stable. $\blacksquare$ 

As we mentioned before, the relevant invariant set concerning the
future asymptotic dynamics is the set $X_2=X_3=X_4=0$ where the dynamics is
given by

\begin{align}
&\frac{dX}{d\tau}=\sqrt{6} Y^2 s \sqrt{1-X^2-Y^2} -3 X \left(2 X^2-1\right)
\left(X^2+Y^2-1\right), \label{system71a}\\
&\frac{dY}{d\tau}=-X Y \left(\sqrt{6} s \sqrt{1-X^2-Y^2} +6 X
\left(X^2+Y^2-1\right)\right). \label{system71b}
\end{align}
According the our center manifold calculation the center manifold of the
origin can be approximated by the graph $x_2=x_3=x_4=0, x=  \sqrt{\frac{2}{3}%
} s y^2+ \frac{4}{3} \sqrt{\frac{2}{3}} s^3 y^4$, or, using the compact
variables, by 
\begin{align}
& X_2=X_3=X_4=0, \notag \\
& X \left(-X^2-Y^2+1\right)^{9/2}= \sqrt{\frac{2}{3}} s
\left(X^2-1\right)^4 Y^2  \notag \\
& -\frac{4}{3} \sqrt{\frac{2}{3}} s \left(s^2-3\right) \left(X^2-1\right)^3
Y^4 \notag \\
& +\frac{2}{3} \sqrt{\frac{2}{3}} s \left(8 s^4-7 s^2+9\right)
\left(X^2-1\right)^2 Y^6  \notag \\
& -\frac{4}{27} \sqrt{\frac{2}{3}} s \left(224 s^6-126 s^4+36 s^2-27\right)
\left(X^2-1\right) Y^8  \notag \\
& +\frac{1}{81} \sqrt{\frac{2}{3}} s \left(2 \left(10880 s^6-5088 s^4+666
s^2-81\right) s^2+81\right)Y^{10}.
\end{align}
This curve is denoted by a solid line in the Figure \ref{fig:Case5HL}, where
it is presented a compact phase portrait of the system \eqref{system71a}, \eqref{system71b} for different
choices of the parameter $s$. For $s=0$ the line $P_{26}(y_c)=%
\left(0,0,0,0,y_c\right)$ is the attractor. For $s\neq 0$ the attractor is
the origin. 

 We see that for small enough $X$-value the curve is a good
approximation of the center manifold of the origin. The advantages of using
Center Manifold theory are unveiled immediately when the procedure is
compared with the Normal Forms Calculations \cite{wiggins} presented by Leon
and Saridakis in \cite{Leon:2009rc}. First it is taken a linear
transformation $(x_2,x_3,x_4,x,y)\rightarrow (x, x_3, x_2, x_4, y)$ to
transform the linear part of the system to its real Jordan form: $\mathbf{%
diag}\left(-6,-4,-3,-2,0\right).$ Taking the quadratic transformation 
\begin{equation*}
\left( 
\begin{array}{c}
x_2 \\ 
x_3 \\ 
x_4 \\ 
x \\ 
y%
\end{array}
\right)\rightarrow \left( 
\begin{array}{c}
x_2 -x_2 (x+x_2+x_3) \\ 
x_3 -x_3 (x+x_2+x_3) \\ 
x_4 + \sqrt{\frac{2}{3}} s y^2-\frac{1}{2} (x+x_2+x_3) x_4 \\ 
x -x (x+x_2+x_3) \\ 
y -\frac{1}{6} \left(3 x+3 x_2+3 x_3-2 \sqrt{6} s x_4\right) y%
\end{array}
\right),
\end{equation*}
are eliminated the non resonant terms of second order. \newpage Finally, it
can be implemented the cubic transformation 
\begin{widetext}
$$\left(
\begin{array}{c}
 x_2\\
 x_3\\
 x_4\\
  x\\
 y
\end{array}
\right)\rightarrow \left(
\begin{array}{c}
x_2+x_2
   \left[(x+x_2+x_3)^2-x_4^2\right]\\
x_3+   x_3
   \left[(x+x_2+x_3)^2-x_4^2\right]\\
x_4+   \frac{1}{24}
   \left[-12 x_4^3+9 (x+x_2+x_3)^2 x_4+96
   s^2 y^2 x_4+8 \sqrt{6} s
   (x_3-2 x) y^2\right]\\
x+   x
   \left[(x+x_2+x_3)^2-x_4^2\right]\\
y+   \frac{y \left\{945
   x^2+42 \left(45 x_2+45 x_3-16 \sqrt{6} s
   x_4\right) x+5 \left\{189 x_2^2+14 \left(27 x_3-8
   \sqrt{6} s x_4\right) x_2+3 \left\{63
   x_3^2-40 \sqrt{6} s x_4 x_3+28
   \left[\left(2 s^2-3\right) x_4^2-12 s^2 y^2\right]\right\}\right\}\right\}}{2520}
\end{array}
\right),$$
\end{widetext}
resulting in the simplified system 
\begin{align}
&\frac{dx_2}{dN}=-6 x_2+\mathcal{O}{(4)}, \\
&\frac{dx_3}{dN}=-4x_3 +\mathcal{O}{(4)}, \\
&\frac{dx_4}{dN}=x_4 \left(-3+4 s^2 y^2\right)+\mathcal{O}{(4)}, \\
&\frac{dx}{dN}=-2x+\mathcal{O}{(4)}, \\
&\frac{dy}{dN}=-2 s^2 y^3+\mathcal{O}{(4)},
\end{align}
where $\mathcal{O}{(4)}$ denotes terms of fourth order in the vector norm.
Therefore, the local center manifold of the origin $W^c_\text{loc}(\mathbf{0}%
),$ is tangent to the $y$-axis at the origin and it can be represented
locally up to fourth order as the graph 
\begin{align}
&W^c_\text{loc}=\left\{(x_2,x_3,x_4,x,y)\in\mathbb{R}^5: x_2=x_{20}e^{-\frac{%
3}{2 s^2 y^2}}, x_3=x_{30}e^{-\frac{1}{s^2 y^2}}, \right.  \notag \\
&\left. x_4=x_{40}{y^{-2}}e^{-\frac{1}{s^2 y^2}}, x=x_{0}e^{-\frac{1}{2 s^2
y^2}}, |y|<\varepsilon\right\},
\end{align}
 where $\varepsilon\ll 1$. An it follows under the initial condition $%
y(0)=y_0$, that $y(N)=y_0 (1+4 s^2 y_0^2 N)^{-1/2}$, therefore $P_{26}$ is
the late-time attractor. 

The center manifold, on the other hand, is more economical in the
use of computing resources and one obtains a system, say \eqref{HLrest1},
with a reduced dimensionality (1D), when compared with normal forms (5D).
Anyway, one can complement both results to find information about the
dynamics of a model at hand as for example in \cite{Leon:2018lnd}. 

\subsection{\protect Powerlaw Potential}
\label{Case5C}

\begin{figure*}[t!]
\centering
\subfigure[]{\includegraphics[width=3.5in,
height=3.5in,angle=-90]{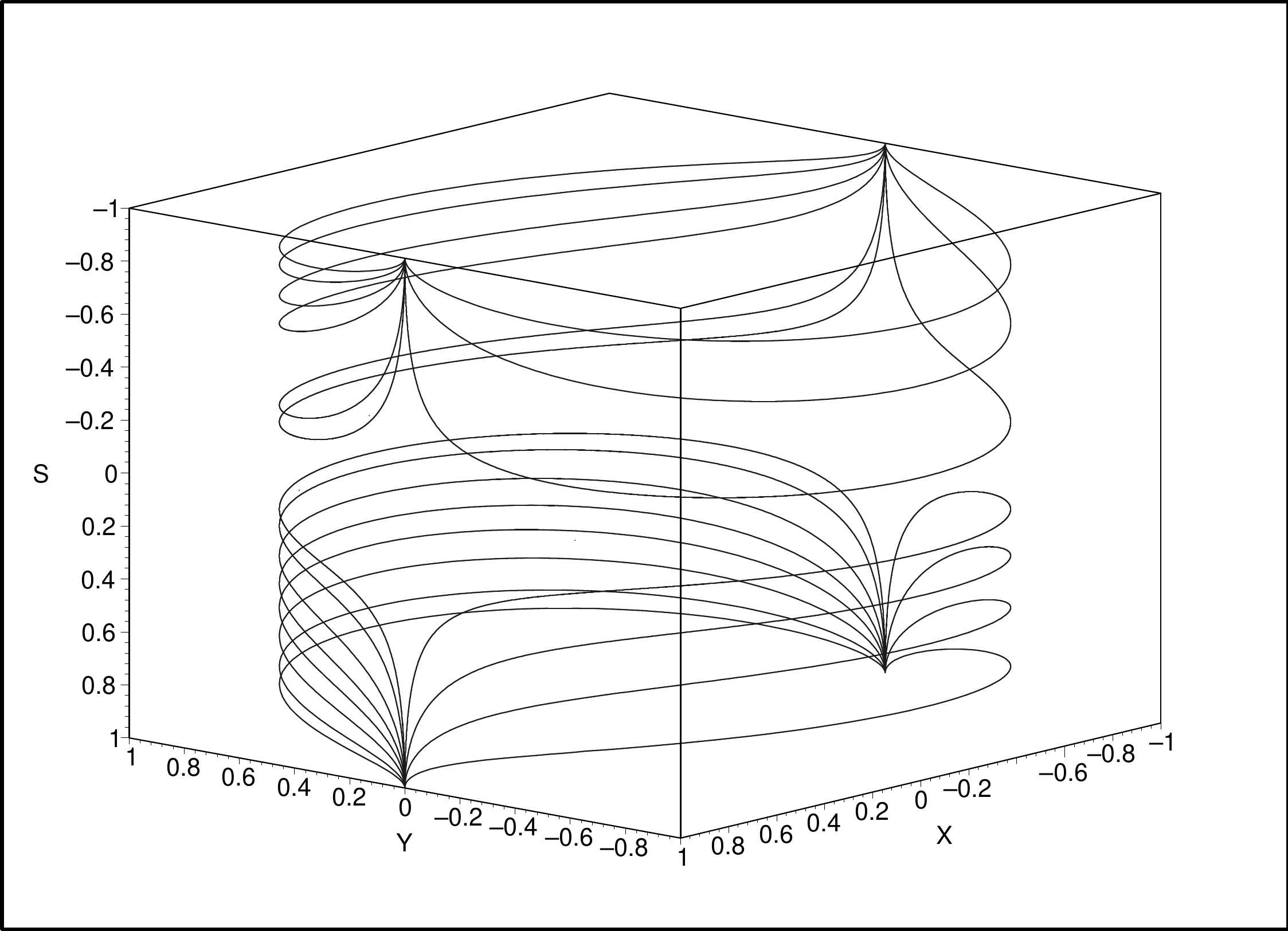}} \hspace{2cm}  \subfigure[]{%
\includegraphics[width=4in, height=4in]{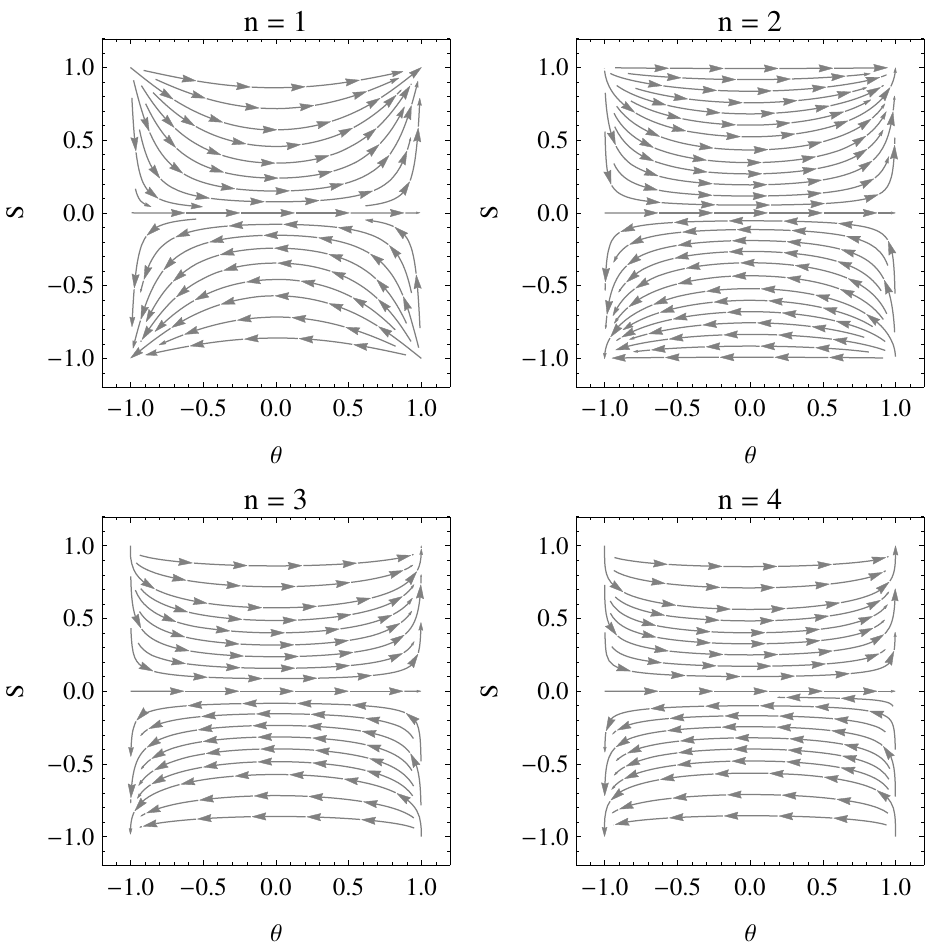}} 
\caption{\textit{{(a) Compact phase portrait of \eqref{SYSTEM88a}, \eqref{SYSTEM88b}, \eqref{SYSTEM88c} for the
choice $n=2$. (b) Dynamics on the cylinder at infinity given by Eqs. 
\eqref{SYST515a}, \eqref{SYST515b} for some values of $n$.}}}
\label{fig:Case5HLPL}
\end{figure*}

The system 
\eqref{GenHLext2}, \eqref{GenHLext3}, \eqref{GenHLext4}, \eqref{GenHLextx}, 
\eqref{GenHLexty}, \eqref{GenHLexts}  becomes: 
\begin{align}
&\frac{dx_2}{dN}=2 x_2 \left(3 x^2+x_2+2 x_3+3 x_4-1\right),
\label{PowerHLext2} \\
&\frac{dx_3}{dN}=2x_3 \left(3 x^2+x_2+2 x_3+3 x_4-2\right),
\label{PowerHLext3} \\
&\frac{dx_4}{dN}=2 x_4 \left(3x^2+x_2+2 x_3+3 x_4-3\right),
\label{PowerHLext4}
\\
&\frac{dx}{dN}= 3 x^3+(x_2+2 x_3+3 x_4-3) x+\sqrt{6} s y^2,
\label{PowerHLextx} \\
&\frac{dy}{dN}=\left(3 x^2-\sqrt{6} s x+x_2+2 x_3+3 x_4\right) y,
\label{PowerHLexty} \\
&\frac{ds}{dN}=\frac{\sqrt{6}}{n} x s^2 ,  \label{PowerHLexts}
\end{align}
defining a dynamical system in $\mathbb{R}^6$. The equilibrium points at the
finite region and their stability conditions are summarized as follows. 

\begin{itemize}
\item $(x_2,x_3,x_4,x,y,s)=(0,0,0,0,y_c,0)$ with eigenvalues \newline
$-6, -4, -3, -2, 0 , 0$.  Nonhyperbolic with 2D center manifold and 4D
stable manifold. 

\item $(x_2,x_3,x_4,x,y,s)=\left(0,0,1-x_c^2,x_c,0,0\right)$ with
eigenvalues
$6, 4, 3, 2, 0, 0$.  Nonhyperbolic with 2D center manifold and 4D unstable
manifold. 

\item $(x_2,x_3,x_4,x,y,s)=(0,0,0,0,0,s_c)$. The eigenvalues are%

$0, 0, -6, -4, -3, -2$.  Nonhyperbolic with 2D center manifold and 4D stable
manifold. 

\item $(x_2,x_3,x_4,x,y,s)=(0,0,1,0,0,s_c)$. The eigenvalues are%

$0, 0, 2, 3, 4, 6$.  Nonhyperbolic with 2D center manifold and 4D unstable
manifold. 

\item $(x_2,x_3,x_4,x,y,s)=(0,1,0,0,0,s_c)$. The eigenvalues are%

$0, -2, -1, 2, 2, 4$. Nonhyperbolic. It behaves as a saddle. 

\item $(x_2,x_3,x_4,x,y,s)=(1,0,0,0,0,s_c)$. The eigenvalues are%

$0, -4, -2, -2, 1, 2$. Nonhyperbolic. It behaves as a saddle. 

\item  $(x_2,x_3,x_4,x,y,s)=(0,0,0,-1,0,0)$. The eigenvalues are%

$6, 4, 3, 2, 0, 0$.  Nonhyperbolic with 2D center manifold and 4D unstable
manifold. 

\item  $(x_2,x_3,x_4,x,y,s)=(0,0,0,0,0,0)$. The eigenvalues are%

$-6, -4, -3, -2, 0, 0$.  Nonhyperbolic with 2D center manifold and 4D stable
manifold. 

\item  $(x_2,x_3,x_4,x,y,s)=(0,1,0,0,0,0)$. The eigenvalues are%

$4, -2, 2, 2, -1, 0$. Nonhyperbolic. It behaves as a saddle. 

\item  $(x_2,x_3,x_4,x,y,s)=(1,0,0,0,0,0)$. The eigenvalues are%

$-4, -2, -2, 2, 1, 0$. Nonhyperbolic. It behaves as a saddle. 

\item  $(x_2,x_3,x_4,x,y,s)=(0,0,0,1,0,0)$. The eigenvalues are%

$6, 4, 3, 2, 0, 0$.  Nonhyperbolic with 2D center manifold and 4D unstable
manifold. 
\end{itemize}

 Concerning the dynamics on the infinity region of the phase space,
we have found that the relevant asymptotic dynamics occurs on the invariant
set $X_2=X_3=X_4=0$, where the evolution equations reduced, after taking a
time rescaling, to 
\begin{align}
&\frac{dX}{d\tau}=\sqrt{6} Y^2 \,\tan\left(\frac{\pi S}{2}\right) +3 X
\left(2 X^2-1\right) \sqrt{1-X^2-Y^2}, \label{SYSTEM88a}\\
&\frac{dY}{d\tau}=-X Y \Big(\sqrt{6} \,\tan\left(\frac{\pi S}{2}\right) - 6
X \sqrt{1-X^2-Y^2}\Big), \label{SYSTEM88b}\\
&\frac{dS}{d\tau}=-\frac{\sqrt{6} X (\cos (\pi S)-1)}{\pi n}. \label{SYSTEM88c}
\end{align}
Summarizing, we have that the equilibrium points/lines with the higher
dimension of the stable manifold are
$(0,0,0,0,y_c,0)$, $(0,0,0,0,0,s_c)$ and $(0,0,0,0,0,0)$. We distinguish the
three cases by assuming $y_c\neq 0, s_c\neq 0$. For all these solutions the
center manifold is 2D. Therefore, one can be interested on the stability
analysis of the corresponding center manifolds, since they might contain the
relevant late-time attractors. On the other hand, we also see that the
equilibrium points/lines with the higher dimension of the unstable manifold
are $\left(0,0,1-x_c^2,x_c,0,0\right)$, $(0,0,1,0,0,s_c)$, $(0,0,0,-1,0,0)$,
and $(0,0,0,1,0,0)$. We distinguish the four cases. As before, one can be
interested on the stability analysis of the corresponding center manifolds,
since they might contain the relevant early-time attractors. 

 To analyze the dynamics on the cylinder at infinity it is more
convenient to use the parametrization 

\begin{equation}
X=\sin\left(\frac{\pi\theta}{2}\right), Y=\cos\left(\frac{\pi\theta}{2}%
\right), \theta \in \left[-1, 1\right].
\end{equation}
We consider only the sector $Y\geq 0$ due to the system is invariant under
the reflection $Y\rightarrow -Y$. Therefore, we obtain the system

\begin{align}
&\frac{d\theta}{d\tau}= \frac{2 \sqrt{6} \cos \left(\frac{\pi \theta }{2}%
\right) \tan \left(\frac{\pi S}{2}\right)}{\pi }, \label{SYST515a}
\\
& \frac{dS}{d\tau}=-\frac{\sqrt{6} \sin \left(\frac{\pi \theta }{2}\right)
(\cos (\pi S)-1)}{\pi n}. \label{SYST515b}
\end{align}

 In the figure \ref{fig:Case5HLPL} it is shown (a) a compact phase
portrait of \eqref{SYSTEM88a}, \eqref{SYSTEM88b}, \eqref{SYSTEM88c} for the choice $n=2$. (b) Dynamics on the
cylinder at infinity given by Eqs. \eqref{SYST515a}, \eqref{SYST515b} for some values of $n$.

\subsection{\protect E-models}
\label{Case5D}

\begin{figure*}[ht!]
\centering
\subfigure[]{\includegraphics[width=3.5in, height=3.5in,angle=-90]{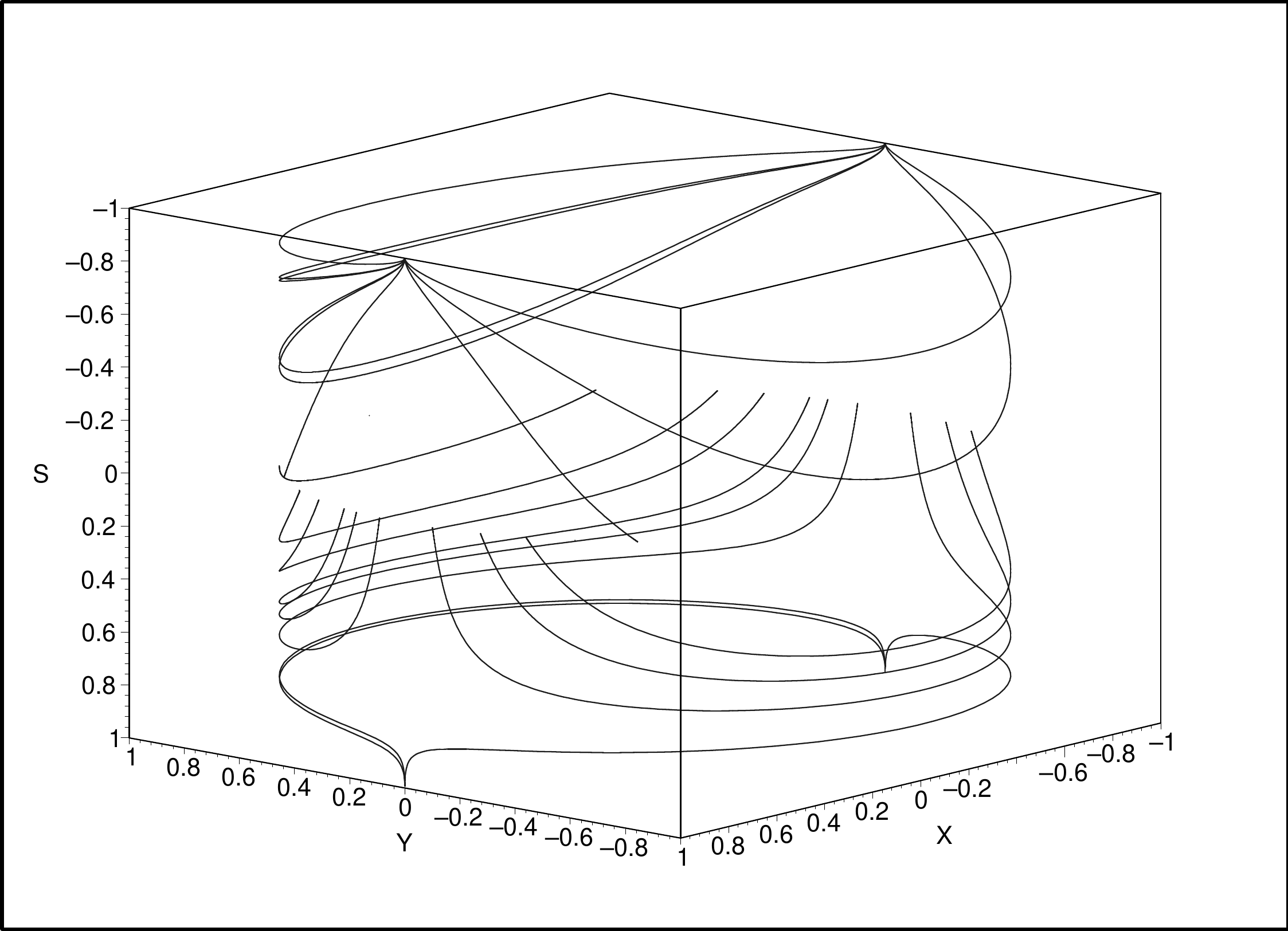}}
\hspace{2cm}  \subfigure[]{\includegraphics[width=4in, height=4in]{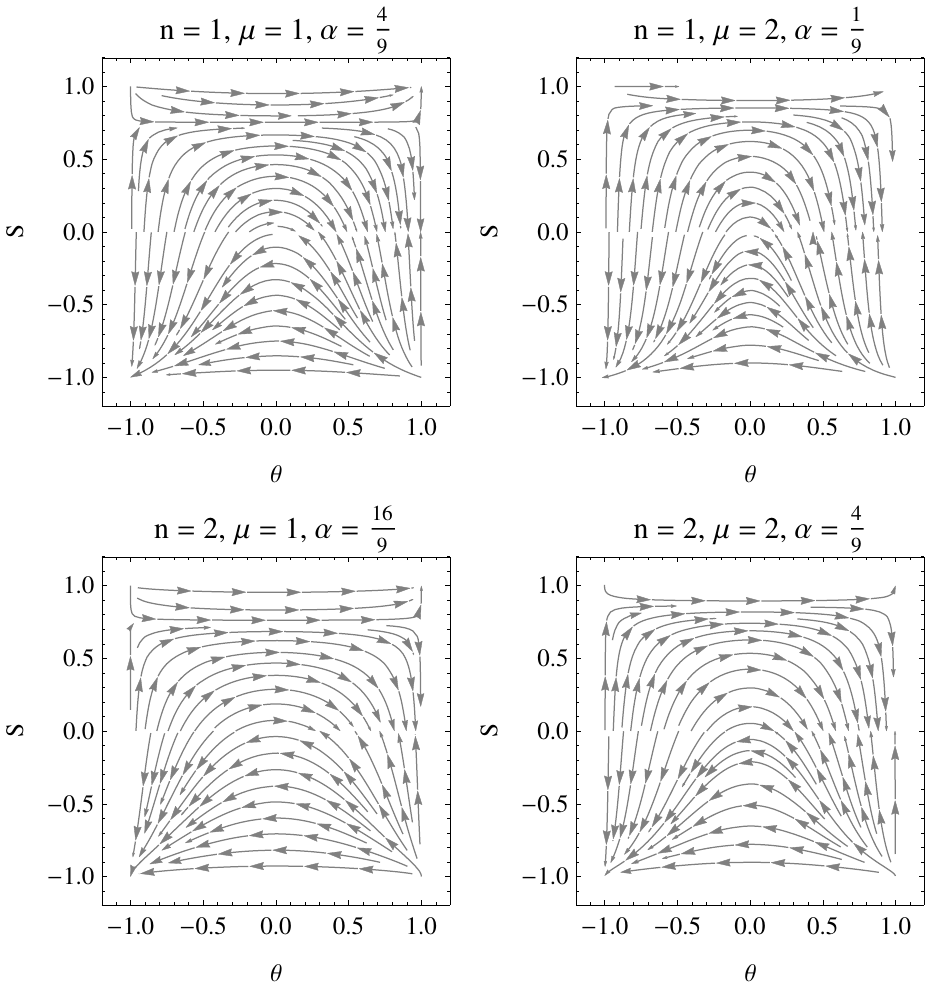}}
\caption{\textit{{(Color online) (a) Compact phase portrait of 
\eqref{SYSTEM105a}, \eqref{SYSTEM105b}, \eqref{SYSTEM105c} for the choice $n=2, \protect\mu=1$. (b) Dynamics on the
cylinder at infinity given by Eqs. \eqref{SYST106a},  \eqref{SYST106b} for some values of $n, 
\protect\mu$.}}}
\label{fig:Case5HLE}
\end{figure*}

 The dynamics on the finite region of the phase space is given by: 
\begin{align}
&\frac{dx_2}{dN}=2 x_2 \left(3 x^2+x_2+2 x_3+3 x_4-1\right),
\label{EmodelHLext2} \\
&\frac{dx_3}{dN}=2x_3 \left(3 x^2+x_2+2 x_3+3 x_4-2\right),
\label{EmodelHLext3} \end{align}
\begin{align}
&\frac{dx_4}{dN}=2 x_4 \left(3x^2+x_2+2 x_3+3 x_4-3\right),
\label{EmodelHLext4} \\
&\frac{dx}{dN}= 3 x^3+(x_2+2 x_3+3 x_4-3) x+\sqrt{6} s y^2,
\label{EmodelHLextx} \end{align}
\begin{align}
&\frac{dy}{dN}=\left(3 x^2-\sqrt{6} s x+x_2+2 x_3+3 x_4\right) y,
\label{EmodelHLexty} \\
&\frac{ds}{dN}=\frac{\sqrt{6}}{n} x s(s-\sqrt{6}\mu),  \label{EmodelHLexts}
\end{align}
defining a dynamical system in $\mathbb{R}^6$. 

The equilibrium points at the finite region and their stability
conditions are summarized as follows. 

\begin{itemize}
\item { $(x_2,x_3,x_4,x,y,s)=(0, 0, 0, 0, y_c, 0)$, with eigenvalues%

$-6,-4,-3,-2,0,0$. Nonhyperbolic with 2D center manifold and 4D stable
manifold. }

\item { $(x_2,x_3,x_4,x,y,s)=\left(0, 0, 1-x_c^2, x_c, 0, \sqrt{6}
\mu\right)$, with eigenvalues 
$0,2,4,6,\frac{6 x_c \mu }{n},3-6 x_c  \mu$. Nonhyperbolic with a 5D
unstable manifold for $0<x_c\mu<\frac{1}{2}$. Nonhyperbolic and behaves as
saddle otherwise. }

\item { $(x_2,x_3,x_4,x,y,s)=\left(0, 0, 1-x_c^2, x_c, 0, 0\right)$,
with eigenvalues 
$0,2,3,4,6,-\frac{6 x_c \mu }{n}$. Nonhyperbolic with a 5D unstable manifold
for $x_c\mu<0$. Nonhyperbolic and behaves as saddle otherwise. }

\item  $(x_2,x_3,x_4,x,y,s)=(0, 0, 0, 0, 0, s_c)$, with eigenvalues%

$0,0,-6,-4,-3,-2$. Nonhyperbolic with 2D center manifold and 4D stable
manifold. 

\item  $(x_2,x_3,x_4,x,y,s)=(0, 0, 1, 0, 0, s_c)$, with eigenvalues%

$0,0,2,3,4,6$. Nonhyperbolic with 2D center manifold and 4D unstable
manifold. 

\item  $(x_2,x_3,x_4,x,y,s)=(0, 1, 0, 0, 0, s_c)$, with eigenvalues%

$0,-2,-1,2,2,4$. Nonhyperbolic, behaves as a Saddle. 

\item  $(x_2,x_3,x_4,x,y,s)=(1, 0, 0, 0, 0, s_c)$, with eigenvalues%

$0,-4,-2,-2,1,2$. Nonhyperbolic, behaves as a Saddle. 

\item  $(x_2,x_3,x_4,x,y,s)=\left(0, 0, 0, -1, 0,\sqrt{6} \mu \right)$%
, with eigenvalues
$0,2,4,6,-\frac{6 \mu }{n},6 \mu  +3$. Nonhyperbolic with a 5D unstable
manifold for $-\frac{1}{2}<\mu<0$. Nonhyperbolic and behaves as saddle
otherwise. 

\item  $(x_2,x_3,x_4,x,y,s)=\left(0, 0, 0, 0, 0, \sqrt{6} \mu \right)$%
, with eigenvalues 
$-6,-4,-3,-2,0,0$. Nonhyperbolic with 2D center manifold and 4D stable
manifold. 

\item $(x_2,x_3,x_4,x,y,s)=\left(0, 1, 0, 0, 0, \sqrt{6} \mu \right)$%
, with eigenvalues 
$4,-2,2,2,-1,0$. Nonhyperbolic, behaves as a Saddle. 

\item $(x_2,x_3,x_4,x,y,s)=\left(1, 0, 0, 0, 0, \sqrt{6} \mu \right)$%
, with eigenvalues 
$-4,-2,-2,2,1,0$. Nonhyperbolic, behaves as a Saddle. 

\item $(x_2,x_3,x_4,x,y,s)=\left(0, 0, 0, 1, 0, \sqrt{6} \mu \right)$%
, with eigenvalues 
$0,2,4,6,\frac{6 \mu }{n},3-6 \mu$. Nonhyperbolic with a 5D unstable
manifold for $0<\mu<\frac{1}{2}$. Nonhyperbolic and behaves as saddle
otherwise. 

\item  $\left(0, 0, 1-\frac{1}{4 \mu^2}, \frac{1}{2 \mu}, 0, \sqrt{6}
\mu\right)$, with eigenvalues
$0,0,\frac{3}{n},2,4,6$. Nonhyperbolic with 2D center manifold and 4D
unstable manifold. 

\item  $(x_2,x_3,x_4,x,y,s)=\left(1-\frac{1}{12 \mu^2}, 0, 0, \frac{1%
}{6\mu}, -\frac{1}{3 \sqrt{2}\mu}, \sqrt{6} \mu\right)$, with eigenvalues%

$-4,-2,\frac{1}{n},2,-\frac{\sqrt{\mu^4-9 \mu^6}}{\sqrt{3} \mu^3}-1,\frac{%
\sqrt{\mu^4-9\mu^6}}{\sqrt{3} \mu^3}-1$. Saddle. 

\item  $\left(0, 1-\frac{1}{6 \mu^2}, 0, \frac{1}{3 \mu}, -\frac{1}{3 
\sqrt{2} \mu }, \sqrt{6} \mu \right)$, with eigenvalues \newline
$-2,\frac{2}{n},2,4,-\frac{\sqrt{\frac{8 \mu^4}{3}-15 \mu^6}}{2 \mu^3}-\frac{%
1}{2},\frac{\sqrt{\frac{8 \mu^4}{3}-15 \mu^6}}{2 \mu ^3}-\frac{1}{2}$.
Saddle. 

\item $(x_2,x_3,x_4,x,y,s)=\left(1-\frac{1}{12 \mu ^2}, 0, 0, \frac{1%
}{6 \mu }, \frac{1}{3 \sqrt{2} \mu }, \sqrt{6} \mu \right)$, with eigenvalues%

$-4,-2,\frac{1}{n},2,-\frac{\sqrt{\mu^4-9 \mu^6}}{\sqrt{3}\mu ^3}-1,\frac{%
\sqrt{\mu^4-9 \mu ^6}}{\sqrt{3} \mu ^3}-1$. Saddle. 

\item  $\left(0, 1-\frac{1}{6 \mu ^2}, 0, \frac{1}{3 \mu }, \frac{1}{%
3 \sqrt{2} \mu }, \sqrt{6} \mu\right)$, with eigenvalues \newline
$-2,\frac{2}{n},2,4,-\frac{\sqrt{\frac{8 \mu ^4}{3}-15 \lambda ^6}}{2 \mu ^3}%
-\frac{1}{2},\frac{\sqrt{\frac{8\mu ^4}{3}-15 \mu^6}}{2 \mu ^3}-\frac{1}{2}$%
. Saddle. 

\item  $(x_2,x_3,x_4,x,y,s)=\left(0, 0, 0, 2 \mu, -\sqrt{1-4 \mu^2}, 
\sqrt{6}\mu\right)$, with eigenvalues
$\frac{12 \mu ^2}{n},24 \mu ^2,24 \mu ^2-6,24 \mu ^2-4,24 \mu ^2-2,12 \mu
^2-3$. Saddle. 

\item  $(x_2,x_3,x_4,x,y,s)=\left(0, 0, 0, 2 \mu, \sqrt{1-4 \mu ^2}, 
\sqrt{6} \mu\right)$, with eigenvalues
$\frac{12 \mu ^2}{n},24 \mu ^2,24 \mu ^2-6,24 \mu ^2-4,24 \mu ^2-2,12 \mu
^2-3$. Saddle. 

\item  $(x_2,x_3,x_4,x,y,s)=(0, 0, 0, -1, 0, 0)$, with eigenvalues%

$0,2,3,4,6,\frac{6 \mu }{n}$. Nonhyperbolic with a 5D unstable manifold for $%
\mu>0$. Nonhyperbolic and behaves as saddle otherwise. 

\item  $(x_2,x_3,x_4,x,y,s)=(0, 0, 0, 0, 0, 0)$, with eigenvalues%

$-6,-4,-3,-2,0,0$. Nonhyperbolic with 2D center manifold and 4D stable
manifold. 

\item  $(x_2,x_3,x_4,x,y,s)=(0, 1, 0, 0, 0, 0)$, with eigenvalues%

$4,-2,2,2,-1,0$. Nonhyperbolic and behaves as saddle. 

\item  $(x_2,x_3,x_4,x,y,s)=(1, 0, 0, 0, 0, 0)$, with eigenvalues%

$-4,-2,-2,2,1,0$. Nonhyperbolic and behaves as saddle. 

\item  $(x_2,x_3,x_4,x,y,s)=(0, 0, 0, 1, 0, 0)$, with eigenvalues%

$0,2,3,4,6,-\frac{6 \mu }{n}$. Nonhyperbolic with a 5D unstable manifold for 
$\mu<0$. Nonhyperbolic and behaves as saddle otherwise. 
\end{itemize}

The relevant asymptotic dynamics occurs on the invariant set $%
X_2=X_3=X_4=0$, where the evolution equations reduced, after taking a time
rescaling, to 
\begin{align}
&\frac{dX}{d\tau}=\sqrt{6} Y^2 \,\tan\left(\frac{\pi S}{2}\right) +3 X
\left(2 X^2-1\right) \sqrt{1-X^2-Y^2}, \label{SYSTEM105a}\\
&\frac{dY}{d\tau}=-X Y \Big(\sqrt{6} \,\tan\left(\frac{\pi S}{2}\right) - 6
X \sqrt{1-X^2-Y^2}\Big), \label{SYSTEM105b}\\
&\frac{dS}{d\tau}=\frac{X \left(-6 \mu \sin (\pi S)-\sqrt{6} \cos (\pi
S)+\sqrt{6}\right)}{\pi n}. \label{SYSTEM105c}
\end{align}
Summarizing, the equilibrium points/lines with the higher dimension of the
stable manifold are $(0, 0, 0, 0, y_c, 0)$, \newline $\left(0, 0, 0, 0, 0, \sqrt{6}
\mu \right)$, and 
$(0, 0, 0, 0, 0, 0)$. For all these solutions the center manifold is 2D.
Therefore, one can be interested on the stability analysis of the
corresponding center manifolds, since they might contain the relevant
late-time attractors. Whereas, the equilibrium points/lines with the higher
dimension of the unstable manifold are \newline $\left(0, 0, 1-x_c^2, x_c, 0, \sqrt{6}
\mu \right)$,
$\left(0, 0, 1-x_c^2, x_c, 0, 0\right)$, \newline $\left(0, 0, 0, -1, 0,\sqrt{6}
\mu \right)$, $\left(0, 0, 0, 1, 0, \sqrt{6} \mu \right)$, 
$(0, 0, 0, -1, 0, 0)$, $(0, 0, 0, 1, 0, 0)$. For all these solutions the
center manifold is 1D and the unstable manifold is 5D. Therefore, one can be
interested on the stability analysis of the corresponding center manifolds,
since they might contain the relevant early-time attractors. 

 To analyze the dynamics on the cylinder at infinity it is more
convenient to use the parametrization 
\begin{equation}
X=\sin\left(\frac{\pi\theta}{2}\right), Y=\cos\left(\frac{\pi\theta}{2}%
\right), \theta \in \left[-1, 1\right].
\end{equation}
We consider only the sector $Y\geq 0$ due to the system is invariant under
the reflection $Y\rightarrow -Y$. Therefore, we obtain the system 
\begin{align}
&\frac{d\theta}{d\tau}= \frac{2 \sqrt{6} \cos \left(\frac{\pi \theta }{2}%
\right) \tan \left(\frac{\pi S}{2}\right)}{\pi }, \label{SYST106a}\\
& \frac{dS}{d\tau}=-\frac{\sqrt{6} \sin \left(\frac{\pi \theta }{2}\right) (%
\sqrt{6} \mu\sin (\pi S) + \cos (\pi S)-1)}{\pi n}. \label{SYST106b}
\end{align}

In the figure \ref{fig:Case5HLE} it is shown (a) a compact phase
portrait of \eqref{SYSTEM105a}, \eqref{SYSTEM105b}, \eqref{SYSTEM105c} for the choice $n=2, \mu=1$. (b) Dynamics on
the cylinder at infinity given by Eqs. \eqref{SYST106a}, \eqref{SYST106b} for some values of $%
n, \mu$. 

\section{\protect Conclusions}

 \label{HLconclusions}

Although the HL gravity theory in the original presentation has
several drawbacks, still there are several attempts to cure some of these
problems. In \cite{dan1} the authors reconsidered the \textquotedblleft
detailed balance\textquotedblright\ as a principle that can be used to
restrict the proliferation of couplings in HL gravity, and were discussed some resolutions 
for all of the usual shortcomings within this framework with the
most persistent related to the projectability. The main issues are quite
tough to be solved, including how to derive the RG flows, the calculation of
higher order quantum corrections, and how to recover Lorentz invariance. The
fact some of these issues are very difficult to be answered, however, does
not spoil the interest in the theory. In this paper we have proceeded to a
very detailed analysis of the stability of several cosmological solutions
that are of interest for inflation in HL gravity, using the Center Manifold
Theory. We restricted our attention to few aspects of the theory, and we
have focused more about the mathematical tools used for the analysis of the
solution space of the models. We studied the dynamics of a scalar field in Ho%
\v{r}ava-Lifshitz cosmology for a wide range of self-interacting potentials
for the scalar field under the detailed - balance condition and without
imposing it, by means of the powerful method of $f$-devisers. By
following this approach one can study the cosmological consequences of the
model at hand without the need of an a priori specification of the
potential, and in the end one just substitutes the specific potential form
in the results, instead of having to repeat the whole dynamical elaboration
from the start. As we have seen, the results are richer and more general,
revealing the full capabilities of HL cosmology. 

Because of the complexity of our analysis we categorized it in four
possible cases: (a) $k=0,~\Lambda =0$, (b)  $k\neq 0,\Lambda =0$, (c)~$k=0,~\Lambda \neq 0$, and (d) $k\neq
0,~\Lambda \neq 0$. For a general potential, the number of equations
for the autonomous dimensionless dynamical system under the detailed-balance
condition in the three first cases is three, and four for case (d), however,
the geometry and the constraint conditions of the phase space change. Due to
in general the designed phase spaces are non compact, we have implemented a
compactification procedure such that the dynamics in both the finite and
infinity region can be visualized with the help of 2D and 3D phase space by
integrating numerically for some specific potentials. 

We considered also the case beyond the detailed-balance condition.
In that analysis we studied the critical points for a six dimensional
dynamical system. For the majority of the critical points the stability
depend on the function form of the $f$-devisers. Hence, in order to
demonstrate our results we considered some specific forms for the functions%
$f\left( s\right) $, which follows from specific potentials$%
V\left( \phi \right)$ proposed before in the literature. 

Specifically, we considered the exponential potential and for
potentials beyond the exponential one, e.g., the powerlaw potential $V(\phi
)=\frac{1}{2n}(\mu \phi )^{2n},\mu >0,n=1,2,\ldots $ \cite{Alho:2015cza}
with $f(s)=-\frac{s^{2}}{2n}$; the potential of the so-called E-model: $%
V(\phi )=V_{0}\left( 1-e^{-\sqrt{\frac{2}{3\alpha }}\phi }\right) ^{2n}$
with $f(s)=-\frac{s\left( s-\sqrt{6}\mu \right) }{2n}$, $\mu =\frac{2n}{3%
\sqrt{\alpha }}$, discussed in \cite{Alho:2017opd} for a conventional scalar
field cosmology. Observe that the dynamics of the latter potential is
equivalent to that of the exponential potential plus a cosmological
constant, $V=V_{0}e^{-\sqrt{6}\mu \phi }+\Lambda $ having $f(s)=-s\left( s-%
\sqrt{6}\mu \right) $, up to a rescaling in the independent variable.
Finally, for the exponential potential since we found that in all cases
critical points which describe de Sitter universes exist. We performed a
thorough analysis on the stability of that solutions with special interests
by applying the center manifold theorem. The previous findings by Leon \&
Saridakis were recovered and extended and new examples were presented and
discussed. The advantages of using Center Manifold theory are unveiled
immediately when the procedure is compared with the Normal Forms
Calculations presented previously in the literature. The results of \cite%
{Alho:2015cza,Alho:2017opd} were recovered as particular cases. We
have presented several results concerning the stability of the de Sitter
solution in Ho\v{r}ava-Lifshitz cosmology using the Center Manifold theory.
We find that in all the cases, with the exception of the model studied in
Section \ref{stabilityP26}, the de Sitter solution is unstable: saddle or
center-saddle. In Section \ref{stabilityP26} we proved that the de Sitter
solution is locally asymptotically stable. These analyzes are of
mathematical relevance for the cosmology. 

\section*{\protect Acknowledgements}

This work was funded by Comisi\'{o}n Nacional de Investigaci\'{o}n
Cient\'{\i}fica y Tecnol\'{o}gica (CONICYT) through FONDECYT Iniciaci\'{o}n
11180126. The author thanks to Departmento de Matem\'{a}tica and to
Vicerrector\'{\i}a de Investigaci\'{o}n y Desarrollo Tecnol\'{o}gico at
Universidad Cat\'{o}lica del Norte for financial support. AP acknowledges
the financial support \ of FONDECYT grant no. 3160121. 

 \appendix

\section{\protect  Center Manifold theory}

 \label{sectionCM} 
In this appendix we summarize the main techniques for the
construction of center manifolds for vector fields in $\mathbb{R}^n$. We
follow the approach of the book \cite{wiggins}, chapter 18. The setup is as
follows. We consider vector fields in the form 
\begin{align}
&\mathbf{x}^{\prime }=\mathbf{A x}+\mathbf{f}(\mathbf{x},\mathbf{y}),  \notag
\\
&\mathbf{y}^{\prime }=\mathbf{B x}+\mathbf{g}(\mathbf{x},\mathbf{y}), \; (%
\mathbf{x},\mathbf{y})\in \mathbb{R}^c\times\mathbb{R}^s,  \label{basiceqs}
\end{align}
where 
\begin{align}
& \mathbf{f(0,0)}=\mathbf{0}, \mathbf{Df(0,0)}=\mathbf{0}, \mathbf{g(0,0)}=%
\mathbf{0}, \mathbf{Dg(0,0)}=\mathbf{0}.
\end{align}
$\mathbf{A}$ is a $c\times c$ matrix having eigenvalues with zero real
parts, $\mathbf{B}$ is an $s\times s$ matrix having eigenvalues with
negative real parts, and $\mathbf{f}$ and $\mathbf{g}$ are $C^r$ functions ($%
r\geq 2$).

\begin{defn}[Center Manifold]
{ \label{CMdef} An invariant manifold will be called a center manifold
for \eqref{basiceqs} if it can locally be represented as follows 
\begin{align}
&W^{c}\left( \mathbf{0}\right) =\left\{ \left( \mathbf{x},\mathbf{y}\right)
\in\mathbb{R}^c\times\mathbb{R}^s:\mathbf{y}=\mathbf{h}\left( \mathbf{x}%
\right) ,\;\left\vert \mathbf{x}\right\vert <\delta\right\};  \notag \\
& \;\;\;\mathbf{h}\left( \mathbf{0}\right) =\mathbf{0},\;D\mathbf{h}\left( 
\mathbf{0}\right) =\mathbf{0},
\end{align}
for $\delta$ sufficiently small (cf. \cite{wiggins} p. 246, \cite{Perko},p.
155). }
\end{defn}
The conditions $\mathbf{h}\left( \mathbf{0}\right) =\mathbf{0},\;D%
\mathbf{h}\left( \mathbf{0}\right) =\mathbf{0}$ imply that $W^{c}\left( 
\mathbf{0}\right)$ is tangent to $E^c$ at $\left(\mathbf{x},\mathbf{y}%
\right)=(\mathbf{0},\mathbf{0}),$ where $E^c$ is the generalized eigenspace
whose corresponding eigenvalues have zero real parts. The following three
theorems (see theorems 18.1.2, 18.1.3 and 18.1.4 in \cite{wiggins} p.
245-248) are the main results to the treatment of center manifolds. The
first two are existence and stability theorems of the center manifold for %
\eqref{basiceqs} at the origin. The third theorem allows to compute the
center manifold to any desired degree accuracy by using Taylor series to
solve a quasilinear partial differential equation that $\mathbf{h}\left( 
\mathbf{x}\right)$ must satisfy. The proofs of these results are given in 
\cite{Carr:1981}. 

\begin{thm}[Existence]
\label{existenceCM} There exists a $C^r$ center manifold for %
\eqref{basiceqs}. The dynamics of \eqref{basiceqs} restricted to the center
manifold is, for $\mathbf{u}$ sufficiently small, given by the following
c-dimensional vector field 
\begin{equation}  \label{vectorfieldCM}
\mathbf{u}^{\prime }=\mathbf{A u}+\mathbf{f}\left(\mathbf{u},\mathbf{h}\left(%
\mathbf{u}\right)\right),\; \mathbf{u}\in\mathbb{R}^c.
\end{equation}
\end{thm}
The next results implies that the dynamics of \eqref{vectorfieldCM}
near $\mathbf{u}=0$ determine the dynamics of \eqref{basiceqs} near $\left(%
\mathbf{x},\mathbf{y}\right)=(\mathbf{0},\mathbf{0})$ (see also Theorem
3.2.2 in \cite{Guckenheimer}).

\begin{thm}[Stability]
\label{stabilityCM} i) Suppose the zero solution of %
\eqref{vectorfieldCM} is stable (asymptotically stable) (unstable); then the
zero solution of \eqref{basiceqs} is also stable (asymptotically stable)
(unstable). Then if $(\mathbf{x}(\tau),\mathbf{y}(\tau))$ is a solution of %
\eqref{basiceqs} with $(\mathbf{x}(0),\mathbf{y}(0))$ sufficiently small,
then there is a solution $\mathbf{u}(\tau)$ of \eqref{vectorfieldCM} such
that, as $\tau\rightarrow\infty$ 
\begin{align*}
& \mathbf{x}(\tau)=\mathbf{u}(\tau)+\mathcal{O}(e^{-r \tau}), \\
& \mathbf{x}(\tau)=\mathbf{h}\left(\mathbf{u}(\tau)\right)+\mathcal{O}(e^{-r
\tau}),
\end{align*}
where $r>0$ is a constant.
\end{thm}

\textbf{Dynamics Captured by the center manifold.} Theorem \ref%
{stabilityCM} says that for initial conditions of the \emph{full system}
sufficiently close to the origin, trajectories through them asymptotically
approach a trajectory on the center manifold. In particular, singular points
sufficiently close to the origin, sufficiently small amplitude periodic
orbits, as well as small homoclinic and heteroclinic orbits are contained in
the center manifold.

 The obvious question now is how to compute the center manifold so
that we can use the result of theorem \ref{stabilityCM}? To answer this
question we will derive an equation that $\mathbf{h}(\mathbf{x})$ must
satisfy in order to its graph to be a center manifold for \eqref{basiceqs}.

Suppose we have a center manifold 
\begin{align}
&W^{c}\left( \mathbf{0}\right) =\left\{ \left( \mathbf{x},\mathbf{y}\right)
\in\mathbb{R}^c\times\mathbb{R}^s:\mathbf{y}=\mathbf{h}\left( \mathbf{x}%
\right) ,\;\left\vert \mathbf{x}\right\vert <\delta\right\};  \notag \\
& \;\;\;\mathbf{h}\left( \mathbf{0}\right) =\mathbf{0},\;D\mathbf{h}\left( 
\mathbf{0}\right) =\mathbf{0},
\end{align}
with $\delta$ sufficiently small. Using the invariance of $W^{c}\left( 
\mathbf{0}\right)$ under the dynamics of \eqref{basiceqs}, we derive a
quasilinear partial differential equation that $\mathbf{h}\left( \mathbf{x}%
\right)$ must satisfy. This is done as follows: 
\begin{enumerate}
\item The $(\mathbf{x},\mathbf{y})$ coordinates of any point on $%
W^{c}\left( \mathbf{0}\right) $ must satisfy 
\begin{equation}
\mathbf{y}=\mathbf{h}(\mathbf{x})  \label{coordCM}
\end{equation}

\item  Differentiating \eqref{coordCM} with respect to time implies
that the $(\mathbf{x}^{\prime },\mathbf{y}^{\prime })$ coordinates of any
point on $W^{c}\left( \mathbf{0}\right) $ must satisfy 
\begin{equation}
\mathbf{y}^{\prime }=D\mathbf{h}\left( \mathbf{x}\right)\mathbf{x}^{\prime }
\label{totalderivative}
\end{equation}

\item Any point in $W^{c}\left( \mathbf{0}\right) $ obey the
dynamics generated by \eqref{basiceqs}. Therefore substituting 

\begin{align}
&\mathbf{x}^{\prime }=\mathbf{A x}+\mathbf{f}\left(\mathbf{x},\mathbf{h}(%
\mathbf{x})\right), \\
&\mathbf{y}^{\prime }=\mathbf{B}\mathbf{h}(\mathbf{x})+\mathbf{g}\left(%
\mathbf{x},\mathbf{h}(\mathbf{x})\right)
\end{align}
into \eqref{totalderivative} gives 

\begin{align}
& \mathcal{N}\left(\mathbf{h}(\mathbf{x})\right)\equiv D\mathbf{h}(\mathbf{x}%
)\left[\mathbf{A x}+\mathbf{f}\left(\mathbf{x},\mathbf{h}(\mathbf{x})\right)%
\right] -\mathbf{B}\mathbf{h}(\mathbf{x})  \notag \\
& -\mathbf{g}\left(\mathbf{x},\mathbf{h}(\mathbf{x})\right)=0 .
\label{MaineqcM}
\end{align}

\end{enumerate}

 Equation \eqref{MaineqcM} is a quasilinear partial differential that 
$\mathbf{h}(\mathbf{x})$ must satisfy in order for its graph to be an
invariant center manifold. To find the center manifold, all we need to do is
solve \eqref{MaineqcM}. Unfortunately, it is probably more difficult to
solve \eqref{MaineqcM} than our original problem; however the following
theorem give us a method for computing an approximated solution of %
\eqref{MaineqcM} to any desired degree of accuracy. 

\begin{thm}[Approximation]
\label{approximationCM} Let $\Phi:\mathbb{R}^c\rightarrow\mathbb{R}^s
$ be a $C^1$ mapping with $\Phi(\mathbf{0})=\mathbf{0}$ and $D\Phi(\mathbf{0}%
)=\mathbf{0}$ such that $\mathcal{N}\left(\Phi(\mathbf{x})\right)=\mathcal{O}%
(\|\mathbf{x}\|^q)$ as $\mathbf{x}\rightarrow \mathbf{0}$ for some $q>1.$
Then, $|\mathbf{h}(\mathbf{x})-\Phi(\mathbf{x})|=\mathcal{O}(\|\mathbf{x}%
\|^q)$ as $\mathbf{x}\rightarrow \mathbf{0}$. 
\end{thm}
This theorem allows us to compute the center manifold to any desired
degree of accuracy by solving \eqref{MaineqcM} to the same degree of
accuracy, and we can use power series expansions.


\begin{thebibliography}{999}
\bibitem{hor3}   P.~Horava, 
Phys.\ Rev.\ D \textbf{79}, 084008 (2009). 


\bibitem{Jacobson:2000xp}  T.~Jacobson and D.~Mattingly, 
Phys.\ Rev.\ D \textbf{64}, 024028 (2001). 



\bibitem{Eling:2004dk}  C.~Eling, T.~Jacobson and D.~Mattingly, 
gr-qc/0410001. 


\bibitem{DJ}  W.~Donnelly and T.~Jacobson, 
Phys.\ Rev.\ D \textbf{82}, 064032 (2010). 


\bibitem{kann} S. Kanno and J. Soda, Phys. Rev. D \textbf{74},
063505 (2006). 

\bibitem{Zlosnik:2006zu} T.~G.~Zlosnik, P.~G.~Ferreira and
G.~D.~Starkman, 
Phys.\ Rev.\ D \textbf{75}, 044017 (2007). 


\bibitem{CarrJ}  I. Carruthers and T. Jacobson, Phys Rev D \textbf{83}
024034 (2011). 



\bibitem{Jacobson} T.~Jacobson, 
PoS QG \textbf{-PH}, 020 (2007). 



\bibitem{Carroll:2004ai}  S.~M.~Carroll and E.~A.~Lim, 
Phys.\ Rev.\ D \textbf{70}, 123525 (2004). 




\bibitem{Garfinkle:2011iw}  D.~Garfinkle and T.~Jacobson, 
Phys.\ Rev.\ Lett.\ \textbf{107} (2011) 191102. 


\bibitem{TJab13}  T. Jacobson, Phys.\ Rev.\ D \textbf{89}, 081501
(2014). 


\bibitem{sot01} T.P. Sotiriou, J. Phys. Conf. Ser. 283, 012034
(2011). 

 


\bibitem{Cai:2009ar}  R.~G.~Cai, Y.~Liu and Y.~W.~Sun, 
JHEP \textbf{0906}, 010 (2009); 
R.~G.~Cai, B.~Hu and H.~B.~Zhang, 
Phys.\ Rev.\ D \textbf{80}, 041501 (2009). 




\bibitem{Charmousis:2009tc} C.~Charmousis, G.~Niz, A.~Padilla and
P.~M.~Saffin, 
JHEP \textbf{0908}, 070 (2009). 



\bibitem{Bogdanos:2009uj} C.~Bogdanos and E.~N.~Saridakis, 
Class.\ Quant.\ Grav.\ \textbf{27}, 075005 (2010); 
M.~Henneaux, A.~Kleinschmidt and G.~Lucena G\'omez, 
Phys.\ Rev.\ D \textbf{81}, 064002 (2010); 
K.~Koyama and F.~Arroja, 
JHEP \textbf{1003}, 061 (2010). 




\bibitem{Bellorin:2018blt} J.~Bellorin and B.~Droguett,  
Phys.\ Rev.\ D \textbf{98}, no. 8, 086008 (2018).  


\bibitem{Bellorin:2017gzj}  J.~Bellorin and A.~Restuccia, 
Gen.\ Rel.\ Grav.\ \textbf{49}, no. 10, 132 (2017). 



\bibitem{Bellorin:2017gab}  J.~Bellorin, A.~Restuccia and
A.~Sotomayor, 
J.\ Phys.\ Conf.\ Ser.\ \textbf{831}, no. 1, 012002 (2017). 



\bibitem{Bellorin:2016hcu}  J.~Bellorin and A.~Restuccia, 
Int.\ J.\ Mod.\ Phys.\ D \textbf{27}, no. 01, 1750174 (2017). 



\bibitem{Bellorin:2016nvh}  J.~Bellorin, A.~Restuccia and
A.~Sotomayor, 
J.\ Phys.\ Conf.\ Ser.\ \textbf{738}, no. 1, 012041 (2016). 




\bibitem{Bellorin:2016wsl} J.~Bellorin and A.~Restuccia, 
Phys.\ Rev.\ D \textbf{94}, no. 6, 064041 (2016). 



\bibitem{Bellorin:2015oja}  J.~Bellorin, A.~Restuccia and
A.~Sotomayor, 
Int.\ J.\ Mod.\ Phys.\ D \textbf{25}, no. 02, 1650016 (2015) 




\bibitem{Bellorin:2014qca}  J.~Bellorin, A.~Restuccia and
A.~Sotomayor, 
Phys.\ Rev.\ D \textbf{90}, no. 4, 044009 (2014). 




\bibitem{Restuccia:2014zda} A.~Restuccia, J.~Bellorin and
A.~Sotomayor, 
J.\ Phys.\ Conf.\ Ser.\ \textbf{490}, 012123 (2014). 



\bibitem{Bellorin:2013zbp} J.~Bellorin, A.~Restuccia and
A.~Sotomayor, 
Phys.\ Rev.\ D \textbf{87}, no. 8, 084020 (2013). 


\bibitem{Bellorin:2012di}  J.~Bellorin, A.~Restuccia and
A.~Sotomayor, 
Phys.\ Rev.\ D \textbf{85}, 124060 (2012). 


\bibitem{Bellorin:2011ff}  J.~Bellorin and A.~Restuccia, 
Phys.\ Rev.\ D \textbf{84}, 104037 (2011). 




\bibitem{Bellorin:2010je}  J.~Bellorin and A.~Restuccia, 
Int.\ J.\ Mod.\ Phys.\ D \textbf{21}, 1250029 (2012). 


\bibitem{dan1}  D.~Vernieri and T.~P.~Sotiriou, 
Phys.\ Rev.\ D \textbf{85} (2012) 064003. 

\bibitem{dan2}  D.~Vernieri, 
Phys.\ Rev.\ D \textbf{91} (2015) no.12, 124029. 




\bibitem{Kiritsis:2009sh} E.~Kiritsis and G.~Kofinas, 
Nucl.\ Phys.\ B \textbf{821}, 467 (2009); 
G.~Calcagni, 
JHEP \textbf{0909}, 112 (2009). 





\bibitem{Blas:2009qj}  D.~Blas, O.~Pujolas and S.~Sibiryakov, 
Phys.\ Rev.\ Lett.\ \textbf{104}, 181302 (2010); 
D.~Blas, O.~Pujolas and S.~Sibiryakov, 
JHEP \textbf{0910}, 029 (2009). 




\bibitem{Christodoulakis:2011np} T.~Christodoulakis and N.~Dimakis, 
J.\ Geom.\ Phys.\ \textbf{62} (2012) 2401. 

\bibitem{dan111}  N. Frusciante, M.\ Raveri, D. Vernieri, B. Hu and
A. Silvestri, Physics of the Dark Universe \textbf{13,} 7 (2016). 

 




\bibitem{Lu:2009em}  H.~Lu, J.~Mei and C.~N.~Pope, 
Phys.\ Rev.\ Lett.\ \textbf{103}, 091301 (2009). 




\bibitem{Leon:2009rc} G.~Leon and E.~N.~Saridakis, 
JCAP \textbf{0911}, 006 (2009). 




\bibitem{Mukohyama:2009zs} S.~Mukohyama, K.~Nakayama, F.~Takahashi
and S.~Yokoyama, 
Phys.\ Lett.\ B \textbf{679}, 6 (2009); 
E.~N.~Saridakis, 
Int.\ J.\ Mod.\ Phys.\ D \textbf{20}, 1485 (2011); 
A.~Ali, S.~Dutta, E.~N.~Saridakis and A.~A.~Sen, 
Gen.\ Rel.\ Grav.\ \textbf{44}, 657 (2012); 
S.~Nojiri and S.~D.~Odintsov, 
Phys.\ Rev.\ D \textbf{81}, 043001 (2010). 




\bibitem{Mukohyama:2009gg} S.~Mukohyama, 
JCAP \textbf{0906}, 001 (2009); 
B.~Chen, S.~Pi and J.~Z.~Tang, 
JCAP \textbf{0908}, 007 (2009); 
Y.~S.~Piao, 
Phys.\ Lett.\ B \textbf{681}, 1 (2009); 
B.~Chen, S.~Pi and J.~Z.~Tang, 
JCAP \textbf{0908}, 007 (2009). 




\bibitem{Danielsson:2009gi}  U.~H.~Danielsson and L.~Thorlacius, 
JHEP \textbf{0903}, 070 (2009); 
R.~G.~Cai, L.~M.~Cao and N.~Ohta, 
Phys.\ Rev.\ D \textbf{80}, 024003 (2009); 
A.~Ghodsi and E.~Hatefi, 
Phys.\ Rev.\ D \textbf{81}, 044016 (2010); 
A.~Kehagias and K.~Sfetsos, 
Phys.\ Lett.\ B \textbf{678}, 123 (2009); E.~N.~Saridakis, 
Gen.\ Rel.\ Grav.\ \textbf{45}, 387 (2013). 




\bibitem{Saridakis:2009bv}  E.~N.~Saridakis, 
Eur.\ Phys.\ J.\ C \textbf{67}, 229 (2010). 




\bibitem{Dutta:2009jn}  S.~Dutta and E.~N.~Saridakis, 
JCAP \textbf{1001}, 013 (2010); 
S.~Dutta and E.~N.~Saridakis, 
JCAP \textbf{1005}, 013 (2010). 




\bibitem{Kim:2009dq}  S.~S.~Kim, T.~Kim and Y.~Kim, 
Phys.\ Rev.\ D \textbf{80}, 124002 (2009). 




\bibitem{Cai:2009qs}  R.~G.~Cai, L.~M.~Cao and N.~Ohta, 
Phys.\ Lett.\ B \textbf{679}, 504 (2009); 
M.~Jamil, E.~N.~Saridakis and M.~R.~Setare, 
JCAP \textbf{1011}, 032 (2010); 
M.~Jamil and E.~N.~Saridakis, 
JCAP \textbf{1007}, 028 (2010). 




\bibitem{Brandenberger:2009yt}  R.~Brandenberger, 
Phys.\ Rev.\ D \textbf{80}, 043516 (2009); 
Y.~F.~Cai and E.~N.~Saridakis, 
JCAP \textbf{0910}, 020 (2009); 
M.~Khodadi, Y.~Heydarzade, F.~Darabi and E.~N.~Saridakis, 
Phys.\ Rev.\ D \textbf{93}, no. 12, 124019 (2016). 




\bibitem{Bramberger:2017tid}  S.~F.~Bramberger, A.~Coates,
J.~Magueijo, S.~Mukohyama, R.~Namba and Y.~Watanabe, 
Phys.\ Rev.\ D \textbf{97}, no. 4, 043512 (2018). 



\bibitem{Bellorin:2018wst}  J.~Bellorin, A.~Restuccia and
F.~Tello-Ortiz,  
Phys.\ Rev.\ D \textbf{98}, no. 10, 104018 (2018).  



\bibitem{Abreu:2018wjg} E.~M.~C.~Abreu, A.~C.~R.~Mendes,
G.~Oliveira-Neto, J.~Ananias Neto, L.~G.~R.~Rodrigues and M. ~Silva De
Oliveira, 
arXiv:1805.11042 [gr-qc]. 



\bibitem{Maier:2017dtb}  R.~Maier and I.~D.~Soares, 
Phys.\ Rev.\ D \textbf{96}, no. 10, 103532 (2017) Addendum: [Phys.\ Rev.\ D 
\textbf{97}, no. 4, 049902 (2018)]. 




\bibitem{Wang:2017brl} A.~Wang,  
Int.\ J.\ Mod.\ Phys.\ D \textbf{26} (2017) no.07, 1730014.  




\bibitem{Nilsson:2018knn}  N.~A.~Nilsson and E.~Czuchry,  
Phys.\ Dark Univ.\ \textbf{23}, 100253 (2019).  


\bibitem{gw01}  A.E. Gumrukcuoglu, M. Saravani and T.P. Sotiriou,
Phys. Rev. D 97, 024032 (2018). 



\bibitem{Wang:2009rw}  A.~Wang and Y.~Wu, 
JCAP \textbf{0907}, 012 (2009). 



\bibitem{hl001}  M. M. Anber and J. F. Donoghue, Phys. Rev. D 83,
105027 (2011). 

\bibitem{hl002}  M. Pospelov and Y. Shang, Phys. Rev. D 85, 105001
(2012). 

\bibitem{hl003}  S. Groot Nibbelink and M. Pospelov, Phys. Rev. Lett.
94, 081601 (2005). 

\bibitem{hl004}  
A.~Coates, C.~Melby-Thompson and S.~Mukohyama,  
arXiv:1805.10299 [hep-th].  


\bibitem{rg01} G. D'Odorico, F. Saueressig and M. Schutten, Phys.
Rev. Lett. 113, 17, 171101 (2014). 

\bibitem{rg04}  T. Griffin, P. Horava, C.M. Merby-Thompson, Phys.
Rev. Lett. 110, 081602 (2013). 

\bibitem{rg02}  A. O. Barvinsky, D. Blas, M. Herrero-Valea, S. M.
Sibiryakov and C. F. Steinwachs, Phys. Rev. D 93, 064022 (2016). 

\bibitem{rg03}  A. O. Barvinsky, D. Blas, M. Herrero-Valea, S. M.
Sibiryakov and C. F. Steinwachs, JHEP 1807, 035 (2018). 

\bibitem{rg05} A. Contillo, S. Rechenberger, and F. Saueressig, JHEP
12, 017 \ (2013). 

\bibitem{ef01}  
S.~Mukohyama,  
Class.\ Quant.\ Grav.\ \textbf{27}, 223101 (2010).  


\bibitem{ef02} I. Kimpton and A. Padilla, Matter in Horava-Lifshitz
gravity, JHEP 04, 133 (2013).





\bibitem{Vernieri:2017dvi}  D.~Vernieri and S.~Carloni, 
EPL \textbf{121}, no. 3, 30002 (2018). 


\bibitem{Vernieri:2019vlh} D.~Vernieri,  
arXiv:1906.07738 [gr-qc].  




\bibitem{Paliathanasis:2019qch} A.~Paliathanasis and G.~Leon,  
arXiv:1903.10821 [gr-qc].  



\bibitem{Vernieri:2018sxd} D.~Vernieri,  
Phys.\ Rev.\ D \textbf{98}, no. 2, 024051 (2018).  


\bibitem{LeBlanc:1994qm}  V.~G. LeBlanc, D.~Kerr, and J.~Wainwright. 
\newblock {\em Class. Quant. Grav.}, \textbf{12}:513--541, 1995. 

\bibitem{Heinzle:2009zb}  J.~Mark Heinzle and Claes Uggla. 
\newblock {\em Class. Quant. Grav.}, \textbf{27}:513--541, 015009, 2010. 

\bibitem{WE}  J.~Wainwright and G.F.R.~Ellis (eds). 
\newblock {Dynamical
Systems in Cosmology}. \newblock Cambridge University Press: Cambridge, UK.
(1997) 343 p. 



\bibitem{Coley:1999uh} A.~A.~Coley, 
gr-qc/9910074. 


\bibitem{Coley:2003mj}  A.~A. Coley. 
\newblock Dordrecht, Netherlands: Kluwer (2003) 200 p. 



\bibitem{Carloni:2009jc} S.~Carloni, E.~Elizalde and P.~J.~Silva, 
Class.\ Quant.\ Grav.\ \textbf{27}, 045004 (2010). 




\bibitem{Setare:2010zd} M.~R.~Setare and D.~Momeni, 
Int.\ J.\ Theor.\ Phys.\ \textbf{50}, 106 (2011). 


\bibitem{Cardoso:2008bp}  V. Cardoso, A.~S. Miranda, E. Berti, H.
Witek, and V.~T. Zanchin. 
\newblock {\em Phys. Rev.}, \textbf{D79}:064016, 2009. 

\bibitem{Lavkin:1990gu}  A.~G. Lavkin. 
\newblock {\em Sov. J. Nucl. Phys.}, \textbf{52}:759--760, 1990. 



\bibitem{Charters:2001hi}  T.~C. Charters, A. Nunes, and J.~P.
Mimoso. 
\newblock {\em Class. Quant. Grav.}, \textbf{18}:1703--1714, 2001. 

\bibitem{Aref'eva:2009xr}  I.~Ya. Aref'eva, N.~V. Bulatov, and S.~Yu.
Vernov. 
\newblock {\em Theor. Math. Phys.}, \textbf{163}:788--803.





\bibitem{Escobar:2013js}  D.~Escobar, C.~R.~Fadragas, G.~Leon and
Y.~Leyva, 
Astrophys.\ Space Sci.\ \textbf{349}, 575 (2014). 




\bibitem{Fadragas:2013ina} C.~R.~Fadragas, G.~Leon and
E.~N.~Saridakis, 
Class.\ Quant.\ Grav.\ \textbf{31} (2014) 075018. 


\bibitem{an1}  G.~Papagiannopoulos, S.~Basilakos, J.~D.~Barrow and
A.~Paliathanasis, 
Phys.\ Rev.\ D \textbf{97} (2018) no.2, 024026.

\bibitem{an2} A.~Paliathanasis, 
JCAP \textbf{1708} (2017) no.08, 027. 




\bibitem{an3}  L.~Karpathopoulos, S.~Basilakos, G.~Leon,
A.~Paliathanasis and M.~Tsamparlis, 
Gen.\ Rel.\ Grav.\ \textbf{50} (2018) 79. 






\bibitem{Alho:2015cza}  A.~Alho, J.~Hell and C.~Uggla, 
Class.\ Quant.\ Grav.\ \textbf{32}, no. 14, 145005 (2015). 


\bibitem{Alho:2017opd}  A.~Alho and C.~Uggla, 
Phys.\ Rev.\ D \textbf{95}, no. 8, 083517 (2017). 


\bibitem{Yearsley:1996yg} J.~Yearsley and J.~D.~Barrow, 
Class.\ Quant.\ Grav.\ \textbf{13}, 2693 (1996). 



\bibitem{Cardenas2003} R.~Cardenas, T.~Gonzalez, Y.~Leiva, O.~Martin
and I.~Quiros, 
Phys.\ Rev.\ D \textbf{67}, 083501 (2003). 




\bibitem{Pavluchenko:2003ge}  S.~A.~Pavluchenko, 
Phys.\ Rev.\ D \textbf{67},103518 (2003). 




\bibitem{Sahni:1999gb}  V.~Sahni and A.~A.~Starobinsky, 
Int.\ J.\ Mod.\ Phys.\ D \textbf{9}, 373 (2000). 





\bibitem{Ratra:1987rm} B.~Ratra and P.~J.~E.~Peebles, 
Phys.\ Rev.\ D \textbf{37}, 3406 (1988). 




\bibitem{Wetterich:1987fm} C.~Wetterich, 
Nucl.\ Phys.\ B \textbf{302}, 668 (1988). 




\bibitem{Copeland:2009be} E.~J.~Copeland, S.~Mizuno and M.~Shaeri, 
Phys.\ Rev.\ D \textbf{79}, 103515 (2009). 




\bibitem{Leyva:2009zz} Y.~Leyva, D.~Gonzalez, T.~Gonzalez, T.~Matos
and I.~Quiros, 
Phys.\ Rev.\ D \textbf{80}, 044026 (2009). 

 



\bibitem{delCampo:2013vka}  S.~del Campo, C.~R.~Fadragas, R.~Herrera,
C.~Leiva, G.~Leon and J.~Saavedra, 
Phys. Rev. D \textbf{88}, 023532 (2013).




\bibitem{Sahni:1999qe} V.~Sahni and L.~-M.~Wang, 
Phys.\ Rev.\ D \textbf{62}, 103517 (2000). 







\bibitem{Lidsey:2001nj} J.~E.~Lidsey, T.~Matos and
L.~A.~Urena-Lopez, 
Phys.\ Rev.\ D \textbf{66}, 023514, (2002). 




\bibitem{Matos:2000ng} T.~Matos and L.~A.~Urena-Lopez, 
Class.\ Quant.\ Grav.\ \textbf{17}, L75 (2000). 



\bibitem{Matos:2009hf} T.~Matos, J.~-R.~Luevano, I.~Quiros,
L.~A.~Urena-Lopez and J.~A.~Vazquez, 
Phys.\ Rev.\ D \textbf{80}, 123521 (2009). 



\bibitem{UrenaLopez:2000aj}  L.~A.~Urena-Lopez and T.~Matos, 
Phys.\ Rev.\ D \textbf{62}, 081302 (2000). 



\bibitem{Barreiro:1999zs} T.~Barreiro, E.~J.~Copeland and
N.~J.~Nunes, 
Phys.\ Rev.\ D \textbf{61}, 127301 (2000). 




\bibitem{Gonzalez:2007hw}  T.~Gonzalez, R.~Cardenas, I.~Quiros and
Y.~Leyva, 
Astrophys.\ Space Sci.\ \textbf{310}, 13 (2007). 



\bibitem{Gonzalez:2006cj} T.~Gonzalez, G.~Leon and I.~Quiros, 
Class.\ Quant.\ Grav.\ \textbf{23},3165 (2006). 







\bibitem{Mukohyama:2009mz}  S.~Mukohyama,  
Phys.\ Rev.\ D \textbf{80}, 064005 (2009).  




\bibitem{Li:2009bg} M.~Li and Y.~Pang,  
JHEP \textbf{0908}, 015 (2009). 



\bibitem{Blas:2009yd} D.~Blas, O.~Pujolas and S.~Sibiryakov,  
JHEP \textbf{0910}, 029 (2009).  




\bibitem{Sotiriou:2009gy}  T.~P.~Sotiriou, M.~Visser and
S.~Weinfurtner,  
Phys.\ Rev.\ Lett.\ \textbf{102}, 251601 (2009).  




\bibitem{Sotiriou:2009bx} T.~P.~Sotiriou, M.~Visser and
S.~Weinfurtner,  
JHEP \textbf{0910}, 033 (2009).  




\bibitem{Appignani:2009dy}  C.~Appignani, R.~Casadio and
S.~Shankaranarayanan,  
JCAP \textbf{1004}, 006 (2010).  








\bibitem{Kiritsis:2009vz}  E.~Kiritsis,  
Phys.\ Rev.\ D \textbf{81}, 044009 (2010). 



\bibitem{wiggins}  S.~Wiggins. 
\newblock {\it Introduction to Applied
Nonlinear Dynamical Systems and Chaos}. \newblock Springer (2003). 




\bibitem{Leon:2018lnd}  G.~Leon, A.~Paliathanasis and
J.~L.~Morales-Mart\'inez, 
Eur.\ Phys.\ J.\ C \textbf{78}, no. 9, 753 (2018). 

\bibitem{Perko}  L.~Perko. \newblock {Differential Equations and
Dynamical Systems }. \newblock Springer, Berlin, (1991).

\bibitem{Carr:1981} J. Carr. \newblock {Applications of Center
Manifold theory}. \newblock New York: Springer-Verlag (1981). 

\bibitem{Guckenheimer}  J.~Guckenheimer and P.~Holmes. 
\newblock {Nonlinear Oscillations, Dynamical Systems and Bifurcations of Vector
  Fields}. \newblock Springer, New York, (1983). 


\end{thebibliography}
\end{document}